\documentclass[a4paper,11pt,openbib,oneside]{memoir} 

\usepackage{sty/envLatex}
\usepackage{sty/layout}

\begin{document}

\renewcommand{\dobib}{}

\frontmatter
\pagenumbering{roman}

\begin{titlingpage}
\begin{SingleSpace}
\calccentering{\unitlength} 
\begin{adjustwidth*}{\unitlength}{-\unitlength}
\begin{center}
\begin{figure}[H]
	\centering
	\begin{subfigure}[t]{0.30\textwidth}
		\centering
		\includegraphics[width=\textwidth]{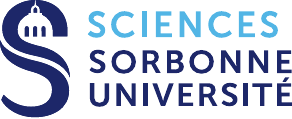}
	\end{subfigure}
	\quad	\quad	\quad	\quad	\quad	\quad	\quad	\quad	\quad	\quad	\quad	\quad	\quad	\quad	\quad	\quad	\quad	\quad	\quad
	\begin{subfigure}[t]{0.18\textwidth}
		\centering
		\includegraphics[width=\textwidth]{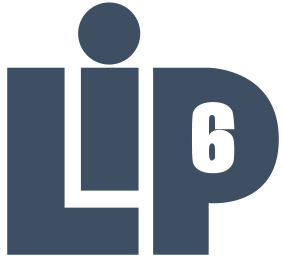}
	\end{subfigure}
\end{figure}
\vspace{5mm}
{\Large \textsc{Sorbonne Université - EDITE de Paris}} \\
\vspace{5mm}
{\large \textsl{Laboratoire d'Informatique de Sorbonne Université (LIP6)}}\\
\vspace*{8mm}
\rule[0.5ex]{\linewidth}{2pt}\vspace*{-\baselineskip}\vspace*{3.2pt}
\rule[0.5ex]{\linewidth}{1pt}\\[\baselineskip]
{\HUGE The interplay between quantum}\\[4mm]
{\HUGE contextuality and Wigner negativity}\\[2mm]
\rule[0.5ex]{\linewidth}{1pt}\vspace*{-\baselineskip}\vspace{3.2pt}
\rule[0.5ex]{\linewidth}{2pt}\\
\vspace{8mm}
{\Large \textsc{By Pierre-Emmanuel Emeriau}}\\
\vspace{5mm}
{\Large \textsc{PhD Thesis in Computer Science}}\\
\vspace{8mm}
{\large \textsc{Under the supervision of Elham Kashefi and Shane Mansfield}}\\
\vspace{8mm}
\end{center}
\begin{flushleft}
{\small \textit{Keywords}: quantum information, contextuality, Wigner negativity, optimisation theory, measure theory,\\ \hspace{1.55cm} continuous-variables}\\
\vspace{5mm}
{\small PhD defence held publicly in November 2021 with the following thesis committee: 
\begin{itemize}[leftmargin=*]
\justifying
\item \textsc{Abramsky} Samson, Christopher Strachey Professor of Computing at Oxford University, United Kingdom, \textit{Président du jury}.
\item \textsc{Ac\'in} Antonio, Group Leader ICREA Professor at ICFO, Spain.
\item \textsc{Arrighi} Pablo, Professeur HDR, Université de Paris-Saclay, France. \textit{Rapporteur}.
\item \textsc{Del Rio} L\'idia, Senior Researcher at ETH Zurich, Switzerland.
\item \textsc{Galv\~ao} Ernesto, Group leader at INL, Portugal. \textit{Rapporteur}.
\item \textsc{Kashefi} Elham, Directrice de Recherche HDR au CNRS, Sorbonne Université, France and Professor at University of Edinburgh, United Kingdom. \textit{PhD supervisor}.
\item \textsc{Mansfield} Shane, Head of Quantum Algorithms, Quandela. France, \textit{PhD supervisor}.
\item \textsc{Parigi} Valentina, Associate Professor at LKB, Sorbonne Université, Paris, France.
\end{itemize}
}
{\scriptsize \noindent This work is licensed under the \href{http://creativecommons.org/licenses/by-nc-nd/4.0/}{Creative Commons Attribution-NonCommercial-NoDerivatives 4.0 International License}.}
\end{flushleft}
\end{adjustwidth*}
\end{SingleSpace}
\end{titlingpage}
\clearemptydoublepage


\chapter*{Abstract}
\begin{SingleSpace}
\lettrine{Q}{uantum} physics has revolutionised our way of conceiving nature and is now bringing about a new technological revolution. The use of quantum information in technology promises to supersede the so-called classical devices used nowadays. Understanding what features are inherently non-classical is crucial for reaching better-than-classical performance.

This thesis focuses on two nonclassical behaviours: quantum contextuality and Wigner negativity. The former is a notion superseding nonlocality that can be exhibited by quantum systems. To date, it has mostly been studied in discrete-variable scenarios, where observables take values in discrete and usually finite sets. In those scenarios, contextuality has been shown to be necessary and sufficient for advantages in some cases. 
On the other hand, negativity of the Wigner function is another unsettling non-classical feature of quantum states that originates from phase-space formulation in continuous-variable quantum optics. Continuous-variable scenarios offer promising candidates for implementing quantum computations and informatic protocols.
Wigner negativity is known to be a necessary resource for quantum speedup with continuous variables. However contextuality has been little understood and studied in continuous-variable scenarios.

We first set out a robust framework for properly treating contextuality in continuous variables. We also quantify contextuality in such scenarios by using tools from infinite-dimensional optimisation theory. This is achieved by a converging hierarchy of finite-dimensional semidefinite programs that approximates the contextual fraction.

Building upon this, we show that Wigner negativity is equivalent to contextuality in continuous variables with respect to Pauli measurements thus establishing a continuous-variable analogue of a celebrated result by Howard \textit{et al.} in discrete variables.

We then introduce experimentally-friendly witnesses for Wigner negativity of single mode and multimode quantum states, based on fidelities with Fock states. They possess a threshold expectation value indicating whether the measured state has a negative Wigner function. We phrase the problem of finding the threshold values as infinite-dimensional linear programs, and we derive two converging hierarchies of semidefinite programs to approximate the threshold values. 

We further extend the range of previously known discrete-variable results linking contextuality and advantage into a new territory of information
retrieval.
We introduce a discrete-variable communication game---called the Torpedo Game---where perfect quantum strategies stem from negativity of the discrete Wigner function. Sequential contextuality is shown not only to be necessary and sufficient for quantum advantage, but also to quantify the degree of advantage for information retrieval tasks.

\end{SingleSpace}
\clearpage

\clearemptydoublepage


\chapter*{Dedication and acknowledgements}
\begin{SingleSpace}

These last three years have been incredibly rich, both scientifically and personally.
I am deeply indebted to my supervisors Elham and Shane who, from the very beginning, have continuously shared their love of Quantum Information in so many topics.
My grateful thanks to Elham for her amazing support and her perfectly adjusted mixture of guidance and freedom throughout the projects.
I cannot begin to express my thanks to Shane, who has been a wonderful supervisor, with incredible insight into quantum foundational related questions.
I can safely say that I learnt more about Quantum Information than on beer related topics by his side (and I did learn a great deal on craft beers!).
N\'i fh\'eidfinn go leor bu\'iochas a ghabhadh le Shane. T\'a an t-eipeagraf ar do shon.

I would also like to extend my deepest gratitude to my coauthors Mark, Rui, Frédéric, Ulysse and Robert.
Mark has always been there despite the distance from the end of the first year onward.
I will be forever grateful to Ulysse for the numerous insightful discussions I had with him and his impressive dedication as he would rather proofread this dissertation than watch a movie or read a book during a 15 hour flight.
Many thanks to Robert and his passion for C*-algebras; his strong mathematical background allowed for fruitful conversations.
Special thanks to Antoine who opened the door to the optimisation theory realm for me.
I must also thank Pablo Arrighi and Anthony Leverrier for their support with the supervising committee; Pablo Arrighi and Ernesto Galv\~ao for agreeing to be `rapporteurs' of this thesis and giving valuable feedbacks; and I am also very thankful to the rest of the jury members Samson Abramsky---who has kindly agreed to preside over the viva, Antonio Ac\'in, L\'idia del Rio, Valentina Parigi.
Thanks to Ricardo and Anna with whom I had interesting discussions on quantum game theory.

Many thanks to Elham, Damian, Eleni, Frédéric and Alex for the wonderful QI team they have built in Paris and again Elham for the amazing Edipar group in Paris and Edinburgh. This was an invaluable asset during my PhD as I could have intense discussions in so many topics with passionate and diverse people. These three years have been a delight because of the numbers of people I could interact with, both in research and outside the lab---mostly at Baker street, Bières Cultes or around the 'quais St Bernard' or during amazing retreats. 
For that I thank the Edipar group and its brilliant members Ellen, Léo, Luka, Niraj, Atul, Brian, Dominik, Harold, Daniel, Mina, Alex, Rawad, Mina, Mahshid, Theodoros, Jonas, Ieva, Yao, Armando.
I also thank the members of the marvelous QI team
Raja, Victor, Federico, Nathan, Clément, Luis, Matthieu, Anu, Andrea, Alisa, Adrien, Simon, Shouvik, Matteo, Francesco, Paul, Ivan. 
I also thank everyone at Quandela; it has been extremely stimulating to work on experimentally related projects.

\medskip

Évidemment un immense merci à tous mes merveilleux amis qui m'ont accompagné en dehors du labo.
À Rémi et Xavier : Bagneux on n'oublie pas.
À Robin parce qu'il est bonnard.
À Eliot mashallah.
À François pour ce but vide que tu m'as offert et que j'ai mis à côté scellant notre amitié.
À Lucie pour ses piggybacks.
À Raja pour sa joie de vivre au quotidien qui réchauffe les c\oe urs.
À Ulysse pour ces nombreuses galettes épinards partagées et également Léonie pour ces soirées jeux de société et escape games brillamment réussis.
À mon ami de plus longue date Johan.
À mon fillot Antoine pour tous ces moments riches depuis Ginette.
Aux Choletais (et alentours) Mathieu, Marie et Thomas qui m'ont soutenu depuis si longtemps déjà.
À Jean-Victor et Raphaëlle pour ces nombreux partages sensibles. 
À Giulia pour son amour partagé du jazz.
Et aussi à Aude, Coline, Tamara, Mathilde G., Mathilde S. et Alice.

Un grand merci au groupe de Ginette Côme, Clément, Victor, Paul, Anne-Claire, Gustave, Pierre H et Peter Battle et également aux coincheurs Léa, Sylvain, Paul Pi Su Sardu.
Au PTRC pour m'avoir fait découvrir ce sport magnifique et ses membres incroyables: un véritable exhutoire pendant ces 3 ans. 
Merci à Damien et Leïla.
Merci à Léa M., Romane et Roméo, Yaelin et nos parties d'échecs, Hortense.
Merci aussi à Ev et Tom.
À Etienne et l'atelier Mala pour ses nouilles veggies. 
Enfin merci à Matthieu et Virginie pour m'avoir aiguillé sur le bon chemin postbac.

À tous les enseignants incroyables que j'ai eu la chance de rencontrer sur mon parcours et qui ont su me passionner; en particulier Cyril Ciaudo, Alexianne Poutevigne, Hamid Seddiq, Bertrand Delahaye, Christophe Collet, Françoise Thebault, Jean Nougareyde, Bertrand Delamotte, Jérôme Beugnon et Patrick Ciarlet.

Un immense merci à Alexandre, Jennifer, Pacôme et Manolie pour leur soutien sur et en dehors des tatamis.

Merci à Caroline, Keat, Emmanuel, Alexandra et Basilou pour tout ce qu'ils ont déjà fait pour moi.

Merci à toute ma famille, grand-parents, oncles, tantes et cousins.

Évidemment une reconnaissance éternelle à mes parents Isabelle et Bruno et ma s\oe ur Mathilde pour m'avoir accompagné jusqu'ici, pour toutes les valeurs qu'ils m'ont transmises et pour leur générosité abyssale. Un immense merci à Mathieu pour sa curiosité sans borne et son partage d'expériences et de connaissances. À Alice pour ses "areu" encourageants. 

À Héloïse pour son soutien immuable et pour l'équilibre apporté dans ma vie tout au long de cette thèse. 
Mes mots ne pourraient se hisser à la hauteur de l’amour et l’affection qu'elle m'a portés. 
Merci d'être dans ma vie ; on a encore tant de choses à découvrir tous les deux. 

\end{SingleSpace}
\clearpage

\clearemptydoublepage


\setlength{\epigraphwidth}{0.5\textwidth}
\epigraph{Sa sagesse résonnait jusque dans ses silences. \\ Jamais ses réponses n'étaient précipitées. \\ Jamais mes questions n'étaient sous-estimées.}{\textit{Le roi n'avait pas ri} \\ Guillaume \textsc{Meurice}}

\clearemptydoublepage

\hypersetup{linkcolor=black}
\makeatletter
\renewcommand{\@pnumwidth}{2em}
\renewcommand{\@tocrmarg}{3em}
\makeatother
\renewcommand{\contentsname}{Table of Contents}
\maxtocdepth{subsection}
\tableofcontents*
\addtocontents{toc}{\par\nobreak \mbox{}\hfill{\bfseries Page}\par\nobreak}
\hypersetup{linkcolor=quantumgreen}
\clearemptydoublepage

\mainmatter


\chapter*{Introduction}
\label{chap:intro}
\addcontentsline{toc}{chapter}{Introduction}
\chaptermark{Introduction}
\pagestyle{intro}

\lettrine{Q}{uantum} mechanics has been one of the major scientific revolutions of the 20$^\text{th}$ century. 
While philosophically puzzling \cite{einsteincan1935,wigner1995remarks,Schlosshauer2005,pusey2012reality,MansfieldPBR2016}, it has profoundly challenged our way of perceiving reality, giving rise to intriguing notions such as \textit{entanglement} or \textit{superposition}. 
It has also allowed us to perform remarkably precise calculation of properties of physical systems---especially at the particle scale. 
The so-called second quantum revolution is often traced to Feynman's idea of a ``probabilistic simulator of a probabilistic nature'' \cite{Feynman1982} to solve difficult problems originating from quantum physics together with the first quantum algorithms outperforming known classical ones \cite{deutsch1992rapid,shor1994algorithms}.

This gave rise to the field of quantum information built around the idea that information may be encoded in quantum degrees of freedom of physical systems. 
Quantum information processing seeks to provide an advantage over classical information processing in various fields such as computing, communication, cryptography or sensing.
To outperform standard classical computers at certain tasks,
one can attempt to harness effects at the microscopic level which have no counterpart in classical physics. 

Studying the fundamentally non-classical features of quantum mechanics is critical for quantum information. If non-classical features can be identified at the very root of quantum information protocols, and their advantage founded on those, then their quantumness can be inherently guaranteed. 
There is a lesson that can be learnt from the dequantization of the quantum recommendation system \cite{kerenidisquantum2016} that was proposed by Ewin Tang \cite{tangquantuminspired2019}. If one wants to prevent against dequantization, it is necessary to ensure that algorithms rely upon some purely quantum phenomenon.\footnotemark
\footnotetext{Note however that first it is still intriguing that a quantum algorithm had to be found in order to derive a classical algorithm with an improved complexity compared to known classical algorithms; secondly purely quantum notions will protect against dequantization but there might still exist classical algorithms that can be as efficient; finally the quantum algorithm still retains some advantage though it was massively diminished by Tang.}
This leads naturally to the following questions:
\begin{center}
    \textit{Which features of quantum physics are inherently non-classical? \\
    Among those, which are relevant to achieving a quantum-over-classical advantage in information processing tasks? \\
    How non-classicality can be related in a quantified way to advantages?}
\end{center}

One of the first influential works indicating how quantum theory may depart from classical analogues was that of Einstein, Podolsky and Rosen \cite{einsteincan1935} in 1935.\index{EPR argument}
They identified early on that if the quantum description of the world is seen as fundamental then entanglement poses a problem of ``spooky action at a distance''.
It was argued that the principle of locality, a key axiom of Einstein's special relativity, was in conflict with the description of the world given by quantum mechanics. 
The conclusion of their paper was that quantum theory should be consistent with a deeper or more complete description of the physical world, in which such problems would disappear.
This is known as \emph{local realism}. 
\emph{Realism} means the introduction of a hidden variable model with hidden states describing the actual physics behind the scene.
Indeed probabilistic predictions of quantum mechanics may result from an incomplete knowledge of the true state of a system rather than being a fundamental feature. 
There are other examples for this in physical theories: for instance statistical mechanics---a probabilistic theory---admits a deeper, very complex description in terms of classical mechanics, which is purely deterministic.
\emph{Local realism} is a constraint on the hidden variables that requires that they cannot be updated from a space-like separated region.
This impossibility was further studied and instantiated as an experimentally verifiable inequality by Bell \cite{belleinstein1964} in 1964 that quantum mechanics is predicted to violate.\index{Bell inequality}
This violation was verified experimentally \eg \cite{Aspect1982,rowe2001experimental,Zeilinger2015,Woo2015,hensen2015loophole}.
This is an example of a \emph{no-go theorem}: quantum physics cannot be described by a local hidden variable model.

\emph{Local realism} was further generalised\footnotemark as \emph{measurement non-contextuality} \cite{KochenSpecker1967,kochen1975problem}. 
\footnotetext{Actually contextuality was originally considered to be similar but still a distinct non-classical feature rather than a generalisation of nonlocality. Fine \cite{Fine1982} proved that nonlocality is a special case of contextuality for the (2,2,2) Bell scenario and this was later generalised to any discrete-variable scenario by Abramsky and Brandenburger \cite{abramsky2011sheaf}. This is a consequence of the Fine--Abramsky-Brandenburger (FAB) theorem which we will present in Section~\ref{sec01:sheaf}.}
Contextuality is a crucial non-classical behaviour that can be exhibited by quantum systems.
The Heisenberg uncertainty principle identified that certain pairs of quantum observables are incompatible, e.g., position and momentum.
In operational terms, observing one will disturb the outcome statistics of the other.
This is sometimes cited as evidence that position and momentum cannot simultaneously be assigned definite values.
However, this is not quite right and a more careful conclusion is that we simply cannot observe these values simultaneously.
To make a stronger statement requires measurement contextuality.\index{Contextuality}
Roughly speaking, the meaning of words can provide an intuition for a contextual behaviour. The meaning of ``I had the appendix removed'' varies greatly whether I am addressing a physician or my supervisors for the present dissertation. Pre-assigning a meaning to this sentence is not possible irrespective of the context \cite{abramsky2014semanticunification,abramsky2014semantic}.
In more operational terms, contextuality is present whenever the behaviour of a system is inconsistent with the basic assumptions that
\begin{enumerate*}[label=(\roman*)]
\item
all of its observable properties may be assigned definite values at all times, and
\item
jointly measuring \textit{compatible} observables does not disturb the global value assignments, or, in other words, these assignments are context-independent.
\end{enumerate*}
Aside from its foundational importance, today contextuality is increasingly studied as an essential ingredient for enabling a range of quantum-over-classical advantages in informatic tasks, which include the onset of universal quantum computing in certain computational models \cite{raussendorf2013contextuality,howard2014contextuality,abramsky2017contextual,bermejo2017contextuality,abramsky2017quantum}.
Moreover, contextuality has been tested experimentally \eg \cite{Xiang2013,Helmut2006,bartosik2009experimental,kirchmair2009state}. 

The importance of seminal foundational results like the Bell \cite{belleinstein1964} and Bell--Kochen--Specker \cite{bell1966,kochen1975problem} theorems is that they identify such non-intuitive behaviours and then rule out the possibility of finding \textit{any} underlying model for them that would not suffer from the same issues.
Incidentally, note that the EPR paradox\index{EPR argument} was originally presented in terms of continuous variables, whereas the CHSH model (which is what was tested experimentally from Aspect onwards) addressed a discrete-variable analogue of it. Also frameworks for formalising contextuality have previously only focused on discrete-variable quantum information \cite{abramsky2011sheaf,csw,Spekkens2005}. In Chapter~\ref{chap:CVcont}, we provide the first robust framework formalising contextuality in continuous variables.\index{Contextuality!CV}

Quantum information with continuous variables~\cite{lloyd1999quantum}---where information is encoded in continuous degrees of freedom of quantum systems---is one of the promising directions for the future of quantum technologies \cite{lloyd1999quantum}.
From a theoretical point of view, quantum information with continuous variables is described via the formalism of infinite-dimensional Hilbert spaces. It offers different perspectives from discrete-variable quantum information.
From a practical point-of-view, continuous-variable quantum systems are emerging
as very promising candidates for implementing
quantum informational and computational tasks \cite{braunstein2005quantum,weedbrook2012gaussian,crespi2013integrated,Bourassa2021blueprintscalable,walschaers2021non,chabaud2021continuous}.
The main reason for this is that they offer
unrivalled possibilities for deterministic generation
of large-scale resource states over millions of modes~\cite{yoshikawa2016invited,yokoyama2013ultra}
and also offer reliable and efficient detection methods, such as homodyne or heterodyne detection~\cite{Leonhardt-essential,Ohliger2012}.
Together they cover many of the basic operations required in the one-way or
measurement-based model of quantum computing \cite{raussendorf2001one}, for example.
Typical implementations are in optical systems where the continuous variables
correspond to the position-like and momentum-like quadratures of
the quantised modes of an electromagnetic field.\index{Quadrature}
Indeed position and momentum,
as mentioned previously in relation to the uncertainty principle,
are the prototypical examples of continuous variables in quantum mechanics.

In order to handily manipulate continuous-variable states, mathematical tools initially inspired by quantum optics have been developed such as the phase-space formalism~\cite{moyal1949quantum} which has then been extended to discrete-variable systems~\cite{connell1984,cohen1986joint,gross2006hudson}. In this framework, quantum states are represented by a quasiprobability distribution over phase space, like the Wigner function~\cite{Wigner1932}.\index{Wigner function} 
These representations provide a geometric intuition for quantum states~\cite{lee1991measure}. 
Quantum states are separated into two categories, Gaussian and non-Gaussian, depending on whether their Wigner function is a Gaussian function or not. 
The set of Gaussian states is well-understood~\cite{ferraro2005gaussian} but has limited power for computation. On the other hand characterising the set of non-Gaussian states is an active research topic~\cite{albarelli2018resource,takagi2018convex,zhuang2018resource,chabaud2020stellar}.
The latter are essential to a variety of quantum information processing tasks such as quantum state distillation~\cite{giedke2002characterization,eisert2002distilling,fiuravsek2002gaussian}, quantum error-correction~\cite{niset2009no}, universal quantum computing~\cite{lloyd1999quantum,ghose2007non} or quantum computational speedup~\cite{Bartlett2002,chabaud2020classical}.
Within those, an important subclass of non-Gaussian states are the states which display negativity in the Wigner function. These two classes of states coincide for pure states---namely, non-Gaussian pure states have a negative Wigner function.\index{Wigner negativity}
Indeed by Hudson's theorem~\cite{hudson1974wigner,soto1983wigner},\index{Hudson's theorem} pure states with a positive Wigner function are necessarily Gaussian states. However, this is not the case for mixed states and the (convex) set of states with a positive Wigner function becomes much harder to characterise~\cite{mandilara2009extending,filip2011detecting}.
In addition to its fundamental relevance as a non-classical property of physical systems~\cite{kenfack2004negativity}, Wigner negativity is also essential for quantum computing as quantum computations described by positive Wigner functions can be simulated efficiently classically~\cite{Mari2012}.
Wigner negativity is thus a necessary resource, though not sufficient~\cite{garcia2020efficient}, for quantum computational speedup. We explore intensively how to witness Wigner negativity, with reliable and accessible detection setup, in Chapter~\ref{chap:witnessing}.

Knowing which characteristic lies at the source of better-than-classical performances can both allow for comparison of quantum systems in terms of their utility,
and offer a heuristic for generating further examples of quantum-enhanced performance. This last point is highlighted in Chapter~\ref{chap:informationretrieval} where we derive an information retrieval game with a quantum-over-classical advantage.
Since contextuality and Wigner negativity both seem to play a fundamental role as non-classical features enabling quantum-over-classical advantages in various tasks, another question that arises is:
\begin{center}
    \textit{What is the precise relationship between Wigner negativity and contextuality?}
\end{center}
For discrete-variable systems of odd power-of-prime dimension, Howard \textit{et al.} \cite{howard2014contextuality} showed that this negativity actually corresponds to contextuality with respect to Pauli measurements, thereby establishing the operational utility of contextuality for the gate-based model of quantum computation (particularly in a fault-tolerant setting). This equivalence was further established for odd dimensions \cite{delfosse2017equivalence} and qubit systems \cite{raussendorf2017contextuality,delfosse2015wigner}. However this link has not previously been exhibited for continuous-variable systems and this is the subject of Chapter~\ref{chap:equivalence}.

\section*{Thesis structure and summary of results}
\label{sec0:summaryofresults}
\addcontentsline{toc}{section}{Summary of results}

\setcounter{figure}{0}
\begin{figure}[ht!]
    \centering
    \scalebox{.7}{\begin{tikzpicture}[scale=1]

\definecolor{redcb}{HTML}{D81B60}
\definecolor{bluecb}{HTML}{1E88E5}
\definecolor{yellowcb}{HTML}{FE6100}
\definecolor{greencb}{HTML}{004D40}

\tikzset{>={Latex[width=3mm,length=3mm]}}

\draw [rounded corners] (-3.75,-1) rectangle (3.75,1);
\node (Chap1) at (0,0) {
\shortstack{ 
\huge \textbf{Chapter \ref{chap:formalisms}} \\[8pt]
\Large \textbf{Frameworks and formalism}
}};

\draw [rounded corners] (-11,3.35) rectangle (-5,8.25);
\node (Chap2) at (-8,5.75) {
\shortstack{ 
\huge \textbf{Chapter \ref{chap:CVcont}} \\[10pt]
\Large \textbf{Continuous-variable} \\
\Large \textbf{contextuality} \\[5pt]
\large \color{redcb} Measure theory \\
\large \color{bluecb} Contextuality \\
\large \color{greencb} Optimisation
}};

\draw [rounded corners] (-3,4) rectangle (3,9.5);
\node (Chap3) at (0,6.7) {
\shortstack{ 
\huge \textbf{Chapter \ref{chap:equivalence}} \\[10pt]
\Large \textbf{Equivalence between} \\
\Large \textbf{contextuality and} \\
\Large \textbf{Wigner negativity} \\[5pt]
\large \color{redcb} Measure theory \\
\large \color{bluecb} Contextuality \\
\large \color{yellowcb} Wigner negativity
}};

\draw [rounded corners] (5,3.35) rectangle (11,8.25);
\node (Chap4) at (8,5.75) {
\shortstack{ 
\huge \textbf{Chapter \ref{chap:witnessing}} \\[10pt]
\Large \textbf{Witnessing Wigner} \\
\Large \textbf{negativity} \\[5pt]
\large \color{redcb} Measure theory \\
\large \color{yellowcb} Wigner negativity \\
\large \color{greencb} Optimisation
}};

\draw [rounded corners] (-3,-7.9) rectangle (3,-3);
\node (Chap5) at (0,-5.5) {
\shortstack{ 
\huge \textbf{Chapter \ref{chap:informationretrieval}} \\[10pt]
\Large \textbf{Quantum advantage in} \\
\Large \textbf{information retrieval} \\[5pt]
\large \color{yellowcb} Wigner negativity \\
\large \color{bluecb} Contextuality \\
\large \color{greencb} Optimisation
}};

\draw [->,very thick] (0,1) to (-8,3.35);
\draw [->,very thick] (0,1) to (0,4);
\draw [->,very thick] (0,1) to (8,3.35);
\draw [->,very thick] (0,-1) to (0,-3);
\draw [->,dashed,very thick] (-3,6.95) to (-5,6);
\draw [->,very thick] (-5,5.8) to (-3,6.75);
\draw [->,very thick,dashed] (3,6.95) to (5,6);
\draw [->,very thick,dashed] (5,5.8) to (3,6.75);

\end{tikzpicture}}
    \caption{
        Dependencies between the chapters of this dissertation. The black arrows enlighten the direct dependencies between the chapters while the dashed arrows indicate a partial dependence. Key topics in the chapters are indicated below the title of the corresponding chapter and are colour coded. They are defined in Chapter~\ref{chap:formalisms}. Note that Chapters \ref{chap:CVcont}, \ref{chap:equivalence} and \ref{chap:witnessing} use the continuous-variable (or infinite-dimensional) notions while Chapter~\ref{chap:informationretrieval} uses the discrete-variable ones. We did not emphasise the common denominator `Quantum Information' as it is a trivial link throughout the dissertation. 
    }
    \label{fig:intro_graphdependency}
\end{figure}
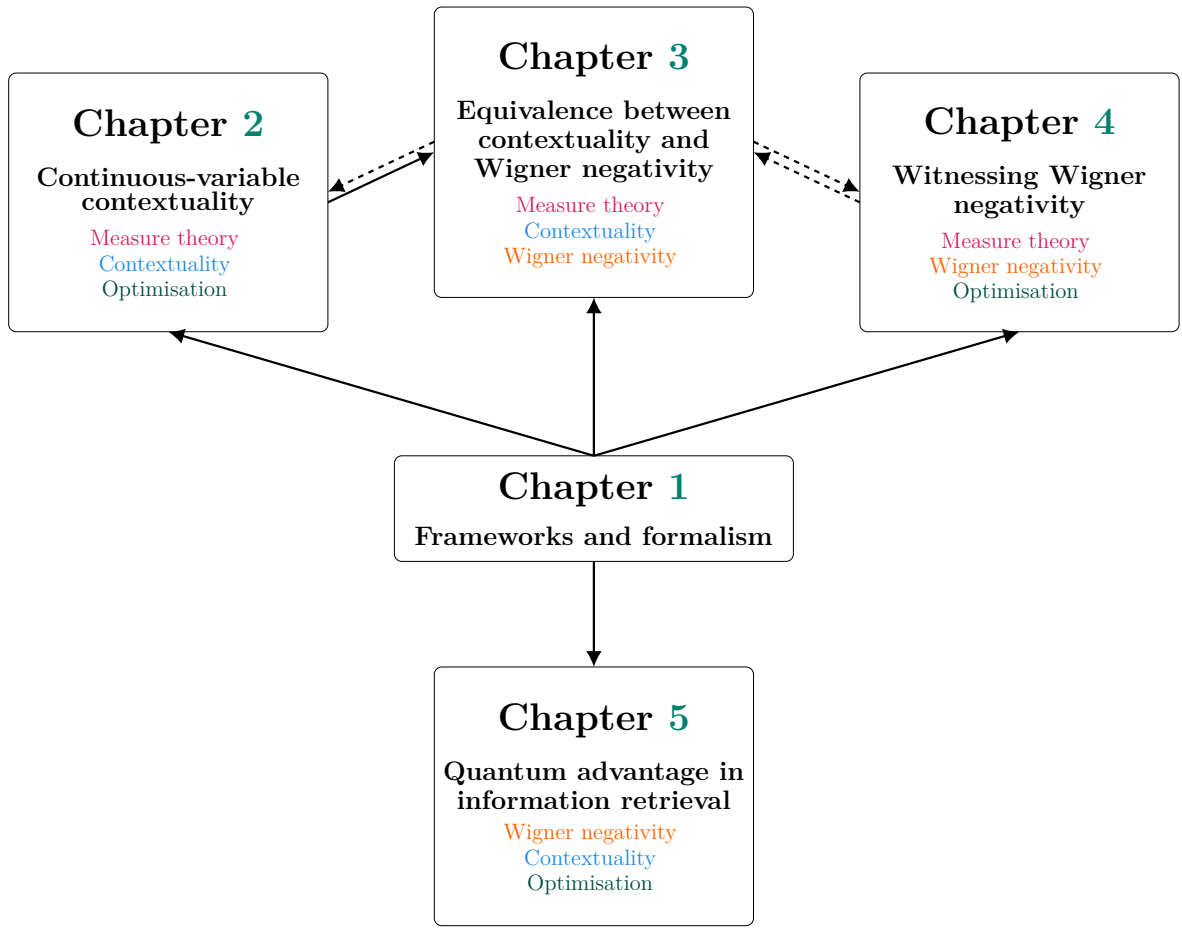

\textbf{Chapter \ref{chap:formalisms}.} This introductory chapter presents the notions that will be used throughout this dissertation. It starts by giving an overview of quantum information in Section~\ref{sec01:preliminaries}; then it gives the necessary tools for understanding the phase-space formulation of quantum mechanics in Section~\ref{sec01:phasespace}; 
it provides some background on measure theory in Section~\ref{sec01:measure};
it gives an overview of optimisation theory with a focus on infinite-dimensional linear programming in Section~\ref{sec01:convexopti}; and finally, it presents the sheaf theoretic framework for contextuality \cite{abramsky2011sheaf} in Section~\ref{sec01:sheaf}.

\noindent \textbf{Chapter \ref{chap:CVcont}.} This chapter investigates what happens for measurement contextuality as presented in \cite{abramsky2011sheaf}---known as the sheaf theoretic approach to contextuality---when one is dealing with continuous-variable systems. 
A robust framework for contextuality in continuous-variable scenarios is presented that follows along the lines of the discrete-variable framework introduced by Abramsky and Brandenburger \cite{abramsky2011sheaf}. Crucially the Fine--Abramsky--Brandenburger (FAB) theorem \cite{Fine1982,abramsky2011sheaf} extends to continuous variables. 
We prove it in scenarios that comprise an uncountable number of measurement labels as it will be essential for the following chapter. 
This establishes that noncontextuality of an empirical behaviour,
originally characterised by the existence of a deterministic hidden-variable model \cite{belleinstein1964,kochen1975problem}, can equivalently be characterised by the existence of a factorisable hidden-variable model, and that ultimately both of these are captured by the notion of extendability.
An important consequence is that nonlocality may be viewed as a special case of contextuality in continuous-variable scenarios just as for discrete-variable scenarios.
The contextual fraction, a quantifiable measure of contextuality that bears a precise relationship to Bell inequality violations and quantum advantages \cite{abramsky2017contextual}, can also be defined in this setting using infinite-dimensional linear programming.
It is shown to be a non-increasing monotone with respect to the free operations of a resource theory for contextuality \cite{abramsky2017contextual,abramsky2019comonadic}.
Crucially, these include the common operation of binning to discretise data which is usually employed in continuous-variable scenarios.
A consequence is that any witness of contextuality on discretised empirical data also witnesses and gives a lower bound on genuine continuous-variable contextuality.
While infinite linear programs are of theoretical importance
and capture exactly the contextual fraction and Bell-like inequalities in which we are interested,
they are not directly useful for actual numerical computations.
To get around this limitation, we introduce a hierarchy of semidefinite
programs \cite{lasserre10} which are relaxations of the original problem and whose values
converge monotonically to the contextual fraction. To use the results from \cite{lasserre10}, we establish that we can reduce the problem to the case of compact outcome spaces. In any case energy bounds of the experimental apparatus provide an argument for this but it is primordial for theoretical considerations where observables might be unbounded. 

\noindent \textbf{Chapter \ref{chap:equivalence}.} Negativity of the Wigner function is a striking non-classical feature of quantum states that is related to contextuality for discrete-variable systems \cite{howard2014contextuality,delfosse2015wigner,bermejo2017contextuality,delfosse2017equivalence}. It has been widely studied
as a resource for quantum speed-up and advantage
\cite{galvao2005discrete,veitch2012negative,pashayan2015estimating,catani2018state,saha2019state}. In this chapter, we extend this result and make a connection between contextuality and Wigner negativity for continuous-variable systems. In particular, we show that contextuality is equivalent to Wigner negativity when we allow the measurement of all possible displacements on 2 or more continuous-variable systems. 

\noindent \textbf{Chapter \ref{chap:witnessing}.} Beyond the fundamental relevance of Wigner negativity, it is also a necessary resource for quantum speedup with continuous variables. As quantum technologies emerge, the need to identify and characterise the resources which provide an advantage over existing classical technologies becomes more pressing. It is therefore desirable to detect Wigner negativity. 
In this chapter, we derive witnesses for Wigner negativity of single-mode and multimode quantum states, based on fidelities with Fock states, which can be reliably measured using standard detection setups. They possess a threshold expectation value indicating whether the measured state has a negative Wigner function. Moreover, the amount of violation provides an operational quantification of Wigner negativity. 
We phrase the problem of finding the threshold value for a given witness as an infinite-dimensional linear optimisation problem. By relaxing and restricting the corresponding linear program, we derive two converging hierarchies of semidefinite programs, which provide numerical sequences of increasingly tighter upper and lower bounds for the sought threshold value. We further show that our witnesses form a complete family---each Wigner negative state is detected by at least one witness---thus providing a reliable method for experimentally witnessing Wigner negativity of quantum states from few measurements. From a foundational perspective, our findings provide insights on the set of positive Wigner functions which still lacks a proper characterisation.

\noindent \textbf{Chapter \ref{chap:informationretrieval}.} Here we focus on discrete-variable quantum information and on expanding the range of applications for which contextuality and negativity
can be linked to and used to directly quantify quantum advantage. 
Random access codes have provided many examples of quantum advantage in communication,
but concern only a specific communication game.
In this chapter, we investigate a broad generalisation of those that are referred to as information retrieval tasks.
We introduce and give a detailed analysis of an information retrieval task---the Torpedo Game---that is distinct from a random access code. We show that it admits a greater quantum advantage than the comparable random access code.
Perfect quantum strategies involving prepare-and-measure protocols with experimentally accessible three-level systems \cite{TavakoliExp2021}
emerge via analysis in terms of the discrete Wigner function.
The example is leveraged to an operational advantage in a pacifist version of the strategy game \textit{Battleship}.
We pinpoint a characteristic of quantum systems that enables quantum advantage  
in any bounded-memory information retrieval task.
While preparation contextuality has previously been linked to advantages in random access coding
we focus here on a different characteristic called
sequential contextuality \cite{mansfield2018quantum}.
It is shown not only to be necessary and sufficient for quantum advantage,
but also to quantify the degree of advantage for any information retrieval task.

\section*{Publications}
\label{sec0:publications}
\addcontentsline{toc}{section}{Publications}

\noindent - The publication, pre-prints and work in progress on which this thesis is built are listed below.

\medskip

\noindent \cite{Chabaud2021witnessingwigner} ``\href{https://doi.org/10.22331/q-2021-06-08-471}{Witnessing Wigner Negativity}'',
U. Chabaud, P-E. Emeriau, F. Grosshans,
\textbf{Quantum, 5:471, June 2021.}

\medskip

\noindent \cite{barbosa2019continuous} ``\href{https://doi.org/10.1007/s00220-021-04285-7}{Continuous-variable nonlocality and contextuality}'',
R. S. Barbosa, T. Douce, P-E. Emeriau, E. Kashefi, S. Mansfield,
\textbf{Communication in Mathematical Physics, 391:1047–1089, March 2022.}

\medskip

\noindent \cite{emeriau2020quantum} ``\href{http://dx.doi.org/10.1103/PRXQuantum.3.020307}{Quantum Advantage in Information Retrieval}'',
P-E. Emeriau, M. Howard, S. Mansfield,
\textbf{PRX Quantum, 3:020307, April 22.}

\medskip

\noindent \cite{booth2021} ``\href{https://arxiv.org/abs/2111.13218}{Contextuality and {W}igner negativity are equivalent for continuous-variable quantum measurements}'',
R. Booth, U. Chabaud, P-E. Emeriau, \textbf{arXiv:2111.13218}.

\section*{Notations}
\label{sec0:notations}
\addcontentsline{toc}{section}{Notations}

\subsubsection*{Sets:}
\begin{itemize}
    \item $\N$ is the set of the natural numbers. For $n \in \N$, $ \llbracket0,n\rrbracket $ denotes the list of natural numbers between $0$ and $n$.
    \item $\R$ is the set of the real numbers.
    \item $\C$ is the set of the complex numbers. An asterisk * adjacent to a complex number denotes that we take its complex conjugate. $\mathbb{S}_1$ is the set of complex numbers of modulus 1.
    \item $\Z_d$ is the set of integers modulo $d$ for $d$ a positive integer. 
\end{itemize}
We add an exponent $*$ to these above sets when $0$ is removed from them. 
For $d\in \N^*$ and $\K \in \{\N,\R,\C,\Z_d\}$, $\K^d$ denotes the Cartesian product with itself $d$ times. For $\bm \alpha \defeq (\alpha_i) \in \K^d$, $\K = \N,\C$ or $\R$ ,$\vert \bm \alpha \vert = \sum_{i=1}^d \alpha_i$. 

\noindent The Kronecker delta is the function of two variables over nonnegative integers that assigns 1 if the variables are equal and 0 otherwise. It is denoted $\delta_{ij}$.
$\bm \delta_k \in \K^d$ designates the vector filled with 0 except for a 1 at the k$^\text{th}$ coordinate. 

\begin{itemize}
    \item $\N^d_k \defeq \setdef{\alpha \in \N^d}{\lvert \alpha \rvert \leq k}$ for $k \in \N$, $d \in \N^*$.
    \item $\Rpolm$ is the ring of real polynomials in the variables $\bm x \in \R^d$.
    \item $\Rpolkm \subset \Rpolm$ is the set of polynomials of total degree at most $k \in \N$.
    \item $\SOSm \subset \Rpolm$ is the set of sum-of-squares polynomials.
    \item $\SOSkm \subset \SOSm$ is the set of sum-of-squares polynomials of total degree at most $2k \in \N$.
\end{itemize}
Below $R$ is a subset of $\R^d$ for some $d \in \N^*$. We implicitly consider real-valued functions.
\begin{itemize}
    \item $C(R)$ is the space of continuous functions on $R$. 
    \item $C_0(R)$ is the space of continuous functions on $R$ that vanish at infinity in the case where $R$ is not bounded.  
    \item $C^{\infty}(R)$ is the space of smooth (infinitely-differentiable) functions over $R$.
    \item $L^1(R)$ is the space of integrable functions over $R$.
    \item $L^2(R)$ is the space of square-integrable functions over $R$.
    \item $L^{2'}(R)$ is the dual space $L^2(R)$ over $R$ which is isomorphic to $L^2(R)$.
    \item $\ell^2$ is the space of square-summable real sequences. 
    \item $\mathcal S(R)$ is the space of Schwartz functions over $R$ (the space of $C^{\infty}$ functions that go to $0$ at infinity faster than any inverse polynomial, as do their derivatives) \cite{schwartz1947theorie}.
    \item $\mathcal S'(R)$ is the dual space of $\mathcal S(R)$ which is the space of tempered distributions over $R$ (\ie distributions that have at most a polynomial growth at infinity).
    \item $\mathcal S(\mathbb N)$ is the space of rapidly decreasing real sequences (\ie that go to $0$ at infinity faster than any inverse polynomial). 
    \item $\mathcal S'(\mathbb N)$ is the dual space of $\mathcal S(\mathbb N)$ which is the space of slowly increasing real sequences (\ie that are upper bounded by a polynomial). 
\end{itemize}
We add a subscript $+$ when we take the nonnegative elements of all of the above sets (\eg $\R_+$ denotes the nonnegative reals, $S_+(R)$ denotes the nonnegative Schwartz functions over $R$, etc). 

\begin{itemize}
    \item $\Matrices{m}{\R}$ is the space of $m \times m$ real matrices for $m \in \N^*$.
    \item $\SymMatrices{m}$ is the set of $m\times m$ real symmetric matrices for $m \in \N^*$. 
\end{itemize}
An exponent $T$ on a matrix denotes its transpose while an exponent $\dagger$ denotes its conjugate transpose. The symbol $\succeq$ stands for semidefinite positiveness.
\begin{itemize}
    \item $\bm U = \tuple{U,\Fc_U}$ denotes a generic measurable space composed of a set $U$ and a $\sigma$-algebra $\Fc_U$ on $U$. We will use the convention of using bold font to refer to the measurable space and regular font to refer to the underlying set.
    \item $\Measures(\bm U)$ is the set of measures on $\bm U$. 
    \item $\FSMeasures(\bm U)$ is the set of finite-signed measures on $\bm U$ (\ie finite measures that are non necessarily nonnegative). 
    \item $\PMeasures(\bm U)$ is the set of probability measures on $\bm U$. 
    \item $\Pc(X)$ denotes the powerset of a set $X$.
    \item $V\setminus U = V - U = \{x \in V | x \notin U\}$ denotes the difference of two sets $U$ and $V$ such that $U\subseteq V$.
    \item for $U \subseteq V$, $U^c$ denotes the complement of $U$ in $V$.
\end{itemize}
For a measurable space $\bm U$, $E \in \Fc_U$, $\bm 1_E$ (resp. $\bm 0_E$) denotes the function that assigns $1$ (resp. $0$) to all elements of $E$. $\bm 1_E$ is the indicator function on $E$.
We use $\cong$ between two sets to mean they are isomorphic. For a vector space $V$ over a field $F$, $V^*$ denotes the set of linear functions $V \to F$.\footnote{This is the same notation than the one used to remove $0$ from a set but it will be clear from context which notation we mean.}

For $k \in \N$, the $k^{\text{th}}$ Laguerre function is $\fdec{\mathcal L_k(x)}{x \in \R_+}{(-1)^k L_k(x) e^{\frac x2}}$
with $L_k(x) \defeq \sum_{l=0}^k \frac{(-1)^l}{l!} \binom kl x^l $ the $k^{\text{th}}$ Laguerre polynomial.

\subsubsection*{Quantum Information}
$\mathscr{H}$ denotes a Hilbert space. When not explicit from the context, $\mathscr{H}$ is a generic separable infinite-dimensional Hilbert space. $\Dc(\mathscr{H})$ is the set of quantum states (\ie positive semidefinite operators with unit trace) over $\mathscr{H}$. $\ket \psi \in \Dc(\mathscr{H})$ represents a generic pure state of $\mathscr H$ and $\rho \in \Dc(\mathscr{H})$ a generic density operator. The identity operator is denoted $\Id$ ($\Id_n$ for $n \in \N^*$ is the usual $n \times n$ identity matrix) and the usual qubit Pauli operators are denoted $\Xp$, $\Yp$ and $\Zp$. $\Xp$ and $\Zp$ also designate the generalisation of their qubit counterpart to the qudit realm. It will be explicit from the context.

\subsection*{Contextuality}
From an operational perspective, an experimental setup can be formalised as a measurement scenario $\tuple{\Xc,\Mc,\Oc}$ where $\Xc$ is the set of measurement labels, $\Mc$ the set of maximal contexts, $\Oc = (\Oc_x)_{x \in \Xc}$ is the collection of outcome sets for each measurement. Empirical models will be usually denoted by $e$ and $\CF(e)$ designates its contextual fraction while $\NCF(e) = 1 - \CF(e)$ designates its noncontextual fraction.

\subsection*{Multi-index notation}
For $M \in \N^*$:
\begin{equation}
    \begin{aligned}
    \bm0&=(0,\dots,0)\in\mathbb N^M\\
    \bm1&=(1,\dots,1)\in\mathbb N^M\\
    m\bm 1&=(m,\dots,m)\\
    m\bm k&=(mk_1,\dots,mk_M)\\
    \pi_{\bm k}&=\prod_{i=1}^M(k_i+1)\\
    s(m) &= \binom{M+m}{m} \\
    \hat D(\bm\alpha)&=\hat D(\alpha_1)\otimes\cdots\otimes\hat D(\alpha_M)\\
    \ket{\bm k}&=\ket{k_1}\otimes\cdots\otimes\ket{k_M}\\
    \bra{\bm k}&=\bra{k_1}\otimes\cdots\otimes\bra{k_M}\\
    \bm k\le\bm n\,&\Leftrightarrow\,k_i\le n_i\quad\forall i=1,\dots,M\\
    L_{\bm k}(\bm\alpha)&=L_{k_1}(\alpha_1)\cdots L_{k_M}(\alpha_M)\\
    \mathcal L_{\bm k}(\bm\alpha)&=\mathcal L_{k_1}(\alpha_1)\cdots\mathcal L_{k_M}(\alpha_M)\\
    |\bm k|&=k_1+\cdots+k_M\\
    \bm\alpha^{\bm k}&=\alpha_1^{k_1}\cdots\alpha_M^{k_M}\\
    \bm k!&=k_1!\cdots k_m!\\
    \binom{\bm n}{\bm k}&=\binom{n_1}{k_1}\cdots\binom{n_M}{k_M}\\
    \bm k+\bm n&=(k_1+n_1,\dots,k_M+n_M)\\
    e^{\bm\alpha}&=e^{\alpha_1}\cdots e^{\alpha_M}.
    \label{eq:ch00_multiindex}
    \end{aligned}
\end{equation}

\dobib

\pagestyle{myvf}


\chapter{Frameworks and formalism}
\label{chap:formalisms}

\lettrine{I}n this chapter, we introduce the frameworks and formalism on which this dissertation is built. Section~\ref{sec01:preliminaries} introduces the basic notions of Quantum Information; Section~\ref{sec01:phasespace} gives the necessary background for phase-space formalism and culminates with the expression of the Wigner function; Section~\ref{sec01:measure} provides the necessary notions of measure theory;
Section~\ref{sec01:convexopti} provides the tools for (infinite-dimensional) optimisation theory; and Section~\ref{sec01:sheaf} introduces the sheaf theoretic approach for contextuality.

\section{Quantum information preliminaries}
\label{sec01:preliminaries}

\subsection{Basics of quantum information theory}
\label{subsec01:QM}

We give a brief and pedestrian introduction to the basics of Quantum Information. We refer readers to \cite{NielsenChuang} for a more extensive introduction.

Quantum information is the field that describes the behaviour of information when encoded in degrees of freedom of particles governed by the laws of quantum physics. 
The properties of a quantum system are represented by its quantum state. 
Formally, \textit{pure} quantum states correspond to normalised state vectors in a separable Hilbert space $\mathscr{H}$. In quantum physics the Dirac notation is ubiquitous: a ket $\ket{\psi}$ is a vector from $\mathscr{H}$ which corresponds to a \textit{pure} quantum state; a bra $\bra{\phi}$ is a linear form $\fdec{\phi}{\mathscr{H}}{\C}$. 
The action of a linear form on a vector $\bra{\phi}(\ket{\psi})$ gives a scalar denoted $\braket{\phi | \psi}$. 
It can be used to construct the projection onto the state $\ket{\psi}$ as $P_{\psi} : \ket{\phi} \in \mathscr{H} \mapsto \braket{\psi | \phi} \ket{\psi} \in \mathscr{H}$ or more concisely $\ket \psi \bra \psi$. 

From an operational point of view, a quantum experiment can be decomposed into three distinct steps: \textit{preparation}, \textit{evolution} and \textit{measurement}. Note that the evolution could always be taken into account in the measurement or the preparation though it is often easier to consider measurements or preparations in a fixed basis. Preparation might be used to encode classical information; evolution to manipulate quantum information; measurements to extract classical information again. 
\paragraph{Quantum states.}
The simplest example is the two-dimensional Hilbert space $\C^2$ where information is encoded in the so-called quantum bits or \textit{qubits}. Given an orthonormal basis $\{\ket{0},\ket{1}\}$ of $\C^2$, a pure qubit state can be expressed as:
\begin{equation}
    \ket{\psi} = a \ket{0} + b \ket{1} = \begin{pmatrix} a \\ b \end{pmatrix}, 
    \label{eq:ch01_qubit}
\end{equation}
where $a,b \in \C$ with the normalisation condition that $|a|^2+|b|^2=1$. 
Physically, it could for instance model the spin of an electron for which the two possible basis states are the spin pointing up and down; or the polarisation of a single photon where the two basis states are vertical and horizontal polarisation. 
Surprisingly, quantum physics allows for superposition, that is, it allows for complex linear combinations. 
That is why quantum theory is inherently probabilistic. 
Quantum randomness occurs when measuring a quantum system in superposition: in general, it is impossible to predict deterministically the outcome of such a measurement. 
This is not because of a lack of knowledge of the system but rather due to the probabilistic nature\footnote{or the incompleteness\dots} of quantum theory. 
When measuring the state defined in \refeq{ch01_qubit} in the computational basis $\{\ket{0},\ket{1}\}$, we will witness outcome $0$ associated to $\ket{0}$ with probability $|a|^2$ and outcome $1$ associated to $\ket{1}$ with probability $|b|^2$ according to the Born rule\index{Born rule} that we detail shortly after. 

Quantum systems may also exhibit classical randomness. 
In that case, we will say that the quantum state is \textit{mixed} rather than \textit{pure}. 
Mathematically, it is represented by a semidefinite, Hermitian operator with unit trace acting on $\mathscr{H}$ called a density matrix or density operator. We write $\Dc (\mathscr H)$ the set of density operators.
Any density operator can be written as a convex combination of pure states. 
For instance, the mixed state $\frac12 \ketbra00+ \frac 12 \ketbra11$ known as the maximally mixed state is classically probabilistic: it corresponds to the state obtained by flipping an unbiased coin; it is different from the pure superposition $\frac 1{\sqrt2} (\ket0 + \ket1)$ whose density matrix is $\frac12 \ketbra00 + \frac12 \ketbra01 + \frac12 \ketbra10 + \frac12 \ketbra11$.

To describe the state of a global system consisting of two independent subsystems $\ket {\psi_1}$ and $\ket {\psi_2}$ lying in respectively Hilbert spaces $\mathscr{H}_1$ and $\mathscr{H}_2$, we will use a tensor product structure. 
We write the global state as $\ket {\psi_1} \otimes \ket {\psi_2} = \ket{\psi_1 \psi_2} \in \mathscr{H}_1 \otimes \mathscr{H}_2$. 
However the two subsystems might not be independent. 
A pure quantum state of several particles which cannot be written as a tensor product of quantum states is said to be \textit{entangled}. 
As an example, the state $\frac 1{\sqrt2}(\ket{00} + \ket{11})$ is entangled while the state $\frac 12 (\ket{00} + \ket{01} + \ket{10} + \ket{11}) = \frac 1{\sqrt 2}(\ket0 + \ket1) \otimes \frac 1{\sqrt 2} (\ket0 + \ket1)$ is separable. 
For ensembles, a mixed state is separable when it can be written as a mixture of separable pure states and entangled otherwise. 
Measurement on entangled states may give rise to correlations unachievable with classical systems. 
For a composite system $\rho_{AB} \in \mathscr H_A \otimes \mathscr H_B$, the reduced state of the first subsystem is obtained by tracing out---that is taking the partial trace---over the second subsystem. It is described by $\Tr_{\mathscr H_B}(\rho)$. Partial trace is the unique operator that describes part of a larger quantum system with the correct description for subsystems of the composite system \cite{NielsenChuang}. 

The \textit{fidelity}\index{Fidelity|textbf}\footnote{Note that in \cite{NielsenChuang}, the fidelity is defined as the square root of Eq~\eqref{eq:ch01_deffidelity}. There is no fundamental difference between the two definitions. } between two arbitrary quantum states $\rho$ and $\sigma$ is defined as 
\begin{equation}
F(\rho,\sigma) \defeq \Tr(\sqrt{\sqrt\rho\sigma\sqrt\rho})^2.
\label{eq:ch01_deffidelity}
\end{equation}
When one of the states is pure, it reduces to $F(\rho,\sigma)=\Tr(\rho\sigma)$. 
The \textit{trace distance}\index{Trace distance|textbf} between two quantum states $\rho$ and $\sigma$ is defined as
\begin{equation}
D(\rho,\sigma) \defeq \frac12\Tr(\sqrt{(\rho-\sigma)^2}).
\label{eq:ch01_deftracedistance}
\end{equation}
The trace distance can be related to the maximum probability of distinguishing between two quantum states. Fidelity and trace distance are related by $1-F\le D\le\sqrt{1-F}$~\cite{NielsenChuang}.

\paragraph{Evolution.}
The simplest evolution of a closed quantum system is described by a unitary transformation with a unitary operator $\hat U$ such that $\hat U \hat U^{\dagger} = \Id$. It is generated by a Hermitian operator $\hat H$---often called Hamiltonian---such that $\hat U = e^{i \hat H}$. The unitary evolution of a pure state $\ket \psi$ results in a pure state $\hat U \ket \psi$ while for a mixed state $\rho$, it results in a mixed state $\hat U \rho \hat U^\dagger$. 
Important 2-dimensional examples are given by the Pauli $\Xp$ gate, the Pauli $\Zp$ gate and the Pauli $\Yp$ gate\footnote{From a computational point of view, they correspond to the more intuitive errors that may occur: respectively \textit{bit flip}, \textit{phase flip} or both.}:
\begin{equation}
    \Xp = \begin{pmatrix}
            0 & 1\\
            1 & 0
        \end{pmatrix}, \quad
    \Zp = \begin{pmatrix}
            1 & 0\\
            0 & -1
        \end{pmatrix}, \quad
    \Yp = \begin{pmatrix}
            0 & -i\\
            i & 0
        \end{pmatrix}.
    \label{eq:ch01_Paulis2}
\end{equation}
Other important single qubit operations are the Hadamard gate $H$ and the $\frac \pi8$-gate $T= \sqrt S$:
\begin{equation}
    \Ha = \frac 1 {\sqrt2} \begin{pmatrix}
            1 & 1\\
            1 & -1
        \end{pmatrix}, \quad
    \Sp = \begin{pmatrix}
            1 & 0\\
            0 & i
        \end{pmatrix}, \quad
    \Tp = \begin{pmatrix}
            1 & 0\\
            0 & e^{i \frac \pi4}
        \end{pmatrix}.
    \label{eq:ch01_singlequbit}
\end{equation}
As an example, applying $\Xp$ on the state given by \refeq{ch01_qubit} yields:
\begin{equation}
    \Xp \ket \psi = b \ket0 + a \ket1.
    \label{eq:ch01_ex_evolution}
\end{equation}

More general evolutions are given by quantum channels known as completely positive trace-preserving (CPTP) maps. 
Stinespring's dilation theorem guarantees that CPTP maps can ultimately be lifted to unitary operations \cite{Stinespring1955} on a higher dimensional Hilbert space.
This can be seen as a purification of CPTP maps. 

\paragraph{Measurement.} Following the approach of von Neumann, a measurement is represented by a self-adjoint operator known as an \textit{observable} $\hat O = \hat O^\dagger$. 
An observable has a family of eigenvectors that form an orthonormal basis for the Hilbert space.
Measuring the observable yields an outcome that corresponds to one of the eigenvalues associated to the eigenvectors comprising the basis. The probability of obtaining each outcome follows from the Born rule\index{Born rule|textbf}. When the state of the system is a pure state $\ket \psi$ it reads:
\begin{equation}
    \text{Pr}(\lambda) = \bra \psi  \Pi_\lambda  \ket \psi,
    \label{eq:ch01_Bornrulepure}
\end{equation}
where $\lambda$ is an eigenvalue of $\hat O$ and $\Pi_\lambda$ is the projector onto the eigenspace\footnote{Or more simply the eigenvector if the observable at hand is not degenerate.} associated to $\lambda$. For qubits, Pauli operators are Hermitian and can thus be measured. 
As already mentioned, measuring $\Zp$ on the state given by \refeq{ch01_qubit} yields two outcomes: $0$ with probability $\bra \psi \Pi_0 \ket \psi = \braket{\psi | 0} \braket{0 | \psi} = |\braket{0 | \psi}|^2 = |a|^2$; and $1$ with probability $\bra \psi \Pi_0 \ket \psi = |b|^2$. 
For a mixed state $\rho$, the Born rule\index{Born rule|textbf} reads:
\begin{equation}
    \text{Pr}(\lambda) = \Tr(\Pi_\lambda \rho).
    \label{eq:ch01_Bornrulemixed}
\end{equation}
By linearity, the expectation value of $\hat O$ can be expressed as:
\begin{equation}
    \langle \hat O \rangle = \Tr(\hat O \rho).
    \label{eq:ch01_expectation}
\end{equation}
This procedure corresponds to a projection-valued measurement (PVM). \index{PVM|textbf}

PVMs are generalised by positive operator-valued measurements (POVMs)\index{POVM|textbf} in the same sense that mixed states generalise the notion of pure states. A POVM is a set of positive semidefinite matrices $\{ F_i \}_{i \in \Ic}$ that sum to the identity operator where $\Ic$ denotes the set of outcomes. When the state of the system is $\rho$, the outcome $i \in \Ic$ is obtained with probability:
 \begin{equation}
    \text{Pr}(i) = \Tr(F_i \rho).
    \label{eq:ch01_POVM}
\end{equation}
A POVM reduces to a PVM when $\{ F_i \}_{i \in \Is}$ are orthogonal projectors.

\section{Phase-space formalism}
\label{sec01:phasespace}

We now turn our attention to the phase-space formulation of quantum physics, motivate its use and introduce the so-called Wigner function. 
We refer readers to a recent and comprehensive overview \cite{rundle2021overview} for further details. 
This formulation is one of many possible alternative formulations of quantum physics, \eg Schrödinger's wave formulation or Heisenberg, Born and Jordan's matrix mechanics both from 1925. 
The phase-space formulation is often favoured in quantum optics applications. 
The Wigner function was first introduced by Wigner in 1932 \cite{Wigner1932} as the Fourier transform of the spatial autocorrelation wavefunction.
It can be thought as the joint (quasi)probability distribution of position and momentum.
The marginals of the Wigner function reproduce the correct probability distribution of quadratures\footnotemark\ \ie integrating the Wigner function over position at a given momentum gives the correct probability associated with the chosen momentum and vice versa.
\footnotetext{Position $\hat q$ and momemtum $\hat p$ are known as quadrature operators since they correspond to two `quadratures' of the electromagnetic field. See Subsection~\ref{subsec01:CV}.}
Crucially, it is a quasiprobability distribution in phase-space: it sums to unity though one has to drop the nonnegativity requirement of probability distributions.
This latter requirement is critical in quantum computing applications.
We will see how Wigner functions emerge as the expectation value of a kernel\footnotemark\ for finite-dimensional as well as infinite-dimensional systems. 
\footnotetext{Kernel is rather versatile in mathematics. Here, by kernel, we mean \textit{integral kernel}.}
The phase-space formulation comes with a remarkable symplectic structure \cite{adesso2014continuous}.

\subsection{Phase-space for \texorpdfstring{$d$}{d}-dimensional quantum states}
\label{subsec01:phasespace_qudits}

It is possible to represent finite-dimensional quantum states as quasi-probability distributions over a phase space of discrete points.
Fix $d \in \N$ an odd power-of-prime.
Wootters \cite{gibbons2004discrete,wootters1987wigner}
introduced  a  method  for constructing discrete Wigner functions (DWF) based on finite fields, wherein vectors from a complete set of mutually unbiased bases in $\mathbb{C}^d$ are put in one-to-one correspondence with the lines of a finite affine plane of order $d$.
This work was based on previous constructions \eg \cite{connell1984,cohen1986joint}.
Two orthonormal bases are said to be mutually unbiased \cite{schwinger1960unitary} if the square of the inner product between an element of the first basis and an element of the second basis equals $\frac 1d$. They are `unbiased' in the following sense: if a system is prepared according to a basis state of one of the basis, all the outcomes of the measurement with respect to the other basis are predicted to occur with the same probability $\frac 1d$.

Gross \cite{gross2006hudson}
singled out one particularly symmetric definition of DWF that obeys the discrete version of Hudson’s Theorem \cite{hudson1974wigner} that we recall after introducing the Wigner function. We introduce the necessary tools for this construction below.

We consider a $d$-dimensional Hilbert space $\Hs$ equipped with the computational basis $\{\ket0, \ket1, \dots, \ket{d-1} \}$. Below all arithmetic will be modulo $d$. The phase-space of a single $d$-level system is $V = \Z_d\times\Z_d$---which can be thought of as an affine plane, or more precisely a toric $d\times d$ grid.\index{Phase-space!DV|textbf} For $m$ particles, it can be straightforwardly extended by associating a phase-space to each particle. We extend the definition of Pauli operators $\Xp$ (translation) and $\Zp$ (boost) to $\Hs$ as follows: for all $k \in \Z_d$,\index{Pauli gates!DV|textbf}  
\begin{equation}
    \begin{aligned}
        \Xp \ket k & = \ket{k+1} \\
        \Zp \ket k & = \omega^k \ket k,
    \end{aligned}
\end{equation}
with $\omega = \exp(i \frac{2 \pi}{d} )$. 
Note that, unlike the qubit case, these are not Hermitian operators anymore. Strictly speaking they should not correspond to physical observational quantities. 
However their non-degenerate eigenstates still form a complete orthonormal basis of the Hilbert space; thus they can be used to define a von Neumann measurement. 
The qudit \textit{Pauli group} is comprised of \textit{Weyl operators} (also known as generalised Pauli operators) which are products of powers of these operators, \eg $\Xp^q \Zp^p$ for $(q,p) \in V$. 
The \textit{Clifford group} is defined as containing the unitaries which map Weyl operators to other Weyl operators under conjugation. 
A unitary $U$ stabilises a state $\ket{\psi}$ if $U\ket{\psi}=\ket{\psi}$. 
A \textit{stabiliser state} is the unique $n$-qudit state stabilised by a subgroup of size $d^n$ of the Pauli group. 
Equivalently, stabiliser states may be understood as the image of computational basis states under the Clifford group.

We introduce the \textit{displacement operator}\index{Displacement operator!DV|textbf} at point $(q,p) \in V$:
\begin{equation}
    \hat D(q,p) \defeq \omega^{2^{-1} qp} \Xp^q \Zp^p.
    \label{eq:ch01_displacementDV}
\end{equation}
These are known as are Gross' Weyl operators \cite{gross2006hudson}. $2^{-1}$ is the multiplicative inverse of 2 in $\Z_d$, that is $2^{-1}=(d+1)/2$. 

The \textit{parity operator}\index{Parity operator!DV|textbf}---also known as the phase point operator at the origin---reads:
\begin{equation}
    \hat \Pi \defeq \sum_{k \in \Z_d} \ketbra{-k}{k}.
    \label{eq:ch01_parityDV}
\end{equation}
Parity transformation---or parity inversion---is essentially a flip in the sign of spatial coordinates. It is physically motivated by its continuous-variable analogue though we indeed have that $\hat \Pi \ket k = \ket {-k \text{ mod d}}$ for $k \in \Z_d$.

\subsection{The case of continuous-variable quantum information}
\label{subsec01:CV}

Before giving the expression of the Wigner function compatible with discrete and continuous systems we need to introduce some elements of quantum information with continuous variables. We refer readers to \cite{adesso2014continuous,braunstein2005quantum,walschaers2021non} for extensive introductions to the material presented here. 

In the continuous-variable (CV) formalism, information is encoded in a mode (or qumode) rather than in a finite dimensional system like a qubit or a qudit. The separable Hilbert space $\Hs$ is infinite-dimensional and is equipped with a countable orthonormal basis $\{\ket n\}_{n \in \N}$ known as the Fock basis, or photon number basis in an optical context. 
The phase-space associated to a single mode is now $V = \R^2 \approxeq \C$ and it can be straightforwardly extended to several modes.\index{Phase-space!CV|textbf} 
A single-mode pure state can be expanded in the Fock basis as:
\begin{equation}
    \ket \psi = \sum_{n\in \N} \psi_n \ket n
    \label{eq:ch01_CVpurestate}
\end{equation}
where for all $n\in \N$, $\psi_n \in \C$ such that $\sum_n \vert \psi_n \vert^2 = 1$. A general mixed state can be expressed as:
\begin{equation}
    \rho = \sum_{k,l \in \N} \rho_{k,l} \ketbra{k}{l},
    \label{eq:ch01_CVmixedstate}
\end{equation}
where for all $k,l \in \N$, $\rho_{k,l} \in \C$, $\rho_{k,l} = \rho_{l,k}^*$ and $\sum_{k} \rho_{k,k} = 1$.

\textit{Ladder operators} are ubiquitous in CV quantum information. Canonical adjoint operators $\hat a$ and $\hat a^\dagger$, known as the annihilation and creation operators, are defined by their action on the Fock basis as:\index{Ladder operators|textbf}
\begin{equation}
    \begin{aligned}
        & \hat a \ket n = \sqrt n \ket{n-1} \quad \text{for } n \in \N^*, \\
        & \hat a \ket 0 = 0, \\
        & \hat a^\dagger \ket n = \sqrt{n+1} \ket{n+1} \quad \text{for } n \in \N.
    \end{aligned}
\end{equation}
These operators obey the commutation relation:
\begin{equation}
    \left[ \hat a,\hat a^\dagger \right] = \Id.
\end{equation}
Note that they do not correspond naturally to physical observations since they are not Hermitian operators. However, from them,  we can define well-known observables. Position and momentum operators---also known as the quadrature operators---are linked to these ladder operators as follows:\index{Quadrature|textbf}
\begin{equation}
    \begin{aligned}
        & \hat q = \frac 1{\sqrt2} (\hat a + \hat a^\dagger), \\
        & \hat p = \frac 1{i\sqrt2} (\hat a - \hat a^\dagger).
    \end{aligned}
\end{equation}
They obey the canonical commutation relation
\begin{equation}
    \left[\hat q, \hat p \right] = i \Id,
\end{equation}
which is one expression of the Heisenberg uncertainty principle \cite{heisenberg1985} \ie that position and momentum cannot be simultaneously measured on the same state with arbitrary precision.

The eigenstates of $\hat a$ are the coherent states $\{\ket \alpha\}_{\alpha \in \C}$:\index{Coherent state|textbf}
\begin{equation}
    \ket \alpha \defeq e^{-\frac 12 |\alpha|^2} \sum_{n \in \N} \frac{\alpha^n}{\sqrt{n!}} \ket n.
    \label{eq:ch01_coherentstate}
\end{equation}
They define an overcomplete basis of $\Hs$.
A central operator in phase-space formulation is the \textit{displacement operator}\index{Displacement operator!CV|textbf} by a value $\alpha \in \C$ \cite{weedbrook2012gaussian} which reads
\begin{equation}
    \hat D(\alpha) \defeq e^{\alpha \hat a^\dagger - \alpha^* \hat a},
    \label{eq:ch01_displacement}
\end{equation}
or similarly for $\alpha = \frac{1}{\sqrt 2}(q + ip) $ with $(q,p) \in V$:
\begin{equation}
    \hat D(q,p) \defeq e^{i p \hat q - i q \hat p } = e^{i \frac {qp}2 } e^{-i q \hat p} e^{i p \hat q} .
    \label{eq:ch01_displacement_pos_mom}
\end{equation}
Compared to the discrete-variable case in Eq.~\eqref{eq:ch01_displacementDV}, this amounts to take the phase factor as $\omega = e^i$ with the continuous-variable generalisation of the $\Xp$ gate as $\Xp(q) \defeq e^{-i q \hat p}$ and of the $\Zp$ gate as $\Zp(p) \defeq e^{i p \hat q}$.\index{Pauli gates!CV|textbf}
This gives that $\hat D(q,p) = e^{i \frac {qp} 2} \Xp(q) \Zp(p)$.

The coherent state of amplitude $\alpha \in \C$ can be obtained by displacing the vacuum state $\ket 0$ by an amount $\alpha$:
\begin{equation}
    \ket \alpha = \hat D(\alpha) \ket 0.
\end{equation}
The coefficients of the displacement operator in Fock basis are given by~\cite{wunsche1998laguerre}:
\begin{equation}
      \braket{k|\hat D(\alpha)|l}=e^{-\frac12|\alpha|^2} \sum_{p=0}^{\min k,l}\frac{\sqrt{k!l!}(-1)^{l-p}}{p!(k-p)!(l-p)!}\alpha^{k-p}\alpha^{*l-p},
      \label{eq:ch01_coefD}
\end{equation}
for all $k,l\in\N$ and all $\alpha\in\C$.

Lastly, the \textit{parity operator}\index{Parity operator!CV|textbf} is defined as:
\begin{equation}
    \hat \Pi \defeq (-1)^{\hat a^\dagger \hat a} = e^{i \pi \hat a^\dagger \hat a}= \sum_{n \in \N} (-1)^n \ketbra{n}{n}.
    \label{eq:ch01_parity}
\end{equation}
On a particle at position $q \in \R$, it acts as $\hat \Pi \psi(q) = \psi(-q)$ where $\psi(q) = \braket{q | \psi}$ and $\ket q$ is an eigenstate of $\hat q$.

\subsection{The Wigner function}
\label{subsec01:Wigner}

We fix $M\in \N^*$ the number of modes. The phase-space is $V^M$ and it is endowed with its vector space structure and its canonical basis.
As mentioned before, the \textit{Wigner function} is a quasiprobability distribution \cite{cahill1969density} on $V^M$ that can be expressed as the expectation value of a kernel---the displaced parity operator. For an operator $\hat O$ acting on $M$ discrete-variable or continuous-variable systems (see Table \ref{tab:ch01_comparisonCV_DV} for the comparison), the Wigner function at point $\bm r = (q_1,\dots,q_M,p_1,\dots,p_M) \in V^M$ reads:\index{Wigner function|textbf}
\begin{equation}
    W_{\hat O}(\bm r) \defeq \left(\frac 2c\right)^{M} \Tr(\hat D(\bm r) \Pi^{\otimes M} \hat D(\bm r)^\dagger \hat O)
    \label{eq:ch01_Wignerdef}
\end{equation}\index{Parity operator}
where $c = 2\pi$ for continuous-variable systems and $c = 2d$ for a $d$-dimensional system and where $\hat D(\bm r) = \hat D(q_1,p_1) \otimes \dots \otimes \hat D(q_M,p_M)$.\index{Displacement operator!CV}

\begin{table}[ht!]
    \centering
        \begin{tabular}{l  c  c}
            \toprule \toprule
             & Discrete variables  & Continuous variables  \\ \midrule
            Phase-space $V^M$ & $(\Z_d^2)^M$ & $(\R^2)^M$ \\
            Phase factor $\omega$ & $e^{i \frac{2 \pi}{d}}$ & $e^i$ \\
            Displacement at $(q,p) \in V^M$ & $e^{\frac{2 i \pi}{d} 2^{-1} qp} \Xp^q \Zp^p$ & $e^{ i \frac{qp}2} \Xp(q) \Zp(p)$  \\
            Parity operator & $\sum_{k \in \Z_d} \ketbra{-k}{k} $ & $\sum_{k \in \N} (-1)^k \ketbra kk$ \\
            Constant $c$ & $2d$ & $2\pi$ \\
            \bottomrule \bottomrule
        \end{tabular}
    \caption{Comparison between quantities comprising the Wigner function of $m$ systems expressed in the discrete-variable setting of odd prime dimension $d$ and in the continuous-variable setting.}
    \label{tab:ch01_comparisonCV_DV}
\end{table}

A useful relation is the link between the Wigner function and the characteristic function that is derived using
the symplectic structure of the phase-space $V^M$. The \textit{characteristic function} of a density operator $\rho$ is defined as \cite{ferraro2005gaussian,Gosson2011symplectic}:\index{Characteristic function|textbf}
\begin{equation}
    \label{eq:ch01_characfunction}
    \begin{aligned}
    \Phi_\rho : V^M & \longrightarrow \R \\
    \bm r & \longmapsto \Tr(\hat D(-\bm r) \rho)
    \end{aligned}
\end{equation}
Due to the commutation relation between position and momentum operators there is a symplectic structure for $V^M$.
We denote $\Omega$ the symplectic form. For $\bm r_1,\bm r_2 \in V^M$,\index{Symplectic form|textbf}
\begin{equation}
    \label{eq:ch01_symplecticform}
    \Omega(\bm r_1,\bm r_2) = \bm r_1 \cdot J \bm r_2 \quad \text{where} \quad J = \begin{pmatrix}
    0 & \Id_M \\
    - \Id_M & 0 
    \end{pmatrix}
\end{equation}
We have that $J^{-1} = J^T = -J$.

Various conventions are used to define the Wigner function in quantum optics. We rely on the one adopted in \cite{Gosson2006symplectic,Gosson2011symplectic}.
The Wigner function is the symplectic Fourier transform of the characteristic function:\index{Wigner function}
\begin{equation}
    \label{eq:ch01_WignerCharacFunction}
    W_\rho(\bm r) = \text{FT} \left[ \Phi_\rho \right] (J \bm r)
\end{equation}
where, for $\bm r \in V^M$, the Fourier transform of a $L^1(V^M)$ function $f$ is defined as (see table~\ref{tab:ch01_comparisonCV_DV}):\index{Fourier transform|textbf}
\begin{equation}
    \text{FT} \left[ f \right](\bm r) \defeq \begin{cases}
    \displaystyle \sum_{\bm p \in V^M} (\omega^*)^{\bm r \cdot \bm p} f(\bm p) \quad \text{ for discrete-variable systems,} \\
    \displaystyle \intg{V^M}{(\omega^*)^{\bm r \cdot \bm p} f(\bm p)}{\bm p} \quad \text{ for continuous-variable systems.}
    \end{cases}
\end{equation}
The inverse Fourier transform is denoted FT$^{-1}$ and it is defined as:
\begin{equation}
    \text{FT}^{-1} \left[ f \right](\bm p) \defeq \begin{cases}
    \displaystyle \frac{1}{c^M} \sum_{\bm r \in V^M} \omega^{\bm r \cdot \bm p} f(\bm r) \quad \text{ for discrete-variable systems,} \\
    \displaystyle \frac{1}{c^M} \intg{V^M}{\omega^{\bm r \cdot \bm p} f(\bm r)}{\bm r} \quad \text{ for continuous-variable systems.}
    \end{cases}
\end{equation}

Now we turn our attention to the celebrated \textit{Hudson's theorem} \cite{hudson1974wigner}.\index{Hudson's theorem|textbf} 
For continuous-variable systems, Hudson's theorem establishes that a pure state which has a nonnegative Wigner function is necessarily a Gaussian state (\ie a state whose Wigner function is a Gaussian distribution). 
Its discrete-variable counterpart \cite{gross2006hudson,Galvao2006ClassicalityDiscreteWigner} says that an odd-dimensional pure state is nonnegatively represented in the DWF if and only if it is a stabiliser state. Hudson’s theorem has remarkable implications, providing large classes of quantum circuit with a local hidden variable model that enables efficient simulation \cite{galvao2005discrete,Galvao2006ClassicalityDiscreteWigner,veitch2012negative,Mari2012}.

The Wigner function is a quasiprobability distribution: a normalised distribution (\ie its sum or integral over phase space equals 1) which can take negative values. 
Whenever it is everywhere nonnegative, it can be interpreted as a kind of probability distribution for the quadrature measurements.
Crucially, Wigner negativity---the fact that the Wigner distribution can be negative in regions of phase space---is a necessary resource for quantum computational speedup \cite{Bartlett2002}. 
Informally, a result from Mari and Eisert \cite{Mari2012} states that if the quantum computation 
\begin{enumerate*}[label=(\roman*)]
\item starts in a product state with a nonnegative Wigner function, 
\item is followed by quantum gates that can be represented by a nonnegative Wigner function (defined through the Choi matrix), 
\item and terminates with measurements associated with a nonnegative Wigner function
\end{enumerate*} 
then there exists an efficient classical algorithm that simulates the computation (assuming that local probability distributions may be sampled efficiently). 
This essentially amounts to sampling directly from the Wigner distribution which, in this case, is a proper probability distribution. This results applies in the DV and CV settings. 

From a resource theoretic point of view, negativity of the Wigner function can only decrease under Gaussian operations~\cite{albarelli2018resource}, i.e., operations that map Gaussian states to Gaussian states. In particular, it is invariant under displacements. It is also a robust property, since two almost indistinguishable quantum states have similar Wigner functions. 
An operational measure of Wigner negativity for a quantum state $\rho\in\mathcal D(\mathscr H)$ is given by its distance to the set of states having a positive Wigner function~\cite{mari2011directly}:\index{Wigner function}
\begin{equation}\label{eq:ch01_eta}
    \eta_\rho=\inf_{\substack{\sigma\in\mathcal D(\mathcal H)\\W_\sigma \geq 0}}D(\rho,\sigma),
\end{equation}
where $D$ denotes the trace distance\index{Trace distance} (see Subsection~\ref{subsec01:QM}). It quantifies the operational distinguishability between the state $\rho$ and any state having a positive Wigner function~\cite{NielsenChuang}.

To fix ideas, the Wigner functions of Fock states $\{\ket k\}_k$ are given by \cite{kenfack2004negativity}:
\begin{equation}
    \forall \alpha \in \C, \, W_k(\alpha) = \frac{2}{\pi} \mathcal L_k(4|\alpha|^2),
    \label{eq:ch01_WignerFockstates}
\end{equation}
where $\mathcal L_k$ is the $k^{\text{th}}$ Laguerre function \cite{szego1959orthogonal} defined as\index{Laguerre!function|textbf} 
\begin{equation}
    \forall x \in \R_+, \; \mathcal L_k(x) \defeq (-1)^k L_k(x) e^{\frac x2}
    \label{eq:ch01_LaguerreFunction}
\end{equation}
with $L_k$ the $k^{\text{th}}$ Laguerre polynomial defined as\index{Laguerre!polynomial|textbf}
\begin{equation}
    L_k(x) \defeq \sum_{l=0}^k \frac{(-1)^l}{l!} \binom kl x^l \Mdot
    \label{eq:ch01_LaguerrePolynomial}
\end{equation}
Except for the vacuum state $\ket 0$, all other Fock states have a Wigner function with a negative part. 

\section{Measure theory in a nutshell}
\label{sec01:measure}

Introduced below are the necessary tools of measure theory that are used throughout this dissertation.
We refer readers to  \cite{billingsley2008probability,tao2011introduction} for a more thorough introduction to measure theory.

To avoid some pathological behaviours when dealing with probability distributions on a continuum, we first need to define $\sigma$-algebras which will result in a good notion of measurability. 
\begin{definition}[$\sigma$-algebras]\index{$\sigma$-algebra|textbf}
A $\sigma$-algebra on a set $U$ is a family $\Bc$ of subsets of U containing the empty set and closed under complementation and countable unions, that is:
\begin{enumerate}[label=(\roman*)]
    \item $\emptyset \in \Bc$.
    \item for all $E \in \Bc$, $E^c \in \Bc$.
    \item for all $E_1,E_2, \dots \in \Bc$, $\cup_{i=1}^\infty E_i \in \Bc$.
\end{enumerate}
\end{definition}

\begin{definition}[Measurable space]\index{Measurable!space|textbf}
A measurable space is a pair $\bm U = \tuple{U,\Fc_U}$ consisting of a set $U$ and a $\sigma$-algebra $\Fc_U$ on $U$.
\end{definition}
\noindent In some sense, the $\sigma$-algebra specifies the subsets of $U$ that can be assigned a `size', and which are therefore called the \emph{measurable sets} of $\bm U$. We will use the convention of using boldface to refer to measurable spaces and regular font to refer to the underlying set. 
A trivial example of a $\sigma$-algebra over any set $U$ is its powerset $\Pc(U)$, which gives the discrete measurable space $\tuple{U,\Pc(U)}$, where every set is measurable.
This is typically used when $U$ is countable (finite or countably infinite), in which case this discrete $\sigma$-algebra is generated by the singletons.
Another example, of central importance in measure theory, is the Borel $\sigma$-algebra $\Bc_\R$ generated from the open sets of $\R$, whose elements are called the Borel sets. 
This gives the measurable space $\tuple{\R,\Bc_\R}$.
Working with Borel sets avoids the problems that would arise if we naively attempted to measure or assign probabilities to points in the continuum.
More generally, any topological space gives rise to a Borel measurable space in this fashion. The product of measurable spaces can be defined as follows:

\begin{definition}[Finite product]\index{Measurable!space!finite product of|textbf}
The product of measurable spaces $\bm U_1 = \tuple{U_1,\Fc_1}$ and $\bm U_2 = \tuple{U_2,\Fc_2}$ is the measurable space
\[\bm U_1 \times \bm U_2 = \tuple{U_1 \times U_2, \Fc_1 \otimes \Fc_2} \Mcomma\]
where $U_1 \times U_2$ is the Cartesian product of underlying sets, and the so-called tensor-product $\sigma$-algebra $\Fc_1 \otimes \Fc_2$ is the $\sigma$-algebra on $U_1 \times U_2$ which is generated by the subsets of the form $E_1 \times E_2$ with $E_1 \in \Fc_1$ and $E_2 \in \Fc_2$.
\label{def:ch01_productmeasurable}
\end{definition}
The product above is given by the \textit{box topology}. In this dissertation, we will also need to deal with measurable spaces formed by taking the product of an uncountably infinite family of measurable spaces. 
In that case the box topology is no longer the most natural choice as enlightened by Tychonoff's theorem \cite{tychonoff1930topologische} and we will use the \textit{product topology}. 
\begin{definition}[Infinite product]\index{Measurable!space!infinite product of|textbf}
Fix a possibly uncountably infinite index set $I$.
The product of measurable spaces $(\bm U_i = \tuple{U_i,\Fc_i})_{i \in I}$ is the measurable space:
\[\prod_{i \in I} \bm U_i = \tuple{\prod_{i \in I} U_i, \prod_{i \in I} \Fc_i} = \tuple{U_I,\Fc_I} \Mcomma\]
where $U_I = \prod_{i \in I} U_i$ is the Cartesian product of the underlying sets, and the $\sigma$-algebra $\Fc_I = \prod_{i \in I} \Fc_i$ is obtained via the product topology \ie it is generated by subsets of $\prod_{i \in I} U_i$ of the form $\prod_{i \in I} B_i$ where for all $i \in I$, $B_i \subseteq U_i$ and $B_i \subsetneq U_i$ only for a finite number of $i \in I$.
\label{def:ch01_productmeasurableinfinite}
\end{definition}
Remarkably, the product topology is the smallest topology that makes the projection maps $\fdec{\pi_k}{\prod_{i \in I} U_i }{U_k}$ continuous.
This definition is more general as it reduces straightforwardly to the case of a finite product.
We can now formally define measurable functions and measures on those spaces. 
\begin{definition}[Measurable function]\index{Measurable!function|textbf}
A measurable function between measurable spaces $\bm U = \tuple{U,\Fc_U}$ and $\bm V = \tuple{V, \Fc_V}$ is a function $\fdec{f}{U}{V}$ whose preimage preserves measurable sets, \ie such that, for any $E \in \Fc_V$, ${f^{-1}(E) \in \Fc_U}$.
\end{definition}
\noindent This is similar to the definition of a continuous function between topological spaces.
Measurable functions compose as expected.

\begin{definition}[Measure]\index{Measure|textbf}
\label{def:ch01_measure}
A measure on a measurable space $\bm U = \tuple{U,\Fc_U}$ is a function $\fdec{\mu}{\Fc_U}{\Rext}$ from the $\sigma$-algebra to the extended real numbers $\Rext = \R \cup \enset{-\infty,+\infty}$ satisfying:
\begin{enumerate}[label=(\roman*)]
\item\label{enum:nonnegativity} (nonnegativity)
$\mu(E)\geq 0$ for all $E\in\Fc_U$;
\item (null empty set)
$\mu(\emptyset)=0$;
\item ($\sigma$-additivity)
for any countable family $\family{E_i}_{i=1}^\infty$ of pairwise disjoint measurable sets, $\mu(\bigcup_{i=1}^\infty E_i) = \sum_{i=1}^\infty \mu(E_i)$.
\end{enumerate}
\end{definition}
\noindent In particular, a measure $\mu$ on $\bm U = \tuple{U,\Fc_U}$ allows one to integrate well-behaved measurable functions $\fdec{f}{U}{\R}$ (where $\R$ is equipped with its Borel $\sigma$-algebra $\Bc_\R$) to obtain a real value, denoted
\begin{equation}
    \intg{\bm U}{f}{\mu} \; \text{ or } \; \intg{x \in U}{f(x)}{\mu(x)}.
\end{equation}
A simple example of a measurable function is the \emph{indicator function} of a measurable set $E \in \Fc_U$:
\[ \bm 1{_E}(x) \defeq \begin{cases} 1 & \text{if $x \in E$} \\ 0 & \text{if $x \not\in E$.}\end{cases}\]
For any measure $\mu$ on $\bm U$, its integral yields the `size' of E:
\begin{equation}\label{eq:ch01_integralindicator}
\intg{\bm U}{\bm 1_{E}}{\mu} = \mu(E) \Mdot 
\end{equation}
A measure $\mu$ on a measurable space $\bm U$ is said to be \emph{finite} if $\mu(U)<\infty$ and it is a \emph{probability measure} if it is of unit mass \ie $\mu(U)=1$.
We will denote by $\Measures(\bm U)$ and $\PMeasures(\bm U)$, respectively, the sets of measures and probability measures on the measurable space $\bm U$. 
Note that neither $\Measures(\bm U)$ nor $\PMeasures(\bm U)$ form a vector space. 
One way to recover this structure is to consider \emph{finite-signed} measures. 
In comparison to the definition of a measure, one drops the nonnegativity requirement \ref{enum:nonnegativity} in Definition~\ref{def:ch01_measure}, but insists that the values be finite. We denote this real vector space $\FSMeasures(\bm U)$. 
Equipped with total variation\index{Total variation distance} distance\footnotemark\, it becomes a normed vector space. 
\footnotetext{For $\mu,\nu \in \FSMeasures(\bm U)$, the total variation distance\index{Total variation distance|textbf} is defined as $\norm{\mu - \nu} = \vert \mu - \nu \vert(U) = \sup_{A \in \Fc_U} \vert \mu(A) - \nu(A) \vert$. When the set is countable, the total variation distance is related to the $L^1$ norm by the identity $\norm{\mu - \nu} = \frac 12 \sum_{x \in U} \vert \mu(x) - \nu(x) \vert$.}

A measurable function $f$ between measurable spaces $\bm U$ and $ \bm V$ carries any measure $\mu$ on $\bm U$ to a measure $f_*\mu$ on $\bm V$. 
This is known as a \emph{push-forward} operation.\index{Push-forward|textbf}
This push-forward measure is given by $f_*\mu(E) = \mu(f^{-1}(E))$ for any set $E$ measurable in $\bm V$.
An important use of push-forward measures is that for any integrable function $g$ between measurable spaces $\bm V$ and $\tuple{\R,\Bc_\R}$, one can write the following change-of-variables formula
\begin{equation}\label{eq:ch01_changeofvariables}
\intg{\bm U}{g \circ f}{\mu} = \intg{\bm V}{g}{f_*\mu} \Mdot
\end{equation}
Importantly, the push-forward operation preserves the total measure, hence it takes $\PMeasures(\bm U)$ to $\PMeasures(\bm V)$.
A case that will be of particular interest to us is the push-forward of a measure $\mu$ on a product space $\bm U_1 \times \bm U_2$ along a projection  $\fdec{\pi_i}{ U_1 \times  U_2}{U_i}$. This yields the \emph{marginal measure} $\mu|_{\bm U_i}={\pi_i}_*\mu$, where for any $E \in \Fc_{U_1}$ measurable,  $\mu|_{\bm U_1}(E) = \mu(\pi_1^{-1}(E)) = \mu(E \times U_2)$.
In the opposite direction, given a measure $\mu_1$ on $\bm U_1$ and a measure $\mu_2$ on  $\bm U_2$, a \emph{product measure} $\mu_1 \times \mu_2$ is a measure on the product measurable space $\bm U_1 \times \bm U_2$ satisfying $(\mu_1 \times \mu_2)(E_1 \times E_2) = \mu_1(E_1)\mu_2(E_2)$ for all $E_1 \in \Fc_1$ and $E_2 \in \Fc_2$. For probability measures, there is a uniquely determined product measure.

The last ingredient we need from measure theory is called a \emph{Markov kernel}. 
\begin{definition}[Markov kernel]\index{Markov kernel|textbf}
A Markov kernel (or probability kernel) between measurable spaces $\bm U = \tuple{U,\Fc_U}$
and $\bm V = \tuple{V, \Fc_V}$ is a function
$\fdec{k}{U \times \Fc_V}{[0,1]}$ (the space $[0,1]$ is assumed to be equipped with its Borel $\sigma$-algebra)
such that:
\begin{enumerate}[label=(\roman*)]
    \item for all $E \in \Fc_V$, $\fdec{k(\dummy,E)}{U}{[0,1]}$ is a measurable function;
    \item for all $x \in U$, $\fdec{k(x,\dummy)}{\Fc_V}{[0,1]}$ is a probability measure.
\end{enumerate}
\end{definition}
\noindent Markov kernels generalise the discrete notion of stochastic matrices.

\section{Optimisation theory}
\label{sec01:convexopti}

Below we give a brief overview of the optimisation problems that arise in this dissertation. 
We begin by recalling some basic definitions \cite{boyd2004convex}.
Let $E$ be a real topological vector space.

\begin{definition}[Convex set]
A set $F \subset E$ is said to be convex if for any $x,y \in F$ and $\lambda \in \left[0,1\right]$, we have that $\lambda x + (1-\lambda) y \in F$. 
\end{definition}
\begin{definition}[Convex function]
Let $F$ be a convex subset of $E$. $\fdec{f}{F}{\R}$ is said to be convex if $f(\lambda x + (1-\lambda) y) \leq \lambda f(x) + (1-\lambda) f(y)$ for any $x,y \in F$ and $\lambda \in \left[0,1\right]$.
\end{definition}
\noindent If $-f$ is convex then we say that $f$ is a \textit{concave} function. 
\begin{definition}[Cone]\index{Cone|textbf}
A set $C \in E$ is a cone if for all $x \in C$ and $\lambda \in \R^+$, $\lambda x \in C$.
\end{definition}
\noindent If $C$ is a convex subset of $E$ and a cone, we say that $C$ is a convex cone. 
\begin{definition}[Proper cone]
A cone $C \subset E$ is a proper cone if:
\begin{itemize}
    \item $C$ is convex. 
    \item $C$ is closed (\ie the limit of vectors with respect to the topology on $E$ in $C$ belong to $C$). 
    \item $C$ is solid (\ie it has nonempty interior). 
    \item $C$ is pointed (\ie if $x \in C$ and $-x \in C$ then $x=0_E$).
\end{itemize}
\end{definition}
\noindent A proper cone $C \subset E$ induces a partial ordering on $E$: $x \geq_C y \iff x-y \in C$ for $x,y \in E$. Thus $x =_C y$ if $x-y \in C$ and $y-x \in C$. In the rest of the manuscript, we will only deal with proper cones that we will refer to as cones. 

A \emph{convex optimisation problem} is an optimisation problem in which the objective function and the feasible set are convex. A detailed introduction to the general problem as well as algorithmic techniques for solving it can be found in \eg \cite{ben2001lectures,boyd2004convex,hiriart2013convex}.
A convex optimisation problem can be generally expressed as:
\begin{equation}
    \Inf{\bm x \in F} f(\bm x).
    \label{eq:ch01_genprobconv}
\end{equation}
$F \subset E$ is called the \textit{feasible set}. It is a convex subset specified by constraints: for instance linear inequalities or semidefinite inequalities.
$\fdec{f}{E}{\R}$ is a convex real-valued function called the \textit{objective function}. Note that the maximisation problem of optimising a concave function $f$ can be expressed in this form with the convex function $-f$. Usually, $E$ can be embedded into $\R^n$ for some $n \in \N^*$. However, we will also consider other cases \eg the space of real valued continuous functions. Such problems will yield \emph{infinite-dimensional} convex optimisation problems. Convex programs have the remarkable property that a local optimum is always a global optimum. This fact is greatly exploited by solvers such as point interior methods. 

A \emph{feasible plan} or \emph{feasible solution} is an element $x\in E$ which belongs to $F$ \ie an element that satisfies all the constraints.
We say that a convex optimisation program (P) is \textit{consistent} if it possesses a feasible solution.
A \emph{strictly feasible solution} strictly satisfies all the constraints. 
If the infimum in \refeq{ch01_genprobconv} is attained, an \emph{optimal plan} or \emph{optimal solution} is a feasible plan that reaches this value. 
The \textit{value} of a program (P), denoted $\text{val(P)}$, is the value returned by the optimisation problem.
It is the value reached by an optimal solution if there exists an optimal plan.   
It can be infinite if the program diverges.  

In particular, we will be interested in a subfield of convex optimisation known as \textit{conic programming}: 
\begin{equation}
    \Inf{\bm Ax \, =_C \, b} f(\bm x) \, ,
    \label{eq:ch01_genprobconic}
\end{equation}
where $A$ is a linear operator acting on $E$, $b \in E$ and $C$ is a convex cone in $E$.\index{Cone}
Conic programming thus consists of the optimisation of a convex function over the intersection of an affine subspace $\{y \in E | y=Ax-b\}$ and a convex cone $C$. This can be extended to constraints of the form $Ax \geq_C b$. We will mainly focus on linear objective functions. 

Importantly conic optimisation reduces to \textit{linear programming} (LP) when the convex cone is a nonnegative orthant\index{Cone}; and to \textit{semidefinite programming} (SDP) when it is the convex cone of positive semidefinite matrices. Remarkably, these two classes come with a duality theory: from a \textit{primal program} expressed as a minimisation problem, one can derive a \textit{dual program} expressed as a maximisation problem. \textit{Weak duality} always holds meaning that the value of the minimisation problem is lowerbounded by the value of the maximisation problem. 
Furthermore when the values of the programs agree, we have \textit{strong duality} between the programs. Equivalently, we say there is \textit{no duality gap}. \index{Strong duality|textbf}
The latter will only hold under specific conditions that we will detail later. There exist robust and efficient classical algorithms for solving linear programs and semidefinite programs \cite{ben2001lectures,boyd2004convex}, in particular via interior point methods.

Finally we will also encounter more exotic optimisation problems where we will have to deal with quadratic terms.
Terms like $x^T Q x$ for some vector $x$ and some symmetric matrix $Q$ will only be convex if $Q$ is positive semidefinite. In that case, we will deal with quadratic SDPs. However we will often encounter optimisation problems where the objective function is no longer convex. The resulting program will not fit in the convex optimisation category. In particular we loose the very desirable property that a local optimum is automatically a global optimum resulting in a numerical resolution which is often much harder.

\subsection{Linear programming}
\label{subsec01:LP}

We present below the generic form of an infinite-dimensional linear program and main results regarding strong duality. Extensive introductions can be found in \cite{barvinok02,dantzig2006linear} and we will follow the form given in \cite[IV--(6.1)]{barvinok02} for such programs.

Let $E_1$, $F_1$ and $E_2$, $F_2$ be pairs of dual topological vector spaces equipped with the weak topology\footnote{The weak topology on $E_1$ induced by $F_1$ and the duality $\langle \dummy , \dummy \rangle_1$ is the weakest (or coarsest) topology on $E_1$ that makes all the maps $\fdec{\langle \dummy , f_1 \rangle_1}{E_1}{\R}$ continuous as $f_1$ ranges over $F_1$. And similarly for $E_2$, $F_1$ and $F_2$.} induced by the dualities $\langle \dummy , \dummy \rangle_{1,2}$. They are defined as $\fdec{\langle \dummy , \dummy \rangle_1}{E_1 \times F_1}{\R}$ and $\fdec{\langle \dummy , \dummy \rangle_2}{E_2 \times F_2}{\R}$. 

We fix convex cones $K_1 \subseteq E_1$ and $K_2 \subseteq E_2$. \index{Cone}
Convex dual cones $K_1^*$ and $K_2^*$ are defined as:
\begin{equation*}
    K_i^* = \{ f \in F_i \text{ such that } \forall e \in K_i, \; \dotp{e}{f}  \ge 0\} \Mdot
\end{equation*}
Furthermore, we fix the linear transformation $\fdec{A}{E_1}{E_2}$. We also fix its dual transformation $\fdec{A^*}{F_2}{F_1}$ such that $\forall e \in E_1, \forall f \in F_2$, $\dotp{A(e)}{f}_2 = \dotp{e}{A^*(f)}_1 $. Finally, we fix $c \in F_1$ and $b \in E_2$. A canonical example of a (potentially infinite-dimensional) primal\footnote{Note that the naming `primal' and `dual' is purely arbitrary and could be exchanged.} linear program can be expressed as:\index{Linear program|textbf}
\leqnomode
\begin{flalign*}
    \label{prog:LPstdform}
    \tag*{(LP)}
    \hspace{3cm} \left\{
    \begin{aligned}
            & \quad \text{Find } e_1 \in E_1 \\
            & \quad \text{minimising } \langle e_1,c \rangle_1 \\
            & \quad \text{subject to:}  \\
            & \hspace{1cm} \begin{aligned}
            & A(e_1) \geq_{K_2} b  \\
            & e_1 \geq_{K_1} 0 \Mdot
            \end{aligned} 
    \end{aligned}
    \right. &&
\end{flalign*}
Its dual problem can be expressed as:
\begin{flalign*}
    \label{prog:DLPstdform}
    \tag*{(D-LP)}
    \hspace{3cm}\left\{
    \begin{aligned}
        & \quad \text{Find } f_2 \in F_2 \\
        & \quad \text{maximising } \langle b,f_2 \rangle_2 \\
        & \quad \text{subject to:}  \\
        & \hspace{1cm} \begin{aligned}
        & A^*(f_2) \leq_{K_1^*} c \\
        & f_2 \geq_{K_2} 0 \Mdot
        \end{aligned}
    \end{aligned}
    \right. &&
\end{flalign*}
\reqnomode

\paragraph{Example.}\label{test} To fix ideas, let us consider an example that will be of particular importance in this manuscript. For some Borel measurable space $\bm U$ of a compact Hausdorff set $U$, $E_1$ and $E_2$ are the set of finite signed measures $\FSMeasures(\bm U)$ and $F_1$ and $F_2$ are the set of continuous real valued functions on $U$, $C(U)$. Note that $\FSMeasures(\bm U)$ is indeed a concrete realisation of the topological dual space of $C(U)$ via the  Riesz–Markov–Kakutani representation theorem \cite{kakutani}.\index{Riesz!Markov-Kakutani representation theorem} The duality is then naturally given by:
\begin{equation}
    \forall \mu \in \FSMeasures(\bm U), \forall f \in C(U), \; \dotp{\mu}{f} = \intg{U}{f}{\mu}.
\end{equation}
$K_1$ and $K_2$ are the convex cone\index{Cone} of positive finite measure. $K_1^*$ and $K_2^*$ are then the set of positive real-valued functions on $\R$.
We further fix $A$ and $A^*$ to be the identity map. $b \in E_2$ is a fixed finite measure and $c \in F_1$ a continuous function on $U$. 
Program \refprog{LPstdform} then seeks the infimum of the real number $\langle e_1,c \rangle_1$ over all finite-signed measures $e_1$ under the constraint that $e_1$ is a nonnegative measure and that it is lower bounded by the finite measure $b$, meaning that $e_1 - b$ is itself a nonnegative finite measure. Likewise, program \refprog{DLPstdform} seeks the supremum of the real number $\langle b,f_2 \rangle_2$ over all real-valued continuous functions $f_2$ under the constraints that $f_2$ is a nonnegative function and that it is upper bounded by the continuous function $c$, meaning that $c - f_2$ is itself a nonnegative continuous function. 

\paragraph{Weak duality.} It always holds that $\text{val\refprog{LPstdform}} \geq \text{val\refprog{DLPstdform}}$.

\paragraph{Strong duality.} \index{Strong duality!of linear programs}This only holds under specific conditions. It is often a very desirable property. For instance, the computation of the contextual fraction \cite{abramsky2017contextual} (see Section~\ref{subsec01:quantifying}) can be phrased as a linear program. Its dual program provides a method for calculating the maximally violated normalised Bell inequality which only holds because strong duality applies in this setting.

When the problem is \textit{finite-dimensional}, a sufficient condition for strong duality is that the program has a finite value and its feasible region has an interior point. The latter is known as Slater's condition. More formally, it requires the existence of a strictly feasible plan \ie a point that satisfies the constraints with strict inequality. For instance, for problem \refprog{LPstdform}, it amounts to the existence of a positive $e_1 \in E_1$ such that $ A(e_1) > b$.

When the problem is \textit{infinite-dimensional}, this becomes slightly more cumbersome. The following theorem, expressed with notation from \refprog{LPstdform}, provides a sufficient condition:
\begin{theorem}[Th.\ (7.2) \cite{barvinok02}]
Suppose that the set $\{ (A(x),\dotp{x}{c}_1) : x \in K_1 \}$ is closed in $E_2 \oplus \R$ and that there exists a primal feasible plan \ie that the primal program is consistent. Then there is no duality gap \ie {\upshape val\refprog{LPstdform} $=$ val\refprog{DLPstdform}}. If {\upshape val\refprog{LPstdform}} is finite then there exists a primal optimal plan. 
\label{th:ch01_strongduality}
\end{theorem}

\subsection{Semidefinite programming}
\label{subsec01:SDP}

Semidefinite programs are a generalisation of linear programs where one optimises in the cone\index{Cone} of semidefinite matrices rather than a positive orthant. Here we give the canonical form of finite-dimensional semidefinite programs \cite{vandenberghe1996semidefinite}.
Let $M,N\in\mathbb N$, $\bm b=(b_1,\dots,b_M)\in\R^M$, $C\in\SymMatrices{N}$ (the set of $N \times N$ symmetric matrices), and $B^{(i)}\in\SymMatrices{N}$ for all $i \in \llbracket 1,M \rrbracket$.\index{Semidefinite program|textbf}
\leqnomode
\begin{flalign*}
   \label{prog:SDPstdform}
    \tag*{(SDP)}
    \hspace{3cm}\left\{
        \begin{aligned}
            & \quad \text{Find } X\in\SymMatrices{N} \\
            & \quad \text{minimising } \Tr(C^TX) \\
            & \quad \text{subject to:}  \\
            & \hspace{1cm} \begin{aligned}
                &  \forall i \in 1,\dots,M, \quad\Tr(B^{(i)}X)=b_i \\
                &  X \succeq 0 \Mdot
            \end{aligned}
        \end{aligned}
    \right. &&
\end{flalign*}
\reqnomode
A matrix $X \in \Matrices{M}{\R}$ is positive semidefinite, denoted $X \succeq 0$, if and only if for all $\bm u \in \R^M$, $\bm u^T X \bm u \geq 0$.
Its dual semidefinite program reads:
\leqnomode
\begin{flalign*}
   \label{prog:DSDPstdform}
    \tag*{(D-SDP)}
    \hspace{3cm}\left\{
        \begin{aligned}
            & \quad \text{Find } \bm y\in\R^M \\
            & \quad \text{maximising } \bm b^T\bm y \\
            & \quad \text{subject to:}  \\
            & \hspace{1cm} \begin{aligned}
                & \sum_{i=1}^My_iB^{(i)} \preceq C \Mdot
            \end{aligned}
        \end{aligned}
    \right. &&
\end{flalign*}
\reqnomode
Compared to the very general programs given in the previous subsection, the programs given above seem more rigid because we only account for finite-dimensional semidefinite programs in this dissertation---a similar simpler form could be given for finite-dimensional LP.
Infinite-dimensional linear programs can either be \textit{relaxed} (for instance when some constraints are relaxed or even suppressed \ie when the optimisation is performed on a larger search space) or \textit{restricted} (for instance when constraints are added). 
It turns out that with a clever choice of restrictions and relaxations, infinite-dimensional linear programs may sometimes be approximated by converging upper and lower hierarchies of finite dimensional semidefinite programs (which can actually be ran on a computer). 
In the case of optimising over measures, it is sometimes possible to rephrase the problem in terms of the infinite sequence of moments of the measure by requiring that the so-called moment matrix is positive semidefinite. 
We then truncate the sequence of moment to obtain a finite dimensional semidefinite program \cite{lasserre10,lasserre2001global}. 
This is known as the \textit{Lasserre hierarchy} (or Lasserre relaxation) that we introduce shortly after. 
Dual to this vision, it is also sometimes possible to find sum-of-squares representation for continuous real-valued nonnegative functions. Then by truncating at a certain degree of the sum-of-squares representation, it can be rephrased as a semidefinite program \cite{parrilo2003semidefinite,lasserre10}. 

\paragraph{Weak duality.} It always holds that $\text{val\refprog{SDPstdform}} \geq \text{val\refprog{DSDPstdform}}$.

\paragraph{Strong duality.} If one of the programs has a finite value and a feasible region with an interior point (a point that strictly satisfies all the constraints) then we have strong duality \ie $\text{val\refprog{SDPstdform}} = \text{val\refprog{DSDPstdform}}$.\index{Strong duality!of semidefinite programs}

\paragraph{Numerical implementation.} There exist nowadays numerous software packages to solve SDPs. A great open-source resource is the library Picos \cite{sagnol2012picos} which allows one to write high-level SDPs in Python. It can be interfaced with robust SDP solvers such as Mosek \cite{mosek} or SDPA \cite{fujisawa2002sdpa}. Note that SDPA features a highly accurate multiple-precision arithmetic SDP solver called SDPA-GMP \cite{nakata2010numerical}. While much slower, this significantly improves accuracy if numerical issues are critical.

\subsection{The Lasserre--Parrilo hierarchy}
\label{subsec01:Lasserrehierarchy}

Below we introduce the Lasserre--Parrilo hierarchy for relaxing infinite-dimensional linear programs known as Generalised Moment Problems \cite{lasserre2001global,lasserre10,parrilo2003semidefinite}.\index{Lasserre hierarchy|textbf}

We start by giving insightful results: Subsection~\ref{subsubsec01:sos} provides results concerning the representation of positive polynomials while Subsection~\ref{subsubsec01:moment} provides results to understand when a sequence can be represented by a Borel measure.

\subsubsection*{Notation, terminology}
We work in $\R^d$ for $d \in \N^*$. We fix $\bm K$ to be a Borel measurable subspace of $\bm \R^d$.
We use the multi-index notation in Eq.~\eqref{eq:ch00_multiindex}, which we recall briefly. Let $\Rpolm$ denote the ring of real polynomials in the variables $\bm x \in \R^d$, and let $\Rpolkm \subset \Rpolm$ contain those polynomials of total degree at most $k$.
The latter forms a vector space of dimension $s(k) \defeq \binom{d+k}{k}$,
with a canonical basis consisting of monomials
$\bm x^{\bm \alpha} \defeq x_1^{\alpha_1}\cdots x_d^{\alpha_d}$\footnote{We also extend this notation to noncommutative variables in Subsection~\ref{subsec01:quadratic}.}
indexed by the set $\N^d_k \defeq \setdef{\bm \alpha \in \N^d}{\lvert \bm \alpha \rvert \leq k}$ where $\lvert\bm \alpha\rvert\defeq\sum_{i=1}^d\alpha_i$. 
For $k \in \N$, $\bm x \in \R^d$, we define $\bm \vrm_k(\bm x) \defeq (\bm x^{\bm \alpha})_{\vert \bm \alpha \vert \leq k} = (1,x_1,\dots,x_n,x_1^2,x_1x_2,\dots,x_n^k)^T $ the vector of monomials of total degree less or equal than $k$.

Any $ p \in \Rpolkm$
is associated with a vector of coefficients  $\bm p \defeq (p_{\bm \alpha}) \in \R^{s(k)}$ via expansion in the canonical basis 
as $p(\bm x) = \sum_{\bm \alpha \in \N^d_k} p_{\bm \alpha}\bm x^{\bm \alpha}$. 

\subsubsection*{Moment problem in probability}

Given a finite set of indices $\Gamma$, a set of reals $\enset{\gamma_j : j \in \Gamma}$ and polynomials $\fdec{h_j}{K}{\R}$, $j \in \Gamma$ (integrable with respect to every measure $\mu \in \FSMeasures(\bm K)$) the corresponding \textit{Global Moment Problem} (GMP) can be expressed as:

\leqnomode
\begin{flalign*}
   \label{prog:GMPstdform}
    \tag*{(GMP)}
    \hspace{3cm}\left\{
        \begin{aligned}
            & \quad \text{Find } \mu \in\FSMeasures(\bm K) \\
            & \quad \text{maximising } \mu(\bm K) \\
            & \quad \text{subject to:}  \\
            & \hspace{1cm} \begin{aligned}
                &  \forall j \in \Gamma,\quad \intg{K}{h_j}{\mu} \leq \gamma_j \Mdot
            \end{aligned}
        \end{aligned}
    \right. &&
\end{flalign*}
\reqnomode
It dual program can be expressed as:
\leqnomode
\begin{flalign*}
   \label{prog:D-GMPstdform}
    \tag*{(D-GMP)}
    \hspace{3cm}\left\{
        \begin{aligned}
            & \quad \text{Find } \bm \lambda \in \R^\Gamma \\
            & \quad \text{minimising } \sum_{j \in \Gamma} \gamma_j \lambda_j \\
            & \quad \text{subject to:}  \\
            & \hspace{1cm} \begin{aligned}
                &  \forall \bm x \in K,\quad \sum_{j \in \Gamma} \lambda_j h_j(\bm x) - \bm x \geq 0 \\
                & \forall j \in \Gamma, \quad \lambda_j \geq 0 \Mdot
            \end{aligned}
        \end{aligned}
    \right. &&
\end{flalign*}
\reqnomode

\subsubsection{Positive polynomials and sum-of-squares}
\label{subsubsec01:sos}

Here we present the link between positive polynomials and sum-of-squares representation so that we can derive a converging hierarchy of restriction problems for program~\refprog{D-GMPstdform}.

\begin{definition}[sum-of-squares polynomial]\index{Sum-of-squares polynomial|textbf}
A polynomial $p \in \Rpolm$ is a sum-of-squares (SOS) polynomial if there exists a finite family of polynomials $\family{q_i}_{i \in I}$ such that
$p = \sum_{i\in I} q_i^2$.
\end{definition}
\noindent SOS polynomials are widely used in convex optimisation.
We will denote by $\SOSm \subset \Rpolm$ the set of (multivariate) SOS polynomials, and $\SOSkm \subset \SOSm$ the set of SOS polynomials of degree at most $2k$. The following proposition hints towards the reason why it is desirable to be able to look for a sum-of-squares decomposition: it can be cast as a semidefinite optimisation problem.
\begin{proposition}[Prop. 2.1, \cite{lasserre10}]
\label{prop:ch01_sosdecomposition}
A polynomial $p \in \Rpolm_{2k}$ has a sum-of-squares decomposition if and only if there exists a real symmetric positive semidefinite matrix $Q \in \SymMatrices{s(k)}$ such that $\forall \bm x \in \R^d$, $p(\bm x) = \bm \vrm_k(\bm x)^T Q \bm \vrm_k(\bm x)$.
\end{proposition}
Then we will be looking at conditions under which a nonnegative polynomial can be expressed as a sum-of-squares polynomial. This is in essence the question raised by Hilbert in his 17$^\text{th}$ conjecture \cite{hilbert1888darstellung}. 
\begin{definition}[quadratic module]
For a family $q = (q_j)_{j\in\enset{1, \ldots, m}}$ of polynomials, the set:
\begin{equation}
    Q(q) \defeq  \setdef{ \sum_{j=0}^m \sigma_j q_j}{(\sigma_j)_{j\in\enset{0,\ldots, m}} \subset \SOSm}
\end{equation}
is a convex cone in $\Rpolm$ called the quadratic module generated by the family $q$ with, for convenience, $q_0 = 1$ added. For $k \in \N$, we define $Q_k(q)$ to be the quadratic module $Q(q)$ where we further impose that $(\sigma_j)_{j\in\enset{0,\ldots, m}} \subset \SOSkm$ \ie we limit the degree of SOS polynomials.
\end{definition}

\begin{assumption} \label{ass:ch01_algebraic}
{\upshape Let $K \subset \R^d$. We make the following three assumptions on $K$.
\begin{enumerate}[label=(\roman*)]
\item \label{ass:ch01_semialgebraic}  Suppose $K$ is a basic semi-algebraic set \ie there exists a family of polynomials $g = (g_j)_{j\in\enset{1, \ldots, m}} \in \Rpolm^m$ of degrees deg($g_j$) respectively such that:
\begin{equation}
    K \defeq \setdef{\bm x \in \R^d}{\forall j = 1,\dots,m, \; g_j(\bm x) \geq 0}.
\end{equation}
\item \label{ass:ch01_compact} Further suppose that $K$ is compact. 
\item \label{ass:ch01_archimidean} Finally suppose that there exists $u \in Q(q)$ such that the level set $\setdef{\bm x \in \R^d}{u(\bm x) \geq 0}$ is compact.
\end{enumerate}
}
\end{assumption}
\noindent In the family of polynomials $(g_j)_{j\in\enset{1, \ldots, m}}$ we add $g_0 = 1$ for convenience. 

The following theorem is the key result that we will exploit for deriving the hierarchy of SDP restrictions for the dual program~\refprog{GMPstdform}. 
\begin{theorem}[Putinar's Positivellensatz \cite{Putinar93}]\index{Putinar's Positivellensatz|textbf}
Let $K \subset \R^d$ satisfy Assumptions~\ref{ass:ch01_algebraic}. If $p \in \Rpolm$ is strictly positive on $K$ then $p \in Q(g)$, that is
\begin{equation}
    p = \sum_{j=0}^m \sigma_j g_j  
\end{equation}
for some sum-of-squares polynomials $\sigma_j \in \SOSm$ for $j=0,1,\dots,m$.
\label{th:ch01_putinar}
\end{theorem}
\noindent A proof can also be found in \cite{Laurent2009}.

Using the result above and Assumption~\ref{ass:ch01_algebraic}, one can derive a hierarchy of SDPs \cite{lasserre10} which provide a converging sequence of optimal values towards the value of program~\refprog{D-GMPstdform}:

\leqnomode
\begin{flalign*}
   \label{prog:DSDP-GMP}
    \tag*{(D-GMP$^k$)}
    \hspace{3cm}\left\{
        \begin{aligned}
            & \quad \text{Find } \bm \lambda = (\lambda_j)_{j \in \Gamma} \in \R^\Gamma \text{ and } \forall j=0,\dots,m, \, f_j \in \SOSm_{k- \lceil \frac{\text{deg}(g_j) }{2}\rceil} \hspace{-3cm}\\
            & \quad \text{minimising } y_0 \\
            & \quad \text{subject to:}  \\
            & \hspace{1cm} \begin{aligned}
                &  \sum_{j \in \Gamma} \lambda_j h_j - \bm 1_{\bm K} = \sum_{j=0}^m f_j g_j  \\
                &  \forall j\in \Gamma, \quad \lambda_j \geq 0 \Mdot
            \end{aligned}
        \end{aligned}
    \right. &&
\end{flalign*}
\reqnomode

\subsubsection{Moment sequences and moment matrices}
\label{subsubsec01:moment}

In this subsection, we want to understand why the program \refprog{LP-CFCV} can be relaxed so that a converging hierarchy of SDPs can be derived. The program \refprog{LP-CFCV} is essentially a maximisation problem on finite-signed Borel measures with additional constraints such as the fact that these are proper measures (\ie they are nonnegative). We will represent a measure by its moment sequence and find conditions for which this moment sequence has a (unique) representing Borel measure. 

\begin{definition}[Riesz functional $L_{\bm y}$]\index{Riesz!functional|textbf}
Given a sequence $\bm y = (y_{\bm \alpha})_{\bm \alpha \in \N^d}\in \R^{\N^d}$,
we define the linear functional $\fdec{L_{\bm y}}{\Rpolm}{\R}$ by 
\begin{equation}
L_{\bm y}(p) \defeq \sum_{\bm \alpha \in \N^d}p_{\bm \alpha} y_{\bm \alpha}.
\end{equation}
\label{def:ch01_RieszFunctional}
\end{definition}

\begin{definition}[Moment sequence]\index{Moment sequence|textbf}
Given a measure $\mu \in \Measures(\bm K)$, its moment sequence $\bm y = (y_{\bm \alpha})_{\bm \alpha \in \N^d} \in \R^{\N^d}$ is given by
\begin{equation}
    y_{\bm \alpha} \defeq \intg{\bm K}{\bm x^{\bm \alpha}}{\mu(\bm x)} \Mdot
    \label{eq:ch0_momentseq}
\end{equation}
We say that $\bm y$ has a unique representing measure $\mu$ when there exists a unique $\mu$ such that Eq.~(\ref{eq:ch0_momentseq}) holds. If $\mu$ is unique then we say it is determinate (\ie determined by its moments).
\end{definition}

\noindent The linear functional $L_{\bm y}$ then gives integration of polynomials with respect to $\mu$ \ie for any $p \in\Rpolm$:
\begin{equation}
\begin{aligned}
L_{\bm y}(p) &=
\sum_{\bm \alpha \in \N^d}p_{\bm \alpha} y_{\bm \alpha}
=
\sum_{\bm \alpha \in \N^d}p_{\bm \alpha} \intg{\bm K}{\bm x^{\bm \alpha}}{\mu(\bm x)}
=
\intg{\bm K}{\sum_{\bm \alpha \in \N^d}p_{\bm \alpha}\bm x^{\bm \alpha}}{\mu(\bm x)} \\
&=
\intg{\bm K}{p(\bm x)}{\mu(\bm x)} \\
&=
\intg{\bm K}{p}{\mu},
\end{aligned}
\end{equation}
where we reversed summation and integration because the sum is finite since $p$ is a polynomial.

The following theorem is often used in optimisation theory over measures as it provides a necessary and sufficient condition for a sequence to have a representing measure. 
\begin{theorem}[Riesz-Haviland \cite{haviland1936momentum}]\index{Riesz!Haviland theorem|textbf}
Let $\bm y = (y_{\bm \alpha})_{\bm \alpha \in \N^d} \in \R^{\N^d}$ and suppose that $K \subseteq \R^d$ is closed. Then $\bm y$ has a representing (nonnegative) measure \ie there exists $\mu$ a measure on $K$ such that:
\begin{equation*}
    \forall \bm \alpha \in \N^d, \; \intg{K}{\bm x^{\bm \alpha}}{\mu} = y_{\bm \alpha}
\end{equation*}
if and only if $L_{\bm y}(p) \geq 0$ for all polynomials $p \in \Rpolm$ nonnegative on $K$.
\label{th:ch01_RieszHaviland}
\end{theorem}

We recall that for $k \in \N$, $s(k) = \binom {d+k}k$.
\begin{definition}[Moment matrix]\index{Moment matrix|textbf}
For each $k \in \N$, the moment matrix of order $k$ $M_k(\bm y) \in \SymMatrices{s(k)}$ of a truncated sequence $(y_{\bm \alpha})_{\bm \alpha \in \N^d_{2k}}$
is the $s(k) \times s(k)$ symmetric matrix with rows and columns indexed by $\N^d_k$ (\ie by the canonical basis for $\Rpolkm$) defined as follows: for any $\bm \alpha, \bm \beta \in \N^d_k$,
\begin{equation}
    \left(M_k(\bm y)\right)_{\bm \alpha,\bm \beta} \defeq L_{\bm y}(\bm x^{\bm \alpha + \bm \beta}) = y_{\bm \alpha + \bm \beta} \Mdot
\end{equation}
\end{definition}

\begin{definition}[Localising matrix]\index{Localising matrix|textbf}
Given a polynomial $p \in \Rpolm$, the localising matrix $M_k(p \bm y) \in \Matrices{s(k)}{\R}$ of a moment sequence $(y_{\bm \alpha})_{\bm \alpha \in \N^d} \in \R^{\N^d}$
is defined by: for all $\bm \alpha, \bm \beta \in \N^d_{k}$,
\begin{equation}
     \left(M_k(p \bm y)\right)_{\bm \alpha, \bm \beta} \defeq  L_{\bm y}(p(x) x^{\bm \alpha + \bm \beta}) = \sum_{\gamma \in \N^d} p_{\gamma} y_{\bm \alpha + \bm \beta + \gamma} \Mdot
\end{equation}
\end{definition}

\noindent The localising matrix reduces to the moment matrix for $p=1$. For well-defined moment sequences, \ie sequences that have a representing finite Borel measure, moment matrices and localising matrices are positive semidefinite, which provides insight on the reason why problem \refprog{LP-CFCV} can be relaxed to a problem with positive semidefiniteness constraints. 

\begin{proposition}
Let $\bm y = (y_{\bm \alpha})_{\bm \alpha \in \N^d} \in \R^{\N^d}$ be a sequence of moments for some finite Borel measure $\mu$ on $\bm K$. Then for all $k\in \N$, $M_k(\bm y) \succeq 0 $. If $\mu$ has support contained in the set $\setdef{\bm x \in K}{g(\bm x) \geq 0}$ for some polynomial $g \in \Rpolm$ then, for all $k \in \N$, $M_k(g \bm y) \succeq 0$.
\end{proposition}

\begin{proof}
Let $\bm y = (y_{\bm \alpha})$ be the moment sequence of a given Borel measure $\mu$ on $\bm K$. Fix $k \in \N$. For any vector $\bm v \in \R^{s(k)}$ (noting that $\bm v$ is canonically associated with a polynomial $v \in \Rpolkm$ in the basis $(\bm x^{\bm \alpha})$):

\begin{align}
	\bm v^T M_k(\bm y) \bm v &= \sum_{\bm \alpha,\bm \beta \in \N^d_k}  v_{\bm \alpha} y_{\bm \alpha + \bm \beta} v_{\bm \beta}  \\
	& = \sum_{\bm \alpha, \bm\beta \in \N^d_k} v_{\bm\alpha} v_{\bm\beta} \intg{K}{\bm x^{\bm\alpha + \bm\beta}}{\mu}  \\
	& = \intg{K}{\left( \sum_{\bm\alpha \in \N^d_k} v_{\bm\alpha} \bm x^{\bm \alpha} \right)^2}{\mu} \\
	& = \intg{K}{v^2(\bm x)}{\mu} \geq 0 \Mdot \label{eq:ch01_momentmatrixpos}
\end{align} 
Thus $M_{k}(\bm y) \succeq 0 $. 

Similarly we can prove that the localising matrix $M_k(g \bm y)$ is positive semidefinite when $g$ is a nonnegative polynomial on the support of $\mu$. Indeed for all $\bm v \in \R^{s(k)}$:
\begin{equation}
	\bm v^T M_k(g \bm y) \bm v = \intg{K}{v^2(\bm x) g(\bm x)}{\mu} \geq 0 \Mcomma
\end{equation}
which concludes the proof.
\end{proof}

The following theorem, which is the dual facet of Theorem~\ref{th:ch01_putinar}, is the key result for deriving the hierarchy of SDP relaxations for the primal problem \refprog{LP-CFCV}. It provides a necessary and sufficient condition for a sequence to have a representing measure.
\begin{theorem}[Th. 3.8 \cite{lasserre10}]
Let $\bm y = (y_{\bm \alpha})_{\bm \alpha \in \N^d} \in \R^{\N^d}$. Let $K \subset \R^d$ satisfy Assumptions~\ref{ass:ch01_algebraic}. Then $\bm y$ has a finite Borel representing measure with support contained in $K$ if and only if:
\begin{align}
    & M_k(\bm y) \succeq 0, \; \forall k \in \N, \\
    & M_k(g_j \bm y) \succeq 0, \; \forall j=1,\dots,m , \; \forall k \in \N.
\end{align}
\label{th:ch01_representingmeasure}
\end{theorem}

Using the above result and Assumption~\ref{ass:ch01_algebraic}, one can derive a hierarchy of SDPs \cite{lasserre10} which provide a converging sequence of optimal values towards the value of program~\refprog{GMPstdform}:
\leqnomode
\begin{flalign*}
   \label{prog:SDP-GMP}
    \tag*{(GMP$^k$)}
    \hspace{3cm}\left\{
        \begin{aligned}
            & \quad \text{Find } \bm y = (y_{\bm \alpha})_{\bm \alpha \in \N^d_{2k}} \in \R^{s(2k)} \\
            & \quad \text{maximising } y_0 \\
            & \quad \text{subject to:}  \\
            & \hspace{1cm} \begin{aligned}
                &  \forall j \in \Gamma,\quad L_{\bm y}(h_j) \leq \gamma_j \\
                &  M_k(\bm y) \succeq 0 \\
                &  \forall i \in 1,\dots,m, \quad M_{k - \lceil \frac{\text{deg}(g_i)}{2} \rceil }(g_i \bm y) \succeq 0 \Mdot
            \end{aligned}
        \end{aligned}
    \right. &&
\end{flalign*}
\reqnomode
We refer readers to \cite{lasserre10} for the proof of convergence of the hierarchies given by programs \refprog{SDP-GMP} and \refprog{DSDP-GMP}.

\subsection{Quadratic optimisation problems}
\label{subsec01:quadratic}

Quadratic problems play an important role in Quantum Information since jointly optimising an expectation value over quantum states $\rho$'s and PVMs $\{ E_i \}_i$ amounts to optimising quadratic terms like $\Tr(\rho E_i)$ under the constraints that these operators are positive semidefinite and that the PVMs sum to the identity. However these terms are not convex so the resulting optimisation problem is challenging to solve (it is NP-hard in general). \index{PVM}

Suppose we have a maximisation problem for a nonconvex bilinear quadratic problem. One obvious strategy to obtain a lower bound is to decompose the optimisation into two SDP sub problems: one when the PVMs are fixed; one when the quantum state is fixed. The solution of one sub program can then be used to fix the quantities in the other program.
This is known as a see-saw iteration algorithm and they usually provide a tight lower bound though convergence is never guaranteed. 

To obtain an upper bound on the value of the nonconvex problem, one can implement the Navascu\'es--Pironio--Ac\'in (NPA) hierarchy \cite{navascues2008convergent} and in particular the finite-dimensional variation known as the Navascu\'es-V\'ertesi (NV) hierarchy \cite{navascues2015characterizing} which will optimise from outside the set of quantum correlations to provide an upper bound on the sought value. Below we briefly introduce noncommutative polynomial optmisation (NPO) and give the NV hierarchy associated. We focus on this variant as the dimension of the Hilbert space is bounded and this will be required for our problems.

Consider the set $S$ of all $d$-tuples of Hermitian operators $\bm X = (X_1,\dots,X_d)$ satisfying $\Rc = \{q_i(\bm X) \succeq 0:i=1,\dots,m\}$
where $(q_i)_{i=1,\dots,m}$ are Hermitian polynomials (\ie $q_i(\bm X)$ is Hermitian for $\bm X$ a $d$-tuple of Hermitian operators).
Given $p$ a  Hermitian polynomial and $D \in \N^*$, a generic NPO problem can be expressed as:\index{NV hierarchy|textbf}
\leqnomode
\begin{flalign*}
   \label{prog:NPO}
    \tag*{(NPO)}
    \hspace{3cm}\left\{
        \begin{aligned}
            & \quad \text{Find a Hilbert space } \mathscr{H}, \psi \in \Dc(\mathscr H)  \\
            & \quad \text{and } \bm X \text{ a $d$-tuple of Hermitian operators on } \mathscr H \\
            & \quad \text{maximising } \braket{\psi | p(X) | \psi} \\
            & \quad \text{subject to:}  \\
            & \hspace{1cm} \begin{aligned}
                &  \text{dim}(\mathscr H) \leq D \\
                &  \forall i = 1,\dots,m, \quad q_i(X) \succeq 0 \Mdot
            \end{aligned}
        \end{aligned}
    \right. &&
\end{flalign*}
\reqnomode
A sequence $\bm y = (y_{\bm \alpha}) \in \R^{s(2k)}$ admits a valid quantum representation if there exists $\bm X \subset \Bc(\mathscr H)$ satisfying $\Rc$ with $\text{dim}(\mathscr H) \leq D$ and a pure state $\ket \psi \in \mathscr H$ such that $y_{\bm \alpha} = \braket{\psi | \bm X^{\bm \alpha} |\psi}$. The moment vectors $\bm y$ that admit a quantum representation with a dimensional constraint satisfy a number of additional linear restrictions which depend on $D$. For instance for $D=1$, there are additional constraints that ensure the variables commute. Denote $S^k_D$ the span of the set of feasible sequences $\bm y = (y_{\bm \alpha}) \in \R^{s(2k)}$. The NV hierarchy is generally expressed as:
\leqnomode
\begin{flalign*}
   \label{prog:SDP-NPO}
    \tag*{(SDP-NPO$^k$)}
    \hspace{3cm}\left\{
        \begin{aligned}
            & \quad \text{Find } \bm y = (y_{\bm \alpha})_{\bm \alpha \in \N^d_{2k}} \in \R^{s(k)} \\
            & \quad \text{maximising } L_{\bm y}(p) \\
            & \quad \text{subject to:}  \\
            & \hspace{1cm} \begin{aligned}
                &  \bm y \in S^k_D \\
                &  y_{\bm 0} = 1 \\
                & M_k(\bm y) \succeq 0 \\
                & \forall i=1,\dots,m,\quad M_k(q_i \bm y) \succeq 0 \Mdot
            \end{aligned}
        \end{aligned}
    \right. &&
\end{flalign*}
\reqnomode
The key to implement this program is to characterise the subspace $S^k_D$ and \cite{navascues2015characterizing} provide methods to do so. It can be achieved by a randomised method where $\bm X$ and $\ket \psi$ are randomly generated to build a basis for $S^k_D$. Another method is deterministic and involve symbolic computations to retrieve a basis for $S^k_D$.

\section{Sheaf-theoretic framework for contextuality}
\label{sec01:sheaf}

In this section, we aim at introducing the traditional notion of contextuality as first considered by Bell \cite{belleinstein1964} and Kochen and Specker \cite{KochenSpecker1967,kochen1975problem}.
We will define it following the later, more general approach developed in the seminal paper of Abramsky and Brandenburger often referred to as the sheaf-theoretic approach to measurement contextuality \cite{abramsky2011sheaf}---though in the present exposition, we will not dive deeply into the sheaf structure itself. The main ingredients are \textit{empirical models} which can be thought of as providing formal descriptions of tables of data specifying the probabilities of joint outcomes for compatible measurements. These empirical models need an underlying abstract description of an experiment which is given by \textit{a measurement scenario}. A crucial result is the Fine--Abramsky--Brandenburger (FAB) theorem which pinpoints a unified structure describing nonlocality and contextuality. Moreover, contextuality can quantified using the contextual fraction \cite{abramsky2017contextual} which can be computed by a linear program. 

\subsection{Measurement scenarios and empirical models}
\label{subsec01:empiricalmodels}

\subsubsection*{Measurement scenarios}
An abstract description of a particular experimental setup is formalised as \textit{a measurement scenario}. 

\begin{definition}[Measurement scenario]\index{Measurement scenario!DV|textbf}
A measurement scenario is a triple $\tuple{\Xc,\Mc,\Oc}$ where:
\begin{itemize}
    \item $\Xc$ is a finite set of measurement labels.
    \item $\Mc$ is a covering family of $\Xc$ \ie it is a set of subsets of $\Xc$ such that $\bigcup_{C \in \Mc} C = \Xc$. We require that each element $C \in \Mc$ is a maximal measurement context in the sense that $\forall C,C' \in \Mc$ such that $C \subseteq C'$ then $C=C'$. It represents a maximal set of compatible measurements. No element of this family is a proper subset of another. 
    \item $\Oc = (\Oc_x)_{x \in \Xc}$ is a finite set of outcomes for each measurement. For some set of measurements $U \subseteq \Xc$, the joint outcome set is given by the product of the respective outcome spaces: $\Oc_U = \prod_{x \in U} \Oc_x$. 
\end{itemize}
\end{definition}

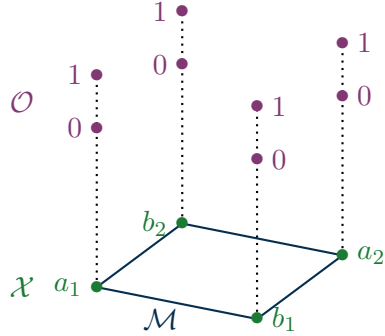
\begin{figure}
    \centering
    \begin{tikzpicture}[x=40pt,y=40pt,thick,label distance=-0.25em,baseline=(current bounding box.center)]

\node (e) at (1.5,-0.3) {};
\node (n) at (0.8,0.6) {};
\node (T) at (0,1.5) {};
\node (u) at (0,0.5) {};
\node (u2) at (-.7,0.25) {};
\node (l) at (-0.7,0) {};
\node (dr) at (0.60,-0.3) {};
\node [inner sep=0em] (a) at (0,0) {};
\node [inner sep=0em] (b) at ($ (a) + (e) $) {};
\node [inner sep=0em] (a') at ($ (a) + (e) + (n) $) {};
\node [inner sep=0em] (b') at ($ (a) + (n) $) {};
\node [inner sep=0em] (a0) at ($ (a) + (T) $) {};
\node [inner sep=0em] (a1) at ($ (a0) + (u) $) {};
\node [inner sep=0em] (b0) at ($ (b) + (T) $) {};
\node [inner sep=0em] (b1) at ($ (b0) + (u) $) {};
\node [inner sep=0em] (a'0) at ($ (a') + (T) $) {};
\node [inner sep=0em] (a'1) at ($ (a'0) + (u) $) {};
\node [inner sep=0em] (b'0) at ($ (b') + (T) $) {};
\node [inner sep=0em] (b'1) at ($ (b'0) + (u) $) {};

\draw [c2] (a) -- (b);
\draw [c2] (a) -- (b');
\draw [c2] (a') -- (b);
\draw [c2] (a') -- (b');

\draw [dotted] (a1) -- (a);
\draw [dotted] (b1) -- (b);
\draw [dotted] (a'1) -- (a');
\draw [dotted] (b'1) -- (b');

\node [inner sep=0.1em,label=left:{\textcolor{c4}{$a_1$}},c4] at (a) {$\bullet$};
\node [inner sep=0.1em,label=right:{\textcolor{c4}{$b_1$}},c4] at (b) {$\bullet$};
\node [inner sep=0.1em,label=right:{\textcolor{c4}{$a_2$}},c4] at (a') {$\bullet$};
\node [inner sep=0.1em,label=left:{\textcolor{c4}{$b_2$}},c4] at (b') {$\bullet$};





\node [inner sep=0.1em,label=left:{\textcolor{c3}{$0$}},c3] at (a0) {$\bullet$};
\node [inner sep=0.1em,label=left:{\textcolor{c3}{$1$}},c3] at (a1) {$\bullet$};
\node [inner sep=0.1em,label=right:{\textcolor{c3}{$0$}},c3] at (b0) {$\bullet$};
\node [inner sep=0.1em,label=right:{\textcolor{c3}{$1$}},c3] at (b1) {$\bullet$};
\node [inner sep=0.1em,label=right:{\textcolor{c3}{$0$}},c3] at (a'0) {$\bullet$};
\node [inner sep=0.1em,label=right:{\textcolor{c3}{$1$}},c3] at (a'1) {$\bullet$};
\node [inner sep=0.1em,label=left:{\textcolor{c3}{$0$}},c3] at (b'0) {$\bullet$};
\node [inner sep=0.1em,label=left:{\textcolor{c3}{$1$}},c3] at (b'1) {$\bullet$};

\node at ($ (a0) + (u2) $) {\textcolor{c3}{$\mathcal{O}$}};
\node at ($ (a) + (l) $) {\textcolor{c4}{$\mathcal{X}$}};
\node at ($ (a) + (dr) $) {\textcolor{c2}{$\mathcal{M}$}};

\end{tikzpicture}
	\caption{The example of the $(2,2,2)$ Bell measurement scenario represented in the bundle diagram format.}
	\label{fig:ch01_bundleexample}
\end{figure}

A very convenient way to picture contextuality is via bundle diagrams as introduced in \cite{abramsky2015contextuality}. The \textcolor{c4}{elements of $\mathcal{X}$} are represented by the vertices of the base of the bundle diagram (see green vertices in figure \ref{fig:ch01_bundleexample}). The \textcolor{c2}{elements of $\mathcal{M}$} are represented by edges (or hyperedges in general) between those vertices (see blue lines). The outcomes (i.e. the \textcolor{c3}{elements of $\mathcal{O}$}) are displayed in fibres above each measurement (see purple vertices). The connection to bundles is very carefully set up in \cite{abramsky2015contextuality}.

\paragraph*{Example.} Let us express the standard (2,2,2) Bell scenario (2 parties, 2 observables each, 2 possible outcomes). It is represented as:
\begin{equation*}
    \mathcal{X}=\{a_1,a_2,b_1,b_2\}, \quad \mathcal{M}=\{\{a_1,b_1\},\{b_1,a_2\},\{a_2,b_2\},\{b_2,a_1\}\}, \quad \forall x \in \Xc, \Oc_x=\{0,1\}
\end{equation*}

\subsubsection*{The language of sheaves}
Different conventions have been used for defining the set of \textit{sections}. We will adopt the one from \cite{barbosa2021closing} and call a \textit{section} an element of $\Oc_U$ for $U \in \Pc(\Xc)$ a set of measurement labels. Sections from $\Oc_U$ are joint outcomes for all measurements in $U$. A \textit{global section} is a global assignment \ie it is an element of $\Oc_\Xc$.
The sheaf $\Ec$\footnotemark\ maps $U \in \Pc(\Xc)$ to $\Ec(U) = \Oc_U$.\index{Sheaf!event} It is called the \emph{event sheaf} as it assigns, to any 
set of measurements, information about the outcome events that could result from jointly performing them.
Note that as well as applying the map to valid contexts $U \in \Mc$ 
we will see that it can also be of interest to consider hypothetical outcome spaces
for sets of measurements that do not necessarily form valid contexts,
in particular $\Ec(X) = \Oc_X$, the joint outcome space for \textit{all} measurements.
\footnotetext{It is a functor $\Ec: \Pc(\Xc) \longrightarrow \text{Set}$}
Sheaves are widely used in modern mathematics. See \cite{maclane1992sheaves} for a comprehensive reference.
They might roughly be thought of as providing a means of assigning information to the open sets of a topological space. 
This is performed in such a way that information can be restricted to smaller open sets and consistent information on family of open sets can be uniquely `glued' on their union.
We are concerned with discrete topological spaces whose points represent measurements,
and the information that we are interested in assigning has to do with outcome spaces for these measurements and probability measures on these outcome spaces.
Sheaves can be defined concisely in category-theoretic terms as contravariant functors (presheaves) satisfying an additional gluing condition, though in what follows we will also give a more concrete description in terms of restriction maps. 

Sheaves come with a notion of \emph{restriction}.
Restriction arises in the following way: whenever $U, V \in \Pc(\Xc)$ with $U \subseteq V$ we have an obvious restriction map $\fdec{\rho^V_U}{\Ec(V)}{\Ec(U)}$ which simply projects from the product outcome space for $V$ to that for $U$.\index{Sheaf!restriction}
Note that $\rho^U_U$ is the identity map for any $U \in \Pc(\Xc)$ and that if $U \subseteq V \subseteq W$ in $\Pc(\Xc)$ then $\rho^V_U \circ \rho^W_V = \rho^W_U$.
Already this is enough to show that $\Ec$ is a \textit{presheaf}. For an inclusion $U \subseteq V$ and a section $o \in \Ec(V) = \Oc_V$, it is often more convenient to use the notation $ o|_U$ to denote $\rho^V_U (o) \in \Ec(U) = \Oc_U$, the restriction of $o$ to $\Oc_U$ which selects outcomes in $o$ that corresponds to measurement labels in $U$.

Additionally, the unique gluing property holds for $\Ec$.
Suppose that $\Nc \subseteq \Pc(X)$ and we have an $\Nc$-indexed family of sections $\family{ o_U \in \Oc_U}_{U \in \Nc}$ that is compatible in the sense that its elements agree on overlaps, \ie that for all $U, V \in \Nc$, $ o_U|_{U \cap V} =  o_V|_{U \cap V}$. Then these sections can always be `glued' together in a unique fashion to obtain a section $ o_{N}$ over $N \defeq \cup \Nc$ such that $ o_N|_U =  o_U$ for all $U \in \Nc$.
This makes $\Ec$ a \emph{sheaf}.
For a concise view on the reason why contextuality might be approached by topological arguments see \cite{Mansfield2020contextualityis}.

\subsubsection*{Empirical model.}
An empirical model is an object that allows one to specify the probabilities of observed joint outcomes on a given measurement scneario.  

\begin{definition}[Empirical model]\index{Empirical model!DV|textbf}
An empirical model $e$ (also called empirical behaviour) on the measurement scenario $\tuple{\Xc,\Mc,\Oc}$ is a family $e = (e_C)_{C \in \Mc}$, where $e_C$ is a probability distribution on the joint outcome space $\Oc_C$ for each maximal context $C$. These probability distributions further satisfy the compatibility conditions\index{Compatibility condition}:  
\begin{equation*}
    \forall C, C' \in \Mc, \quad e_C|_{C \cap C'} = e_{C'}|_{C \cap C'}.
\end{equation*}
\end{definition}
\noindent The notation $e_C|_U$ for $C \in \Mc$ and $U \subset C$ is shorthand for the standard marginalisation of a probability distribution: for $t \in \Oc_U$, $e_C|_U(t) := \sum_{s \in \Oc_C, s|_U=t} e_C(s)$. 
The compatibility condition, which requires that the marginals of these distributions agree on overlapping contexts, is sometimes referred to as the \textit{no-disturbance} condition or as the \textit{generalised no-signalling} condition. A special case is no-signalling which occurs when measurements are performed at space-like separated regions. Compatibility then ensures that the choice of which measurement to perform at one location does not affect the empirical prediction at another location. 
The no-disturbance principle is satisfied by all empirical models that arise from quantum predictions. Note that due to the structure inherited from probability distributions, empirical models are closed under convex combinations.

\paragraph*{Example.} We express an empirical model for the Clauser--Horne--Shimony--Holt (CHSH) experiment \cite{CHSH1969} on the (2,2,2) Bell scenario as the table \ref{tab:ch01_CHSH}.\index{CHSH model}
Note that 
\begin{equation}
    e^{\text{CHSH}} = \frac \eta 4 e^\Id + (1-\eta) e^{\text{PR box}} 
\end{equation}
where $\eta = 1 - \frac {\sqrt 2} 2$. $e^{\text{CHSH}}$ is an empirical model corresponding to the CHSH experiment, $e^\Id$ the one corresponding to a maximally mixed state and $e^{\text{PR box}}$ an empirical model for a PR box.\index{PR box} 
\begin{table}[!ht]
	\centering
	\begin{tabular}{cc||cccc}
		 A & B & $00$ & $01$ & $10$ & $11$ \\ \hline
		 $a_1$ & $b_1$ & $\eta_1$ & $\eta_2$ & $\eta_2$ & $\eta_1$ \\
		 $a_1$ & $b_2$ & $\eta_1$ & $\eta_2$ & $\eta_2$& $\eta_1$ \\
		 $a_2$ & $b_1$ & $\eta_1$ & $\eta_2$ & $\eta_2$ & $\eta_1$ \\
		 $a_2$ & $b_2$ & $\eta_2$ & $\eta_1$ & $\eta_1$ & $\eta_2$
	\end{tabular}
	\caption{Empirical model on the $(2,2,2)$ Bell scenario specifying the probabilities of the joint outcomes for the CHSH model with $\eta_1 = \frac{2+\sqrt 2}{8}$ and $\eta_2 = \frac{2-\sqrt 2}{8}$. Here $a_i$ and $b_i$ for $i=1,2$ represent quantum observables for CHSH with eigenvalues relabelled as $0$ and $1$. Joint probabilities are obtained by the Born rule.\index{CHSH model}}
	\label{tab:ch01_CHSH}
\end{table}

For instance the first row indicates that for the context $C=\{a_1,b_1\}$, the empirical model assigns probability $\frac12$ to the local sections $(0,0)$ and $(1,1)$ in context $\{a_1,b_1\}$. 
To understand the compatibility condition, let us focus on the first and second rows of the table. 
Fix $C = \{a_1,b_1\}$ and $C' = \{a_1,b_2\}$. 
Then $U = C \cap C' = \{a_1\}$. 
Consider any $t \in \Oc_U$: for instance the measurement associated to $a_1$ gives the outcome 0. 
For the first row, the only local sections $s \in \Oc_C$ such that $s|_U = t$ are $(0,0)$ and $(0,1)$ which, marginalised on the second outcome corresponding to the measurement of $a_2$, gives the outcome 0 for $a_1$. Summing the probabilities associated to these sections: $e_C|_U(0) = e_C((0,0)) + e_C((0,1)) = \eta_1 + \eta_2 = \frac12$. With the same procedure for the second row: $e_{C'}|_U(0) = e_{C'}((0,0)) + e_{C'}((0,1)) = \eta_2 + \eta_1 = \frac12$ as required.

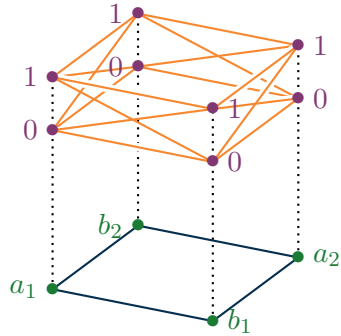
\begin{figure}[!ht]
    \centering
    \begin{tikzpicture}[x=40pt,y=40pt,thick,label distance=-0.25em,baseline=(current bounding box.center)]

\node (e) at (1.5,-0.3) {};
\node (n) at (0.8,0.6) {};
\node (T) at (0,1.5) {};
\node (u) at (0,0.5) {};
\node [inner sep=0em] (a) at (0,0) {};
\node [inner sep=0em] (b) at ($ (a) + (e) $) {};
\node [inner sep=0em] (a') at ($ (a) + (e) + (n) $) {};
\node [inner sep=0em] (b') at ($ (a) + (n) $) {};
\node [inner sep=0em] (a0) at ($ (a) + (T) $) {};
\node [inner sep=0em] (a1) at ($ (a0) + (u) $) {};
\node [inner sep=0em] (b0) at ($ (b) + (T) $) {};
\node [inner sep=0em] (b1) at ($ (b0) + (u) $) {};
\node [inner sep=0em] (a'0) at ($ (a') + (T) $) {};
\node [inner sep=0em] (a'1) at ($ (a'0) + (u) $) {};
\node [inner sep=0em] (b'0) at ($ (b') + (T) $) {};
\node [inner sep=0em] (b'1) at ($ (b'0) + (u) $) {};

\draw [c2] (a) -- (b);
\draw [c2] (a) -- (b');
\draw [c2] (a') -- (b);
\draw [c2] (a') -- (b');

\draw [dotted] (a1) -- (a);
\draw [dotted] (b1) -- (b);
\draw [dotted] (a'1) -- (a');
\draw [dotted] (b'1) -- (b');

\node [inner sep=0.1em,label=left:{\textcolor{c4}{$a_1$}},c4] at (a) {$\bullet$};
\node [inner sep=0.1em,label=right:{\textcolor{c4}{$b_1$}},c4] at (b) {$\bullet$};
\node [inner sep=0.1em,label=right:{\textcolor{c4}{$a_2$}},c4] at (a') {$\bullet$};
\node [inner sep=0.1em,label=left:{\textcolor{c4}{$b_2$}},c4] at (b') {$\bullet$};


\draw [line width=3.2pt,white] (a'0) -- (b'0);
\draw [line width=3.2pt,white] (a'0) -- (b'1);
\draw [line width=3.2pt,white] (a'1) -- (b'0);
\draw [line width=3.2pt,white] (a'1) -- (b'1);

\draw [c1] (a'0) -- (b'0);
\draw [c1] (a'0) -- (b'1);
\draw [c1] (a'1) -- (b'0);
\draw [c1] (a'1) -- (b'1);

\draw [line width=3.2pt,white] (a0) -- (b'0);
\draw [line width=3.2pt,white] (a0) -- (b'1);
\draw [line width=3.2pt,white] (a1) -- (b'0);
\draw [line width=3.2pt,white] (a1) -- (b'1);

\draw [c1] (a0) -- (b'0);
\draw [c1] (a0) -- (b'1);
\draw [c1] (a1) -- (b'0);
\draw [c1] (a1) -- (b'1);

\draw [line width=3.2pt,white] (a'0) -- (b0);
\draw [line width=3.2pt,white] (a'0) -- (b1);
\draw [line width=3.2pt,white] (a'1) -- (b0);
\draw [line width=3.2pt,white] (a'1) -- (b1);

\draw [c1] (a'0) -- (b0);
\draw [c1] (a'0) -- (b1);
\draw [c1] (a'1) -- (b0);
\draw [c1] (a'1) -- (b1);

\draw [line width=3.2pt,white] (a0) -- (b0);
\draw [line width=3.2pt,white] (a1) -- (b1);

\draw [c1] (a0) -- (b0);
\draw [c1] (a1) -- (b1);
\draw [c1] (a0) -- (b1);
\draw [c1] (a1) -- (b0);

\node [inner sep=0.1em,label=left:{\textcolor{c3}{$0$}},c3] at (a0) {$\bullet$};
\node [inner sep=0.1em,label=left:{\textcolor{c3}{$1$}},c3] at (a1) {$\bullet$};
\node [inner sep=0.1em,label=right:{\textcolor{c3}{$0$}},c3] at (b0) {$\bullet$};
\node [inner sep=0.1em,label=right:{\textcolor{c3}{$1$}},c3] at (b1) {$\bullet$};
\node [inner sep=0.1em,label=right:{\textcolor{c3}{$0$}},c3] at (a'0) {$\bullet$};
\node [inner sep=0.1em,label=right:{\textcolor{c3}{$1$}},c3] at (a'1) {$\bullet$};
\node [inner sep=0.1em,label=left:{\textcolor{c3}{$0$}},c3] at (b'0) {$\bullet$};
\node [inner sep=0.1em,label=left:{\textcolor{c3}{$1$}},c3] at (b'1) {$\bullet$};

\end{tikzpicture}
	\caption{Bundle diagram for the CHSH model presented in Table~\ref{tab:ch01_CHSH} on the $(2,2,2)$ Bell scenario.\index{CHSH model}}
	\label{fig:ch01_bundle_CHSH}
\end{figure}

We can now complete the bundle diagram \ref{fig:ch01_bundleexample} with the joint outcomes specified by the \textcolor{c1}{empirical model $e$} (see orange edges in figure \ref{fig:ch01_bundle_CHSH}). Note that bundle diagrams are often used to witness possibilistic contextuality (\ie logical or strong contextuality that we will define shortly) so that it is not always necessary to represent the probabilities associated to joint outcomes. Rather we only represent an edge between outcomes whenever the associated probability is non zero.\footnote{Note, however, that for the CHSH model it would be necessary to take account of the precise probabilities in order to witness contextuality.}

\subsubsection*{Extendability and Contextuality} 

\begin{figure}[ht!]
    \centering
    \includegraphics[width=.8\linewidth]{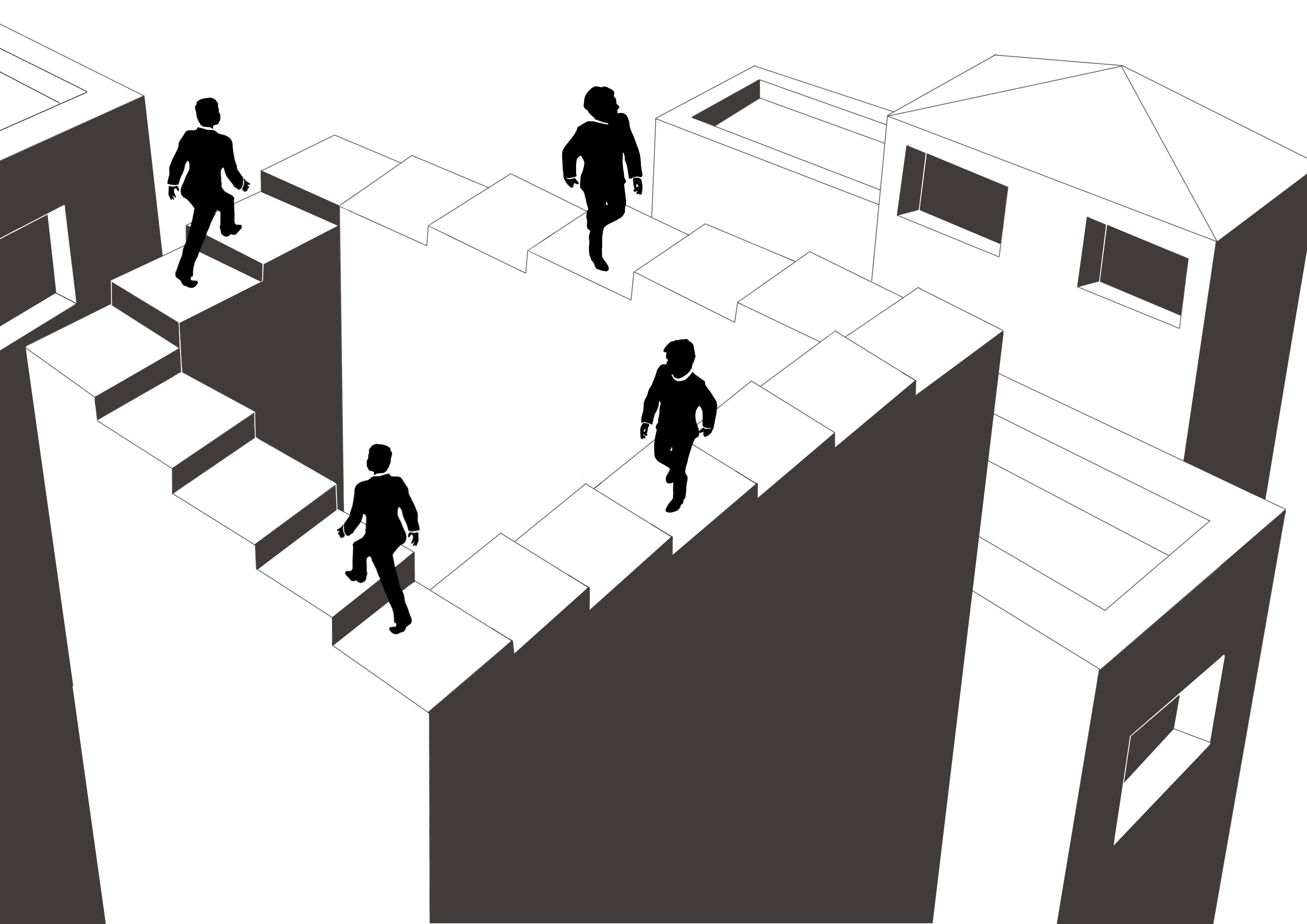}
    \caption{A drawing inspired by M.C. Escher's lithograph \textit{Klimmen en dalen}.}
    \label{fig:ch01_escher}
\end{figure}

We are now able to state the notion of measurement contextuality (or rather noncontextuality) which is a characteristic of the empirical behaviors. Informally noncontextuality can be thought of as an extendability property: local sections can be glued together consistently so that the empirical data can be described from global sections only.
Local predictions can then be obtained from marginalising a probability distribution on global sections. 

\begin{definition}[Noncontextuality or extendability] An empirical model $e$ on a measurement scenario $\tuple{\Xc,\Mc,\Oc}$ is said to be noncontextual (or extendable) if there exists a global probability distribution $\drm$ on global assignments $\Oc_\Xc$ such that $\forall C \in \Mc, e_C = \drm|_C$.
\label{def:ch01_NC}
\end{definition}\index{Extendability!DV|textbf}\index{Contextuality!DV|textbf}

When such a global distribution $\drm$ cannot be found then we say the empirical model $e$ is \textit{contextual}. 
Equivalently, it is contextual when compatible local sections cannot be glued into global sections. 
Contextuality thus naturally arises as the tension between local and global consistency \cite{abramsky2011sheaf,barbosa2021closing}. 
Such tension might give rise to apparent paradoxes. This is beautifully illustrated by M.C. Escher’s lithograph \textit{Klimmen en dalen} (ascending and descending). A drawing inspired by this is given in Figure~\ref{fig:ch01_escher}. Locally looking at the staircase, every character and its adjacent neighbours are going up the stairs consistently. However, taking a global look at the staircase gives an impossible figure in 3 dimensional Euclidean geometry. 
Notions of partial extendability have also been considered in \cite{mansfield2014extendability,simmons2017maximally}.

A stronger version of contextuality is referred to as \textit{logical contextuality}. 
It arises when there exists a local section in the empirical model that cannot be obtained from a global section that is consistent with the support of the model. 
This can easily be pictured on a bundle diagram. 
On figure \ref{fig:ch01_bundleHardy} which illustrates the Hardy paradox \cite{Hardy1993}, the yellow path shows a local section ($\{1,1\}$ in context $a_1,b_1$) that cannot be extended consistently. 
Finally, \textit{strong contextuality}---a stronger form of logical contextuality---arises when no local sections can be extended to a global one consistently \ie when there is no global section $g \in \Oc_\Xc$ such that $\forall C \in \Mc, \, e_C(g|_C) > 0$. The Popescu-Rohrlich (PR) box\footnote{Interestingly the example of these local distributions that cannot be explained by a global distribution was already introduced in \cite{vorob1962consistent}} \cite{popescu1994quantum} is such an example (see figure~\ref{fig:ch01_bundlePR} where no path can be extended consistently). These stronger forms of contextuality can be rather easily witnessed on bundle diagrams as emphasised below.

\begin{figure}[ht!]
\centering
	\begin{minipage}{.45\linewidth}
		\centering
		\begin{tikzpicture}[x=40pt,y=40pt,thick,label distance=-0.25em,baseline=(current bounding box.center)]

\node (e) at (1.5,-0.3) {};
\node (n) at (0.8,0.6) {};
\node (T) at (0,1.5) {};
\node (u) at (0,0.5) {};
\node [inner sep=0em] (a) at (0,0) {};
\node [inner sep=0em] (b) at ($ (a) + (e) $) {};
\node [inner sep=0em] (a') at ($ (a) + (e) + (n) $) {};
\node [inner sep=0em] (b') at ($ (a) + (n) $) {};
\node [inner sep=0em] (a0) at ($ (a) + (T) $) {};
\node [inner sep=0em] (a1) at ($ (a0) + (u) $) {};
\node [inner sep=0em] (b0) at ($ (b) + (T) $) {};
\node [inner sep=0em] (b1) at ($ (b0) + (u) $) {};
\node [inner sep=0em] (a'0) at ($ (a') + (T) $) {};
\node [inner sep=0em] (a'1) at ($ (a'0) + (u) $) {};
\node [inner sep=0em] (b'0) at ($ (b') + (T) $) {};
\node [inner sep=0em] (b'1) at ($ (b'0) + (u) $) {};

\draw [c2] (a) -- (b);
\draw [c2] (a) -- (b');
\draw [c2] (a') -- (b);
\draw [c2] (a') -- (b');

\draw [dotted] (a1) -- (a);
\draw [dotted] (b1) -- (b);
\draw [dotted] (a'1) -- (a');
\draw [dotted] (b'1) -- (b');

\node [inner sep=0.1em,label=left:{\textcolor{c4}{$a_1$}},c4] at (a) {$\bullet$};
\node [inner sep=0.1em,label=right:{\textcolor{c4}{$b_1$}},c4] at (b) {$\bullet$};
\node [inner sep=0.1em,label=right:{\textcolor{c4}{$a_2$}},c4] at (a') {$\bullet$};
\node [inner sep=0.1em,label=left:{\textcolor{c4}{$b_2$}},c4] at (b') {$\bullet$};

\draw [line width=3.2pt,white] (a0) -- (b'0);
\draw [line width=3.2pt,white] (a1) -- (b'0);
\draw [line width=3.2pt,white] (a0) -- (b'1);

\draw [c1] (a0) -- (b'0);
\draw [c1] (a1) -- (b'0);
\draw [c1,preaction={draw,yellow,-,double distance=1\pgflinewidth}] (a0) -- (b'1);

\draw [line width=3.2pt,white] (a'1) -- (b'0);
\draw [line width=3.2pt,white] (a'1) -- (b'1);
\draw [line width=3.2pt,white] (a'0) -- (b'1);

\draw [c1] (a'1) -- (b'0);
\draw [c1] (a'1) -- (b'1);
\draw [c1,preaction={draw,yellow,-,double distance=1\pgflinewidth}] (a'0) -- (b'1);

\draw [line width=3.2pt,white] (a'0) -- (b0);
\draw [line width=3.2pt,white] (a'1) -- (b0);
\draw [line width=3.2pt,white] (a'0) -- (b1);

\draw [c1] (a'0) -- (b0);
\draw [c1] (a'1) -- (b0);
\draw [c1,preaction={draw,yellow,-,double distance=1\pgflinewidth}] (a'0) -- (b1);

\draw [line width=3.2pt,white] (a0) -- (b0);
\draw [line width=3.2pt,white] (a0) -- (b1);
\draw [line width=3.2pt,white] (a1) -- (b0);
\draw [line width=3.2pt,white] (a1) -- (b1);

\draw [c1] (a0) -- (b0);
\draw [c1] (a0) -- (b1);
\draw [c1] (a1) -- (b0);
\draw [c1,preaction={draw,yellow,-,double distance=1\pgflinewidth}] (a1) -- (b1);

\node [inner sep=0.1em,label=left:{\textcolor{c3}{$0$}},c3] at (a0) {$\bullet$};
\node [inner sep=0.1em,label=left:{\textcolor{c3}{$1$}},c3] at (a1) {$\bullet$};
\node [inner sep=0.1em,label=right:{\textcolor{c3}{$0$}},c3] at (b0) {$\bullet$};
\node [inner sep=0.1em,label=right:{\textcolor{c3}{$1$}},c3] at (b1) {$\bullet$};
\node [inner sep=0.1em,label=right:{\textcolor{c3}{$0$}},c3] at (a'0) {$\bullet$};
\node [inner sep=0.1em,label=right:{\textcolor{c3}{$1$}},c3] at (a'1) {$\bullet$};
\node [inner sep=0.1em,label=left:{\textcolor{c3}{$0$}},c3] at (b'0) {$\bullet$};
\node [inner sep=0.1em,label=left:{\textcolor{c3}{$1$}},c3] at (b'1) {$\bullet$};

\end{tikzpicture}
		\caption{Logical contextual empirical model on the $(2,2,2)$ Bell scenario (Hardy model).}
		\label{fig:ch01_bundleHardy}
	\end{minipage}
	\hfill\vline\hfill
	\begin{minipage}{.45\linewidth}
		\centering
		\begin{tikzpicture}[x=40pt,y=40pt,thick,label distance=-0.25em,baseline=(current bounding box.center)]

\node (e) at (1.5,-0.3) {};
\node (n) at (0.8,0.6) {};
\node (T) at (0,1.5) {};
\node (u) at (0,0.5) {};
\node [inner sep=0em] (a) at (0,0) {};
\node [inner sep=0em] (b) at ($ (a) + (e) $) {};
\node [inner sep=0em] (a') at ($ (a) + (e) + (n) $) {};
\node [inner sep=0em] (b') at ($ (a) + (n) $) {};
\node [inner sep=0em] (a0) at ($ (a) + (T) $) {};
\node [inner sep=0em] (a1) at ($ (a0) + (u) $) {};
\node [inner sep=0em] (b0) at ($ (b) + (T) $) {};
\node [inner sep=0em] (b1) at ($ (b0) + (u) $) {};
\node [inner sep=0em] (a'0) at ($ (a') + (T) $) {};
\node [inner sep=0em] (a'1) at ($ (a'0) + (u) $) {};
\node [inner sep=0em] (b'0) at ($ (b') + (T) $) {};
\node [inner sep=0em] (b'1) at ($ (b'0) + (u) $) {};

\draw [c2] (a) -- (b);
\draw [c2] (a) -- (b');
\draw [c2] (a') -- (b);
\draw [c2] (a') -- (b');

\draw [dotted] (a1) -- (a);
\draw [dotted] (b1) -- (b);
\draw [dotted] (a'1) -- (a');
\draw [dotted] (b'1) -- (b');

\node [inner sep=0.1em,label=left:{\textcolor{c4}{$a_1$}},c4] at (a) {$\bullet$};
\node [inner sep=0.1em,label=right:{\textcolor{c4}{$b_1$}},c4] at (b) {$\bullet$};
\node [inner sep=0.1em,label=right:{\textcolor{c4}{$a_2$}},c4] at (a') {$\bullet$};
\node [inner sep=0.1em,label=left:{\textcolor{c4}{$b_2$}},c4] at (b') {$\bullet$};

\draw [line width=3.2pt,white] (a'0) -- (b'1);
\draw [line width=3.2pt,white] (a'1) -- (b'0);

\draw [c1] (a'0) -- (b'1);
\draw [c1,preaction={draw,yellow,-,double distance=1\pgflinewidth}] (a'1) -- (b'0);

\draw [line width=3.2pt,white] (a0) -- (b'0);
\draw [line width=3.2pt,white] (a1) -- (b'1);

\draw [c1,preaction={draw,yellow,-,double distance=1\pgflinewidth}] (a0) -- (b'0);
\draw [c1] (a1) -- (b'1);

\draw [line width=3.2pt,white] (a'0) -- (b0);
\draw [line width=3.2pt,white] (a'1) -- (b1);

\draw [c1] (a'0) -- (b0);
\draw [c1,preaction={draw,yellow,-,double distance=1\pgflinewidth}] (a'1) -- (b1);

\draw [line width=3.2pt,white] (a0) -- (b0);
\draw [line width=3.2pt,white] (a1) -- (b1);

\draw [c1] (a0) -- (b0);
\draw [c1,preaction={draw,yellow,-,double distance=1\pgflinewidth}] (a1) -- (b1);

\node [inner sep=0.1em,label=left:{\textcolor{c3}{$0$}},c3] at (a0) {$\bullet$};
\node [inner sep=0.1em,label=left:{\textcolor{c3}{$1$}},c3] at (a1) {$\bullet$};
\node [inner sep=0.1em,label=right:{\textcolor{c3}{$0$}},c3] at (b0) {$\bullet$};
\node [inner sep=0.1em,label=right:{\textcolor{c3}{$1$}},c3] at (b1) {$\bullet$};
\node [inner sep=0.1em,label=right:{\textcolor{c3}{$0$}},c3] at (a'0) {$\bullet$};
\node [inner sep=0.1em,label=right:{\textcolor{c3}{$1$}},c3] at (a'1) {$\bullet$};
\node [inner sep=0.1em,label=left:{\textcolor{c3}{$0$}},c3] at (b'0) {$\bullet$};
\node [inner sep=0.1em,label=left:{\textcolor{c3}{$1$}},c3] at (b'1) {$\bullet$};

\end{tikzpicture}

		\caption{Strongly contextual empirical model on the $(2,2,2)$ Bell scenario (PR box).\index{PR box}}
		\label{fig:ch01_bundlePR}
	\end{minipage}
\end{figure}

\subsection{The FAB theorem}
\label{subsec01:FAB}

Previously, we characterised noncontextuality of an empirical model by the extendability property. Global sections are sufficient to capture noncontextual empirical behaviours via deterministic global states that assign predefined outcomes to all measurements. This is precisely the model referred to in the Kochen-Specker theorem \cite{KochenSpecker1967,kochen1975problem}. On the other hand, Bell's theorem---focused on a multi-party experiment in which the parties may be spacelike separated---identifies another classical feature: factorisability rather than determinism. Fine unified these two notions in the case of the (2,2,2) Bell scenario \cite{Fine1982}. Later, Abramsky and Brandenburger \cite{abramsky2011sheaf} showed that this existential equivalence holds for any measurement scenario with observables with a discrete spectrum. It establishes an unambiguous, unified treatment of locality and noncontextuality, which is captured in a canonical way by the notion of extendability.

We begin by introducing the notion of hidden-variable models (HVM). Note that HVMs are often referred to as ontological models \cite{Spekkens2005} nowadays. 
The latter has become widely used in quantum foundations in recent years. It indicates that the hidden variable---or the ontic state---is supposed to provide an underlying description of the physical world at perhaps a more fundamental level than the empirical-level description via the quantum state for example.
The idea behind the introduction of HVMs is that there exists some space $\Lambda$ of hidden variables predetermining the empirical behaviour. 
The motivation is that hidden variables could \textit{explain away} some of the more non-intuitive aspects of the empirical predictions of quantum mechanics,
which would then arise as resulting from an incomplete knowledge of the true state of a system rather than
being a fundamental feature.
There is some precedent for this in physical theories: for instance, statistical mechanics---a probabilistic theory---admits a deeper,
albeit usually very complex, description in terms of classical mechanics,
which is purely deterministic.
It is desirable to further
impose constraints on hidden-variable models which will restrict the set of achievable empirical behaviours and require that it behaves \textit{classically} in some sense.
In the case of Bell locality, we require that the hidden-variable model must be local \ie factorisable (in a sense made precise below). 
However, hidden variables may not be directly accessible themselves so we allow that we only have probabilistic information about which hidden variable pertains in the form of a probability distribution $p$ on $\Lambda$. 
The empirical behaviour should then be obtained as an average over the hidden-variable behaviours.

\begin{definition}[Hidden-variable model]\index{Hidden-variable model!DV|textbf}
A hidden-variable model on a measurement scenario $\tuple{\Xc,\Mc,\Oc}$ consists of the triple $\tuple{\Lambda, p, (h^\lambda)_{\lambda \in \Lambda}}$ where:
\begin{itemize}
    \item $\Lambda$ is the finite space of hidden variables; 
    \item $p$ is a probability distribution on $\Lambda$;
    \item for each $\lambda \in \Lambda$, $h^\lambda$ is empirical model \ie $h^\lambda = (h_C^\lambda)_{C \in \Mc}$ is a family where, for each context $C \in \Mc$, $h_C^\lambda$ is a probability distribution on the joint outcome space $\Oc_C$. For each $\lambda \in \Lambda$, $h_\lambda$ satisfies the following compatibility condition:\index{Compatibility condition}
    \begin{equation*}
        \forall C,C' \in \Mc, \quad h_C^\lambda|_{C \cap C'} = h_{C'}^\lambda|_{C \cap C'}.
    \end{equation*}
\end{itemize}
\end{definition}
\noindent Crucially, our definition of hidden-variable model assumes \textit{$\lambda$-independence} \cite{Dickson1998} and \textit{parameter-independence} \cite{Jarrett1984,Shimony1986}.
$\lambda$-independence corresponds to the requirement that the probability distribution $p$ is independent of the measurement context. It is crucial for otherwise every behaviour (including contextual ones) could be modelled trivially by HVMs.
Parameter-independence corresponds to the compatibility condition,
which also ensures that the corresponding hidden-variable behaviour satisfies no-disturbance \cite{brandenburger2013use}.

A hidden-variable model $\tuple{\Lambda, p, (h^\lambda)_{\lambda \in \Lambda}}$ gives rise to an empirical behaviour as follows:
\begin{equation}
    \forall C \in \Mc, \forall s \in \Oc_C, \quad e_C(s) = \sum_{\lambda \in \Lambda} p(\lambda) h_C^\lambda(s).
    \label{eq:ch01_hidden_arising_empirical}
\end{equation}
This is a valid empirical behaviour since it is a convex sum of empirical behaviours $(h_\lambda)_{\lambda\in \Lambda}$.

As mentioned, to ensure that the hidden-variable model behaves \textit{classically} we need to impose further constraints. This motivates the notions of deterministic and of factorisable hidden-variable models.

\begin{definition}[Deterministic HVM]\index{Hidden-variable model!DV!deterministic|textbf}
A hidden-variable model $\tuple{\Lambda, p, (h^\lambda)_{\lambda \in \Lambda}}$ is said to be deterministic if for every $\lambda \in \Lambda$ and for every maximal context $C \in \Mc$ the probability distribution $h_C^\lambda$ is a Kronecker delta
\ie there exists an assignment
$o\in \Oc_C$ such that $h_C^\lambda = \delta_{o}$.
\end{definition}
\noindent Note that global sections $\Oc_\Xc$ can be seen as a canonical form of deterministic hidden variables, which assign a definite outcome to each measurement, independent of the measurement context. Each global section $g \in \Oc_\Xc$ gives raise to a Kronecker delta distribution $\delta_g$. This induces a distribution on each context that can be obtained by marginalisation; this is again a Kronecker delta at the restriction of the global section $g$ to the given context: $\delta_g|_C = \delta_{g|_C}$. Because the set of global assignments is finite, we can always restrict the hidden-variable space to be of finite dimension as an infinite-dimensional space will not capture more behaviours. This fact will become clear from the statement of the FAB theorem.

\begin{definition}[Factorisable HVM]\index{Hidden-variable model!DV!factorisable|textbf}
A hidden-variable model $\tuple{\Lambda, p, (h^\lambda)_{\lambda \in \Lambda}}$ is factorisable if the probability assigned to a joint outcome factors as the product of the probabilities assigned to the individual measurements \ie for every $\lambda \in \Lambda$ and for every maximal context $C \in \Mc$, $h_C^\lambda$ factorises as a product probability distribution.
That is, for $\lambda \in \Lambda$, $C \in \Mc$ and for every $s \in \Oc_C$:
\begin{equation*}
    h_C^\lambda(s) = \prod_{m \in C} h_C^\lambda|_{\{m\}}(s|_{\{m\}}) = \prod_{m \in C} h_m^\lambda(s|_{\{m\}}).
\end{equation*}
\end{definition}
\noindent Due to the assumption of parameter-independence, we can unambiguously write $h_m^\lambda$ for $h_C^\lambda|_{\{m\}}$ as the marginalisation is independent of the context $C$.

We can now state the FAB theorem.
\begin{theorem}[FAB theorem \cite{abramsky2011sheaf}]\index{FAB theorem!DV|textbf}
Let $\tuple{\Xc,\Mc,\Oc}$ a measurement scenario with finite number of measurement labels and finite outcome sets. Let $e$ be an empirical behaviour on $\tuple{\Xc,\Mc,\Oc}$. Then the following propositions are equivalent:
\begin{enumerate}[label=(\arabic*)]
    \item $e$ is extendable.\index{Extendability!DV}
    \item $e$ admits a realisation by a deterministic hidden-variable model. \index{Hidden-variable model!DV!deterministic}
    \item $e$ admits a realisation by a factorisable hidden-variable model.\index{Hidden-variable model!DV!factorisable}
\end{enumerate}
\end{theorem}

\subsection{Quantifying contextuality}
\label{subsec01:quantifying}

A more refined question than asking whether a given empirical behaviour is contextual or not is to ask:
\begin{center}
    \textit{To what extent is a given empirical behaviour contextual?}
\end{center}
Indeed some fraction of the empirical model may admit a noncontextual explanation. 
This allows for a quantitative statement in terms of a measure known as the \textit{contextual fraction}, originally introduced in the seminal work \cite{abramsky2011sheaf} and further developed in \cite{abramsky2017contextual}.
Instead of asking for a proper probability distribution on global assignments that allows the retrieval of the empirical behaviour $e$ by marginalisation at each context, the idea is to ask for a subprobability distribution $\bm \brm$ (a distribution that sums to less than 1) on global sections $\Oc_\Xc$ explaining some fraction of the empirical behaviour \ie we require that $\forall C\in \Mc, \, \bm \brm|_C \leq e_C$. 

We can create an empirical model by taking a convex sum of empirical behaviours \ie for two empirical models $e$ and $e'$ on the same measurement scenario and for $\lambda \in \left[ 0,1 \right]$, $\lambda e + (1-\lambda) e'$ is again a well-defined empirical model. 
An equivalent way of explaining some fraction of the empirical behaviour with a noncontextual behaviour is to ask for a convex decomposition of the form:
\begin{equation}
    e = \lambda e^{NC} + (1-\lambda) e',
    \label{eq:ch01_convexempirical}
\end{equation}
where $\lambda \in \left[ 0,1 \right]$, $e^{NC}$ is a noncontextual empirical model and $e'$ is some other (no-signalling) empirical model. 
The maximum weight $\lambda$ on the noncontextual part or, equivalently, the maximum mass (\ie the quantity $\bm 1 . \bm \brm = \sum_{i = 1}^{\vert \Oc_\Xc \vert} b_i$) of a subprobability distribution is called \textit{the noncontextual fraction} of $e$ and denoted $\NCF(e) \in \left[0,1\right]$. 
It generalises the local fraction \cite{Elitzur1992LF,Barrett2006LF,aolita2012LF}.
The contextual fraction $\CF(e)$ is defined as: $\CF(e) \defeq 1 - \NCF(e)$.
It was originally introduced in \cite{abramsky2011sheaf} where it was proven that a model with contextual fraction of 1 is necessarily strongly contextual. A convex decomposition as in Eq.~(\ref{eq:ch01_convexempirical}) with maximal weight on the noncontextual part necessarily results in the following:
\begin{equation}
    e = \NCF(e) e^{NC} + \CF(e) e^{SC},
\end{equation}
where $e^{SC}$ is strongly contextual. Note that this decomposition might not be unique. Thus the contextual fraction might be formally defined as:

\begin{definition}[Noncontextual fraction]\index{Contextual fraction!DV|textbf}
Let $e$ be an empirical model on the scenario $\tuple{\Xc,\Mc,\Oc}$.
The \emph{noncontextual fraction} of $e$, written $\NCF(e)$, is defined as
\begin{equation}
    \sup\setdef{\bm 1 . \bm \brm}{\bm \brm \; \text{\upshape a subprobability distribution on } \Oc_\Xc  \; \text{\upshape such that } \forall C \in \Mc, \bm \brm|_C \leq e_C } \Mdot 
\end{equation}
\label{def:ch01_NCF}
\end{definition}

\subsubsection*{Computing the contextual fraction as a linear program} 

A very convenient property is that the computation of the noncontextual fraction can be phrased as a linear program (see Subsection \ref{subsec01:LP}) \cite{abramsky2017contextual}. Fix a measurement scenario $\tuple{\Xc,\Mc,\Oc}$ (with a finite number of observables and finite spaces of outcomes). Let $e$ be an empirical model on that scenario. Let $n \defeq \vert \Oc_\Xc \vert$ be the number of possible global assignments and $m \defeq \sum_{C \in \Mc} \vert \Oc_C \vert$ be the number of total local assignments ranging over all contexts. It can also be expressed as $m \defeq \vert \{ \langle C,s \rangle \text{ s.t. } C \in \Mc \text{ and } s \in \Oc_C \} \vert$. We define the incidence matrix $\Mrm$ as the $m \times n$ (0,1)-matrix such that:
\begin{equation}
    \Mrm[\langle C,s \rangle,g] \defeq 
    \begin{cases}
    1 \text{ if } g|_C = s \\
    0 \text{ otherwise},
    \end{cases}
    \label{eq:ch01_incidencemat}
\end{equation}
ranging over contexts $C \in \Mc$, local sections $s \in \Oc_C$, and global sections $g \in \Oc_\Xc$. 
Each column corresponds to a global section and each row to a local section. 
To understand its action,
read the matrix column-by-column by fixing a global section $g \in \Oc_\Xc$. For that column, 
the incidence matrix assigns a 1 in a row corresponding to a local section $s$ in context $C$ whenever the global section under consideration marginalises to $s$ \ie if $g|_C = s$. Thus it can be seen as a table of possible restrictions from global to local assignments. 

The empirical model $e$ can be represented as a vector $ \bm \vrm^e  \in \left[0,1\right]^m$ where for a given context $C \in \Mc$ and a local assignment $s \in \Oc_C$, $\bm \vrm^e[\langle C, \, s \rangle] = e_C(s)$. It is a flattened version of the tables that are usually used to represent empirical models (see table \ref{tab:ch01_CHSH}). Then $e$ is noncontextual (see Definition \ref{def:ch01_NC}) if there exists a global probability distribution $\bm \drm \in \left[0,1\right]^n$ such that $\Mrm \bm \drm = \bm \vrm^e$. The computation of the noncontextual fraction of $e$ defined in Definition \ref{def:ch01_NCF} is given by the value of the following program:\index{Linear program}
\leqnomode
\begin{flalign*}
    \label{prog:LP-CF}
    \tag*{(P-CF$^\text{DV}$)}
    \hspace{4cm}\left\{
    \begin{aligned}
        & \quad \text{Find } \bm \brm \in \R^n \\
        & \quad \text{maximising } \bm 1.\bm \brm \\
        & \quad \text{subject to:}  \\
        & \hspace{1cm} \begin{aligned}
            & \Mrm \bm \brm \leq \bm \vrm^e \\
            & \bm \brm \geq \bm 0 \Mdot
        \end{aligned}
    \end{aligned}
    \right. &&
\end{flalign*}
\reqnomode

\paragraph*{Computing generalised Bell inequality}
\begin{definition}[Generalised Bell inequality]\index{Bell inequality!DV|textbf}
A generalised Bell inequality for the measurement scenario $\tuple{\Xc,\Mc,\Oc}$ is given by $\tuple{\bm \arm,R}$ where $\bm \arm \in \R^m$ is a real vector indexed by local assignments $\langle C,s \rangle$ for $C \in \Mc$ and $s \in \Oc_C$, and $R \in \R$ is a bound. For all noncontextual empirical model $e$ on $\tuple{\Xc,\Mc,\Oc}$, it must hold that $\bm \arm . \bm \vrm^e \leq R$. $\tuple{\bm \arm,R}$ is said to be tight if there exists a noncontextual behaviour $e$ that saturates the bound \ie if $\bm \arm . \bm \vrm^e = R$.
\end{definition}
For no-signalling empirical models, the quantity $\bm \arm . \bm \vrm^e$ is upper bounded by 
\begin{equation}
\norm{a} \defeq \sum_{C \in \Mc} \max \{ a\left[ \langle C,s \rangle \right] | s \in \Oc_C  \},
\label{eq:ch01_boundBI}  
\end{equation}
which amounts, for each context $C \in \Mc$, to having a behaviour which is a Kronecker delta at the maximal component of $\bm \arm$ for each context. Note that this behaviour might be signalling (\ie it might not respect the compatibility condition). We will only consider inequalities for which $\norm{a} > R$ for otherwise the inequality would be trivially satisfied by all models.

Now the amount by which an empirical model $e$ violates a generalised Bell inequality is $\max\{0,\bm \arm . \bm \vrm^e - R\}$. It is usually desirable to normalise the previous quantity in order to have a quantity lying in the unit interval. The \textit{normalised violation} of a generalised Bell inequality $\tuple{\bm \arm, R}$ by a model $e$ is:
\begin{equation}
\frac{\max\{0,\bm \arm . \bm \vrm^e - R\}}{\norm{a} - R}.
\label{eq:ch01_normalisedviolation}
\end{equation}

Importantly, an optimal solution of the dual program of \refprog{LP-CF} (with a clever change of variables) describes a generalised Bell inequality with maximal normalised violation by the empirical model with bound 0. Thus a generalised Bell inequality with maximal normalised violation for $e$ is described by $\tuple{\bm \arm^*,0}$ where $\bm \arm^* \in \R^m$ is an optimal feasible plan of the following program with value $\CF(e)$:
\leqnomode
\begin{flalign*}
    \label{prog:DLP-CF}
    \tag*{(B-CF$^\text{DV}$)}
    \hspace{4cm}\left\{
    \begin{aligned}
        & \quad \text{Find } \bm \arm \in \R^m \\
        & \quad \text{maximising } \bm \arm.\bm \vrm^e \\
        & \quad \text{subject to:}  \\
        &  \hspace{1cm} \begin{aligned}
            & \Mrm^T \bm \arm \leq \bm 0 \\
            & \bm \arm \geq \bm \vert \Mc \vert^{-1} \bm 1 \Mdot
        \end{aligned}
    \end{aligned}
    \right. &&
\end{flalign*}
\reqnomode

We now recall the main theorem from \cite{abramsky2017contextual}:

\begin{theorem}[Th.1 \cite{abramsky2017contextual}]
Let $e$ be an empirical model on a measurement scenario $\tuple{\Xc,\Mc,\Oc}$. 
\begin{enumerate*}[label=(\roman*)] \item The normalised violation of any Bell inequality is at most $\CF(e)$; \item if $\CF(e) > 0$ then there exists a Bell inequality whose normalised violation is exactly $\CF(e)$; \item for any convex decomposition of a contextual model $e$ as $e = \NCF(e) e^{NC} + \CF(e) e^{SC}$, a Bell inequality, for which the normalised violation is $\CF(e)$, is tight at $e^{NC}$ (provided $\NCF(e) > 0$) and maximally violated at $e^{SC}$.
\end{enumerate*}
\label{th:ch01_Belltheorem}
\end{theorem}

What happens for all the results presented above if we now wish to extend the support of empirical models to spaces with continuous outcomes? 
This is the question we address in the next chapter. 
For instance our set of measurement labels $\Xc$ may refer to observables like position and momentum which in principle have a continuous spectrum (being the reals $\R$). 
For a context $C$ including one of these observables, $e_C$ naturally becomes a probability measure over the joint outcome set $\Oc_C$ which necessarily comprises a continuum. Hence the need for the frugal dose of measure theory introduced in Subsection~\ref{sec01:measure}.

\dobib

\clearemptydoublepage


\chapter{Continuous-variable nonlocality and contextuality}
\label{chap:CVcont}

\lettrine{T}o date the study of contextuality has largely focused on discrete-variable scenarios. As a consequence, the main frameworks and approaches to contextuality are tailored to modelling these, e.g.~\cite{abramsky2011sheaf,csw,afls,dzhafarov2015contextuality,Spekkens2005}.
In such scenarios, observables can only take values in discrete, finite sets.
Discrete-variable scenarios typically arise in finite-dimensional quantum mechanics, \eg when modelling quantum computations with qubit systems.

Since quantum mechanics itself is infinite-dimensional,
it also makes sense from a foundational perspective to extend
analyses of the key concept of contextuality to the continuous-variable setting.
Furthermore, continuous variables can be useful when dealing with iteration,
even when attention is restricted to finite-variable actions at discrete time steps,
as is traditional in informatics.
An interesting question, for example, is whether contextuality arises and is of interest in such situations
as the infinite behaviour of quantum random walks.\index{Contextuality!CV}

\begin{figure}[ht!]
    \centering
    {\footnotesize
        \begin{tikzpicture}[scale=.5]

\tikzset{>=latex}

\node (Ra1) at (0,0) {};
\node (Ra2) at (4,2) {};
\node (Rac) at ($(Ra1)!.5!(Ra2)$) {};
\node (up) at (0,1) {};
\node (right) at (2,0) {};

\node (Rb1) at (-4,4) {};
\node (Rb2) at (0,6) {};
\node (Rbc) at ($(Rb1)!.5!(Rb2)$) {};

\node (Rc1) at (4,4) {};
\node (Rc2) at (8,6) {};
\node (Rcc) at ($(Rc1)!.5!(Rc2)$) {};

\node (pp) at (Rac) {\begin{tabular}{c} preparation \\ device \end{tabular}};
\draw [->,in=0,out=90] ($(Rac)+(up)$) to ($(Rbc)+(right)$);
\filldraw [fill=green!20,fill opacity=0.5] (Ra1) rectangle (Ra2);

\node (md1) at (Rbc) {\begin{tabular}{c} measurement \\ device \end{tabular}};
\draw [->,in=180,out=90] ($(Rac)+(up)$) to ($(Rcc)-(right)$);
\draw [->] ($(Rbc)-1.75*(up)$) -- ($(Rbc)-(up)$);
\draw [->] ($(Rbc)+(up)$) -- ($(Rbc)+1.75*(up)$);
\node (md1tt) at ($(Rbc)-2*(up)$) {$m_A \in \{a,a'\}$};
\node (md1tu) at ($(Rbc)+2*(up)$) {$o_A \in \bm R$};
\filldraw [fill=red!20,fill opacity=0.5] (Rb1) rectangle (Rb2);

\node (md2) at (Rcc) {\begin{tabular}{c} measurement \\ device \end{tabular}};
\draw [->] ($(Rcc)-1.75*(up)$) -- ($(Rcc)-(up)$);
\draw [->] ($(Rcc)+(up)$) -- ($(Rcc)+1.75*(up)$);
\node (md2tt) at ($(Rcc)-2*(up)$) {$m_B \in \{b,b'\}$};
\node (md2tu) at ($(Rcc)+2*(up)$) {$o_B \in \bm R$};
\filldraw [fill=blue!20,fill opacity=0.5] (Rc1) rectangle (Rc2);
 
\end{tikzpicture}
    }
    \quad\quad\quad
    \begin{tikzpicture}[scale=.3]

\def\rad{1.15}

\node [inner sep=.0pt] (a11) at (0,0) {};
\node [inner sep=.0pt] (a12) at (5,0) {};
\node [inner sep=.0pt] (a13) at (10,0) {};
\node [inner sep=.0pt] (a21) at (0,5) {};
\node [inner sep=.0pt] (a22) at (5,5) {};
\node [inner sep=.0pt] (a23) at (10,5) {};
\node [inner sep=.0pt] (a31) at (0,10) {};
\node [inner sep=.0pt] (a32) at (5,10) {};
\node [inner sep=.0pt] (a33) at (10,10) {};

\node (left) at (-1,0) {};
\node (up) at (0,1) {};

\node (A') at ($.5*(a11)+.5*(a21)+(left)$) {$a'$};
\node (A) at ($.5*(a21)+.5*(a31)+(left)$) {$a$};
\node (B) at ($.5*(a31)+.5*(a32)+(up)$) {$b$};
\node (B') at ($.5*(a32)+.5*(a33)+(up)$) {$b'$};

\draw ($(a11)+(left)$) -- (a13);
\draw ($(a21)+(left)$) -- (a23);
\draw ($(a31)+(left)$) -- (a33);

\draw ($(a31)+(up)$) -- (a11);
\draw ($(a32)+(up)$) -- (a12);
\draw ($(a33)+(up)$) -- (a13);

\filldraw[white,inner color=blue,outer color=blue!5] ($.75*(a11)+.25*(a22)$) circle (\rad);
\filldraw[white,inner color=blue,outer color=blue!5] ($.25*(a11)+.75*(a22)$) circle (\rad);

\filldraw[white,inner color=blue,outer color=blue!5] ($.75*(a12)+.25*(a23)$) circle (\rad);
\filldraw[white,inner color=blue,outer color=blue!5] ($.25*(a12)+.75*(a23)$) circle (\rad);

\filldraw[white,inner color=blue,outer color=blue!5] ($.75*(a22)+.25*(a33)$) circle (\rad);
\filldraw[white,inner color=blue,outer color=blue!5] ($.25*(a22)+.75*(a33)$) circle (\rad);

\filldraw[white,inner color=blue,outer color=blue!5] ($.75*(a22)+.25*(a31)$) circle (\rad);
\filldraw[white,inner color=blue,outer color=blue!5] ($.25*(a22)+.75*(a31)$) circle (\rad);

\end{tikzpicture}
    \caption{
        [Left] operational depiction of a typical bipartite experimental scenario. [Right] Hypothetical probability density plots for empirical data arising from such an experiment for a continuous generalisation of the PR box~\cite{ketterer2018continuous}. Cf.~the discrete-variable probability tables of \cite{mansfield2012hardy,mansfield2017consequences}.\index{PR box}
    }
    \label{fig:ch02_scenario}
\end{figure}
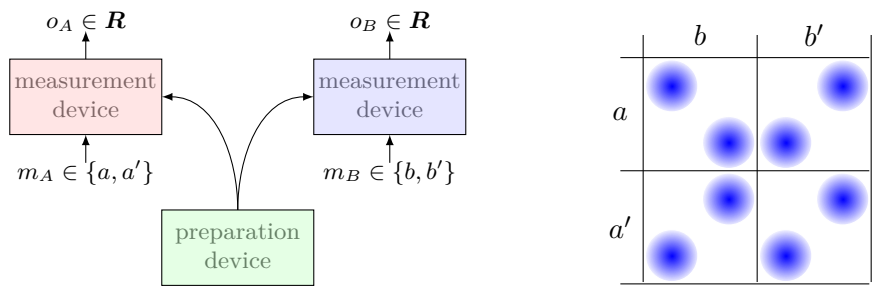

What kind of experiments are we trying to model? 
Suppose that we can interact with a system by performing measurements on it and observing their outcomes.
A feature of quantum systems is that not all observables commute, so that certain combinations of measurements are incompatible.
At best, we can obtain empirical data for contexts in which only compatible measurements are performed, which can be collected by running the experiment repeatedly.
Contextuality then arises when the empirical data obtained is inconsistent with the assumption that for each run of the experiment the system has a global and context-independent assignment of values to all of its observable properties.
To take an operational perspective, a typical example of an experimental setup or scenario that we consider in this chapter is the one depicted in Figure~\ref{fig:ch02_scenario} [left].
In this standard scenario extended to the continuous-variable realm, a system is prepared in some fixed bipartite state, following which parties $A$ and $B$ may each choose between two measurement settings, $m_A \in \enset{ a, a'}$ for $A$ and $m_B \in \enset{ b, b'}$ for $B$.
We assume that outcomes of each measurement live in $\bm R$, which typically will be a measurable subspace of the real numbers $\R$ (with its Borel $\sigma$-algebra).
Depending on which choices of inputs---which specify a context---were made, the empirical data might for example be distributed according to one of the four hypothetical probability density plots in $\bm R^2$ depicted in Figure~\ref{fig:ch02_scenario} [right].
This scenario and hypothetical empirical behaviour has been considered elsewhere \cite{ketterer2018continuous} as a continuous-variable version of the Popescu--Rohrlich (PR) box \cite{popescu1994quantum}.

\paragraph{Related work.}
Note that we are specifically interested in scenarios involving observables with continuous spectra, or in more operational language, measurements having an outcome space which is a continuum.
For quantifying contextuality, we still consider scenarios featuring only discrete sets of observables or measurements, as is typical in continuous-variable quantum computing, but we prove the Fine--Abramsky--Brandenburger (FAB) theorem\index{FAB theorem!CV} for uncountable sets of measurement settings. This will be useful for Chapter~\ref{chap:equivalence}.
The possibility of considering contextuality in settings with outcome spaces being a continuum has also been evoked in \cite{cunha2019measures}.
We also note that several prior works have explicitly considered contextuality in continuous-variable systems \cite{plastino2010state,he2010bell,mckeown2011testing,su2012quantum,asadian2015contextuality,laversanne2017general,ketterer2018continuous}. In particular \cite{CavalcantiCVBI2007} presents a genuine continuous-variable Bell inequality that uses a second-order moment approach.

Our approach is distinct from these in that it provides a genuinely continuous-variable treatment of contextuality itself as opposed to embedding discrete-variable contextuality arguments into, or extracting them from, continuous-variable systems. This chapter is based on \cite{barbosa2019continuous}.

\section{Framework}\label{sec02:formalism}

In this section, we follow closely the discrete-variable framework of \cite{abramsky2011sheaf} which is presented in Section~\ref{sec01:sheaf} to develop a continuous-variable framework that allows to describe and analyse the empirical behaviours that arise with continuous-variable systems, although some extra care is required for dealing with continuous variables.

\subsubsection*{Measurement scenarios}

We recall that an abstract description of an experimental setup is formalised as a measurement scenario. Below we give its description for continuous-variable systems.\index{Measurement scenario!CV|textbf}
\begin{definition}[Measurement scenario] \label{def:ch02_measurementscenario}
A measurement scenario is a triple $\tuple{\Xc,\Mc,\bm \Oc}$ whose elements are specified as follows.
\begin{itemize}
    \item  $\Xc$ is a (possibly infinite) set of measurement labels.
    
    \item $\Mc$ is a covering family of subsets of $\Xc$, \ie such that $\bigcup\Mc = \Xc$.
    The elements $C \in \Mc$ are called maximal contexts and represent maximal sets of compatible observables. We therefore require that $\Mc$ be an anti-chain with respect to subset inclusion, \ie that no element of this family is a proper subset of another.
    Any subset of a maximal context also represents a set of compatible measurements.

    \item
    $\bm \Oc = \family{\bm \Oc_x}_{x \in \Xc}$ specifies a measurable space of outcomes $\bm \Oc_x = \tuple{\Oc_x,\Fc_x}$ for each measurement $x \in \Xc$. If some set of measurements $U \subseteq \Xc$ is considered together, there is a joint outcome space given by the product of the respective outcome spaces $\bm \Oc_U \defeq \prod_{x \in U} \bm \Oc_x = \tuple{\Oc_U, \Fc_U}$ (see definition~\ref{def:ch01_productmeasurable} for the finite case and definition~\ref{def:ch01_productmeasurableinfinite} for the infinite case).\index{Measurable!space}
    \end{itemize}
\end{definition}

\noindent We consider $\Xc$ as general as possible (since we will need $\Xc$ of the form $\R^M \times \R^M$ for some $M \in \N^*$ in Chapter~\ref{chap:equivalence}). Note that, later in the current chapter, we will restrict to finite set of measurement labels for the quantification of contextuality. 

\paragraph*{Example.} The setup represented in Figure~\ref{fig:ch02_scenario} is described by the measurement scenario:
\begin{equation}\label{eq:scenario}
\Xc=\enset{a,a',b,b'} \, , \quad\quad \Mc = \enset{ \, \enset{a,b}, \, \enset{a,b'}, \, \enset{a',b}, \, \enset{a',b'} \, } \, , \quad\quad \forall x \in \Xc, \; \bm \Oc_x = \bm R \, ,
\end{equation}
where $\bm R$ is a measurable subspace of $\tuple{\R,\Bc_\R}$.

As for the discrete-variable case, the \textit{sheaf} $\Ec$ that maps $U \subseteq \Xc$ to $\Ec(U) = \bm \Oc_U$
is called the \emph{event sheaf}.\index{Sheaf!event} We recall that it comes with the notion of restriction, \ie, for $U \subseteq V \in \Pc(\Xc)$ there is a restriction map $\rho^V_U : \bm \Oc_V \rightarrow \bm \Oc_U $ which is continuous in the product topology.\index{Sheaf!restriction}

\subsubsection*{Empirical models}

\begin{definition}[Empirical model] \label{def_ch02_empiricalmodel} \index{Empirical model!CV|textbf}
An empirical model on a measurement scenario $\tuple{\Xc,\Mc,\Oc}$ is a family $e = \family{e_C}_{C \in \Mc}$, where $e_C$ is a probability measure on the space $\Ec(C)=\bm \Oc_C$ for each maximal context $C \in \Mc$. It satisfies the compatibility conditions:
\[\forall C, C' \in \Mc, \quad e_C|_{C \cap C'} = e_{C'}|_{C \cap C'} \Mdot \]
\end{definition}

Empirical models capture in a precise way the probabilistic \emph{behaviours} that may arise upon performing measurements on physical systems.
The compatibility condition ensures that the empirical behaviour of a given measurement or compatible subset of measurements is independent of which other compatible measurements might be performed along with them.
As mentioned for the discrete-variable framework in Subsection~\ref{subsec01:empiricalmodels}, this is known as the \emph{no-disturbance condition}.
A special case is \emph{no-signalling}, which applies in multi-party or Bell scenarios such as that of Figure~\ref{fig:ch02_scenario} and Eq.~\eqref{eq:scenario}.
In that case, contexts consist of measurements that are supposed to occur in  space-like separated locations, and compatibility ensures for instance that the choice of performing $a$ or $a'$ at the first location does not affect the empirical behaviour at the second location, \ie $e_{\enset{a,b}}|_{\enset{b}} = e_{\enset{a',b}}|_{\enset{b}}$.

Note also that while empirical models may arise from the predictions of quantum theory, their definition is theory-independent.
This means that empirical models can just as well describe hypothetical behaviours beyond what can be achieved by quantum mechanics such as the well-studied Popescu--Rohrlich box \cite{popescu1994quantum}.
This can be useful in probing the limits of quantum theory
and in singling out what distinguishes and characterises quantum theory within larger spaces of probabilistic theories,
both well-established lines of research in quantum foundations.

\subsubsection*{Extendability and contextuality}

\begin{definition}[Noncontextuality]\index{Contextuality!CV|textbf}\index{Sheaf!event}\index{Measure}
\label{def:nc}
An empirical model $e$ on a scenario $\tuple{\Xc,\Mc,\bm \Oc}$ is \emph{extendable}\index{Extendability!CV|textbf} or \emph{noncontextual}
if there is a probability measure $\mu$ on the space $\Ec(\Xc)=\bm \Oc_\Xc$
such that $\mu|_C = e_C$ for every $C \in \Mc$.
\end{definition}

Recall that $\bm \Oc_\Xc$ is the space of global assignments.
Of course, it is not always the case that $\Xc$ is a valid context, and if it were then $\mu = e_\Xc$ would trivially extend the empirical model.
The question of the existence of such a probability measure that recovers the context-wise empirical content of $e$ is particularly significant. When it exists, it amounts to a way of modelling the observed behaviour as arising stochastically from the behaviours of underlying states, identified with the elements of $\Oc_\Xc$.
These elements deterministically assigns outcomes to \textit{all} the measurements in $\Xc$ independently of the measurement context that is actually performed.
If an empirical model is not extendable it is said to be \emph{contextual}.

\section{A continuous-variable FAB theorem}\label{sec02:fine}

As in the discrete-variable setting, we characterise contextuality of an empirical model by the absence of a global section for that empirical model. 
In this section, we prove a Fine--Abramsky--Brandenburger (FAB) theorem in the continuous-variable setting.
It establishes that in this setting there is also an unambiguous, unified description of locality and noncontextuality, which is captured in a canonical way by the notion of extendability.

We will begin by introducing hidden-variable models in a more precise way following the treatment given in Section~\ref{subsec01:FAB}.

\begin{definition}[Hidden-variable model]
\index{Hidden-variable model!CV|textbf}
\index{Measurable!space}
\index{Measurable!function}
\label{def:ch02_hvmodel}
A hidden-variable model on a measurement scenario $\tuple{\Xc,\Mc,\Oc}$ consists of the triple $\tuple{\bm \Lambda, p, (k_C)_{C \in \Mc}}$ where:
\begin{itemize}
    \item $\bfLambda = \tuple{\Lambda,\Fc_\Lambda}$ is the measurable space of hidden variables, 
    \item $p$ is a probability distribution on $\bfLambda$,
    \item for each context $C \in \Mc$, $k_C$ is a probability kernel\footnotemark\ between the measurable spaces $\bfLambda$ and $\Ec(C)=\bm \Oc_C$ satisfying the following compatibility condition:
    \begin{equation}\label{eq:compatibility_hiddenvar}
        \forall C,C' \in \Mc, \forall \lambda \in \Lambda, \quad k_C(\lambda,-)|_{C \cap C'} = k_{C'}(\lambda,-)|_{C \cap C'} \Mdot
    \end{equation}
\end{itemize}
\end{definition}

\footnotetext{Recall from Section~\ref{sec01:measure} 
that a probability kernel $k_C$ between $\bfLambda$ and $\Ec(C)$ is a function
$\fdec{k_C}{\Lambda \times \Fc_C}{[0,1]}$ 
which is:\index{Markov kernel}
\begin{enumerate*}[label=(\roman*)]
    \item 
a measurable function in the first argument, \ie $\fdec{k_C(\dummy,B)}{\Lambda}{[0,1]}$ is measurable for every $B \in \Fc_C$; and
    \item
 a probability measure in the second argument,
\ie $\fdec{k_C(\lambda,\dummy)}{\Fc_C}{[0,1]}$ is a probability measure for all $\lambda \in \Lambda$.
\end{enumerate*}
}
\noindent The last ingredient can be understood as defining a function $\underline{k}$ from the set of hidden variables $\Lambda$ to the set of empirical models over $\tuple{\Xc,\Mc,\bm \Oc}$.
The function assigns to each $\lambda \in \Lambda$ the empirical model
$\underline{k}(\lambda) \defeq \family{\underline{k}(\lambda)_C}_{C \in \Mc}$, where the correspondence with the definition above is via $\underline{k}(\lambda)_C = k_C(\lambda,\dummy)$.
This function must be `measurable' in $\bfLambda$
in the sense that  $\fdec{\underline{k}(\dummy)_C(B)}{\Lambda}{[0,1]}$
is a measurable function for all $C \in \Mc$ and $B \in \Fc_C$. 
Hence the need for Markov kernels. 

A hidden-variable model $\tuple{\bfLambda,p,(k_C)_{C \in \Mc}}$ on a measurement scenario $\tuple{\Xc,\Mc,\bm \Oc}$ gives rise to an empirical behaviour. 
The corresponding empirical model $e$ is such that for any maximal context $C \in \Mc$ and measurable set of joint outcomes $B \in \Fc_C$,
\begin{equation}
    \label{eq:ch02_eval_HVM_emp}
    e_C(B) = \intg{\Lambda}{k_C(\dummy,B)}{p}  = \intg{\lambda \in \Lambda}{k_C(\lambda,B)}{p(\lambda)}\Mdot
\end{equation}

Note that this definition of hidden-variable model still assumes $\lambda$-independence and parameter-independence. As in the discrete-variable case, we may impose further constraints to ensure that the HVM behaves \textit{classically}.

\begin{definition}[Deterministic HVM]\index{Hidden-variable model!CV!deterministic|textbf}
A hidden-variable model $\tuple{\bfLambda,p,(k_C)_{C \in \Mc}}$ is \emph{deterministic} if
$\fdec{k_C(\lambda,\dummy)}{\Fc_C}{[0,1]}$ is a Dirac measure
for every $\lambda \in \Lambda$
and for every maximal context $C \in \Mc$;
in other words, there is an assignment
$\bm o\in \Oc_C$ such that $k_C(\lambda,\dummy) = \delta_{\bm o}$.
\end{definition}\index{Measure}

In general discussions on hidden-variable models (\eg \cite{BrandenburgerYanofsky2008}), the condition above, requiring that each hidden variable determines a unique joint outcome for each measurement context, is sometimes referred to as weak determinism. This is contraposed to strong determinism,
which demands not only that each hidden variable fix a deterministic outcome to each individual measurement,
but that this outcome be independent of the context in which the measurement is performed.
Note, however, that since our definition of hidden-variable models assumes the compatilibity condition \eqref{eq:compatibility_hiddenvar}, \ie parameter-independence, 
both notions of determinism coincide \cite{brandenburger2013use}.

\begin{definition}[Factorisable HVM]\index{Hidden-variable model!CV!factorisable|textbf}
A hidden-variable model $\tuple{\bfLambda,p,(k_C)_{C \in \Mc}}$ is \emph{factorisable} if
$\fdec{k_C(\lambda,\dummy)}{\Fc_C}{[0,1]}$ factorises as a product measure
for every $\lambda \in \Lambda$
and for every maximal context $C \in \Mc$.
That is, for any family of
measurable sets $\family{B_x \in \Fc_x}_{x \in C}$,
\[k_C(\lambda,\prod_{x\in C} B_x) = \prod_{x\in C} k_C|_{\enset{x}}(\lambda,B_x)\]
where
$k_C|_{\enset{x}}(\lambda,\dummy)$ is the marginal of the probability measure
$k_C(\lambda,\dummy)$ on $\bm \Oc_C=\prod_{x \in C}\bm \Oc_x$ to the space $\bm \Oc_{\enset{x}} = \bm \Oc_x$.\footnotemark
\end{definition}

\footnotetext{
Note that, due to the assumption of parameter independence (Eq.~\eqref{eq:compatibility_hiddenvar}),
we can unambiguously write $k_x(\lambda,\dummy)$ for
$k_C|_{\enset{x}}(\lambda,\dummy)$, as this marginal is independent of the context $C$ from which one is restricting.
}

In other words, if we consider elements of $\Lambda$ as inaccessible `empirical' models -- \ie if we use the alternative definition of hidden-variable models using the map $\underline{k}$ (see 
the remark below Definition \ref{def:ch02_hvmodel}) -- then factorisability is the requirement that each of these be factorisable in the sense that
 \[\underline{k}_C(\lambda)\left(\prod_{x\in C} B_x\right) = \prod_{x\in C} \, \underline{k}_{\enset{x}}(\lambda)(B_x)
 \]
 where $\underline{k}_{\enset{x}}(\lambda)$ is the marginal of the probability measure
 $\underline{k}_C(\lambda)$ on $\bm \Oc_C=\prod_{x \in C}\bm \Oc_x$ to the space $\bm \Oc_x$.\index{Measure}

We now prove the continuous-variable analogue of the theorem proved in the discrete probability setting by Abramsky and Brandenburger \cite[Proposition 3.1 and Theorem 8.1]{abramsky2011sheaf} (see Section~\ref{subsec01:FAB}),
generalising the result of Fine \cite{Fine1982} to arbitrary measurement scenarios with a possibly infinite number of measurement labels.\index{FAB theorem!CV|textbf}

\begin{theorem}
\label{th:ch02_FAB}
Let $e$ be an empirical model on a measurement scenario $\tuple{\Xc,\Mc,\bm \Oc}$. The following are equivalent:
\begin{enumerate}[label=(\arabic*)]
\item\label{it:ext} $e$ is extendable;\index{Extendability!CV}
\item\label{it:det} $e$ admits a realisation by a deterministic hidden-variable model;\index{Hidden-variable model!CV!deterministic}
\item\label{it:fac} $e$ admits a realisation by a factorisable hidden-variable model.\index{Hidden-variable model!CV!factorisable}
\end{enumerate}
\end{theorem}
\begin{proof}
We prove the sequence of implications \ref{it:ext} $\Rightarrow$ \ref{it:det} $\Rightarrow$ \ref{it:fac} $\Rightarrow$ \ref{it:ext}.

\textbf{\ref{it:ext} $\Rightarrow$ \ref{it:det}.}
The idea is that $\Ec(\Xc)=\bm \Oc_\Xc$ provides a canonical deterministic hidden-variable space. Suppose that $e$ is extendable to a global probability measure $\mu$. Let us set $\bfLambda \defeq \bm \Oc_\Xc$,
set $p \defeq \mu$, and set
$k_C(\bm g,\dummy) \defeq \delta_{\bm g|_C}$
for all global outcome assignments $\bm g \in  \Oc_\Xc$.\index{Extendability!CV}
This is by construction a deterministic hidden-variable model,
which we claim gives rise to the empirical model $e$. Let us verify it.\index{Hidden-variable model!CV!deterministic}\index{Measure}\index{Measurable!space}

Let $C \in \Mc$ and write $\fdec{\rho^\Xc_C}{\bm \Oc_\Xc}{\bm \Oc_C}$ for the  measurable projection\index{Sheaf!restriction}\index{Measurable!function}
which, in the event sheaf, is the restriction map $\fdec{\rho^\Xc_C = \Ec(C \subseteq \Xc)}{\Ec(\Xc)}{\Ec(C)}$.
For any $E \in \Fc_C$, we have
\begin{equation}\label{eq:deltaindicator_aux}
k_C(\bm g,E) = \delta_{\bm g|_C}(E) = \delta_{\rho^\Xc_C(\bm g)}(E) = \bm 1_{E}(\rho^\Xc_C(\bm g))
=
(\bm 1_E \circ \rho^\Xc_C)(\bm g)
\end{equation}
which will assign 1 if $\rho^\Xc_C(\bm g) = \bm g|_C \in E$.
Then, as required

\begin{calculation}
\intg{\bfLambda}{k_C(\dummy,E)}{p}
\just={$\bfLambda = \bm \Oc_\Xc$; $p = \mu$; $k_C(\dummy,E) = \bm 1_E \circ \rho^\Xc_C$ by Eq.~\eqref{eq:deltaindicator_aux}}
\intg{\bm \Oc_\Xc}{\bm 1_E \circ \rho^\Xc_C}{\mu}
\just={change of variables, Eq.~\eqref{eq:ch01_changeofvariables}}
\intg{\bm \Oc_C}{\bm 1_E}{\rho^\Xc_{C*}\mu}
\just={marginalisation for probability measures}
\intg{\bm \Oc_C}{\bm 1_E}{\mu|_C}
\just={integral of indicator function, Eq.~\eqref{eq:ch01_integralindicator}}
\mu|_C(E)
\just={$\mu$ extends the empirical model $e$}
e_C(E) \Mdot
\end{calculation}%

\textbf{\ref{it:det} $\Rightarrow$ \ref{it:fac}.}
It is enough to show that if a hidden-variable model $\tuple{\bfLambda,p,k}$ is deterministic then it is also factorisable.\index{Hidden-variable model!CV!factorisable}
For this, it is sufficient to notice that a Dirac measure
$\delta_{\bm o}$ with $\bm o \in \Oc_C$ on a product space $\bm \Oc_C=\prod_{x \in C}\bm \Oc_x$
factorises as the product of Dirac measures
\begin{equation}
    \delta_{\bm o} = \prod_{x \in C}\delta_{\bm o|_{\enset{x}}} = \prod_{x \in C}\delta_{\rho^C_{\{x\}}(\bm o)} \, ,
\end{equation}
as the projection map $\rho^C_{\{x\}}$ is well-defined and continuous. 
Then for any $\prod_{x \in C} B_x \in \Fc_C$ (such that $B_x \subsetneq \Oc_x$ only for a finite number of $x \in C$) we have:
\begin{equation}
    \delta_{\bm o}\left(\prod_{x \in C} B_x\right) = \prod_{x \in C} \delta_{\bm o |_{\{x\}}}(B_x) \Mdot
\end{equation}

\textbf{\ref{it:fac} $\Rightarrow$ \ref{it:ext}.}\index{Measure}\index{Measurable!space}\index{$\sigma$-algebra}
Suppose that $e$ is realised by a factorisable hidden-variable model $\tuple{\bfLambda,p,k}$.\index{Hidden-variable model!CV!factorisable}
Write $k_x$ for $k_C|_{\enset{x}}$.
Define a measure $\mu$ on $\bm \Oc_\Xc$ as follows:
for any measurable set $\prod_{x \in \Xc} E_x \in \Fc_\Xc$ where $E_x \subsetneq \Oc_x$ only for a finite number of $x \in \Xc$,
the value of $\mu$ on it is given by
\begin{equation}\label{eq:mudef} \index{Extendability!CV}
\mu\left(\prod_{x\in \Xc}E_x\right) \defeq
\intg{\Lambda}{\left(\prod_{x\in \Xc} k_x(\dummy,E_x)\right)}{p}
=
\intg{\lambda \in \Lambda}{\left(\prod_{x\in \Xc} k_x(\lambda,E_x)\right)}{p(\lambda)}
\Mcomma
\end{equation}
where the product $\prod$ on the right-hand side is of real numbers in the interval $[0,1]$.
Note that the $\sigma$-algebra of $\bm \Oc_\Xc$ is given by the product topology and it is generated by such measurable sets; hence the equation above uniquely determines $\mu$ as a measure on $\bm \Oc_\Xc$.

Now, we show that this is a global section for the empirical probabilities.
Let $C \in \Mc$ and consider a measurable set $F = \prod_{x\in C}F_x \in \Fc_C$ with $F_x \subsetneq \Oc_x$ only for a finite number of $x \in C$. Then

\begin{calculation}
\mu|_{C}(F)
\just={definition of marginalisation}
\mu(F \times \Oc_{\Xc \setminus C})
\just={definition of $F$ and $\Oc_{U}$}
\mu\left(\prod_{x \in C}F_x \times \prod_{x \in \Xc \setminus C}\Oc_{x}\right)
\just={definition of $\mu$, Eq.~\eqref{eq:mudef}}
\intg{\Lambda}{\left(\prod_{x\in C} k_x(\dummy,F_x)\right)\left(\prod_{x \in \Xc \setminus C} k_x(\dummy,\Oc_x)\right)}{p}
\just={$k_x(\lambda,\dummy)$ probability measure so $k_x(\lambda,\Oc_x)=1$}
\intg{\Lambda}{\left(\prod_{x\in C} k_x(\dummy,F_x)\right)}{p}
\just={factorisability of the hidden-variable model}
\intg{\Lambda}{k_C(\dummy,\prod_{x\in C} F_x)}{p}
\just={definition of $F$}
\intg{\Lambda}{k_C(\dummy,F)}{p}
\just={$e$ is the empirical model corresponding to $\tuple{\Lambda,p,k}$}
e_C(F) \Mdot
\end{calculation}

\noindent Since the $\sigma$-algebra $\Fc_C$ of $\bm \Oc_C$ is generated by such measurable sets $F$ above and we have seen that $\mu|_C$ agrees with $e_C$ on these sets, we conclude that $\mu|_C = e_C$ as required. \index{$\sigma$-algebra}
\end{proof}

\section{Quantifying contextuality}
\label{sec02:quantifying}

Beyond questioning whether a given empirical behaviour is contextual or not,
it is also interesting to ask to what \textit{degree} it is contextual.
In discrete-variable scenarios, a very natural measure of contextuality
is the contextual fraction \cite{abramsky2011sheaf} (see Section~\ref{subsec01:quantifying}).
This measure was shown in \cite{abramsky2017contextual} to have
a number of very desirable properties.
It can be calculated using linear programming,
an approach that subsumes the more traditional approach to quantifying
nonlocality and
contextuality using Bell and noncontextual inequalities.
Indeed we
can understand the (dual) linear program as optimising over \textit{all} such inequalities for the scenario in question and returning
the maximum normalised
violation of \textit{any} Bell or noncontextuality inequality achieved by the given empirical model.
Crucially, the contextual fraction was also shown to \textit{quantifiably} relate to
quantum-over-classical advantages in specific informatic tasks \cite{abramsky2017contextual,mansfield2018quantum}.
Moreover it has been demonstrated to be a monotone with respect to the free
operations of resource theories for contextuality
\cite{abramsky2017contextual,duarte2018resource,abramsky2019comonadic}.

In this section, we consider how to carry those ideas
to the continuous-variable setting. 
The formulation as a linear optimisation problem and the attendant correspondence with Bell inequality violations
requires special care as one needs to use infinite-dimensional linear programming,
necessitating some extra assumptions on the measurable outcome spaces.

\subsection{The continuous-variable contextual fraction}

Asking whether a given behaviour is noncontextual amounts to
asking whether the empirical model is extendable, or in other words whether
it admits a deterministic hidden-variable model.
However, a more refined question to pose is:
\begin{center}
\textit{what fraction of the behaviour admits a deterministic hidden-variable model?}
\end{center}
This quantity is what we call the noncontextual fraction.
Similarly, the fraction of the behaviour that is left over
and that can thus be considered \textit{irreducibly} contextual
is what we call the contextual fraction.

\begin{definition}[Noncontextual fraction]
Let $e$ be an empirical model on the measurement scenario $\tuple{\Xc,\Mc,\bm \Oc}$.
The noncontextual fraction of $e$, written $\NCF(e)$, is defined as
\begin{equation}
    \sup\setdef{\mu(\Oc_\Xc)}{\mu \in \Measures(\bm \Oc_\Xc), \, \forall C \in \Mc, \mu|_C \leq e_C} \Mdot 
\end{equation}
Note that since $e_C$ is a probability measure on $\bm \Oc_C$ for all $C \in \Mc$ it follows that $\NCF(e) \in [0,1]$. The \emph{contextual fraction} of $e$, written $\CF(e)$, is given by $\CF(e) \defeq 1 - \NCF(e)$.
\end{definition}
\noindent This is the continuous-variable analogue of Definition \ref{def:ch01_NCF}.

\subsection{Monotonicity under free operations including binning}

In the discrete-variable setting, the contextual fraction was shown to be a monotone under a number of natural classical operations that 
transform and combine empirical models and control their use as resources, 
therefore constituting the `free' operations of a resource theory of contextuality \cite{abramsky2017contextual,duarte2018resource,abramsky2019comonadic}.

All of the operations on empirical models defined for discrete variables in \cite{abramsky2017contextual} -- viz. translations of measurements, transformation of outcomes, probabilistic mixing, product, and choice -- carry almost verbatim to our current setting.
One detail is that one must insist that the coarse-graining of outcomes be achieved by (a family of) measurable functions.
A particular example of practical importance is \emph{binning}, which is widely used in continuous-variable quantum information as
a method of discretising data by partitioning the outcome space $\bm \Oc_x$ for each measurement $x \in \Xc$ into
a finite number of `bins', \ie measurable sets.
Note that a binned empirical model is obtained by pushing forward along a family $\family{t_x}_{x\in \Xc}$ of outcome translations $\fdec{t_x}{\bm \Oc_x}{\bm \Oc'_x}$ where $\bm \Oc'_x$ is finite for all $x \in X$.

\begin{proposition}
If $e$ is an empirical model, and $e^\text{bin}$ is any discrete-variable empirical model obtained from $e$ by binning, then contextuality of $e^\text{bin}$ witnesses contextuality of $e$, and quantifiably gives a lower bound $\CF(e^\text{bin}) \leq \CF(e)$.
\end{proposition}

For the conditional measurement operation introduced in \cite{abramsky2019comonadic}, which allows for adaptive measurement protocols such as those used in measurement-based quantum computation \cite{raussendorf2001one},
one must similarly insist that the map determining the next measurement to perform based on the observed outcome of a previous measurement 
be a measurable function. Since we are, for now on, only considering scenarios with a finite number of measurements, this amounts to a partition of the outcome space $\bm \Oc_x$ of the first measurement, $x$, into measurable subsets labelled by measurements compatible with $x$, indicating which will be subsequently performed depending on the outcome observed for $x$.

The inequalities establishing monotonicity from
\cite[Theorem 2]{abramsky2017contextual}
also hold for continuous variables.
There is a caveat for the equality formula for the product of two empirical models: 
\begin{equation*}
\NCF(e_1 \otimes e_2) = \NCF(e_1)\NCF(e_2).
\end{equation*} 
Whereas the inequality establishing monotonicity ($\geq$) still holds in general, the proof establishing the other direction ($\leq$) makes use of duality of linear programs. Therefore, it only holds under the assumptions we will impose in the remainder of this section.

\subsection{Assumptions on the outcome spaces}
\label{subsec02:assumptions}

In order to phrase the problem of contextuality as a linear programming problem and establish the connection with violations of Bell inequalities, we need to impose some conditions on the measurable spaces
of outcomes.\index{Bell inequality!CV}
First we suppose that we have a \textit{finite number of measurement labels} \ie that $\Xc$ is finite.

From now on, 
we restrict attention to the case where
the outcome space $\bm \Oc_x$ for each measurement $x \in \Xc$
is the Borel measurable space for a \textit{compact} Hausdorff space,
\ie that the set $\Oc_x$ is a compact space
and $\Fc_x$ is the $\sigma$-algebra generated by its open sets, written $\Borel(\Oc_x)$.
Note that this includes most situations of interest in practice. 
In particular, it includes the case of measurements with \textit{outcomes in a bounded subspace of $\R$ or $\R^n$}. 
This is experimentally motivated since measurement devices are \textit{energetically bounded}.

One may wish to relax this assumption to allow for the case of a locally compact outcome space in order to include the case of measurements with outcomes in $\R$ or $\R^n$. Indeed $\R$ would be the canonical outcome space for the quadratures of the electromagnetic field. We address this issue in Subsection~\ref{subsec02:localcompact} and show that it can be reduced to the case of a compact outcome space. Thus the assumptions laid out here are not restrictive for the cases of interest, even theoretical ones. To summarise we essentially have the two following assumptions for the rest of this chapter:
\begin{enumerate}[label=(\roman*)]
    \item $\Xc$ is a finite set of measurement labels,
    \item for each $x \in \Xc$, the outcome set $\Oc_x$ is compact.
\end{enumerate} 
%
%

To obtain an infinite-dimensional linear program,\index{Linear program} we need to work with vector spaces (see Section~\ref{sec01:convexopti}). However as explained in Section~\ref{sec01:measure}, probability measures, or even finite or arbitrary measures, do not form one. We will therefore consider the set 
$\FSMeasures(\bm Y)$ of \emph{finite signed measures} on a measurable space ${ \bm Y = \tuple{Y,\Fc_Y} }$. Recall that these are functions $\fdec{\mu}{\Fc_Y}{\R}$ such that $\mu(\emptyset)=0$ and $\mu$ is $\sigma$-additive.
The set $\FSMeasures(\bm Y)$ forms a real vector space which includes the probability measures $\PMeasures(\bm Y)$.
When $Y$ is a compact Hausdorff space and $\bm Y = \tuple{Y,\Borel(Y)}$,
the Riesz--Markov--Kakutani representation theorem \cite{kakutani}\index{Riesz!Markov-Kakutani representation theorem} says that $\FSMeasures(\bm Y)$
is a concrete realisation of the
topological dual space of $C(Y)$,
the space of continuous real-valued functions on $Y$.
The duality is given by
$\tuple{\mu,f} \defeq \intg{\bm Y}{f}{\mu}$
for $\mu \in \FSMeasures(\bm Y)$ and $f \in C(Y)$.\index{Measure}
%

\subsection{Infinite-dimensional linear programming for computing the contextual fraction}

Consider an empirical model $e = \family{e_C}_{C \in \Mc}$ on a scenario $\tuple{\Xc,\Mc,\bm \Oc}$ satisfying the assumptions discussed above.
Calculation of its noncontextual fraction can be expressed as the infinite-dimensional linear programming problem \refprog{LP-CFCV}.\index{Empirical model!CV} \index{Measurement scenario!CV}
This is our primal linear program; its dual linear program is given by \refprog{DLP-CFCV}.\index{Linear program}
In what follows, we will see how to derive the dual and show that the optimal values of both programs coincide.

\leqnomode \index{Contextual fraction!CV}
\begin{flalign*}
    \label{prog:LP-CFCV}
    \tag*{(P-CF$^\text{\upshape CV}$)}
    \hspace{3cm}\left\{
    \begin{aligned}
        & \quad \text{Find } \mu \in \FSMeasures(\bm \Oc_\Xc) \\
        & \quad \text{maximising } \mu(\Oc_\Xc) \\
        & \quad \text{subject to:}  \\
        &  \hspace{1cm} \begin{aligned}
            & \forall C \in \Mc,\; \mu|_C \;\leq\; e_C \\
            & \mu \;\geq\; 0 \Mdot
        \end{aligned}
    \end{aligned}
    \right. &&
\end{flalign*}
%
\begin{flalign*}
    \label{prog:DLP-CFCV}
    \tag*{(D-CF$^\text{\upshape CV}$)}
    \hspace{3cm}\left\{
    \begin{aligned}
        & \quad \text{Find } \family{f_C}_{C \in \Mc} \in \prod_{C \in \Mc} C(\Oc_C) \\
        & \quad \text{minimising } \sum_{C \in \Mc} \intg{\Oc_C}{f_C}{e_C} \\
        & \quad \text{subject to:}  \\
        & \hspace{1cm} \begin{aligned}
            & \sum_{C \in \Mc} f_C \circ \rho^X_C \;\geq\; \mathbf{1}_{\Oc_\Xc} \\
            & \forall C \in \Mc,\; f_C \;\geq\; \mathbf{0}_{\Oc_C} \Mdot
        \end{aligned}
    \end{aligned}
    \right. &&
\end{flalign*}
\reqnomode

\noindent
We have written $\rho^X_C$ for the projection ${\Oc_\Xc}\longrightarrow{\Oc_C}$ as before,
and $\mathbf{1}_D$ (resp. $\mathbf{0}_D$)
for the constant function $D \longrightarrow \R$ that assigns the number $1$ (resp. $0$) to all elements of its domain $D$; in the above instance, to all $\bm g \in \Oc_\Xc$ (resp. all $\bm o \in \Oc_C$).
%

Analogues of these programs have been studied in the discrete-variable setting \cite{abramsky2017contextual}.
Note however that, in general, these continuous-variable linear programs are over infinite-dimensional spaces and thus not practical to compute directly.
For this reason, in Section~\ref{sec02:sdp} 
we will introduce a hierarchy of finite-dimensional
semidefinite programs that approximate the solution of \refprog{LP-CFCV} to arbitrary precision.

\subsubsection*{Retrieving the standard form of infinite-dimensional LPs}

Here we express the programs \refprog{LP-CFCV} and \refprog{DLP-CFCV} in the standard form of infinite-dimensional linear programs as introduced in Section~\ref{subsec01:LP} following \cite{barvinok02}.
We define the following spaces:
\begin{itemize}
    \item $\displaystyle E_1 \defeq \FSMeasures(\bm \Oc_\Xc)$.
    \item $\displaystyle F_1 \defeq C(\Oc_\Xc)$, the dual space of $E_1$.
    \item $\displaystyle E_2 \defeq \prod_{C \in \Mc} \FSMeasures(\bm \Oc_C)$.
    \item $\displaystyle F_2 \defeq \prod_{C \in \Mc} C(\Oc_C) $, the dual space of $E_2$.
\end{itemize}
The  dualities $\langle \dummy , \dummy \rangle_1 : E_1 \times F_1 \longrightarrow \R$ and $\langle \dummy , \dummy \rangle_2 : E_2 \times F_2 \longrightarrow \R$ are defined as follows:
\begin{align*}
    & \forall \mu \in E_1, \; \forall f \in F_1, & & \hspace{-2cm} \langle \mu,f \rangle_1 \defeq \intg{\Oc_\Xc}{f}{\mu} \\
    & \forall (\nu_C) \in E_2, \; \forall (f_C) \in F_2, & & \hspace{-2cm} \langle  (\nu_C),(f_C) \rangle_2 \defeq \sum_{C \in \Mc} \intg{\Oc_C}{f_C}{\nu_C}\, ,
\end{align*}
where, for simplicity, we have omitted $C \in \Mc$ as a subscript for the families of functions.
We fix $K_1$ to be the convex cone\index{Cone} of positive measures in $E_1 = \FSMeasures(\bm \Oc_\Xc)$ and $K_2$ to be the convex cone of families of positive measures in $E_2 = \prod_{C \in \Mc} \FSMeasures(\bm \Oc_C)$. Then $K_1^*$ is the convex cone of positive function in $F_1 = C(\Oc_\Xc)$ and $K_2^*$ is the convex cone of families of positive functions in $F_2 = \prod_{C \in \Mc} C(\Oc_C)$.

Let $\fdec{A}{E_1}{E_2}$ be the following linear transformation:
\begin{equation*}
    \forall \mu \in E_1, \quad A(\mu) \defeq (\mu|_C)_{C \in \Mc} \in E_2 \Mdot
\end{equation*}
We also define the linear transformation $\fdec{A^*}{F_2}{F_1}$ as:
\begin{equation*}
    \forall (f_C) \in F_2, \quad A^*((f_C)) \defeq \sum_{C \in \Mc} f_C \circ \rho^X_C \in F_1 \Mdot
\end{equation*}
We can verify that $A^*$ is the dual transformation of $A$: $\forall \mu \in E_1, \forall (f_C) \in F_2$, we have
\begin{align}
\langle A(\mu), (f_C) \rangle_2 & = \langle  (\mu|_C), (f_C)  \rangle_2 \\
& = \sum_{C \in \Mc} \intg{\Oc_C}{f_C}{\mu|_C} \\
& = \intg{\Oc_\Xc}{\sum_{C \in \Mc} f_C \circ \rho^X_C }{\mu} \\
& = \langle  \mu , \sum_{C \in \Mc} f_C \circ \rho^X_C  \rangle_1 \\
& = \langle  \mu, A^*((f_C))  \rangle_1 \Mdot
\end{align}

Now fixing the vector function in the objective to be $c \defeq - \bm 1_{\Oc_\Xc} \in F_1$ and the vector in the constraints to be $b \defeq (-(e_C)_{C \in \Mc}) \in E_2$, the program \refprog{LP-CFCV} (resp. \refprog{DLP-CFCV})  can be expressed as in the standard form given in \refprog{LPstdform} (resp. \refprog{DLPstdform}). Note that the minus sign in the vectors $c$ and $b$ was added because we choose the primal program in the standard form to be a minimisation problem while the primal program at hand is a maximisation problem.\index{Linear program}

\subsubsection*{Deriving the dual via the Lagrangian}\index{Lagrangian method}

We now give an explicit derivation of \refprog{DLP-CFCV} as the dual of \refprog{LP-CFCV} via the Lagrangian method which is widely used in optimisation theory.
We do not take into account positivity constraints as constraints in the Lagrangian because they are already satisfied in the convex cones\index{Cone} in which one optimises.
To derive the dual program we introduce $\lvert \Mc \rvert$ dual variables---one continuous map $f_C \in C(\Oc_C)$ for each $C\in \Mc$---to account for the constraints $\mu|_{C} \leq e_C$.
We then define the Lagrangian $\Lc : K_1 \times K_2^* \longrightarrow \R$ as
\begin{equation}
    \Lc\left(\mu,(f_C)\right) \defeq \underbrace{\mu(\Oc_\Xc)\vphantom{\sum_{C \in \Mc}}}_{\text{objective}} + \underbrace{\sum_{C \in \Mc} \intg{\Oc_C}{f_C}{(e_C - \mu|_C)}}_{\text{constraints}} \Mdot
\end{equation}
The primal program \refprog{LP-CFCV} corresponds to 
\begin{equation}
    \sup_{\mu \in K_1} \; \inf_{(f_C) \in K_2^*} \; \Lc(\mu,(f_C)) \Mdot
\end{equation}
The infimum here imposes the constraints that $\mu|_C \leq e_C$ for all $C \in \Mc$, for otherwise the associated optimisation problem is not well-defined (or rather it necessarily has a value of minus infinity).
If these constraints are satisfied, then because of the infimum over $(f_C) \in K_2^*$, the second term of the Lagrangian vanishes yielding, as expected, the objective of the primal problem.
Expressing the dual problem amounts to permuting the infimum and the supremum. Thus we need to rewrite the Lagrangian as:
\begin{align}\index{Lagrangian method}
    \Lc(\mu,(f_C)) \quad&=\quad \mu(\Oc_\Xc) + \sum_{C \in \Mc} \intg{\Oc_C}{f_C}{(e_C - \mu|_C)} \\
                     \quad&=\quad \intg{\Oc_\Xc}{\mathbf{1}_{\Oc_\Xc}}{\mu} + \sum_{C \in \Mc} \intg{\Oc_C}{f_C}{e_C}  - \sum_{C \in \Mc} \intg{\Oc_C}{f_C}{\mu|_C} \\
                \quad&=\quad \intg{\Oc_\Xc}{\mathbf{1}_{\Oc_\Xc}}{\mu} + \sum_{C \in \Mc} \intg{\Oc_C}{f_C}{e_C}  - \sum_{C \in \Mc} \intg{\Oc_\Xc}{f_C \circ \rho^X_C}{\mu} \label{eq:ch02_lag1}\\      
                     \quad&=\quad \intg{\Oc_\Xc}{\mathbf{1}_{\Oc_\Xc}}{\mu} + \sum_{C \in \Mc} \intg{\Oc_C}{f_C}{e_C} - \intg{\Oc_\Xc}{\left( \sum_{C \in \Mc} f_C \circ \rho^X_C \right)}{\mu}  \label{eq:ch02_lag2}\\ 
                     \quad&=\quad \sum_{C \in \Mc} \intg{\Oc_C}{f_C}{e_C} + \intg{\Oc_\Xc}{\left( \mathbf{1}_{\Oc_\Xc} - \sum_{C \in \Mc} f_C \circ \rho^X_C \right)}{\mu} \Mdot
\end{align}
where Eq.~(\ref{eq:ch02_lag1}) is obtained by the push-forward operation and Eq.~(\ref{eq:ch02_lag2}) is obtained by linearity of the integral (and the finiteness of $\Mc$).
The dual program \refprog{DLP-CFCV} indeed corresponds to
\begin{equation}
    \inf_{(f_C) \in K_2^*} \; \sup_{\mu \in K_1} \; \Lc(\mu,(f_C)) \Mdot
\end{equation}
The supremum imposes that $\sum_{C \in \Mc} f_C \circ \rho^X_C \geq \mathbf{1}_{\Oc_\Xc}$ on $\Oc_\Xc$ since otherwise the Lagrangian diverges. If this constraint is satisfied, then because of the supremum the second term vanishes yielding, as expected, the objective of the dual problem.

\subsection{Zero duality gap}\index{Strong duality!of linear programs}
A key result about the noncontextual fraction, which is essential in establishing the connection to Bell inequality violations, is that \refprog{LP-CFCV} and \refprog{DLP-CFCV} are strongly dual, in the sense that no gap exists between their optimal values \ie $\val{\text{P-CF}^{\text{CV}}} = \val{\text{D-CF}^{\text{CV}}}$.
Strong duality holds in finite linear programming when Slater's conditions are met---which will be the case for examples of interest. However it does not hold in general in the infinite-dimensional case. Below we prove a strong duality result for our programs. 

\begin{proposition}
\label{prop:zero_duality}
Problems {\upshape \refprog{LP-CFCV}} and {\upshape \refprog{DLP-CFCV}} have zero duality gap. Thus their optimal values satisfy:
{\upshape
\begin{equation}
    \val{\text{P-CF}^{\text{CV}}}  = \val{\text{D-CF}^{\text{CV}}} = \NCF(e)
\end{equation}
}and there exists a primal optimal plan for the primal program {\upshape \refprog{LP-CFCV}}.
\end{proposition}
\begin{proof}
This proof relies on \cite[Theorem 7.2]{barvinok02} (recalled in Section~\ref{subsec01:LP}, Theorem~\ref{th:ch01_strongduality}).
Because $\mu_0= \bm 0_{\bm \Oc_\Xc}$---the measure that assigns $0$ to every measurable set of $\bm \Oc_\Xc$--- is a feasible solution for \refprog{LP-CFCV} and the noncontextual fraction lies between 0 and 1, \refprog{LP-CFCV} is consistent with finite value. Thus it suffices to show that the following cone\index{Cone}
\begin{equation}
    \Kc = \{ \left( A(\mu), \langle \mu,c \rangle_1 \right) \, : \, \mu \in K_1 \} = \{ \left(\, (\mu|_C)_C, \mu(\Oc_\Xc)\, \right) \, : \, \mu \in K_{1} \}
\end{equation}
is weakly closed in $E_2 \oplus \R$ (\ie closed in the weak topology of $K_1$) where we recall that $K_1$ is the convex cone of positive measures in $E_1 = \FSMeasures(\bm \Oc_\Xc)$.

We first notice that the linear transformation $A$ is a bounded linear operator and thus continuous. Boundedness comes from the fact that, $\forall \mu \in K_1$,
\begin{align}
    \norm{A(\mu)}_{E_2} &= \norm{(\mu|_C)_C}_{E_2} \\
    & = \sum_{C \in \Mc} \norm{\mu|_C}_{\FSMeasures(\bm \Oc_C)} \\
    & \leq \sum_{C \in \Mc}  \norm{\mu}_{E_1} \label{eq:ch02_boundedness}\\
    &  = \vert \Mc \vert \norm{\mu}_{E_1} \Mcomma
\end{align}
where we take the strong topology---\ie the norm induced by the total variation\index{Total variation distance} distance---on finite-signed measure spaces. It is defined as:
\begin{equation*}
    \norm{\mu}_{\FSMeasures(\bm U)} = \vert \mu \vert(U)\Mdot
\end{equation*}
We also equip the finite product space $E_2 = \prod_{C \in \Mc} \FSMeasures(\bm \Oc_C)$ with the norm obtained by summing\footnotemark\ the individual total variation\index{Total variation distance} norms. Eq.~(\ref{eq:ch02_boundedness}) is due to the fact that $\mu \in K_1$ so this is a positive measure and thus $\norm{\mu_C}_{\FSMeasures(\bm \Oc_C)} \leq \norm{\mu}_{E_1}$. This, of course, extends to the weak topology.

\footnotetext{Categorically, this is a coproduct.}
Secondly, we consider a sequence 
$(\mu^k)_{k \in \N}$ in $K_1$
and we want to show that the accumulation point $((\Theta_C)_C,\lambda) = \lim_{k \rightarrow \infty} \left( A(\mu^k), \langle \mu^k,c \rangle_1 \right)$ 
belongs to $\Kc$, where $\Theta = (\Theta_C)_C \in E_2$ and $\lambda \in \R$. 
If we consider the product of indicator functions $(\mathbf{1}_{\Oc_C})_C \in F_2$ then 
$\langle A(\mu^k), (\mathbf{1}_{\Oc_C}) \rangle_2 = \sum_{C \in \Mc} \mu|_C^k(\Oc_C) \longrightarrow_{k} \sum_{C \in \Mc} \Theta_C(\Oc_C) < \infty$ as  $\Theta_C$ is a finite measure for all maximal contexts $C \in \Mc$. Then because $\Mc$ is a covering family of $\Xc$, $\forall k \in \N, \,\mu^k(\Oc_\Xc) \leq \sum_{C \in \Mc} \mu^k|_C(\Oc_C) < \infty$.
Since $(\mu^k) \in K_1^\N$ is a sequence of positive measures, this implies that $(\mu^k)$ is bounded.
Next, by weak-$*$ compactness of the unit ball (Alaoglu's theorem \cite{luenberger97}), there exists a subsequence $(\mu^{k_i})_{k_i}$ that converges weakly to an element $\omega \in K_1$. 
By continuity of $A$, it yields that the accumulation point is such that $((\Theta_C)_C,\lambda) = \left( A(\omega), \langle \omega,c \rangle_1 \right) \in \Kc $.
\end{proof}

\subsection{The case of local compactness}
\label{subsec02:localcompact}

We now focus on cases where the outcome space might be only locally compact which include most theoretical situations that are of interest in practice; for instance $\R$ could be the outcome space for the position and momentum operators. 

For each measurement $x \in \Xc$, $\bm \Oc_x$
is supposed to be the Borel measurable space for a second-countable locally compact Hausdorff space,
\ie that the set $\Oc_x$ is equipped with a second-countable locally compact Hausdorff topology
and $\Fc_x$ is still the $\sigma$-algebra generated by its open sets, written $\Borel(\Oc_x)$.
Second countability and Hausdorffness of two spaces $Y$ and $Z$ suffice to show that
the Borel $\sigma$-algebra of the product topology is the tensor product of the Borel $\sigma$-algebras, \ie $\Borel(Y \times Z) = \Borel(Y) \otimes \Borel(Z)$
\cite[Lemma 6.4.2 (Vol.~2)]{Bogachev}. Hence, these assumptions
guarantee that $\bm \Oc_U$ for $U \in \Pc(\Xc)$ is the Borel $\sigma$-algebra of the product topology on $\Oc_U= \prod_{x \in U}\Oc_x$. These product spaces are also second-countable, locally compact, and Hausdorff as all three properties are preserved by finite products.
When $Y$ is a second-locally compact Hausdorff space and $\bm Y = \tuple{Y,\Borel(Y)}$,
the Riesz--Markov--Kakutani representation theorem \cite{kakutani} says that $\FSMeasures(\bm Y)$
is a concrete realisation of the
topological dual space of $C_0(Y)$,
the space of continuous real-valued functions on $Y$
that vanish at infinity.\footnote{A function $\fdec{f}{Y}{\R}$ on a locally compact space $Y$ is said to \emph{vanish at infinity} if the set $\setdef{y \in Y}{\|f(x)\|\geq\epsilon}$ is compact for all $\epsilon>0$.}
The duality is given by
$\tuple{\mu,f} \defeq \intg{\bm Y}{f}{\mu}$
for $\mu \in \FSMeasures(\bm Y)$ and $f \in C_0(Y)$\footnotemark. Note that when $Y$ is compact (as treated above), we only need to consider functions in $C(Y)$.
\footnotetext{This theorem holds more generally for locally compact Hausdorff spaces if one considers only (finite signed) Radon measures, which are measures that play well with the underlying topology. However, second-countability, together  with local compactness and Hausdorffness, guarantees that every Borel measure is Radon \cite[Theorem 7.8]{Folland}.}

Next, we show that we can approximate the linear program \refprog{LP-CFCV}\footnotemark\ by a slightly modified linear program defined on the space of finite measures on a measurable compact subspace of $\bm \Oc_\Xc$. 
The idea is to approximate to any desired error the mass of a finite measure on a locally compact set by the mass of the same measure on a compact subset. This naturally comes from the notion of \textit{tightness} of a measure.

\footnotetext{Here we will still use the form of the program \refprog{LP-CFCV} though throughout this subsection one has to keep in mind that it is defined over finite-signed measures on a \textit{locally compact} space rather than a compact space.}

\begin{definition}[Tightness of a measure] \index{Measure}
A measure $\mu$ on a metric space $U$ is said to be \textit{tight} if for each $\epsilon > 0$ there exists a compact set $U_{\epsilon} \subseteq U$ such that $\mu(U \setminus U_{\epsilon}) < \epsilon$.
\end{definition}

\noindent Then we need to argue that every measure we will consider is tight. This is a result of the following theorem:

\begin{theorem}[\cite{parthasarathyprobability1967}]
If $S$ is a complete separable metric space, then every finite measure on $S$ is tight.
\label{th:tightness}
\end{theorem}

\noindent For $x \in \Xc$, $\bm \Oc_x$ is a second-countable locally compact Hausdorff space, thus a Polish space \ie a separable completely metrisable topological space. For this reason, the above theorem applies. We are now ready to state and prove the main theorem of this subsection.

\begin{theorem} \index{Linear program}
The linear program \refprog{LP-CFCV} defined over finite-signed measures on a locally compact space can be approximated to any desired precision $\varepsilon$ by a linear program $(\text{\upshape P-CF}^{\text{\upshape CV},\varepsilon})$ defined over finite signed measures on a compact space.
\label{th:approximate_local}
\end{theorem}

\begin{proof} \index{Measure}
Fix $\epsilon > 0$. Let $C \in \Mc$ be a given context and $x \in C$ a given measurement label within that context. 
Because $e_C$ is a probability measure on $\Oc_C$, the marginal measure $e_C|_{\{x \}}$ is a finite measure on $\Oc_x$. 
Following Theorem~\ref{th:tightness}, $e_C|_{\{x \}}$ is tight and there exists a compact subset $K_x^{\epsilon,C} \subseteq \Oc_x$ such that: $e_C|_{\{x \}}(\Oc_x \setminus K_x^{\epsilon,C}) \leq \epsilon$. We apply this procedure\footnotemark\ for every context and for all measurements in a context. 
We now define the compact set: 
\begin{equation*}
\Oc_x^{\epsilon} \defeq \bigcup\limits_{C | x \in C} K_x^{\epsilon,C}.    
\end{equation*}
The previous definition is essential to ensure a noncontextual cutoff of the outcome set which, in turn, ensures the good definition of a compact subset for each measurement label independent of the context. For some subset of measurement labels $U \subseteq \Xc$, we define the compact set $\Oc_U^{\epsilon} \defeq \prod_{x \in U} \Oc_x^{\epsilon}$. For every context $C \in \Mc$ and for every measurement label $x \in C$, we now have that $K_x^{\epsilon,C} \subseteq \Oc_x^{\epsilon}$ and thus $e_C|_{\{x \}}(\Oc_x \setminus \Oc_x^{\epsilon}) \leq \epsilon$. Note that due to the compatibility condition, we can write $e_C|_{\{x\}}$ as $e_{\{x\}}$ for any context. 

\footnotetext{Note that there exists proofs that explicitly constructs the approximating sets $K_x^{\epsilon,C}$ (see \cite{Orbanz2011ProbabilityTI}) based on the separability of the underlying spaces. It makes this construction feasible in practice and justifies this approach.}

Let $\mu$ be any feasible solution of \refprog{LP-CFCV} defined over finite-signed measures on a locally compact space. 
Due to the constraints of \refprog{LP-CFCV} we have that $\forall x \in \Xc, \, \mu|_{\{x\}} \leq e_{\{x\}}$.
Then:
\begin{align}
    \mu(\Oc_\Xc \setminus \Oc^{\varepsilon}_\Xc) & = \mu \left( \prod_{x \in \Xc } \Oc_{x} \setminus \prod_{x \in \Xc } \Oc_{x}^{\epsilon}  \right) \\
    & = \mu \left( \prod_{x \in \Xc } ( \Oc_{x} \setminus \Oc_{x}^{\epsilon} ) \right) \\
    & = \prod_{x \in \Xc} \mu|_{\{x\}} (\Oc_x \setminus \Oc_x^\varepsilon) \\
    & \leq \prod_{x \in \Xc} e_{\{x\}} (\Oc_x \setminus \Oc_x^\varepsilon) \\
    & \leq \epsilon^{\vert \Xc \vert} \Mdot
\end{align}

We now define the linear program $(\text{P-CF}^{\text{CV},\varepsilon})$ which has the same form as \refprog{LP-CFCV} though the unknown measures are taken from $\FSMeasures(\bm \Oc_\Xc^{\epsilon})$ where $\bm \Oc_\Xc^{\epsilon} = \tuple{\Oc_\Xc^{\epsilon},\Borel(\Oc_\Xc^{\epsilon})}$. 
We would like to state that $(\text{P-CF}^{\text{CV},\varepsilon})$ approximates \refprog{LP-CFCV} up to $\epsilon$; \ie that their values are $\varepsilon$-close. 
The missing ingredient from the previous chain of inequalities is that given an optimal measure $\mu^*$ satisfying \refprog{LP-CFCV}, we do not know whether an optimal solution $\mu^*_{\epsilon}$ of $(\text{P-CF}^{\text{CV},\varepsilon})$ is necessarily the restriction of $\mu^*$ to $\Oc_\Xc^{\epsilon}$. In fact, it is possible that we do not even have a unique optimal solution. However we only need to prove that they have the same mass on $\Oc_\Xc^\varepsilon$, \ie $\mu_{\epsilon}^*(\Oc_\Xc^{\epsilon}) = \mu^*|_{\Oc_\Xc^{\epsilon}}(\Oc_\Xc^{\epsilon}) $.
By contradiction, suppose it does not hold. Then because $\mu^*_{\epsilon}$ is an optimal value of $(\text{P-CF}^{\text{CV},\varepsilon})$, we must have $\mu_{\epsilon}^*(\Oc_\Xc^{\epsilon}) > \mu^*|_{\Oc_\Xc^{\epsilon}}(\Oc_\Xc^{\epsilon}) $. From this we construct a new measure $\tilde{\mu}$ on $\bm \Oc_\Xc$ which equals to $\mu^*_{\epsilon}$ on $\bm \Oc_\Xc^{\epsilon}$ and $\mu^*$ on $\bm \Oc_\Xc \setminus \Oc_\Xc^{\epsilon}$. It satisfies all constraints and furthermore $\tilde{\mu}(\Oc_\Xc) > \mu^*(\Oc_\Xc)$. This contradicts that $\mu^*$ is an optimal solution of \refprog{LP-CFCV}. Thus necessarily $\mu_{\epsilon}^*(\Oc_\Xc^{\epsilon}) = \mu^*|_{\Oc_\Xc^{\epsilon}}(\Oc_\Xc^{\epsilon})$.

The linear program $(\text{P-CF}^{\text{CV},\varepsilon})$ defined on a compact space has indeed a value $\varepsilon$-close to the original program \refprog{LP-CFCV}. 
\end{proof}

In conclusion of this subsection, we can approximate the problem of finding the noncontextual fraction with the assumption that the outcome spaces are locally compact spaces by the same problem defined on compact subspaces and it suffices to restrict the study to the case of compact outcome spaces. \index{Linear program} \index{Measure}
Thus for the rest of this chapter, we indeed restrict ourselves to compact outcome spaces.

\section{Continuous generalisation of Bell inequalities}\label{sec02:bellinequality} \index{Bell inequality!CV} 

The dual program \refprog{DLP-CFCV} is of particular interest in its own right.
As we now show, it can essentially be understood as
computing a continuous-variable
`Bell inequality' that is optimised to the empirical data.
Making the change of variables $\beta_C \defeq |\Mc|^{-1} \mathbf{1}_{\Oc_C}-f_C \in C(\Oc_C)$ for each
$C \in \Mc$, the dual program \refprog{DLP-CFCV} transforms to:
\leqnomode
\begin{flalign*}
    \label{prog:B-CFCV}
    \tag*{(B-CF$^\text{\upshape CV}$)}
    \hspace{3cm}\left\{
    \begin{aligned}
        & \quad \text{Find } \family{\beta_C}_{C \in \Mc} \in \prod\limits_{C \in \Mc} C(\Oc_C) \\
        & \quad \text{maximising } \sum_{C \in \Mc} \intg{\Oc_C}{\beta_C}{e_C} \\
        & \quad \text{subject to:}  \\
        & \hspace{1cm} \begin{aligned}
            & \sum_{C \in \Mc} \beta_C \circ \rho^\Xc_C \;\leq\; \mathbf{0}_{\Oc_\Xc} \\
            & \forall C \in \Mc,\; \beta_C \;\leq\; \vert \Mc \vert^{-1} \mathbf{1}_{\Oc_C} \Mdot
        \end{aligned}
    \end{aligned}
    \right. &&
\end{flalign*}
\reqnomode
This program directly computes the contextual fraction $\CF(e)$\index{Contextual fraction!CV} instead of the noncontextual fraction. 
It maximises, subject to constraints,
the total value obtained by integrating
these functionals context-wise against the empirical model in question.
The first set of constraints---a generalisation of a system of linear inequalities determining a Bell inequality ---
ensures that, for noncontextual empirical models, the value of the program is
at most $0$, since any such model extends to a measure $\mu$ on $\bm \Oc_\Xc$ such
that $\mu(\Oc_\Xc) = 1$.
The final set of constraints acts as a normalisation condition on the
value of the program, ensuring that it takes values in the interval $[0,1]$ for any empirical model.
Any family of functions $\beta = (\beta_C) \in F_2$ satisfying the constraints will thus result in what can be regarded as a
generalised Bell inequality,
\begin{equation}
    \sum_{C \in \Mc} \intg{\Oc_C}{\beta_C}{e_C} \leq 0 \, ,
\end{equation}
which is satisfied by all noncontextual empirical models.

\begin{definition}[Form on a measurement scenario]
A form $\beta$ on a measurement scenario $\tuple{\Xc,\Mc,\bm \Oc}$ is a family $\beta = (\beta_C)_{C \in \Mc}$ of functions $\beta_C \in C(\Oc_C)$ for all $C \in \Mc$.
Given an empirical model $e$ on $\tuple{\Xc,\Mc,\bm \Oc}$,
the value of $\beta$ on $e$ is 
\begin{equation}
    \langle e,\beta \rangle_{_2} \defeq \sum_{C \in \Mc} \intg{\Oc_C}{\beta_C}{e_C} \Mdot
\end{equation}
The norm of $\mathbf{\beta}$
is given by
\begin{equation}
    \|\beta\| \defeq \sum_{C \in \Mc}\|\beta_C\| = \sum_{C\in\Mc}\sup \setdef{\beta_C(\bm o)}{\bm o\in \Oc_C}\Mdot
\end{equation}
\end{definition}

\begin{definition}[Inequality]
An inequality $(\mathbf{\beta},R)$ on a measurement scenario $\tuple{\Xc,\Mc,\bm \Oc}$ is given by $\tuple{\beta,R}$ where $\beta$ is a form on $\tuple{\Xc,\Mc,\bm \Oc}$ together with a bound $R \in \R$.
An empirical model $e$ is said to satisfy the inequality $\tuple{\beta,R}$ if the value of $\beta$ on $e$ is below the bound, 
\ie $\langle e,\beta \rangle_{2} \leq R$.
\end{definition}

\begin{definition}[Continuous Bell inequality]\index{Bell inequality!CV}
An inequality $(\beta,R)$ is said to be a continuous Bell inequality if it is satisfied by all noncontextual empirical models, \ie if for any noncontextual model $d$ on $\tuple{\Xc,\Mc,\Oc}$, it holds that 
$\langle d,\beta \rangle_{_2} \leq R$.
\end{definition}
\noindent A continuous Bell inequality $(\beta,R)$ establishes a bound $\langle e,\beta \rangle_{2}$ amongst noncontextual models $e$. For more general models, the value of $\beta$ on $e$ is only limited by the algebraic bound $\norm{\beta}$. A model $e$ reaching $\norm{\beta}$ is likely to be signalling. In the following, we will only consider inequalities $(\beta,R)$ for which $R < \norm{\beta}$ excluding inequalities trivially satisfied by all empirical models.
\begin{definition}[Normalised violation of a continuous Bell inequality]
The normalised violation of a generalised Bell inequality $(\beta,R)$ by an empirical model $e$ is
\begin{equation}
\frac{\max\enset{0,\langle e,\beta \rangle_{_2} - R}}{\|\beta\|-R} \Mcomma
\end{equation}
the amount by which its value $\langle \beta, e \rangle_{_2}$ exceeds the bound $R$ normalised by the maximal `algebraic' violation.
\end{definition}

An optimal plan $\beta \in F_2$ for \refprog{B-CFCV} defines a continuous Bell inequality $\tuple{\beta,0}$ that is maximally violated by the empirical model $e$. Note however that an optimal solution might not be feasible---for instance the optimum might be achieved by a discontinuous function---in which case there will exist a sequence of feasible solutions converging to such a discontinuous function. 

The above definition restricts to the usual notions of Bell inequality and noncontextual inequality in the discrete-variable case
and is particularly close to the presentation in \cite{abramsky2017contextual}.
The following theorem also generalises to continuous variables the main result of \cite{abramsky2017contextual} (see Theorem \ref{th:ch01_Belltheorem}).

\begin{theorem}
Let e be an empirical model on a measurement scenario $\tuple{\Xc,\Mc,\bm \Oc}$.
\begin{enumerate*}[label=(\roman*)]
\item
The normalised violation by $e$ of any Bell inequality is at most $\CF(e)$;
\item \label{it:ch02_BIth}
if $\CF(e) > 0$ then for every $\varepsilon > 0$ there exists a Bell inequality whose normalised violation by $e$ is at least $\CF(e)-\epsilon$.
\end{enumerate*}
\end{theorem}
Item \ref{it:ch02_BIth} is slightly modified compared to the discrete analogue version because there is no guarantee that there exists an optimal plan for the dual program \refprog{DLP-CFCV}. In particular, its optimal solution might be achieved by a discontinuous function that can be approximated by continuous ones. Hence the modification of \ref{it:ch02_BIth} with a normalised violation $\varepsilon$-close to $\CF(e)$.
\begin{proof}
As in the discrete-variable case (see \cite{abramsky2017contextual} for a detailed proof), the proof follows directly from the definitions of the linear programs, and from strong duality, \ie the fact that their optimal values coincide (Proposition \ref{prop:zero_duality} above).
\end{proof}

\section{Approximating the contextual fraction with SDPs} \index{Semidefinite program} \index{Linear program}
\label{sec02:sdp}

In Section~\ref{sec02:quantifying}, 
we presented the problem of computing the noncontextual fraction\index{Contextual fraction!CV} as an infinite-dimensional linear program.
Although of theoretical importance, it cannot be run to perform the actual, numerical computation of this quantity.
Here we exploit the link between measures and their sequence of moments to derive a hierarchy of truncated finite-dimensional semidefinite programs which are a relaxation of the original primal problem \refprog{LP-CFCV}. Dual to this vision, we can equivalently exploit the link between positive polynomials and their sum-of-squares representation to derive a hierarchy of semidefinite programs which are a restriction of the dual problem \refprog{DLP-CFCV}. 
We further prove that the optimal values of the truncated programs
converge monotonically to the noncontextual fraction.
This makes use of global optimisation techniques developed by Lasserre and Parrilo \cite{lasserre10,parrilo2003semidefinite} and further developed in \cite{lasserre2011new}.
These techniques were introduced in Subsection~\ref{subsec01:Lasserrehierarchy} and we will use the same notations throughout this section. 
Another extensive and well presented reference on the subject is \cite{Laurent2009}.
In Subsection~\ref{subsec02:hierarchy}, we derive a hierarchy of SDPs to approximate the contextual fraction and we show its convergence in Subsection~\ref{subsec02:convergence}.

\subsection{Hierarchy of semidefinite relaxations for computing \texorpdfstring{$\NCF (e)$}{NCF(e)}}\index{Lasserre hierarchy}
\label{subsec02:hierarchy}

We fix a measurement scenario $\tuple{\Xc,\Mc,\bm \Oc}$ and an empirical model $e$ on this scenario. 
We will restrict our attention to outcome spaces of the form detailed in subsection~\ref{subsec02:assumptions}.
Let $d = \vert \Xc \vert \in \N^*$ so that $\Oc_\Xc$ is a Borel subset of $\R^d$.
As a prerequisite, we first need to compute the sequences of moments associated to measures $(e_C)_{C \in \Mc}$ derived from the empirical model. For $C \in \Mc$, let $\bm y^{e,C} = (y^{e,C}_{\bm \alpha})_{\bm \alpha \in \N^d}$ be the sequence of all moments of $e_C$. \index{Moment sequence}
For a given $k \in \N$ which will fix the level of the hierarchy, we only need to compute a finite number $s(k)$ of moments for all contexts. These will be the inputs of the program. 

Below, we derive a hierarchy of SDP relaxations for the primal program \refprog{LP-CFCV} such that their optimal values converge monotonically to $\val{\text{P-CF}^{\text{CV}}} = \NCF(e)$. We start by discussing the assumptions we have to make on the outcome space. Then we derive the hierarchy based on, first, the primal program and then the dual program and we further show that these formulations are indeed dual. Finally we prove the convergence of the hierarchy.

\subsubsection*{Further assumptions on the outcome space?}
We already made the assumptions mentioned in Subsection~\ref{subsec02:assumptions} for the outcome spaces $\bm \Oc = (\bm \Oc_x)_{x \in \Xc}$ noting that they are not restrictive when considering actual experimental or theoretical applications. However we would like to meet the assumptions detailed in Assumption~\ref{ass:ch01_algebraic} for the global outcome space $\Oc_\Xc$ so that both Theorems \ref{th:ch01_putinar} and \ref{th:ch01_representingmeasure} apply in our setting. 

Assumption~\ref{ass:ch01_algebraic}~\ref{ass:ch01_compact} is de facto met because we already assumed that for all $x \in \Xc$, $\Oc_x \subset \R$ is compact. Remember that, at worst, $\Oc_x$ is locally compact and we saw that this can be reduced to the compact case in Subsection~\ref{subsec02:localcompact}.

Let us discuss Assumption~\ref{ass:ch01_algebraic}~\ref{ass:ch01_semialgebraic}. We have that $\Oc_\Xc = \prod_{x \in \Xc} \Oc_x$ with $\Oc_x \subset \R$ compact. If $\Oc_x$ is disconnected, we can always complete it into a connected space by attributing a measure zero to the added parts for all measures $e_C$ whenever $x \in C$. Then because $\Oc_x$ is compact, it is bounded and it can be described by two constant polynomials: there exists $a_x,b_x \in \R$ such that $\Oc_x = \left[a_x,b_x\right]$. This makes $\Oc_\Xc$ a polytope so in particular, it is semi-algebraic. We write it as 
\begin{equation}
\Oc_\Xc = \setdef{\bm x \in \R^d}{\forall j = 1,\dots,m, \; g_j(\bm x) \geq 0}
\end{equation}
for some polynomials $g_j \in \Rpolm$ of degree 1.

As noted in \cite{lasserre10}, Assumption~\ref{ass:ch01_algebraic}~\ref{ass:ch01_archimidean} is not very restrictive. For instance, it is satisfied when the set is a polytope. This is the case for $\Oc_\Xc$.

Thus there is no need for further assumptions than what we already assumed in Subsection~\ref{subsec02:assumptions} to apply the main results we presented in Section~\ref{subsec01:Lasserrehierarchy}. 

\subsubsection*{Relaxation of the primal program}\index{Linear program} \index{Semidefinite program}

The program \refprog{LP-CFCV} can be relaxed so that a converging hierarchy of SDPs can be derived. The program \refprog{LP-CFCV} is essentially a maximisation problem on finite-signed Borel measures with additional constraints such as the fact that these are proper measures (\ie they are nonnegative). We will represent a measure by its moment sequence and use conditions for which this moment sequence has a (unique) representing Borel measure (see Subsection~\ref{subsubsec01:moment}). 
We recall the expression of the primal program \refprog{LP-CFCV}:
\leqnomode\index{Linear program} \index{Semidefinite program}
\begin{flalign*}
    \tag*{(P-CF$^\text{\upshape CV}$)}
    \hspace{3cm}\left\{
    \begin{aligned}
        & \quad \text{Find } \mu \in \FSMeasures(\bm \Oc_\Xc) \\
        & \quad \text{maximising } \mu(\Oc_\Xc) \\
        & \quad \text{subject to:}  \\
        &  \hspace{1cm} \begin{aligned}
            & \forall C \in \Mc,\; \mu|_C \;\leq\; e_C \\
            & \mu \;\geq\; 0 \Mdot
        \end{aligned}
    \end{aligned}
    \right. &&
\end{flalign*}
\reqnomode
From Subsection~\ref{subsubsec01:moment} which culminates at Theorem~\ref{th:ch01_representingmeasure}, it can be relaxed for $k\in \N^*$ as:
\leqnomode
\begin{flalign*}
    \label{prog:SDPk-CFCV}
    \tag*{(PS-CF$^{\text{\upshape CV},k}$)}
    \hspace{3cm}\left\{
    \begin{aligned}
        & \quad \text{Find } \bm y \in \R^{s(2k)} \\
        & \quad \text{maximising } y_{\bm 0} \\
        & \quad \text{subject to:}  \\
        & \hspace{1cm} \begin{aligned}
            & \forall C \in \Mc,\; M_k(\bm y^{e,C} - \bm y|_C ) \succeq 0, \\
            & \forall j=1,\dots,m,\; M_{k-1}(g_j \bm y) \succeq 0, \\
            & M_k(\bm y) \succeq 0 \Mdot
        \end{aligned}
    \end{aligned}
    \right. &&
\end{flalign*}
\reqnomode
We consider localising matrices of order $k-1$ rather than $k$ because all $g_j$'s are of exact degree 1. In this way, the maximum degree with the moment matrices matches. In general we have to deal with localising matrices of order $k- \lceil \frac{\text{deg}(g_j)} 2 \rceil$. If $\mu$ is a representing measure on $\bm \Oc_\Xc$ for $\bm y$ then for all contexts $C \in \Mc$, $\bm y|_C$ can be defined through $\bm y$ by requiring that $\bm y|_C$ has representing measure $\mu|_C$. The two last constraints state necessary conditions on the variable $\bm y$ to be moments of some finite Borel measure supported on $\bm \Oc_\Xc$. The first constraint is a relaxation of the constraint $\mu|_C \leq e_C$ for $C \in \Mc$. As expected, \refprog{SDPk-CFCV} is a semidefinite relaxation of problem \refprog{LP-CFCV} so that $\forall k \in \N^*$, $\NCF(e) = \val{\text{LP-CF}^\text{CV}} \leq \val{\text{PS-CF}^{\text{CV},k}} $. Moreover $(\val{\text{PS-CF}^{\text{CV},k}})_k$ form a monotone nonincreasing sequence because more constraints are added as $k$ increases (so that the relaxations are tighter and tighter).

\subsubsection*{Restriction of the dual program}

The program \refprog{DLP-CFCV} can be restricted so that we can derive a converging hierarchy of SDPs. It is essentially the minimisation of continuous functions for which we require additional constraints such as the fact that they are nonnegative. We will exploit the link between positive polynomials and sum-of-squares representation that was presented in Subsection~\ref{subsubsec01:sos}. 
We recall the expression of the dual program \refprog{DLP-CFCV}:
\leqnomode
\begin{flalign*}\index{Linear program} \index{Semidefinite program}
    \tag*{(D-CF$^\text{\upshape CV}$)}
    \hspace{3cm}\left\{
    \begin{aligned}
        & \quad \text{Find } \family{f_C}_{C \in \Mc} \in \prod_{C \in \Mc} C(\Oc_C) \\
        & \quad \text{minimising } \sum_{C \in \Mc} \intg{\Oc_C}{f_C}{e_C} \\
        & \quad \text{subject to:}  \\
        & \hspace{1cm} \begin{aligned}
            & \sum_{C \in \Mc} f_C \circ \rho^X_C \;\geq\; \mathbf{1}_{\Oc_\Xc} \\
            & \forall C \in \Mc,\; f_C \;\geq\; \mathbf{0}_{\Oc_C} \Mdot
        \end{aligned}
    \end{aligned}
    \right. &&
\end{flalign*}
\reqnomode
As this point we could derive the dual of program \refprog{SDPk-CFCV} and show that this is indeed a restriction of the above program. For a more symmetric treatment, we restrict the dual program building on Subsection~\ref{subsubsec01:sos} and Theorem~\ref{th:ch01_putinar}. Instead of optimising over positive continuous functions, we restrict them to belong to the quadratic module $Q(g)$ and then $Q_k(g)$ for some $k \in \N^*$ further requiring that the degrees of SOS polynomials\index{Sum-of-squares polynomial} are fixed. 
For $k \in \N^*$, we have
\leqnomode
\begin{flalign*}
    \label{prog:DSDPk-CFCV}
    \tag*{(DS-CF$^{\text{\upshape CV},k}$)}
    \hspace{3cm}\left\{
    \begin{aligned}
        & \quad \text{Find } (p_C)_{C \in \Mc} \subset \SOSkm \text{ and } \family{\sigma_j}_{j=1,\dots,m} \subset \SOSm_{k-1} \hspace{-2cm}\\
        & \quad \text{maximising } \sum_{C \in \Mc} \intg{\Oc_C}{p_C}{e_C} \\
        & \quad \text{subject to:}  \\
        &  \hspace{1cm} \begin{aligned}
            & \sum_{C \in \Mc} p_C \circ \rho^\Xc_C  - \mathbf{1}_{\Oc_\Xc} = \sum_{j=0}^{m} \sigma_j g_j \Mdot
        \end{aligned}
    \end{aligned}
    \right. &&
\end{flalign*}
\reqnomode
\refprog{DSDPk-CFCV} is a restriction of \refprog{DLP-CFCV} so that for all $k \in \N^*$, we have that $\NCF(e) = \val{\text{D-CF}^{\text{CV}}} \leq \val{\text{DS-CF}^{\text{CV},k}} $. Furthermore, $(\val{\text{DS-CF}^{\text{CV},k}})_k$ form a monotone nonincreasing sequence.

\subsubsection*{Proving duality for the semidefinite programs} \index{Semidefinite program}

As mentioned above, we chose to derive programs \refprog{SDPk-CFCV} and \refprog{DSDPk-CFCV} using dual arguments. These programs should therefore be dual to one another, which will immediately provide weak duality. We prove this for completeness. 

\begin{proposition}
The program  {\upshape \refprog{DSDPk-CFCV}} is the dual formulation of the program {\upshape \refprog{SDPk-CFCV}}. 
\label{prop:ch02_weakduality}
\end{proposition}

\begin{proof}
We start by rewriting $M_k(\bm y)$ as $\sum_{\bm \alpha \in \N^d_k} \bm y_{\bm \alpha} A_{\bm \alpha}$ and $M_{k-1}(g_j \bm y)$ as $\sum_{\bm \alpha \in \N^d_k} y_{\bm \alpha} B^j_{\bm \alpha}$ for $1 \leq j \leq m$ and for appropriate real symmetric matrices $A_{\bm \alpha}$ and $(B^j_{\bm \alpha})_j$. For instance, in the basis $(\bm x^{\bm \alpha})$:
\begin{equation*}
   ( A_{\bm \alpha} )_{\bm s, \bm t} = \bigg( 
    \begin{cases}
        1 \text{ if } \bm s+\bm t = \bm \alpha \\
        0 \text{ otherwise}
    \end{cases}
    \bigg)_{\bm s, \bm t} \Mdot
\end{equation*}
From $A_{\bm \alpha}$, we also extract $A_{\bm \alpha}^C$ for $C \in \Mc$ in order to rewrite $M_k(\bm y|_C)$ as $\sum_{\alpha \in \N^d_k} \bm y_{\bm \alpha} A_{\bm \alpha}^C$. This amounts to identifying which matrices $(A_{\bm \alpha})$ contribute to a given context $C \in \Mc$. Then \refprog{SDPk-CFCV} can be rewritten as:\index{Moment matrix}\index{Localising matrix}
\leqnomode
\begin{flalign*}\index{Semidefinite program}
    \tag*{(SDP-CF$^{\text{\upshape CV},k}$)}
    \hspace{4cm}\left\{
    \begin{aligned}
        & \quad \text{Find } \bm y \in \R^{s(2k)} \\
        & \quad \text{maximising } y_{\bm 0} \\
        & \quad \text{subject to:}  \\
        &  \hspace{1cm} \begin{aligned}
            & \forall C \in \Mc,\; M_k(\bm y^{e,C}) - \sum_{\bm \alpha \in \N^d_k} y_{\bm \alpha} A_{\bm \alpha}^C \succeq 0, \\
            & \forall j=1,\dots,m,\; \sum_{\bm \alpha \in \N^d_k} y_{\bm \alpha} B^j_{\bm \alpha} \succeq 0, \\
            & \sum_{\bm \alpha \in \N^d_k} y_{\bm \alpha} A_{\bm \alpha} \succeq 0 \Mdot
        \end{aligned}
    \end{aligned}
    \right. &&
\end{flalign*}
\reqnomode
Via the Lagrangian, this is equivalent to:
\begin{equation}
	\sup_{\bm y \in \R^{s(k)}} \; \inf_{ \substack{(X^C), (Y_j), Z \\ \text{ SDP matrices}}} \; \Lc(\bm y,(X_C),Y,(Z_j)),
\end{equation}
with
\begin{equation}
    \begin{aligned}
    \Lc(\bm y,(X^C),(Y_j),Z) &= y_{\bm 0} \\
    & + \sum_{C \in \Mc} \Tr(M_k(\bm y^{e,C}) X^C) - \sum_{C \in \Mc} \sum_{\bm \alpha \in \N^d_k} y_{\bm \alpha} \Tr(A_{\bm \alpha}^C X^C) \\
    & + \sum_{j=1}^m \sum_{\bm \alpha \in \N^d_k} y_{\bm \alpha} \Tr(B_{\bm \alpha}^j Y_j) \\
    & + \sum_{\bm \alpha \in \N^d_k} y_{\bm \alpha} \Tr(A_{\bm \alpha} Z) \Mdot
    \end{aligned}
\end{equation}
The dual program corresponds to 
\begin{equation}
	\inf_{ \substack{(X^C), (Y_j), Z \\ \text{ SDP matrices}}} \; \sup_{\bm y \in \R^{s(k)}} \; \Lc(\bm y,(X^C),(Y_j),Z) \Mdot
\end{equation}
We rewrite the Lagrangian as:\index{Lagrangian method}
\begin{equation}
    \begin{aligned}
    \Lc(\bm y,(X_C),Y,(Z_j)) &= \sum_{C \in \Mc} \Tr(M_k(\bm y^{e,C}) X^C) \\
    & + \sum_{\bm \alpha \in \N^d_k} y_{\bm \alpha} \left( \delta_{\bm \alpha, \bm 0}  - \sum_{C \in \Mc} \Tr(A_{\bm \alpha}^C X^C) + \sum_{j=1}^m \Tr(B_{\bm \alpha}^j Y_j) + \Tr(A_{\bm \alpha} Z) \right) \Mdot
    \end{aligned}
\end{equation}
Then the dual program of \refprog{SDPk-CFCV} reads:
\leqnomode
\begin{flalign*}\index{Semidefinite program}
    \hspace{2cm}\left\{
    \begin{aligned}
        & \quad \text{Find } \family{X^C}_{C \in \Mc}, \family{Y_j}_{j=1,\dots,m} \text{ and } Z \text{ SDP matrices } \\
        & \quad \text{maximising } \sum_{C \in \Mc} \Tr(M_k(\bm y^{e,C}) X^C) \\
        & \quad \text{subject to:}  \\        
        & \hspace{1cm} \begin{aligned}
            &  \forall \bm \alpha \in \N^d_k, \; \sum_{C \in \Mc} \Tr(A_{\bm \alpha}^C X^C) - \sum_{j=1}^m \Tr(B_{\bm \alpha}^j Y_j) - \Tr(A_{\bm \alpha} Z) = \delta_{\bm \alpha, \bm 0} \Mdot
        \end{aligned}
    \end{aligned}
    \right. &&
\end{flalign*}
\reqnomode
We finally show that the above program exactly corresponds to \refprog{DSDPk-CFCV}. We start with the objective. For all $C \in \Mc$, with $X^C$ a positive semidefinite matrix:
\begin{align}\index{Moment matrix}
    \Tr(M_k(\bm y^{e,C}) X^C) & = \sum_{\bm \alpha} \sum_{\bm \beta} (M_k(\bm y^{e,C}))_{\bm \alpha \bm \beta} (X^C)_{\bm \beta \bm \alpha} \\
    & = \sum_{\bm \alpha} \sum_{\bm \beta} \bm y^{e,C}_{\bm \alpha + \bm \beta} (X^C)_{\bm \alpha \bm \beta} \\
    & = \sum_{\bm \alpha} \sum_{\bm \beta} \intg{\Oc_C}{\bm x^{\bm \alpha + \bm \beta}}{e_C} (X^C)_{\bm \alpha \bm \beta} \\
    & = \intg{\Oc_C}{ \bm \vrm_k(\bm x)^T X^C \bm \vrm_k(\bm x)}{e_C} \\
    & = \intg{\Oc_C}{ p_C }{e_C},
\end{align}
with for all $C \in \Mc$, $p_C \in \SOSkm$ a sum-of-squares polynomial\index{Sum-of-squares polynomial} via Proposition~\ref{prop:ch01_sosdecomposition} and where we used $\bm \vrm_k(\bm x)$ the vectors of monomials of maximal total degree $k$.

Now, to retrieve the constraint, we multiply each side by $\bm x^{\bm \alpha}$ and we sum for all $\bm \alpha$:
\begin{equation}
    \sum_{C \in \Mc} \Tr(\sum_{\bm \alpha} \bm x^{\bm \alpha} A^C_{\bm \alpha} X^C) - \bm 1 = \sum_{j=1}^m \Tr(\sum_{\bm \alpha} \bm x^{\bm \alpha} B_{\bm \alpha}^j Y_j) + \Tr(\sum_{\bm \alpha} \bm x^{\bm \alpha} A_{\bm \alpha} Z)
\end{equation}
Recalling the definition of moment and localising matrices:
\begin{align}
    & \sum_{\bm \alpha} A^C_{\bm \alpha} \bm x^{\bm \alpha} = \bm \vrm_k(\bm x) \bm \vrm_k(\bm x)^T \\
    & \sum_{\bm \alpha} B_{\bm \alpha}^j \vrm_k(\bm x) \vrm_k(\bm x)^T = g_j(\bm x) \bm \vrm_{k-1}(\bm x) \bm \vrm_{k-1}(\bm x)^T, \; \forall j=1,\dots,m 
\end{align}
Thus, by Proposition~\ref{prop:ch01_sosdecomposition}, for appropriate sum-of-squares polynomials\index{Sum-of-squares polynomial} $(\sigma_j)_{j=0,1,\dots,m} \subset \Rpolm_{k-1}$:
\begin{align}
    & \Tr(\sum_{\bm \alpha} \bm x^{\bm \alpha} A^C_{\bm \alpha} X^C) = p_C \circ \rho_C^\Xc (\bm x) \\
    & \Tr(\sum_{\bm \alpha} \bm x^{\bm \alpha} B_{\bm \alpha}^j Y_j) = g_j(\bm x) \sigma_j(\bm x) \\
    & \Tr(\sum_{\bm \alpha} \bm x^{\bm \alpha} A_{\bm \alpha} Z) = \sigma_0(\bm x)
\end{align}
This is exactly the constraint in \refprog{DSDPk-CFCV} with, for convenience, $g_0 = 1$.

\end{proof}

\subsection{Convergence of the hierarchy of SDPs}\index{Semidefinite program}
\label{subsec02:convergence}

Finally, we prove that the constructed hierarchy provides a sequence of objective values that converges monotonically to the noncontextual fraction $\NCF(e)$.

\begin{theorem} \label{th:ch2_SDPconvergence}\index{Contextual fraction!CV}
The optimal values of the hierarchy of semidefinite programs {\upshape\refprog{SDPk-CFCV}} (resp. {\upshape\refprog{DSDPk-CFCV}}) provide monotonically decreasing upper bounds converging to the noncontextual fraction $\NCF(e)$ which is the value of {\upshape\refprog{LP-CFCV}}. That is 
    {\upshape
    \begin{equation}
        \val{\text{PS-CF}^\text{CV,k}} \; \downarrow \; \val{\text{P-CF}^{\text{CV}}} = \NCF(e) \quad \text{ as } k \rightarrow \infty \, ,
    \end{equation}
        \begin{equation}
        \val{\text{DS-CF}^\text{CV,k}} \; \downarrow \; \val{\text{D-CF}^{\text{CV}}} = \NCF(e) \quad \text{ as } k \rightarrow \infty \Mdot
    \end{equation}
    }
\end{theorem}

\begin{proof}
Because of the strong duality between the original infinite-dimensional linear programs we have:
\begin{equation}
    \val{\text{P-CF}^{\text{CV}}} = \val{\text{D-CF}^{\text{CV}}} = \NCF(e).
\end{equation}
Moreover, $\forall k \geq 1$, \refprog{SDPk-CFCV} is a relaxation of \refprog{LP-CFCV}:
\begin{equation}
    \val{\text{PS-CF}^\text{CV,k}} \geq \val{\text{P-CF}^{\text{CV}}}.
\end{equation}
And $\forall k \geq 1$, \refprog{DSDPk-CFCV} is a restriction of \refprog{DLP-CFCV}:
\begin{equation}
    \val{\text{DS-CF}^\text{CV,k}} \geq \val{\text{D-CF}^{\text{CV}}}.
\end{equation}
Also $\forall k \geq 1$, we have weak duality between \refprog{SDPk-CFCV} and \refprog{DSDPk-CFCV} (by Proposition~\ref{prop:ch02_weakduality}):
\begin{equation}
    \val{\text{DS-CF}^\text{CV,k}} \geq \val{\text{PS-CF}^\text{CV,k}}.
\end{equation}
Thus for all $k \geq 1$:
\begin{equation}
    \val{\text{DS-CF}^\text{CV,k}} \geq \val{\text{PS-CF}^\text{CV,k}} \geq \NCF(e).
\end{equation}
We already saw that $(\val{\text{PS-CF}^\text{CV,k}})_k$ and $(\val{\text{DS-CF}^\text{CV,k}})_k$ form monotone nonincreasing sequences. We now show that $(\val{\text{DS-CF}^\text{CV,k}})_k$ converges to $\NCF(e)$. This is equivalent to showing that we can approximate any feasible solution\footnotemark of program \refprog{DLP-CFCV} with a solution of \refprog{DSDPk-CFCV} for a high enough rank $k$. 
\footnotetext{Note that program \refprog{DLP-CFCV} might not have an optimal solution in which case it only has an optimal solution in the closure of the feasible set. In that case, we can always find a sequence of feasible solutions converging to an optimal solution.}

Fix $\varepsilon > 0$ and a feasible solution $(f_C)_{C \in \Mc} \in \prod_{C \in \Mc} C(\Oc_C)$ of \refprog{DLP-CFCV}. Then $\forall C \in \Mc$, $f_C + \frac{\varepsilon}{\vert \Mc \vert}$ is a positive continuous function on $\Oc_C$. Because $\Oc_C$ is compact (see Subsection~\ref{subsec02:assumptions}) by the Stone-Weierstrass theorem, $f_C +  \frac{\varepsilon}{\vert \Mc \vert}$ can be approximated by a positive polynomial. 
Thus there exist positive polynomials $p_C^{\varepsilon} \in \Rpolm$ such that for all contexts $C \in \Mc$ (in sup norm):
\begin{equation}
    \norm{f_C + \frac{\varepsilon}{\vert \Mc \vert} - p_C^{\varepsilon}} \leq
    \frac{\varepsilon}{\vert \Mc \vert} 
    \label{eq:ch02_approx1}
\end{equation}
and also:
\begin{equation}
    \norm{ \left(f_C + \frac{\varepsilon}{\vert \Mc \vert} - p_C^{\varepsilon} \right) \circ \rho^\Xc_C } < \frac{1}{\vert \Mc \vert} \Min{\bm x \in \Oc_\Xc} \left( \sum_{C \in \Mc} \left(f_C + \frac{\varepsilon}{\vert \Mc \vert} \right) \circ \rho^\Xc_C(\bm x) - \bm x \right)
    \label{eq:ch02_approx2}
\end{equation}
where the minimum is strictly positive as $\sum_C (f_C +\frac{\varepsilon}{\vert \Mc \vert}) \circ \rho^\Xc_C > \bm 1_{\Oc_\Xc}  $.

From Eq.~(\ref{eq:ch02_approx1}), the objective derived with $(p_C^{\varepsilon})_C$ is $\varepsilon$-close to the original objective:
\begin{align}
    \left| \sum_{C \in \Mc} \intg{\Oc_C}{f_C}{e_C} - \sum_{C \in \Mc} \intg{\Oc_C}{p_C^{\varepsilon}}{e_C}\right| & \leq \sum_{C \in \Mc}  \intg{\Oc_C}{\left| f_C + \frac{\varepsilon}{\vert \Mc \vert} - p_C^{\varepsilon}\right|}{e_C} \\
    & \leq \varepsilon\Mdot
\end{align}
Also from Eq.~(\ref{eq:ch02_approx2}):
\begin{align}
    \sum_{C \in \Mc} & p_C^{\varepsilon} \circ \rho^\Xc_C - \bm 1\\
    & > \sum_{C \in \Mc} \left(f_C + \frac{\varepsilon}{\vert \Mc \vert} \right) \circ \rho^\Xc_C - \Min{\bm x \in \Oc_\Xc} \left( \sum_{C \in \Mc} \left(f_C + \frac{\varepsilon}{\vert \Mc \vert} \right) \circ \rho^\Xc_C(\bm x) - \bm x \right) - \bm 1 \\
    & \geq 0,
\end{align}
so that $\sum_{C \in \Mc} p_C^{\varepsilon} \circ \rho^\Xc_C - \bm 1_{\Oc_\Xc}$ is a positive polynomial on $\Oc_\Xc$.
Next, because $\Oc_\Xc$ is of the form required in Assumption~\ref{ass:ch01_algebraic}, by Putinar's Positivellensatz\index{Putinar's Positivellensatz} (see Theorem~\ref{th:ch01_putinar}), $\sum_{C \in \Mc} p_C^{\varepsilon} \circ \rho^\Xc_C - \bm 1_{\Oc_\Xc}$ belongs to the quadratic module $Q(g)$. Therefore, for a high enough rank $k \in \N$, it is a feasible solution of \refprog{DSDPk-CFCV} and thus:
\begin{equation}
    \lvert \NCF(e) - \val{\text{DS-CF}^\text{CV,k}} \rvert \leq \varepsilon \Mdot
\end{equation}

\end{proof}

\section{Discussion and open problems}

In this chapter, we have presented a framework formalising measurement contextuality for continuous-variable systems. We further extended the FAB theorem\index{FAB theorem!CV} to those kinds of scenarios. We also studied how to quantify contextuality via the extension of the contextual fraction to the continuous-variable realm. Its computation can be phrased as an infinite-dimensional linear program that we can relax to derive a converging hierarchy of finite dimensional semidefinite programs.\index{Contextual fraction!CV}

\textit{Logical} forms of contextuality, which are present
at the level of the \textit{possibilistic} rather than \textit{probabilistic}
information contained in an empirical model, remain to be considered
(e.g. in discrete-variable scenarios \cite{fritz2009possibilistic,abramsky2011sheaf,abramsky2013relational,mansfield2012hardy}).
In the discrete setting, these can be treated by analysing
`possibilistic' empirical models obtained by considering the supports of the discrete-variable probability distributions \cite{abramsky2011sheaf}, which indicate the
elements of an outcome space that occur with non-zero probability. It can often be easily pictured on a bundle diagram.
However the notion of support of a measure is not as straightforward,
and the na\"ive approach is not viable since typically all singletons have measure $0$.
Nevertheless, supports can be defined in the setting of Borel measurable spaces,
for instance, which in any case
are the kind of spaces in which we are practically interested in Sections 
\ref{sec02:quantifying}
and \ref{sec02:sdp}.\index{Measurable!space}

Approaches to contextuality that characterise obstructions to global sections using cohomology have had considerable success
\cite{abramsky2012cohomology,abramsky2015contextuality,caru2015detecting,caru2017,raussendorf2016cohomological,roumen2017cohomology,okay2017topological,caru2018towards,okay2018cohomological,mansfield2013mathematical}
and typically apply to logical forms of contextuality.
An interesting prospect is to explore how the present framework may be employed to these ends, and to see whether the continuous-variable setting can open the door to new techniques that can be applied, or whether qualitatively new forms of contextual behaviour may be uncovered.
A related direction to be developed is to understand how our treatment of contextuality can be further extended to continuous measurement spaces as proposed in \cite{cunha2019measures}. \index{Contextuality!CV}

Another direction to be explored is how our continuous-variable framework for contextuality can be extended to apply to more general notions of contextuality that relate not only to measurement contexts but also more broadly to contexts of preparations and transformations as well \cite{Spekkens2005,mansfield2018quantum}, noting that these also admit quantifiable relationships to quantum advantage \eg \cite{mansfield2018quantum,henaut2018tsirelson}.
Indeed, a major motivation to study contextuality is for its connections to quantum-over-classical advantages in informatic tasks.
An important line of questioning is to ask what further connections can be found in the continuous-variable setting, and whether continuous-variable contextuality might
offer advantages that outstrip those achievable with discrete-variable contextual resources.
Note that it is known that infinite-dimensional quantum systems can offer certain
additional advantages beyond finite-dimensional ones \cite{slofstra2016tsirelson}, though the empirical
model that arises in that example is still a discrete-variable one in our sense.

The present work sets the theoretical basis for computational exploration
of continuous-variable contextuality in quantum-mechanical empirical
models.
This, we hope, can provide new insights and inform all other avenues to be developed in future work.
It can also be useful in verifying the non-classicality of empirical models.
Numerical implementation of the programs of Section \ref{sec02:sdp} 
is of particular interest.
The hierarchy of semidefinite programs can be used numerically to witness contextuality in continuous-variable experiments.
Even if the time-complexity of the semidefinite program may increase drastically with its degree, a low-degree program can already provide a first witness of contextual behaviour as it will provide a lower bound on the contextual fraction.

Since our framework for continuous-variable
contextuality is independent of quantum theory itself, it can equally
be applied to `empirical models' that arise in other, non-physical settings.\index{Empirical model!CV}
The discrete-variable framework of \cite{abramsky2011sheaf} has led to a number of
surprising connections and cross-fertilisations with other fields \cite{abramsky2015contextualsemantics},
including natural language \cite{abramsky2014semantic}, relational databases \cite{abramsky2012databases,barbosa2015contextuality}, logic \cite{abramsky2012logical,abramsky2015contextuality,kishida2016logic}, constraint satisfaction \cite{AbramskyGottlob2013robust,abramsky2017quantum} and social systems \cite{dzhafarov2016there}.
It may be hoped that similar connections and applications can be found for the present framework to fields in which continuous-variable data is of
central importance.
For instance probability kernels\index{Markov kernel} of the kind we have used are also widely employed in machine learning (\eg \cite{hofmann2008kernel}), inviting intriguing questions about how our framework might be used or what advantages contextuality may confer in that setting.

\dobib

\clearemptydoublepage


\chapter{Equivalence of Wigner negativity and contextuality for continuous-variable Pauli measurements}
\chaptermark{Equivalence of Wigner negativity and contextuality}
\label{chap:equivalence}

\lettrine{F}{or} discrete-variable systems of odd power-of-a-prime dimension, Howard \textit{et al.} \cite{howard2014contextuality} showed that negativity of the Wigner function\index{Wigner function} actually corresponds to contextuality with respect to Pauli measurements (with an additional ancilla system), thereby establishing the operational utility of contextuality for the gate-based model of quantum computation---particularly in a fault-tolerant setting. The equivalence of Wigner negativity and contextuality was established by deriving a noncontextuality inequality using the graph-theoretic techniques of Cabello, Severini and Winter \cite{csw} which extends Kochen-Specker type state-independent proofs to the state-dependent realm. 
CSW inequalities have been shown to be equivalent to the logical Bell inequalities appearing in the sheaf theoretic approach \cite{abramsky2011sheaf} in \cite{silva2017csw}.\index{Wigner negativity}
This proof of equivalence \cite{howard2014contextuality} and subsequent alternate proof requires \cite{delfosse2015wigner,delfosse2017equivalence} that, as well as the system displaying Wigner negativity, a second ancillary system must be present in order to have a sufficiently rich set of available \textit{Pauli measurements}.\index{Pauli measurement}
The equivalence was generalised to odd dimensions in \cite{delfosse2017equivalence} and also established for qubit systems \cite{raussendorf2017contextuality,delfosse2015wigner}.

Still the Wigner function and the phase-space formulation\index{Phase-space!CV} associated were initially introduced for continuous-variable systems and, to date, there is no link between Wigner negativity and contextuality in the continuous-variable setting. In \cite{bertrand1987tomographic,blass2020negative} it is proven that the Wigner function is the \textit{unique} phase-space quasiprobability distribution yielding the correct marginals for every quadrature. However the link to contextuality\index{Contextuality!CV} remains unclear as it is delicate to exhibit the right measurement scenario \cite{Banaszek1998}.
In fact, generalising the discrete-variable approach to the sheaf-theoretic framework for continuous-variable contextuality runs into several problems of a functional-analytic nature. For example, since we only assume access to measurable properties of the quantum system, already the results of \cite{blass2020negative} are of no use since they implicitly assume point wise equality of marginals whereas we can only guarantee equality almost-everywhere. 
On the other hand, we assume a minimal structure in the nature of our model. In fact, assuming only the (pre-)sheaf structure for a set of measurements on quantum systems, we prove that there is a natural hidden-variable model which relates to the Wigner function on the phase space.
This allows us to properly establish that Wigner negativity\index{Wigner negativity} is equivalent to contextuality with respect to generalised quadrature\index{Quadrature} measurements \ie continuous-variable Pauli measurements.\index{Pauli measurement} These are the most commonly used measurements in continuous-variable quantum information, in particular in quantum optics \cite{adesso2014continuous,walschaers2021non}.
Our result directly relies on continuous-variable contextuality as presented in the previous chapter which generalises the well-established notion presented in~\cite{abramsky2011sheaf}.\index{Contextuality!CV}

We start by properly defining the measurement scenario we are focusing on in Section~\ref{sec03:measurementscenario}. We provide the suitable empirical models and crucially prove that they are families of probability measures on linear functions rather than any functions in Section~\ref{sec03:empiricalmodels}.
Then we prove the main theorem in Section~\ref{sec03:equivalence} and conclude with a few open questions in Section~\ref{sec03:conclusion}. This chapter is based on \cite{booth2021}.

\section{Measurement scenario under consideration}
\label{sec03:measurementscenario}

Hereafter we fix $M \in \N^*$ to be the number of modes---that is, $M$ continuous-variable systems---and $\mathscr{H}$ the corresponding Hilbert space. As introduced in Section~\ref{sec01:phasespace}, the phase space for a single continuous-variable mode is $V = \R^2$. 
The Hilbert space associated with $M$ modes is $V^M$. This a $2M$-dimensional symplectic $\R$-linear space. See \cite{Sudarshan1988} for a concise introduction to the symplectic structure of the phase space and \cite{Gosson2006symplectic} for a detailed review.

From Subsection~\ref{subsec01:Wigner}, the Wigner function\index{Wigner function} representing the total system is a real-valued quasiprobability distribution $V^M \to \R$. It was already established in \cite{albini2009quantum} that the Wigner function must be at least a $L^1$-function for the generalised quadrature measurements to provide a good characterisation.
The proof strategy is to use the extendability\index{Extendability!CV} property of a suitable noncontextual empirical model from Definition~\ref{def:nc} to exhibit a global probability measure on global value assignments. Then we show that it corresponds to the Wigner function. 
This step is delicate since it requires careful functional-analytic considerations.
Having achieved this, it proves that the Wigner function must be everywhere nonnegative since it corresponds to a global probability measure. For the other direction, a nonnegative Wigner function can be used directly as a hidden-variable model so that the corresponding empirical model is noncontextual \cite{Son2009positive}.

We recall some background on the symplectic structure of $V^M$\index{Phase-space!CV} (see Subsection~\ref{subsec01:Wigner}). The symplectic form is denoted $\fdec{\Omega}{V^M \times V^M}{\R}$. For $\bm r_1,\bm r_2 \in V^M$,
\begin{equation}\index{Symplectic form}
    \label{eq:ch03_symplecticform}
    \Omega(\bm r_1,\bm r_2) = \bm r_1 \cdot J \bm r_2 \quad \text{where} \quad J = \begin{pmatrix}
    0 & \Id_M \\
    - \Id_M & 0 
    \end{pmatrix}
\end{equation}
We have that $J^{-1} = J^T = -J$ and of course $J$ can be seen as linear map from $V^M$ to  $V^M$.
A Lagrangian vector subspace\index{Lagrangian!subspace|textbf} is a maximal isotropic subspace, that is, a maximal subspace on which the symplectic form $\Omega$ vanishes. For a symplectic space of dimension $2M$, Lagrangian subspaces are of dimension $M$.
The Lagrangian Grassmannian\index{Lagrangian!Grassmannian|textbf} is the set of all Lagrangian subspaces.

\subsection*{Measurement scenario}
The Wigner function bears a close relationship to displacement operators as emphasised via its link with the characteristic function in Eq.~\eqref{eq:ch01_WignerCharacFunction}.
Wigner negativity will be shown to be equivalent to contextuality with respect to Pauli measurements\index{Pauli measurement} in the same spirit as \cite{howard2014contextuality,delfosse2017equivalence}. Following Definition~\ref{def:ch02_measurementscenario}, this measurement scenario corresponds to the setting described below.
\begin{definition}
\label{def:ch03_measscen}
We fix the measurement scenario $\tuple{\Xc,\Mc,\bm \Oc}$ as follows:\index{Measurement scenario!CV}
\begin{itemize}
    \item the set of measurement labels is $\Xc \defeq V^M$;\index{Phase-space!CV}
    \item the maximal contexts are Lagrangian subspaces\index{Lagrangian!subspace}
    of $V^M$ so that the set of maximal contexts $\Mc$ is the Lagrangian Grassmannian\index{Lagrangian!Grassmannian}
    of $\Xc$;
    \item for each $\bm x \in \Xc$, $\bm \Oc_{\bm x} \defeq \langle \R,\Bc_\R \rangle$ so that for any set of measurement labels $U \subseteq \Xc$, $\Oc_U \cong \R^U$ can be seen as the set of functions from $U$ to $\R$ with its product $\sigma$-algebra $\Fc_U$\footnotemark.\index{$\sigma$-algebra}\index{Measurable!space}
    \footnotetext{It is generated by collection of functions $E$ from $L$ to $\R$ such that $\pi_{\bm x}(E)$ is a real interval for a finite number of $\bm x \in \Xc$ and $\R$ for the rest where $\pi_{\bm x}$ is the projection given later in Eq~\eqref{eq:ch03_catprojection}. }
\end{itemize}
\end{definition}
\noindent Each $\bm x = (q_1,\dots,q_M,p_1,\dots,p_M) \in \Xc$ specifies a point in phase-space which corresponds to measuring the associated displacement operator $\hat D(\bm x) = \hat D(q_1,p_1) \otimes \dots \otimes \hat D(q_M,p_M)$. 
Since a displacement operator is not self-adjoint (\ie Hermitian) we detail below what the precise meaning of such a ``measurement'' is. 

\subsection*{Measuring a displacement operator}\index{Displacement operator!CV}

We will focus on a single mode since it can be straightforwardly extended to multimode quantum states.
A quadrature operator\index{Quadrature} such as the position operator $\hat q$ (see Subsection~\ref{subsec01:CV}) is self-adjoint and it can be expanded via the spectral theorem \cite{hall2013quantum} as:
\begin{equation}
    \hat q = \intg{x \in \mathrm{sp}(\hat q)}{x}{P_{\hat q}(x)} \, ,
\end{equation}
where the spectrum of $\hat q$ is $\mathrm{sp}(\hat q) = \R$ and $P_{\hat q}$ is the spectral measure of $\hat q$ \cite[Th. 8.10]{hall2013quantum}.
For $E \in \Bc(\R)$, $P_{\hat q}(E)$ is given by \cite[Def. 8.8]{hall2013quantum}:
\begin{equation}
    P_{\hat q}(E) = \bm 1_E (\hat q) \Mdot
\end{equation}
Informally, it assigns 1 whenever measuring $\hat q$ yields an outcome that belongs to $E$.
We can view $P_{\hat q}(E)$ as the formal version of the projector $\intg{x \in E}{\ketbra{x}{x}}{x}$ (with $\ket x$ a non-normalisable eigenvector of $\hat q$).
Its functional calculus \cite{hall2013quantum} can be expressed as:
\begin{equation}
     f(\hat q) = \intg{x \in \mathrm{sp}(\hat q)}{f(x)}{P_{\hat q}(x)} \, ,
\end{equation}
for $f$ a bounded measurable function.
Then we can write the spectral measure of $f(\hat q)$ via the push-forward operation: 
\begin{equation}
    \forall E \in \Bc(\R), \; P_{f(\hat q)}(E) = P_{\hat q}(f^{-1}(E)) \Mdot
\end{equation}
It follows immediately that, for $s \in \R$, the spectral measure for the diagonal phase operator $e^{is\hat q}$ is given, for $E \in \mathcal{B}(\mathbb{S}_1)$, by
\begin{equation}
    P_{\exp(is \hat q)}(E) \coloneqq P_{\hat q}(\{x \in \R \mid e^{isx} \in E\}).
\end{equation}

Define the rotated quadrature $\hat q_\theta \defeq \cos(\theta) \hat q + \sin(\theta) \hat p$ for $\theta \in \left[0,2\pi\right]$ and the phase-shift operator $\hat R(\theta)
\defeq \exp(i \frac{\theta}{2} (\hat q^2 + \hat p^2))$. Then
\begin{equation}
    \hat q_\theta = \hat R(\theta) \hat q \hat R(-\theta),
\end{equation}
so that the spectral measure of $\hat q_{\theta}$ is given by\index{Quadrature}
\begin{equation}
    P_{\hat q_\theta}(E) = \hat R(\theta) P_{\hat q}(E) \hat R(-\theta).
\end{equation}

Let $(q,p) \in \R^2$. We can find $r \in \R_+$, $\theta \in \left[ 0,2\pi \right]$ such that $(q,p) = (-r\sin(\theta),r\cos(\theta))$.
Then:
\begin{equation}\index{Displacement operator!CV}
    \hat D(q,p) = e^{i(p \hat q - q \hat p)} = e^{ir(\cos(\theta) \hat q - \sin(\theta) \hat p)} = e^{ir \hat q_{\theta}}.
\end{equation}
This form allows us to deduce spectral measures for the displacement operators. For any $E \in \mathcal{B}(\mathbb{S}_1)$, we have
\begin{align}
    P_{\hat D(q,p)}(E) &= P_{\exp(ir \hat q_{\theta})}(E) \\
      &= P_{\hat q_\theta}(\{x \in \R \mid e^{irx} \in E\})\\
      &= \hat R(\theta) P_{\hat q}(\{x \in \R \mid e^{irx} \in E\}) \hat R(-\theta).
\end{align}

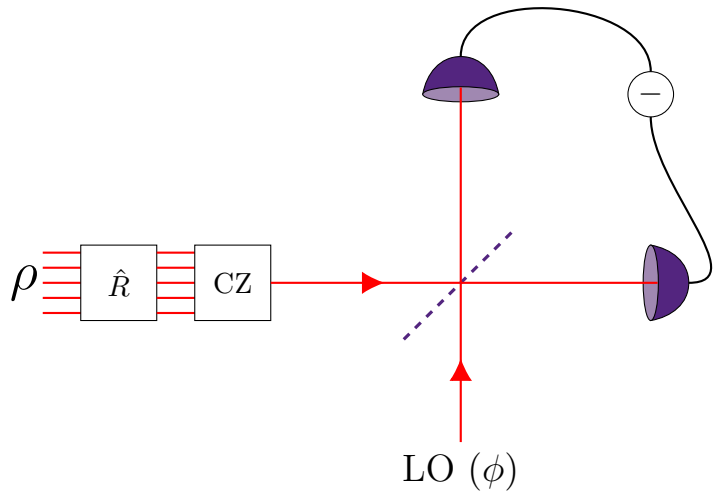
\begin{figure}[ht!]
    \centering
    \begin{tikzpicture}[scale=1 	]

\tikzset{>={Latex[width=3mm,length=3mm]}}

\node (rho) at (-3.25,0) {\huge $\rho$};
\draw[red,thick] (-3,0) -- (-2.5,0);
\draw[red,thick] (-3,0.4) -- (-2.5,0.4);
\draw[red,thick] (-3,0.2) -- (-2.5,0.2);
\draw[red,thick] (-3,-0.2) -- (-2.5,-0.2);
\draw[red,thick] (-3,-0.4) -- (-2.5,-0.4);
\draw (-2.5,-.5) rectangle (-1.5,.5);
\node (R) at (-2,0) {$\hat R$};
\draw[red,thick] (-1.5,0) -- (-1,0);
\draw[red,thick] (-1.5,0.4) -- (-1,0.4);
\draw[red,thick] (-1.5,0.2) -- (-1,0.2);
\draw[red,thick] (-1.5,-0.2) -- (-1,-0.2);
\draw[red,thick] (-1.5,-0.4) -- (-1,-0.4);
\draw (-1,-.5) rectangle (0,.5);
\node (CZ) at (-.5,0) {CZ};
\draw[->,red,thick] (0,0) -- (1.5,0);
\draw [red,thick] (1.4,0) -- (2.5,0);

\node (LO) at (2.5,-2.5) {\Large LO ($\phi$)};
\draw[->,red,thick] (LO) -- (2.5,-1);
\draw [red, thick] (2.5,-1.1) -- (2.5,0);

\draw[dashed,very thick,quantumviolet] (1.75,-.75) -- (3.24,.74);

\draw [fill=quantumviolet] (2.5,2.5) to (3,2.5) to [out=100,in=0] (2.5,3) to  [out=180,in=80](2,2.5) to (2.5,2.5);
\draw [fill=quantumlightviolet] (2.5,2.5) ellipse (.5cm and .1cm);

\draw [fill=quantumviolet] (5,0) to (5,-.5) to [out=10,in=-90] (5.5,0) to  [out=90,in=-10](5,.5) to (5,0);
\draw [fill=quantumlightviolet] (5,0) ellipse (.1cm and .5cm);

\draw [red, thick] (2.5,0) -- (5.09,0);
\draw [red, thick] (2.5,0) -- (2.5,2.59);

\draw [align=center,minimum size=1cm,inner sep=0pt] (5,2.5) circle (.3cm) node {\large $-$};
\draw [thick] (2.5,3) to [out=90,in=90] (5,2.8);
\draw [thick] (5.5,0) to [out=0,in=-90] (5,2.2);

\end{tikzpicture}
    \caption{Experimental protocol corresponding to the measurement scenario of this chapter. It permits measurement of any linear combination of quadratures. After phase-shift operations on individual modes and CZ gates on pairs of modes, homodyne detection one mode of the state is implemented.
    The dashed line represents a balanced beamsplitter. The local oscillator (LO) is a strong coherent state. At the hand of each arm are photodiode detectors.
    The difference in the intensity yields a value proportional to a quadrature of the input mode, which can be rotated depending on the phase of the local oscillator.
    }
    \label{fig:ch03_homodyne}
\end{figure}

In conclusion, ``measuring a displacement operator'' can be implemented with quadrature measurements\index{Quadrature}. Now measuring any multimode displacement amounts to being able to measure arbitrary multimode quadratures; that is, any linear combination of quadratures, \eg $\hat q_1 + 2 \hat p_{2\theta} + 5 \hat q_{M\alpha}$ for arbitrary angles $\theta, \alpha$. First, we apply phase-shift operators $\hat R(\theta)$ for each individual mode to obtain the right rotated quadratures.
Then we apply CZ gates of the form $e^{i g \hat q_i \hat q_j}$ for $g \in \R$ to pairs of modes $i$ and $j$ to sum them. 
This permits construction of the desired linear combinations in a quadrature of a mode. It remains to measure it. This can be implemented with standard homodyne detection.\index{Homodyne detection|textbf} 
It consists of a Gaussian measurement of a quadrature of the field, by mixing the state with a strong coherent state.
Then the intensities of both output arms are measured with photodiode detectors. Their difference yields a value proportional to a quadrature of the input mode, which can be rotated depending on the phase of the local oscillator. The POVM elements for homodyne\index{POVM} detection are given by $\ket{x}_{\phi}\bra{x}$ for all $x \in \R$ where $\ket x_\phi$ is the eigenstate of the rotated quadrature operator $\hat q_\phi$ with eigenvalue $x$.
This is represented in Figure~\ref{fig:ch03_homodyne}. All of these steps can be implemented experimentally \cite{ferraro2005gaussian,su2013gate}.

\subsection*{Maximal contexts}

This measurement scenario is to be interpreted as follows. The measurement corresponding to the label $\bm r \in \Xc$ is described by the spectral measure $P_{\bm {\hat r}}$ for the quadrature corresponding to $\bm r$. 
A pair of spectral measures of self-adjoint operators is compatible, in the sense that they admit a joint spectral measure if and only if they commute, which in turn is true if and only if the operators themselves commute \cite{hall2013quantum}. 
In the case of our measurement scenario, two spectral measures associated to $\bm x,\bm y\in \Xc$ commute if and only if $\Omega(\bm x,\bm y)=0$.\index{Symplectic form}
Thus, measurement labels are compatible only when they both belong to some
isotropic subspace of $\Xc$. The maximal isotropic subspaces are the Lagrangian
subspaces,\index{Lagrangian!subspace} so each context $L \in \Mc$ corresponds to a Lagrangian subspace. $\Mc$ must be the Lagrangian Grassmanian of $\Xc$.\index{Lagrangian!Grassmannian}

\section{Admissible empirical models}
\label{sec03:empiricalmodels}

\subsection*{Empirical models}

We are interested in experiments arising from quadrature measurements of a quantum system. 
For each context $L \in \Mc$, the set $\Oc_L = \prod_{\bm x \in L} \R$ can be seen as the set of functions from $L$ to $\R$ with the corresponding $\sigma$-algebra constructed with the product topology. For $\bm x \in \Xc$, the categorical projections are:
\begin{equation}
    \begin{aligned}
    \pi_{\bm x} : \Oc_L &\longrightarrow \R \\
    f &\longmapsto f(\bm x) \Mdot
    \end{aligned}
    \label{eq:ch03_catprojection}
\end{equation}

We restrict our attention to empirical models $e = (e_L)_{L \in \Mc}$ which satisfy the Born rule;\index{Born rule} \ie there exists some quantum state $\rho \in \Dc(\mathscr{H})$ such that for all contexts $L \in \Mc$ and measurable sets $U \in \Fc_L$:\index{Empirical model!CV}
\begin{equation}
    e_L(U) = \Tr\left(\rho \prod_{\bm x \in L} P_{\bm{\hat{x}} } \circ \pi_{\bm x} (U) \right) \Mdot
\end{equation}
We will therefore use the notation $e^{\rho} = (e_L^\rho)_{L \in \Mc}$ to make explicit the dependence with $\rho$.
Because of the compatibility condition, we may unambiguously write, for each $\bm x \in \Xc$ and for each $U \in \Fc_{\bm x}$ (we write $\Fc_{\bm x}$ for simplicity though we mean $\Fc_{\left\{\bm x\right\}}$):
\begin{equation}
\label{eq:ch03_empiricalx}
    e^{\rho}_{\bm x}(U) = \Tr\left(\rho P_{\bm{\hat{x}} } \circ \pi_{\bm x} (U) \right) \Mdot
\end{equation}
This comes from the marginalisation $e^{\rho}_L|_{\bm x}$ for each $L \in \Mc$ such that $\bm x \in L$.

At this stage there is a mismatch. The Wigner function\index{Wigner function} is a quasiprobability distribution over $V^M = \Xc$, while the extendability\index{Extendability!CV} property of a noncontextual empirical model in the measurement scenario presented above provides a global probability measure on $\Oc_\Xc$, which can be seen as the set of functions $\Xc \to \R$. 
In general, the latter is much larger than the former.
To solve this issue, we show that we can restrict to linear value assignments so that $\Oc_\Xc$ can be taken as $\Xc^*$, the linear dual of $\Xc$. 
Because $\Xc^* \cong \Xc$ then there is no more mismatch between the Wigner function and the global probability measure that exists for a noncontextual empirical behaviour.

We first show that we can restrict to linear value assignments on Lagrangian subspaces\index{Lagrangian!subspace} in Lemma~\ref{lemma:ch03_outcomes_linear}. We do so by showing that the empirical model assigns a nonzero mass only to the linear functions $L \to \R$.
We then lift this property to global value assignments in Proposition~\ref{prop:ch03_linearglobal} in the same spirit as \cite{delfosse2017equivalence}. 

\begin{lemma}
Let $L \subseteq \Xc$ be a Lagrangian subspace. 
Let $U \in \Fc_L$ be a Lebesgue measurable set of functions $L \to \R$ such that $\pi_{\bm x}(U)$ is distinct from $\R$ for a finite number of $\bm x \in L$.
Then there exists a subset $U_{\mathrm{lin}}$ of linear functions $L \to \R$ such that
  \begin{equation}
    e_L^\rho(U_{\mathrm{lin}}) = e_L^\rho(U) \Mdot
  \end{equation}
 \label{lemma:ch03_outcomes_linear}
\end{lemma}
\begin{proof}

First let $(\bm e_k)_{k=1,\dots,M}$ be a basis of $L \cong \R^M$. Let $P$ be the joint spectral measure of $\{P_{\bm{ \hat e_1}},\dots,P_{\bm{ \hat e_M}}\}$.
For any $ \bm y \in L$, define the function
\begin{equation}
    \begin{aligned}
        f_{\bm y} : L & \longrightarrow \R \\
              \bm x & \longmapsto \bm x \cdot \bm y \, ,
    \end{aligned}
\end{equation} 
where $\dummy \cdot \dummy $ is the usual Euclidean scalar product on $L \cong \R^M$.
For any $\bm x \in L$, $P_{\bm {\hat x}}$ is the push-forward of $P$ by the measurable function $f_{\bm x}$ by definition of the functional calculus on $M$ variables.\index{Measurable!function} \index{Push-forward}
Recall that for $\bm x \in L$, $\pi_{\bm x}(U) = \left\{f(\bm x) \mid f \in U \right\} \subseteq \R$.

Then,
\begin{align}
    \Tr\left(\rho \prod_{\bm x \in L} P_{\bm {\hat x}} \circ \pi_{\bm x}(U)\right) 
    &= \Tr\left(\rho \prod_{\bm x \in L} P \left(f_{\bm x}^{-1}\left( \pi_{\bm x}(U)\right)\right)\right) \\ 
    &= \Tr\left(\rho  P\left(\bigcap_{\bm x \in L} f_{\bm x}^{-1}\left(\pi_{\bm x}(U)\right)\right) \right)
\end{align}
with
\begin{align}
    \bigcap_{\bm x \in L} f_{\bm x}^{-1}(\pi_{\bm x}(U)) = \left\{ \bm y \in L \mid \forall \bm x \in L, \, \bm x \cdot \bm y \in \pi_{\bm x}(U) \right\}.
\end{align}
 
Now define
\begin{equation}
    U_{\mathrm{lin}} \coloneqq \left\{
      \begin{aligned}
        L &\longrightarrow \R \\
        \bm x &\longmapsto \bm x \cdot \bm y
      \end{aligned}
      \bigm\vert \bm y \in \bigcap_{\bm x \in L} f_{\bm x}^{-1}(\pi_{\bm x}(U)) 
    \right\} \Mdot
\end{equation}
By construction,
\begin{align}
    \bigcap_{\bm x \in L} f_{\bm x}^{-1}&(\pi_{\bm x}(U_{\mathrm{lin}})) \\
    &= \bigcap_{\bm x \in L} f_{\bm x}^{-1} \left( \left\{ \bm x \cdot \bm y \mid \bm y \in L \text{ s.t. } \forall \bm z \in L, \, \bm y \cdot \bm z \in \pi_{\bm z}(U) \right\} \right)\\
    &= \bigcap_{\bm x \in L}  \left\{ \bm \alpha \in L \mid \bm x \cdot \bm \alpha = \bm x \cdot \bm y \text{ with } \bm y \in L \text{ s.t. } \forall \bm z \in L, \, \bm y \cdot \bm z \in \pi_{\bm z}(U) \right\} \\
    &= \left\{ \bm \alpha \in L \mid \forall \bm x \in L, \, \bm x \cdot \bm \alpha = \bm x \cdot \bm y \text{ with } \bm y \in L \text{ s.t. } \forall \bm z \in L, \, \bm y \cdot \bm z \in \pi_{\bm z}(U) \right\} \\
    &= \left\{ \bm \alpha \in L \mid \forall \bm z \in L, \, \bm \alpha \cdot \bm z \in \pi_{\bm z}(U) \right\} \label{eq:ch03_equality}\\
    &= \bigcap_{\bm x \in L} f_{\bm x}^{-1}(\pi_{\bm x}(U))\, , 
\end{align}
where Eq.~\eqref{eq:ch03_equality} follows from that fact that $\forall \bm x \in L$, $\bm x \cdot  \bm \alpha = \bm x \cdot \bm y$ implies $\bm \alpha = \bm y$.
Also for all $\bm x \in L$, $\pi_{\bm x}(U_{\mathrm{lin}}) \subseteq \pi_{\bm x}(U)$ so that we are indeed reproducing all value assignments from linear functions of $U$.

Then, as claimed,
  \begin{align}
    e_L^\rho (U_{\mathrm{lin}})
    &= \Tr\left(\rho \prod_{\bm x \in L} P_{\bm {\hat x}} \circ \pi_{\bm x}(U_{\mathrm{lin}})\right) \\
    &= \Tr\left(\rho \prod_{\bm x \in L} P \left( f_{\bm x}^{-1} \left( \pi_{\bm x}(U_{\mathrm{lin}})\right)\right) \right)\\
    &= \Tr\left(\rho  P\left(\bigcap_{\bm x \in L} f_{\bm x}^{-1}\left( \pi_{\bm x}(U_{\mathrm{lin}} \right)\right)\right) \\
    &= \Tr\left(\rho  P\left(\bigcap_{\bm x \in L} f_{\bm x}^{-1}\left(\pi_{\bm x}(U)\right)\right)\right) \\
    &= \Tr\left(\rho \prod_{\bm x \in L} P_{\bm {\hat x}}\left(\pi_{\bm x}(U)\right)\right)\\
    &= e_L^\rho (U)\Mdot
  \end{align}
\end{proof}

Now we prove that the set of global value assignments can be identified with $\Xc^*$, the linear dual space of
$\Xc$.
\begin{proposition}
  If $M \geqslant 2$ (\ie for more than 2 modes), global value assignments are linear functions
  $\Xc \to \R$, and the set of global value assignments forms a
  $\R$-linear space of dimension $2M$, namely $\mathscr{E}(\Xc) = \Xc^*$.
  \label{prop:ch03_linearglobal}
\end{proposition}
\begin{proof}
  The sheaf-theoretic framework for contextuality describes value assignments as
  a sheaf\index{Sheaf!event} $\mathscr{E} : \mathcal{P}(\Xc)^\op \to \mathsf{Meas}$, where
  $\mathscr{E}(U)$ is the set of value assignments for the measurement labels
  in $U$, which can be viewed as a set of functions $U \to \R$. For any Lagrangian $L
  \in \Mc$, there is a restriction map $\rho_L^\Xc = \mathscr{E}(\Xc) \to
  \mathscr{E}(L) : f \mapsto f|_L$ that simply restricts the domain of any function from $\Xc$ tp $L$. Then $\mathscr{E}(L)$ must coincide with
  the set of possible value assignments $\Oc_L$.\index{Sheaf!restriction}

  By Lemma~\ref{lemma:ch03_outcomes_linear}, $\mathscr{E}(L)$ consists in linear
  functions $L \to \R$ so that $\mathscr{E}(\Xc)$ contains only functions $\Xc
  \to \R$ whose restriction to any Lagrangian subspace\index{Lagrangian!subspace} is $\R$-linear. Then,
  following \cite[Lemma 1]{delfosse2017equivalence} (the lemma is proven for the discrete phase-space $\Z_d^M \times \Z_d^M$ but its proof extends directly to $\R^M \times \R^M$), we conclude that if $M \geqslant 2$,
  $\mathscr{E}(\Xc)$ contains only $\R$-linear functions $\Xc \to \R$, i.e.
  $\mathscr{E}(\Xc) = \Xc^*$.
\end{proof}

Therefore, without loss of generality, for any $U \subset \Xc$, we can restrict $\Oc_U$ to be the set of linear functions from $U \to \R$. Thus, an empirical model will be a collection of probability measures on $L^*$ for each $L \in \Mc$. For a noncontextual empirical model $e^\rho$, the extendability property yields a global probability measure on $\Xc^*$ that we will identify with the Wigner function of $\rho$ in the following section. \index{Extendability!CV} \index{Measure}

\section{Equivalence between Wigner negativity and contextuality}
\label{sec03:equivalence}

We are now ready to tackle the main proof that Wigner negativity is equivalent to contextuality in our measurement scenario.\index{Wigner function}\index{Wigner negativity}
We prove it by essentially identifying the Wigner function with the probability density of a carefully constructed hidden-variable model.\index{Hidden-variable model!CV}
We first set up the hidden-variable model via Proposition~\ref{proposition:ch03_density} and we prove the equivalence in Theorem~\ref{th:ch03_equivalence}. Crucially, for the identification with the Wigner function, we need to ensure that the hidden-variable model is realisable by a probability measure over hidden variables that has a density. We further require that this density is a $L^1$ function.

\begin{proposition}\index{Empirical model!CV}
  If an empirical model $e^\rho$ for the continuous-variable measurement scenario in Definition~\ref{def:ch03_measscen} is
  noncontextual, then $e^\rho$ admits a realisation by a deterministic hidden-variable model
  $\tuple{\bm \Xc,\mu_r,(k_L)_{L \in \Mc}}$ such that $\mu_r$ has density
  $w_\mu \in L^1(\Xc)$ with respect to the Lebesgue measure.
  \label{proposition:ch03_density}
\end{proposition}
\begin{proof}
  By our extension of the FAB theorem (see Theorem~\ref{th:ch02_FAB}) and Lemma~\ref{lemma:ch03_outcomes_linear}, $e^\rho$ is realised by a canonical hidden-variable model (HVM) $(\bfLambda,\mu,k)$ (see Definition~\ref{def:ch02_hvmodel}), for which
  \begin{itemize}
  \item $\bfLambda = \bm \Oc_\Xc = \bm \Xc^*$ \ie hidden variables are linear value assignments;
  \item each probability kernel $k_L : \Xc^* \to L^*$ is deterministic and
    factorisable;
  \item $\mu$ is a probability measure on $\Xc^*$.
  \end{itemize}

Through the Riesz representation theorem \cite{riesz1909operations}\index{Riesz!representation theorem}, we pick the following isomorphism constructed with the Euclidean scalar product to identify elements from $\Xc$ to elements from $\Xc^*$:
    \begin{equation}
      \begin{aligned}
        \alpha :
        \Xc &\longrightarrow \Xc^* \\
        \bm x &\longmapsto (\bm x \cdot -) \Mdot
      \end{aligned}
    \end{equation}
This will be essential to take the hidden-variable space to be $\Xc$ rather than $\Xc^*$ (so it matches with the Wigner function).    
    
For all $L\in \Mc$, let \index{Markov kernel}
  \begin{equation}
    \begin{aligned}
    \tilde k_L:  \Xc \times \Fc_L & \longrightarrow \R  \\
    (\bm x,E) & \longmapsto k_L( \alpha(\bm x),E) = \delta_{\alpha(\bm x)|_L}(E) \Mdot
    \end{aligned} 
  \end{equation}
Note that for both $(\tilde k_L)_{L \in \Mc}$ and $(k_L)_L$ we can unambiguously write $\tilde k_{\bm x}$ and $k_{\bm x}$ for a measurement label $\bm x \in \Xc$ because of the compatibility condition (see Eq.~\eqref{eq:compatibility_hiddenvar}).
    
Fix $\bm x \in \Xc$. 
For $E \in \Bc(\R)$, let 
\begin{equation}\index{Measure}
    \label{eq:ch03_prho}
    p^\rho_{\bm x}(E) \defeq \Tr \left( \rho P_{\bm{\hat x}} (E) \right) \Mdot
\end{equation}

Fix $U \in \Fc_{\bm x}$ (a measurable subset of linear functions from $\{\bm x\}$ to $\R$).
We first evaluate $p^\rho_{\bm x} \circ \pi_{\bm x} (E) =e^\rho_{\bm x}(E)$ (see Eq.~\eqref{eq:ch03_empiricalx}) with the HVM above (see Eq.~\eqref{eq:ch02_eval_HVM_emp}). 
\begin{align}
      p^\rho_{\bm x} \left(\pi_{\bm x}(U) \right) &= e^\rho_{\bm x}(U) \\
      &= \intg{\bfLambda}{k_{\bm x}(\dummy,U)}{\mu} \\
      &= \intg{f \in \Xc^*}{k_{\bm x}(f,U)}{\mu(f)} \\
      &= \intg{f \in \Xc^*} {\delta_{f|_{\{\bm x\}}}(U)} {\mu(f)} \\
      &= \intg{f \in \Xc^*} {\delta_{f(\bm x)}(\pi_{\bm x}(U))}{\mu(f)} \\
      &= \intg{f \in \Xc^*} {\delta_{\alpha^{-1}(f) \cdot \bm x}(\pi_{\bm x}(U))} {\mu(f)} \\
      &= \intg{\bm y \in (\bm x,-)^{-1} (\pi_{\bm x}(U))}{}{\mu \circ \alpha (\bm y)} \Mdot
\end{align}
Thus $p^\rho_{\bm x}$ is the push-forward\index{Push-forward} of the measure\index{Measure} $\mu \circ \alpha $ on $\Xc$ by the linear functional $(x \cdot -)$ where $J$ is defined in Eq.~\eqref{eq:ch03_symplecticform}. 
By the Lebesgue decomposition theorem \cite{billingsley2008probability}, there is a decomposition $\mu \circ \alpha = \mu_r + \mu_s$ where $\mu_r$ is absolutely continuous with respect to the Lebesgue measure $\mathrm{d} \bm x$ on $\Xc$ and $\mu_s$ is singular with respect to $\mathrm{d}\bm x$. It follows that, that for any $\bm x \in \Xc$, since $A = (\bm x\cdot-)^{-1}(E)$ has non-zero $\mathrm{d} \bm x$-measure for any Borel-measurable $E \subseteq \R$ of non-zero Lebesgue measure,\index{Measure}
\begin{equation}
    \mu_r (A) = \mu \circ \alpha  (A) - \mu_s  (A) = \mu \circ \alpha(A) = p^\rho_{\bm x}(E).
    \label{eq:ch03_pushforward}
\end{equation}
  
Then $(\bm \Xc,\mu_r,(\tilde k_L)_L)$ is a deterministic and factorisable hidden-variable model for the empirical model $e^\rho$\index{Hidden-variable model!CV!deterministic}\index{Hidden-variable model!CV!factorisable}. By the Radon-Nikodym theorem \cite{Nikodym1930}\index{Radon-Nikodym theorem}, $\mu_r$ has a density $w_\mu$ with respect to the Lebesgue measure $\mathrm d \bm x$ on $\Xc$. Since $\mu_r$ is a probability measure, $w_\mu \in L^1(\Xc)$.
\end{proof}

The main result follows from identifying $w_\mu$ (Proposition~\ref{proposition:ch03_density}) and the Wigner function $W_\rho$ as $L^1(\Xc)$
functions:\index{Wigner function}
\begin{theorem}
  Assume $\rho$ is a density operator such that its Wigner function $W_\rho \in L^1(\Xc)$ with respect to the Lebesgue measure.
  Let $e^\rho$ be an empirical model\index{Empirical model!CV} on the measurement scenario\index{Measurement scenario!CV} in Definition~\ref{def:ch03_measscen} for $\rho$ according to the Born rule \ie for any $L \in \Mc$, for $U \in \Fc_L$, $e_L(U) = \Tr(\rho \prod_{\bm x \in L} P_{\bm{\hat x}} \circ \pi_{\bm x}(U))$.
  Then $e^\rho$ is noncontextual if and only if the Wigner function $W_\rho$ of
  $\rho$ is nonnegative, and in that case $W_\rho$ describes a hidden
  variable model for $e^\rho$.\index{Hidden-variable model!CV}
  \label{th:ch03_equivalence}
\end{theorem}
\begin{proof}
  The result holds by identifying the characteristic function of $\rho$, denoted $\Phi_\rho$ (see Eq.~\eqref{eq:ch01_characfunction}) with the Wigner function and with the density $w_\mu$ from Proposition~\ref{proposition:ch03_density}.
  For $\bm x \in \Xc$, we have that $\Phi_\rho(\bm x) = \text{FT}^{-1} \left[W_\rho \right](- J \bm x)$\index{Fourier transform} by taking the inverse Fourier transform of Eq.~\eqref{eq:ch01_WignerCharacFunction} with the change of variables $\bm x \to - J \bm x$. 
  
  On the other hand, fix a noncontextual empirical model $e^\rho$ satisfying the Born rule associated to $\rho$ and the measurement scenario in Definition~\ref{def:ch03_measscen}.
  By Proposition~\ref{proposition:ch03_density} we have:
  \begin{align}
    \Phi(\bm x) &= \Tr\left(\hat D(- \bm x)\rho\right) \\
    &= \Tr\left(\rho \intg{\lambda\in\R}{e^{- i\lambda}}{P_{\hat{J \bm x}}(\lambda)}\right) \\
    &= \intg{\R} {e^{- i\lambda}} {p_{\hat{J \bm x}}^\rho(\lambda)} \\
    &= \intg{\Xc} {e^{- i J \bm x \cdot \bm y}} {\mu_r(\bm y)} \\
    &= \intg{\Xc} {e^{- i J \bm x \cdot \bm y} w_\mu(y)} {\bm y} \\
    &= \text{FT}^{-1} [w_\mu](-J\bm x).
  \end{align}
  where the second line comes from the spectral theorem; the third line via Eq.~\eqref{eq:ch03_prho} and the fact that the integral and the trace may be inverted by the definition of the integral with respect to the spectral measure \cite{hall2013quantum}; the fourth line via the push-forward operation in Eq.~\eqref{eq:ch03_pushforward}\index{Push-forward}; and the two last lines comes from Proposition~\ref{proposition:ch03_density} and the definition of the inverse Fourier transform.\footnote{The missing factor $(\frac{1}{2 \pi})^M$ can always be taken into account in the measure.}
  As a result, we have $ \text{FT}^{-1}[w_\mu](J\bm x) = \text{FT}^{-1} [W_\rho](J\bm x)$
  and since $w_\mu,W_\rho \in L^1(X)$, $w_\mu = W_\rho$
  almost everywhere \cite[Corollary 7.1]{folland2009fourier}. We have that $w_\mu$ is a density function of a probability measure, so it
  follows that both functions must be almost everywhere nonnegative. Because the Wigner function\index{Wigner function} is
  a continuous function from $\Xc$ to $\R$ \cite{cahill1969density}, $W_\rho$
  must be nonnegative.\index{Wigner negativity} 
  
  Conversely, a nonnegative Wigner function provides a canonical noncontextual hidden variable model for $e^\rho$ \cite{Son2009positive}. Explicitly, the corresponding hidden-variable model satisfying Eq.~\eqref{eq:ch02_eval_HVM_emp} is $(\bm \Xc,W_{\rho} \dd{\bm x},(\tilde k_L)_L)$.
\end{proof}

\section{Discussion and open problems}
\label{sec03:conclusion}

We have seen that Wigner negativity\index{Wigner negativity} is equivalent to contextuality\index{Contextuality!CV} with respect to continuous-variable Pauli measurements\index{Pauli measurement}. This raises important questions. First the present argument requires considering a measurement scenario\index{Measurement scenario!CV} that comprises an uncountable family of measurement labels (the entire phase-space $V^M$). From an experimental perspective, it is crucial to wonder what happens if we restrict to a finite family of measurement labels and see whether we can derive a robust version of this theorem. 

Another question concerns a quantifiable relationship between contextuality and Wigner negativity. In the previous chapter we saw that can formalise the contextual fraction \cite{abramsky2017contextual} in continuous variables. In the next chapter, we will derive witnesses for Wigner negativity whose violation lower bounds the distance to the set of states with a positive Wigner function. It would be highly desirable to establish a precise and quantified link between those two measures of nonclassicality.\index{Contextual fraction!CV}

Homodyne detection\index{Homodyne detection} is a standard detection method in continuous variables \cite{yokoyama2013ultra} and it is the basis of several computational models in continuous-variable quantum information \cite{DouceCVIQP2017,Chabaud2017hom,BShomodyne2017,GKP2001}. 
In \cite{Baragiola2019}, a continuous-variable model for fault-tolerant, universal quantum computation using only homodyne detection is developed. It combines Gottesman--Kitaev--Preskill \cite{GKP2001} Clifford quantum computation and Gaussian quantum computation. 
This justifies the measurement scenario under consideration. 
Homodyne detection is a Gaussian measurement, therefore any quantum advantage is due to Wigner negativity being present before the detection setup. This result shows that, just like in the discrete-variable case \cite{howard2014contextuality}, continuous-variable contextuality supplies the necessary ingredients for continuous-variable quantum computing.\index{Contextuality!DV}

\dobib

\clearemptydoublepage


\chapter{Witnessing Wigner negativity}
\label{chap:witnessing}

\lettrine{I}{n} Chapter~\ref{chap:CVcont} we formalised a framework for defining properly contextuality in continuous-variable systems. In Chapter~\ref{chap:equivalence}, we showed that contextuality has a close connection with Wigner negativity, namely that they are equivalent in the setting we presented. We now turn to a more in-depth analysis of Wigner negativity. \index{Wigner function} \index{Wigner negativity}

As the Wigner function is a real-valued quasi-probability distribution~\cite{cahill1969density}---a normalised distribution which can take negative values---it cannot be sampled directly experimentally in general. However, its marginals are proper probability distributions which can thus be sampled using homodyne detection in the optical setting~\cite{lvovsky2009continuous}. \index{Homodyne detection}
Alternatively, heterodyne detection (also called double homodyne detection) allows for sampling from a smoothed version of the Wigner function~\cite{husimi1940some,richter1998determination}. 
In both cases, displacing a state in phase space before the detection is equivalent to measuring the undisplaced state directly with homodyne or heterodyne detection and then applying a classical post-processing procedure---namely, a translation of the classical outcome according to the displacement amplitude~\cite{chabaud2020efficient,chabaud2020certification}.

Reconstructing the Wigner function to assess whether it is negative or not is possible via full tomographic reconstruction \cite{d2003quantum}. However, such reconstructions are very costly in terms of the number of measurements needed, and require performing a tomographically complete set of measurements---thus usually involving multiple measurement settings. In the continuous-variable setting the task is even more daunting, since the Hilbert space of quantum states is infinite-dimensional~\cite{lvovsky2009continuous,chabaud2020building}. 

\begin{figure}[ht]
	\begin{center}
		\includegraphics[width=0.7\columnwidth]{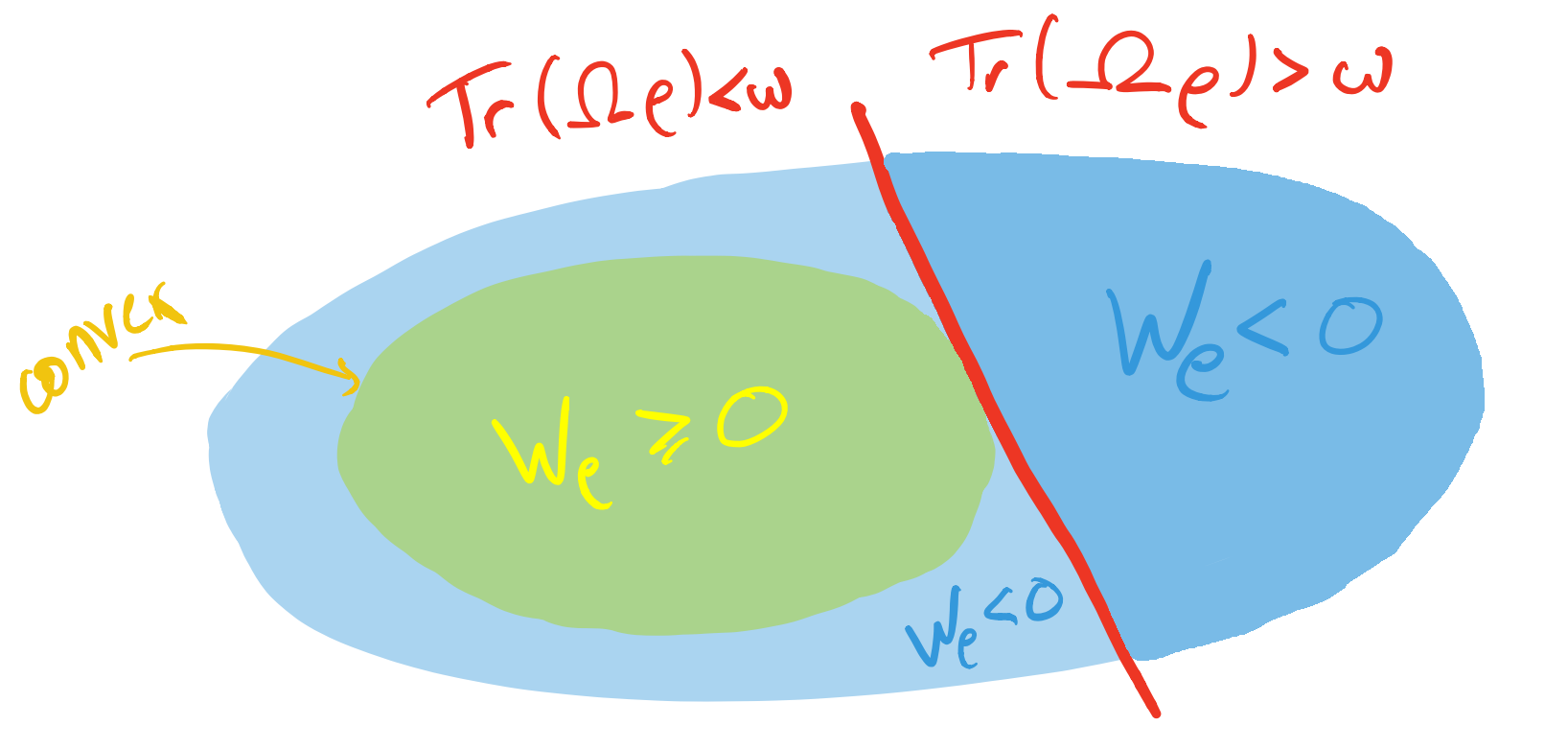}
		\caption{Pictorial representation of a witness $\hat\Omega$ with threshold value $\omega$ for a given property. In green: states without the property. In blue: states with the property. In red: witness threshold value. In light blue: states with the property undetected by the witness.}
		\label{fig:ch04_witness}
	\end{center}
\end{figure}

Instead, another strategy is to introduce \textit{witnesses} for specific properties of quantum states~\cite{terhal2001family,lewenstein2000optimization,mari2011directly,kiesel2012universal,Galvao21Witnesses,chabaud2020certification} that are more accessible experimentally.
These witnesses should be observables that possess a \textit{threshold expectation value} indicating whether the measured state exhibits the desired property or not.  
Intuitively, a witness for a given property can be thought of as a separating hyperplane in the set of quantum states, such that any state on one side of this hyperplane has this property (see~\refig{ch04_witness}). In particular, some states with the sought property may remain unnoticed by the witness. In that regard, it is desirable to have a \textit{complete} set of witnesses such that for each state exhibiting the desired property, there exists at least one witness in the set that captures it.

A natural choice for a witness of Wigner negativity is the fidelity\index{Fidelity} with a pure state having a Wigner function with negative values, since it is a quantity that can be accessed experimentally by direct fidelity estimation~\cite{d2003quantum,chabaud2020building}. Building on this intuition, and given that all Fock states---with the exception of the (Gaussian) vacuum state $\ket0$---have a negative Wigner function, we introduce a broad family of  Wigner negativity witnesses for single-mode and multimode continuous-variable quantum states based on fidelities with Fock states.
Furthermore this family is \textit{complete}, that is, for each state exhibiting Wigner negativity, there exists at least one witness that will detect it. 
The expectation values of the witnesses are linear functions of the state which may be \textit{efficiently estimated experimentally} using standard homodyne or heterodyne detection, thus providing a reliable method for detecting Wigner negativity with certifiable bounds. \index{Homodyne detection}
Additionally, we show that the amount by which the measured expectation value exceeds the threshold value of the witness provides an operational measure of Wigner negativity as it directly lower bounds the distance between the measured state and the set of states with positive Wigner function.\index{Wigner function}

We cast the computation of the threshold values of the witnesses as infinite-dimensional linear programs\index{Linear program}. To perform an actual numerical implementation, we can either relax or restrict these programs. \textit{Upper and lower bounds for the threshold values} of our witnesses are then given by two converging hierarchies of finite-dimensional semidefinite programs\index{Semidefinite program}. While we first study the case of single-mode quantum states and witnesses for simplicity, we then show that our results extend to the multimode setting---when there are more than one quantum system. 

This chapter may be of interest for physicists interested in characterising Wigner negativity\index{Wigner negativity} of quantum states---either theoretically or experimentally---and mathematicians interested in infinite-dimensional convex optimisation theory.
It is structured as follows: we give a detailed exposition of our witnesses in Section~\ref{sec04:results}. Section~\ref{sec04:protocol} describes the experimental procedure for witnessing Wigner negativity of a quantum state using these witnesses, together with use-case examples. The following Section~\ref{sec04:opti}---which deals with infinite-dimensional optimisation techniques of independent interest---is devoted to estimating the threshold values of the witnesses: after some technical background in Section~\ref{subsec04:functionspaces}, Section~\ref{subsec04:LP} reformulates the problem of finding the threshold value of a witness as an infinite-dimensional linear optimisation\index{Linear program}, while Section~\ref{subsec04:SDP} derives two hierarchies of semidefinite relaxations and restrictions for this linear program\index{Semidefinite program}, yielding numerical upper and lower bounds for the threshold value. Section~\ref{subsec04:CVproof} establishes the proof of convergence of these hierarchies of upper and lower bounds to the threshold values in their respective optimisation spaces. 
We introduce the generalisation to the multimode case in Section~\ref{sec04:multi} and conclude with a few open questions in Section~\ref{sec04:conclusion}. This chapter is based on~\cite{Chabaud2021witnessingwigner}.

\section{Wigner negativity witnesses}
\label{sec04:results}

We introduce the following Wigner negativity witnesses:
\begin{equation}\label{eq:ch04_witnessOmega}
    \hat\Omega_{\bm a,\alpha}:=\sum_{k=1}^na_k\hat D(\alpha)\ket k\!\bra k\hat D^\dag(\alpha),
\end{equation}
for $n\in\mathbb N^*$, $\bm a=(a_1,\dots,a_n)\in[0,1]^n$, 
with $\max_ka_k=1$, and $\alpha\in\mathbb C$. These operators are \textit{finite weighted sums of displaced Fock states projectors}. 
They can be thought of as Positive Operator-Valued Measure (POVM) elements,\index{POVM} and their expectation value for a quantum state $\rho$ is given by\index{Displacement operator!CV}
\begin{equation}
    \Tr\!\left(\hat\Omega_{\bm a,\alpha}\,\rho\right)=\sum_{k=1}^na_kF\!\left(\hat D^\dag(\alpha)\rho\hat D(\alpha),\ket k\right),
\end{equation}
where $F$ is the fidelity\index{Fidelity} (see Subsection~\ref{subsec01:QM}). This quantity can be (efficiently) directly estimated from homodyne or heterodyne detection\index{Homodyne detection} of multiple copies of the state $\rho$ by 
translating the samples obtained by the amplitude $\alpha$ in the classical postprocessing and performing fidelity estimation with the Fock states $\ket1,\dots,\ket n$~\cite{d2003quantum,chabaud2020building}.

For $n\in\mathbb N^*$, each choice of $(\bm a,\alpha)\in[0,1]^n\times\mathbb C$ yields a different Wigner negativity witness. In particular, when $\alpha=0$ and only one entry of the vector $\bm a$ is non-zero, the expectation value of the witness is given by the fidelity with a single Fock state.

To each witness $\hat\Omega_{\bm a,\alpha}$ is associated a threshold value defined as:
\begin{equation}\label{eq:ch04_threshold}
    \omega_{\bm a}:=\sup_{\substack{\rho\in\mathcal D(\mathscr H)\\W_\rho \geq 0}}\Tr\!\left(\hat\Omega_{\bm a,\alpha}\,\rho\right).
\end{equation}
Since negativity of the Wigner function is invariant under displacements, the threshold values do not depend on the value of the displacement amplitude $\alpha$ and we thus write $\omega_{\bm a}$ (rather than $\omega_{\bm a,\alpha}$) for the threshold value associated to the witness $\hat\Omega_{\bm a,\alpha}$. This is sensible, given that the threshold value asks for nonnegativity anywhere in phase space, so a displacement in phase space should not change its value. 
Combining \begin{enumerate*}[label=(\roman*)]\item that the threshold value associated to a witness does not depend on the displacement parameter $\alpha$ and \item that we can always take into account displacement\index{Displacement operator} via classical post-processing if one uses homodyne or heterodyne detection associated to $\hat \Omega_{\bm a,0}$~\cite{chabaud2020efficient,chabaud2020certification}, we can restrict the analysis to witnesses of the form $\hat \Omega_{\bm a,0}$ that will generate the family $\{ \hat \Omega_{\bm a,\alpha} \}_{\alpha \in \C}$.\end{enumerate*}
Note however that the choice of displacement amplitude can still play an important role for certifying negativity of certain quantum states. For instance, the most suitable witness for detecting negativity of the displaced Fock state $\hat D(e^{i \frac{\pi}4}) \ket 1$ is itself \ie $\hat \Omega_{(1,0,\dots,0),e^{i \frac{\pi}4}}$ though $\hat \Omega_{(1,0,\dots,0),0}$ will be used experimentally and then classical post-processing will be applied on the samples to retrieve the displaced witness.

If the measured expectation value for an experimental state is higher than the threshold value given by \refeq{ch04_threshold}, this implies by definition that its Wigner function takes negative values.
Moreover, the following result shows that the amount by which the expectation value exceeds the threshold value directly provides an operational quantification of Wigner negativity for that state:\index{Wigner negativity}

\begin{lemma}\label{lemma:ch04_operational}
Let $\rho\in\mathcal D(\mathscr H)$ Wigner negative, and fix a witness $\hat\Omega_{\bm a,\alpha}$ defined in~\refeq{ch04_witnessOmega}, for $n\in\N^*$, $\bm a=(a_1,\dots,a_n)\in[0,1]^n$, and $\alpha\in\C$, with threshold value $\omega_{\bm a}$. Let us further assume that that it violates the threshold value of the witness \ie $\Tr(\hat\Omega_{\bm a,\alpha}\,\rho) > \omega_{\bm a}$ and denote the amount of violation as
\begin{equation}
    \delta_{\bm a,\alpha}(\rho):=\Tr\!\left(\hat\Omega_{\bm a,\alpha}\,\rho\right)-\omega_{\bm a}.
\end{equation}
Then,
\begin{equation}
    \eta_\rho\ge\delta_{\bm a,\alpha}(\rho),
\end{equation}
where $\eta_\rho$ is the distance between $\rho$ and the set of states having a positive Wigner function, defined in~\refeq{ch01_eta}.
\end{lemma}

\begin{proof}
We use the notations of the lemma. Let us consider the binary POVM\index{POVM} $\{\hat\Omega_{\bm a,\alpha},\mymathbb 1-\hat\Omega_{\bm a,\alpha}\}$. For all $\sigma\in\mathcal D(\mathscr H)$, we write $P^\sigma_{\bm a,\alpha}$ the associated probability distribution: $P^\sigma_{\bm a,\alpha}(0)=1-P^\sigma_{\bm a,\alpha}(1)=\Tr(\hat\Omega_{\bm a,\alpha}\,\sigma)$. 

Let $\sigma$ be a state with a positive Wigner function, so that $\Tr(\hat\Omega_{\bm a,\alpha}\,\sigma)\le\omega_{\bm a}$, by definition of the threshold value. We have:
\begin{equation}
    \begin{aligned}
        \delta_{\bm a,\alpha}(\rho)&=\Tr(\hat\Omega_{\bm a,\alpha}\rho)-\omega_{\bm a}\\
        &\le \Tr(\hat\Omega_{\bm a,\alpha}\rho)-\Tr(\hat\Omega_{\bm a,\alpha}\sigma) \\
        &=|P^\rho_{\bm a,\alpha}(0)-P^\sigma_{\bm a,\alpha}(0)|\\
        &=\|P^\rho_{\bm a,\alpha}-P^\sigma_{\bm a,\alpha}\|\\
        &\le D(\rho,\sigma),
    \end{aligned}
\end{equation}
where we used $\delta_{\bm a,\alpha}(\rho)\ge0$ in the second line, $\| P-Q\| =\frac12\sum_x|P(x)-Q(x)|$ denotes the total variation\index{Total variation distance} distance, and we used the operational property of the trace distance\index{Trace distance} in the last line~\cite{NielsenChuang}. We finally take the infimum over $\sigma$  and with the definition of $\eta_\rho$ (see~\refeq{ch01_eta}) that concludes the proof.
\end{proof}

\noindent This results directly extends to the case where only an upper bound of the threshold value is known: the amount by which the expectation value exceeds this upper bound is also a lower bound of the distance to the set of states having a positive Wigner function.

Importantly, the family of Wigner negativity\index{Wigner negativity} witnesses $\{\hat\Omega_{\bm a,\alpha}\}$ is \textit{complete}, i.e., for any quantum state with negative Wigner function there exists a choice of witness $(\bm a,\alpha)$ such that the expectation value of $\hat\Omega_{\bm a,\alpha}$ for this state is higher than the threshold value. Indeed, by taking $\bm a=(1,0,1,0,1,\dots)$, this family includes as a subclass the complete family of witnesses from~\cite{chabaud2020certification}. Indeed from Eq.~\eqref{eq:ch01_parity} (definition of the parity operator\index{Parity operator!CV}), Eq.~\eqref{eq:ch01_Wignerdef} (definition of the Wigner function\index{Wigner function}) and the completeness relation $\sum_{n \in \N} \ketbra{n}{n} = \Id$, the Wigner function of any density operator $\rho \in \mathcal D(\mathscr H)$ reads:
\begin{equation*}
    W_{\rho}(\alpha) = \frac 2 \pi \left(1 - 2 \Tr\left( \hat \Omega_{(1,0,1,0,\dots),\alpha} \rho \right) \right) \Mdot
\end{equation*}
Thus for any state with a negative Wigner function, there exists a choice of $\alpha \in \C$ such that the witness $\hat \Omega_{(1,0,1,0,\dots),\alpha}$ with threshold value $\frac 12$ can detect its negativity.

The threshold value in \refeq{ch04_threshold} is given by an optimisation problem over quantum states having a positive Wigner function. This is a convex subset of an infinite-dimensional space that does not possess a well-characterised structure. While solving this optimisation problem thus seems unfeasible in general, it turns out that we can obtain increasingly good numerical upper and lower bounds for the threshold value using semidefinite programming. This is developed in Section~\ref{sec04:opti}. 

The relevant programs are derived in Section~\ref{subsec04:SDPupper} and ~\ref{subsec04:SDPlower} and their respective convergence is proven in \ref{subsec04:CVproof}. Since these proofs introduces several intermediate forms of the programs,  we explicitly give them below to avoid confusion on which programs to implement numerically.
For  $n\in\mathbb N^*$, $\bm a=(a_1,\dots,a_n)\in[0,1]^n$, and $m\ge n$,
the hierarchies of semidefinite programs that respectively provide lower bounds and upper bounds for the threshold value $\omega_{\bm a}$ associated to the witnesses $\{ \hat \Omega_{\bm a,\alpha} \}_{\alpha \in  \C}$ are:
\leqnomode
\begin{flalign*}
    \label{prog:lowerSDP}
    \tag*{$(\text{SDP}^{m,\leq}_{\bm a})$}
    \hspace{3cm} \left\{
    \begin{aligned}\index{Semidefinite program}
            & \quad \text{Find } Q \in \SymMatrices{m+1} \text{ and } \bm{F} \in \R^{m+1} \\
            & \quad \text{maximising } \textstyle \sum_{k=1}^na_kF_k \\
            & \quad \text{subject to}  \\
            & \hspace{1cm} \begin{aligned}
            & \sum\limits_{k=0}^{m} F_k = 1  \\
            & \forall k \in \llbracket 0,m \rrbracket, \; F_k \geq 0\\
            & \forall l \in \llbracket 1,m \rrbracket, \; \sum_{i+j=2l-1} Q_{ij} = 0 \\
            & \forall l \in \llbracket 0,m \rrbracket, \; \sum_{i+j=2l} Q_{ij} = \sum_{k=l}^{m}  \frac{(-1)^{k+l}}{l!}  \binom kl F_k  \\
            & Q \succeq 0,
            \end{aligned} 
    \end{aligned}
    \right. &&
\end{flalign*}
and
\begin{flalign*}
    \label{prog:upperSDP}
    \tag*{$(\text{SDP}^{m,\geq}_{\bm a})$}
    \hspace{3cm} \left\{
    \begin{aligned}
            & \quad \text{Find } Q \in \SymMatrices{m+1} \text{ and } \bm{F} \in \R^{m+1} \\
            & \quad \text{maximising } \textstyle \sum_{k=1}^na_kF_k \\
            & \quad \text{subject to}  \\
            & \hspace{1cm} \begin{aligned}
            & \sum_{k=0}^{m} F_k = 1  \\
            & \forall k \in \llbracket 0,m \rrbracket, \; F_k \geq 0\\
            & \forall l \in \llbracket 1,m \rrbracket, \forall i+j=2l-1, \; A_{ij} = 0 \\
            & \forall l \leq m, \forall i+j = 2l, \;  A_{ij} = \sum_{k=0}^{l} F_k  \binom lk l!  \\
            & A \succeq 0,
            \end{aligned} 
    \end{aligned}
    \right. &&
\end{flalign*}
\reqnomode
Let $\omega_{\bm a}^{m,\geq} =$ val\refprog{upperSDP} be the optimal value of \refprog{upperSDP}. We show in Section~\ref{sec04:opti} that the sequence $(\omega_{\bm a}^{m,\geq})_{m\ge n}$ is a decreasing sequence of upper bounds of $\omega_{\bm a}$, which converges to $\omega_{\bm a}$. Similarly, let $\omega_{\bm a}^{m,\leq}$ be the optimal value of \refprog{lowerSDP}. We show that the sequence $(\omega_{\bm a}^{m,\leq})_{m\ge n}$ is an increasing sequence of lower bounds of $\omega_{\bm a}$, which converges to $\omega_{\bm a}^\mathcal S$---a modified threshold value computed with Schwartz functions rather than square-integrable functions; we have $\omega_{\bm a}^\mathcal S\le\omega_{\bm a}$, and the equality between the two values is still open.

In particular, the numerical upper bounds $\omega_{\bm a}^{m,\geq}$ can be used instead of the threshold value $\omega_{\bm a}$ to witness Wigner negativity\index{Wigner negativity}, while the numerical lower bounds $\omega_{\bm a}^{m,\leq}$ may be used to control how much the upper bounds differ from the threshold value. We give a detailed procedure in the following section, together with use-case examples and details on the numerical implementation.

A few comments regarding the state-of-the-art. 
The idea of measuring witnesses of Wigner negativity using
displacements and photon number parity measurements was already proposed in \cite{Lutterbach97} and implemented experimentally in photonic cavities in \cite{Nogues2000,Bertet2002}. Our Wigner negativity witnesses provide much more flexibility as it does not require the knowledge of a phase-space point $\alpha \in \C$ where the Wigner function might be negative. 
Our witnesses also outperform existing ones~\cite{mari2011directly,fiuravsek2013witnessing} in terms of generality and practicality, since they form a complete family and provide much more flexibility with the choice of $n\in\N^*$, $\bm a\in[0,1]^n$ and $\alpha\in\mathbb C$. They are accessible with optical homodyne or heterodyne measurement, and do not require making any assumption on the measured state, unlike other existing methods~\cite{Lutterbach97,Mattia2017PRL}. Moreover, our witnesses generalise those of~\cite{chabaud2020certification}, and may provide simpler alternatives to detect Wigner negativity. We also provide two converging hierarchies to approximate the threshold values associated to these witnesses. 
Proving convergence is important 
and was not considered in the other approaches mentioned above. Finally, contrary to all approaches detailed above, ours also generalises to the multimode setting, as detailed in section~\ref{sec04:multi}.

\section{Procedure for witnessing Wigner negativity}
\label{sec04:protocol}

\subsection{Procedure}
\label{subsec04:procedure}

In this section, we describe an experimental procedure to check whether a continuous-variable quantum state exhibits Wigner negativity using our witnesses.\index{Wigner negativity}
\begin{figure}[ht!]
	\begin{mybox}{Witnessing Wigner negativity:}\label{box:ch04_protocol}
    \justifying{
        \begin{enumerate}
            \item Choose a suitable fidelity\index{Fidelity}-based witness $\hat\Omega_{\bm a,\alpha}$ defined in \refeq{ch04_witnessOmega} by picking $n\in\mathbb N^*$, $\bm a\in[0,1]^n$ and $\alpha\in\mathbb C$.
            \item Run the upper bounding semidefinite program \refprog{upperSDP} for $m\ge n$ to get a numerical estimate $\omega^{m,\geq}_{\bm a}$.
             \item Run the lower bounding semidefinite\index{Semidefinite program} program \refprog{lowerSDP} for $m \ge n$ to get a numerical estimate $\omega^{m,\leq}_{\bm a}$. 
             \item Estimate the expectation value for that witness of the experimental state from samples of homodyne or heterodyne detection\index{Homodyne detection} by performing fidelity\index{Fidelity} estimation with the corresponding Fock states and translating the samples by $\alpha$. This yields an experimental witness value denoted $\omega_{\text{exp}}$.
            \item Compare the value obtained experimentally with the numerical bounds:
            if it is greater than the numerical upper bound, i.e.
            $\omega_{\text{exp}}\ge\omega^{m,\geq}_{\bm a}$, then the state displays Wigner negativity, and its distance to the set of Wigner positive states is lower bounded by $\omega_{\text{exp}}-\omega^{m,\geq}_{\bm a}$. 
            Otherwise, if it is lower than the numerical lower bound, i.e. $\omega_{\text{exp}}\leq \omega^{m,\leq}_{\bm a}$, then the witness cannot detect Wigner negativity for this state. 
        \end{enumerate}
    }
	\end{mybox}
\end{figure}

The main subroutine of this procedure is to estimate fidelities with displaced Fock states using classical samples from homodyne or heterodyne detection\footnote{Actually, using a fidelity witness rather than a fidelity estimate is sufficient for our purpose.}, in order to compute the experimental value for a Wigner negativity witness. As already mentioned, displacement can be achieved with classical post-processing by translating the classical samples according to the displacement amplitude, and performing direct fidelity estimation with Fock states \cite{lvovsky2009continuous,d2003quantum,chabaud2020building}.

Upper and lower bounds on the threshold value of the witness are then obtained using semidefinite programming, and comparing the experimental witness value to these bounds gives insight about the Wigner negativity of the measured quantum state.

We give a detailed procedure for using our witnesses for detecting Wigner negativity in the framed box. This procedure starts by the choice of a specific witness, and we explain hereafter a heuristic method for picking a good witness.

If the experimental state is anticipated to have negativity at $\alpha$, then one may use the witness with parameters $(n,\bm a,\alpha)$ with $\bm a=(1,0,1,0,\dots)$, which will detect negativity for $n$ large enough~\cite{chabaud2020certification}. However, this may imply having to estimate fidelities with Fock states having a high photon number with homodyne or heterodyne detection, which requires a lot of samples, while simpler witnesses can suffice for the task and be more efficient, as we show in the next section. Moreover, there are cases where the state to be characterised is fully unknown. 
Instead, a simple heuristic for picking a good witness for Wigner negativity is the following: 
\begin{itemize}
\item From samples of homodyne or heterodyne detection of multiple copies of an experimental state, estimate the expected values of witnesses in \refeq{ch04_witnessOmega} for a small value of $n$ and a large set of values $\bm a$ and $\alpha$, using the same samples for all witnesses. 
\item Based on these values, pick the simplest witness possible---with the smallest value of $n$---that is able to witness Wigner negativity with a reasonable violation. This is done by comparing the estimated expected values with the upper bounds on the corresponding threshold values computed numerically. 
These bounds depend only the choice of the witness parameters $n,\bm a$ and can even be precomputed using~\refprog{upperSDP}. To facilitate the use of our methods, we have collected such bounds for $\bm a=(0,\dots,0,1)$ and $n\le10$ in Table~\ref{tab:ch04_Fockbounds}. We also precomputed these bounds for $n=3$ and a large number of values of $\bm a$ \cite{codes}. 
\item Then, estimate the expected value for that witness using a new collection of samples---thus obtaining proper error bars and avoiding the accumulation of statistical errors.
\end{itemize}

\noindent In what follows, we give a few theoretical examples for using our witnesses to detect negativity of the Wigner function of single-mode quantum states.\index{Wigner negativity}

\subsection{Examples}
\label{subsec04:examples}

We identify three levels of generality within our family of single-mode witnesses in \refeq{ch04_witnessOmega}:\index{Fidelity} \begin{enumerate*}[label=(\roman*)]\item fidelities with single Fock states,\item linear combinations of fidelities with Fock states, and \item displaced linear combinations of fidelities with Fock states.\end{enumerate*} 

Fidelities with Fock states are the most practical of our witnesses, since they require the estimation of only one diagonal element of the density matrix of the measured state.
The corresponding values in Table~\ref{tab:ch04_Fockbounds} can be used directly by experimentalists: if an estimate of $\braket{n | \rho | n}$ for appropriate $n$ is above one of these numerical upper bounds then it ensures that $\rho$ has a Wigner function with negative values. Moreover, by Lemma~\ref{lemma:ch04_operational}, the amount by which the estimate of $\braket{n | \rho | n}$ exceeds the numerical upper bound directly provides a lower bound on the distance between $\rho$ and the set of states having a positive Wigner function.

\begin{figure}[ht!]
    \centering
    \includegraphics[width=.6\linewidth]{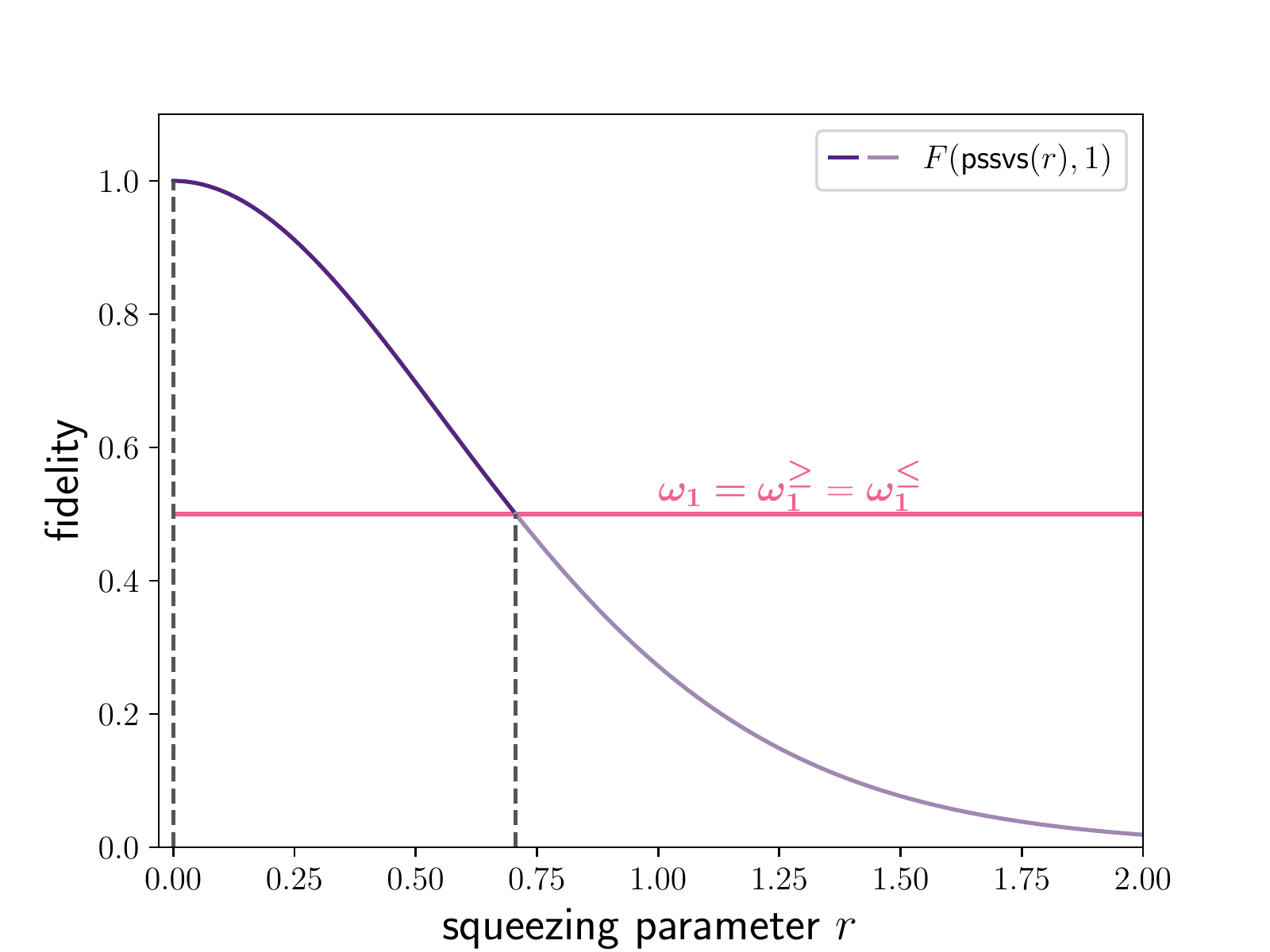}
    \caption{Fidelities of photon-subtracted squeezed vacuum states $\ket{\text{p-ssvs}(r)}$ with the Fock state $\ket1$ with respect to the squeezing parameter $r\in\R$. The dashed grey lines delimit the intervals of parameter values where the witness $\ketbra{1}{1}$ can be used to detect Wigner negativity, \ie when the fidelity\index{Fidelity} (violet curve) is above the witness upper bound (pink line). The height difference when the violet curve is above the pink line directly provides a lower bound on the distance between the corresponding state and the set of states having a positive Wigner function.} 
    \label{fig:ch04_ex_pssvs}
\end{figure}

For instance, if we focus on $n=3$ in Table~\ref{tab:ch04_Fockbounds}, the threshold value $\omega_3$ satisfies $0.378 \le \omega_3 \le 0.427$. Having a state $\rho$ such that $\braket{3|\rho|3} > 0.427$ guarantees that $\rho$ has Wigner negativity. If $\braket{3|\rho|3} < 0.378$ then we conclude that the witness cannot detect negativity for this state.
\begin{table}[htp]
    \centering
        \begin{tabular}{@{}l|cc@{}}
            \toprule
            $n$ & Lower bound & Upper bound \\ \midrule
            1 & 0.5 & 0.5   \\
            2 & 0.5 & 0.5   \\
            3 & 0.378 & 0.427 \\
            4 & 0.375 & 0.441 \\
            5 & 0.314 & 0.385 \\
            6 & 0.314 & 0.378 \\
            7 & 0.277 & 0.344 \\
            8 & 0.280 & 0.348 \\
            9 & 0.256 & 0.341 \\
            10 & 0.262 & 0.334 \\
            \bottomrule
        \end{tabular}
    \caption{Table of numerical upper and lower bounds for the threshold value $\omega_n$ of various Wigner negativity witnesses obtained using our hierarchies of semidefinite programs at rank $m$ up to around $30$. The witnesses considered here are Fock states projectors (photon-number states $\ket n$) from $1$ to $10$. Note that the gap between the lower and upper bounds never exceeds $0.1$. Additionally, the bounds in the first two lines are analytical (see Section~\ref{subsec04:LP}) and for the upper bounds, the corresponding values obtained numerically are $0.528$ and $0.551$, respectively. See Section~\ref{subsec04:numerical} and \cite{codes} for the numerical implementation.}
    \label{tab:ch04_Fockbounds}
\end{table}
When the experimental state is close to a Fock state (different from the vacuum), a natural choice for the witness thus is the fidelity with the corresponding Fock state. 
For instance, consider a photon-subtracted squeezed vacuum state~\cite{ourjoumtsev2006generating}
\begin{equation}
    \ket{\text{p-ssvs}(r)}=\frac1{\sinh r}\hat a\hat S(r)\ket 0,
\end{equation}
where $\hat S(r)=e^{\frac r2(\hat a^2-\hat a^{\dag2})}$ is a squeezing operator with parameter $r\in\mathbb R$. Its fidelity with the single-photon Fock state $\ket1$ is given by:
\begin{equation}\label{eq:ch04_fidepssvs}
    \frac1{(\sinh r)^2}\left|\braket{1|\hat a\hat S(r)|0}\right|^2=\frac1{(\cosh r)^3}.
\end{equation}
When the squeezing parameter is small, this state is close to a single-photon Fock state. In particular, for $0<r<0.70$, the fidelity $F(\text{p-ssvs}(r),1)$ in~\refeq{ch04_fidepssvs} is greater than $\omega_1^\ge=\frac12$ and our witness can be used to detect Wigner negativity of this state (see~\refig{ch04_ex_pssvs}).

Another example is given by superpositions of coherent states: we consider the cat state~\cite{sanders1992entangled}
\begin{equation}
    \ket{\text{cat}_2(\alpha)}=\frac{\ket\alpha+\ket{-\alpha}}{\sqrt{2(1+e^{-2|\alpha|^2})}},
\end{equation}
and the compass state~\cite{zurek2001sub}
\begin{equation}
 \ket{\text{cat}_4(\alpha)}=\frac{\ket\alpha+\ket{-\alpha}+\ket{i\alpha}+\ket{-i\alpha}}{2\sqrt{1+e^{-|\alpha|^2}(2\cos(|\alpha|^2)+1)}},   
\end{equation}
where $\ket\alpha=e^{-\frac12|\alpha|^2}\sum_{k\ge0}\frac{\alpha^k}{\sqrt{k!}}\ket k$ is the coherent state of amplitude $\alpha\in\mathbb C$. We have
\begin{equation}\label{eq:ch04_fidecat}
    \left|\braket{2|\text{cat}_2(\alpha)}\right|^2=\frac{|\alpha|^4}{2\cosh(|\alpha|^2)},
\end{equation}
and 
\begin{equation}\label{eq:ch04_fidecompass}
    \left|\braket{4|\text{cat}_4(\alpha)}\right|^2=\frac{|\alpha|^8/12}{\cosh(|\alpha|^2)+\cos(|\alpha|^2)}.
\end{equation}

\begin{figure}
\centering
\hfill
\begin{subfigure}[b]{0.49\textwidth}
   \includegraphics[width=1\linewidth]{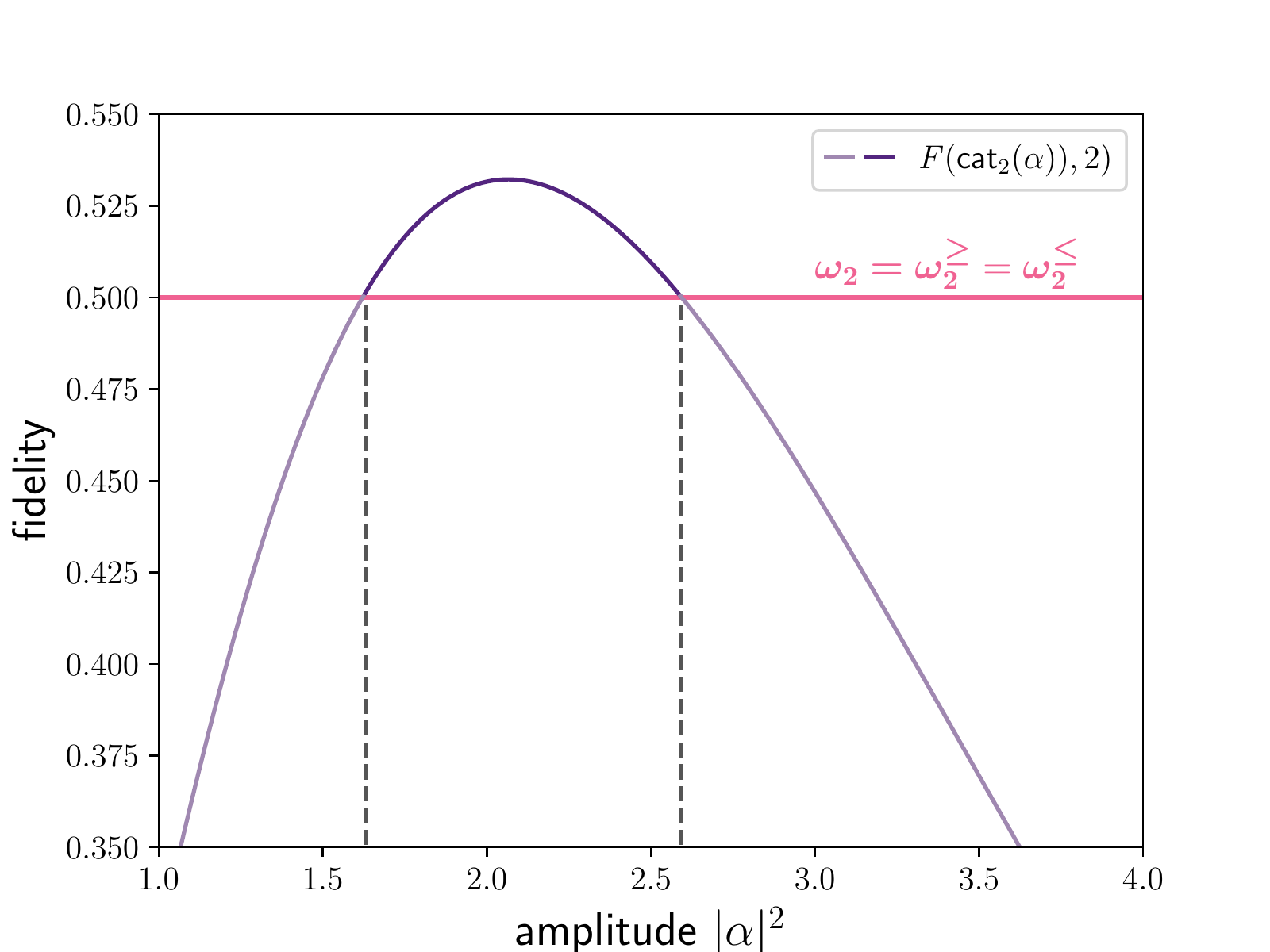}
   \caption{}
\end{subfigure}
\hfill
\begin{subfigure}[b]{0.49\textwidth}
   \includegraphics[width=1\linewidth]{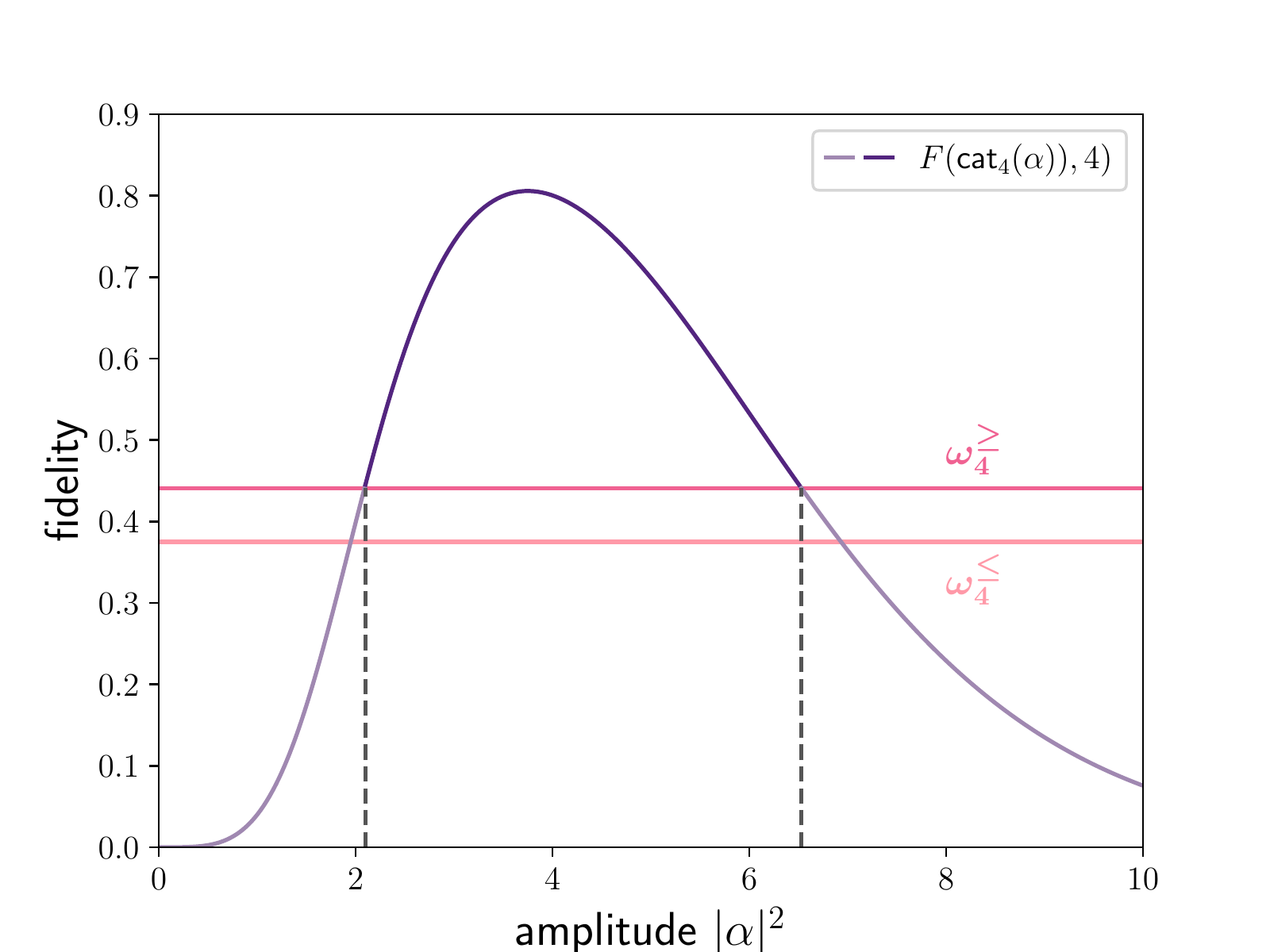}
   \caption{}
\end{subfigure}
\caption{(a) Fidelities of cat states $\ket{\text{cat}_2(\alpha)}$ with amplitude $\alpha\in\mathbb C$ with the Fock state $\ket2$. (b) Fidelities of compass states $\ket{\text{cat}_4(\alpha)}$ with amplitude $\alpha\in\mathbb C$ with the Fock state $\ket4$. The dashed grey lines delimit the intervals of amplitude where our witnesses from Table~\ref{tab:ch04_Fockbounds} can be used to detect Wigner negativity of the corresponding state. When it is below the witness lower bound (light pink line), we are guaranteed that the witness (here $\ket4\!\bra4$) cannot be used to detect Wigner negativity of the state.}
\label{fig:ch04_ex_cats}
\end{figure}

For $1.63\le|\alpha|^2\le2.59$, the fidelity $F(\text{cat}_2(\alpha),2)$ in~\refeq{ch04_fidecat} is greater than $\omega_2^\ge=\frac12$ and our witness corresponding to $n=2$ can be used to detect Wigner negativity of this state. Similarly, for $2.10\le|\alpha|^2\le6.53$, the fidelity\index{Fidelity} $F(\text{cat}_4(\alpha),4)$ in~\refeq{ch04_fidecompass} is greater than $\omega_4^\ge=0.441$ and our witness corresponding to $n=4$ can be used to detect Wigner negativity of this state (see~\refig{ch04_ex_cats}).\index{Wigner negativity}

Some quantum states will remain unnoticed by all single Fock state negativity witnesses. For example, the state $\rho_{0,1,2} := \frac19 \ket0\!\bra0 + \frac49 \ket1\!\bra1+ \frac49 \ket2\!\bra2$ has a negative Wigner function but is not detected by any of the single Fock state negativity witnesses, since the lower bounds for $n=1,2$ in Table~\ref{tab:ch04_Fockbounds} are higher than $\frac49$, and this state has fidelity $0$ with higher Fock states. However, it is detected by simple witnesses based on linear combinations of fidelities. For example, with $n=2$ and $\bm a=(1,1)$ we find numerically that the threshold value of the witness $\ket1\!\bra1+\ket2\!\bra2$ is less than $0.875$ when running the corresponding program \refprog{upperSDP} for $m=7$. 
And since $\Tr(\rho_{0,1,2} (\ket1\!\bra1+\ket2\!\bra2)) = \frac89 > 0.875$, this linear combination of Fock state fidelities can indeed detect Wigner negativity for this state.

However, some quantum states with a negative Wigner function will always go unnoticed with witnesses of the previous form because these witnesses are invariant under phase-space rotations while the Wigner function of those states becomes positive under random dephasing. Consider for instance the superposition $\sqrt{1-\frac1s}\ket0+\frac1{\sqrt s}\ket1$, for $s>2$ (which under random dephasing is mapped to $(1-\frac1s)\ket0\!\bra0+\frac1s\ket1\!\bra1$). In that case, the Wigner negativity of such states can still be witnessed by using the displaced version of our witnesses. In particular, if any single-mode quantum state has a Wigner function negative at $\alpha\in\mathbb C$, then there is a choice of $n\in\mathbb N^*$ such that the witness in \refeq{ch04_witnessOmega} defined by $\bm a=(1,0,1,0,1,\dots)$ and the displacement amplitude $\alpha$ detects its negativity~\cite{chabaud2020certification}.
In practice, simpler witnesses may suffice to detect negativity, and the choice of witness will ultimately depend on the experimental state at hand.
\begin{figure}[ht]
	\begin{center}
		\includegraphics[width=.6\linewidth]{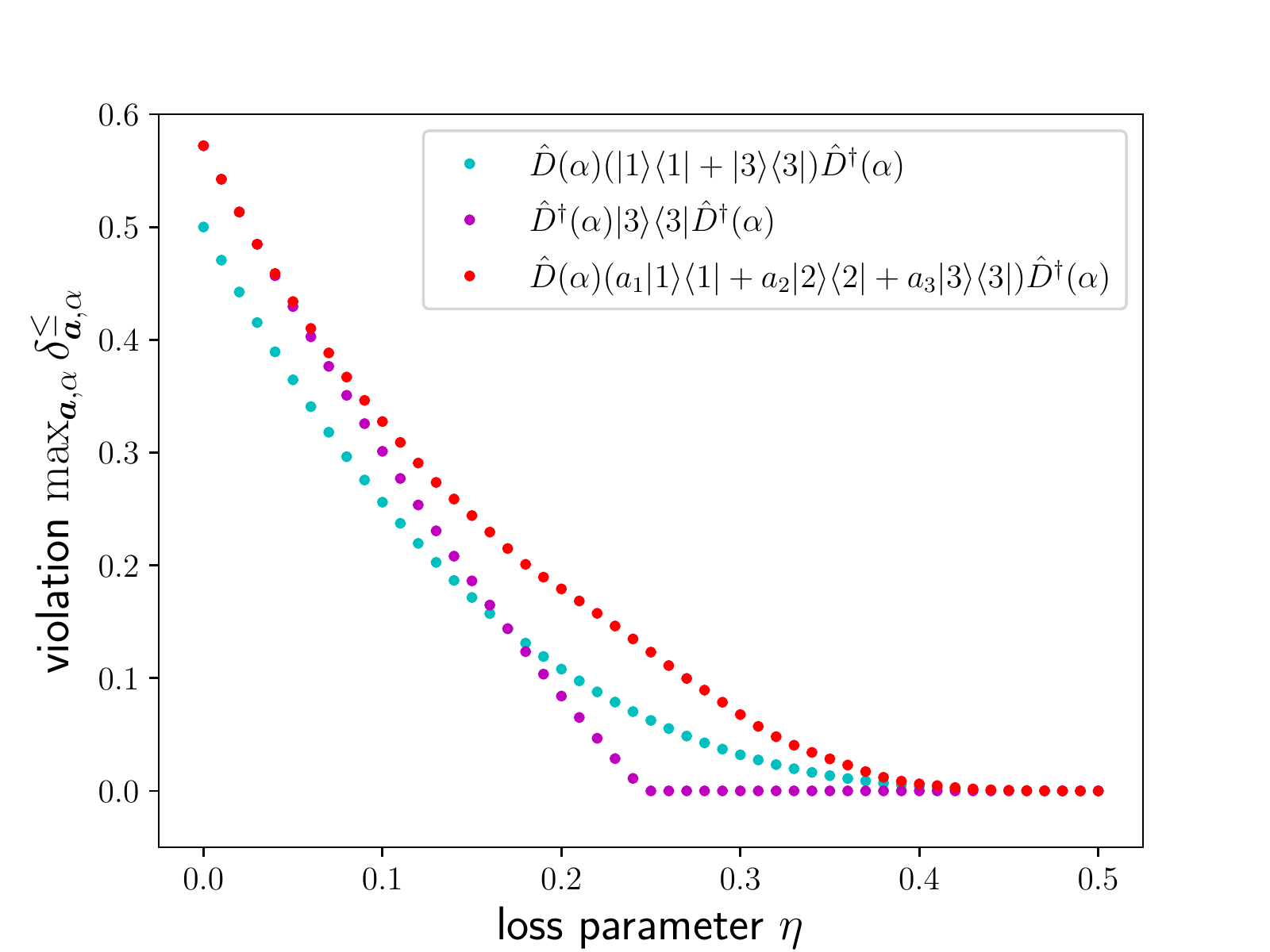}
		\caption{Lower bounds on the violation $\delta_{\bm a,\alpha} = \Tr(\hat\Omega_{\bm a,\alpha}\,\rho_{3,\eta})-\omega_{\bm a}$ for a lossy 3-photon Fock state $\rho_{3,\eta} = \eta^3\ket3\!\bra3+3\eta^2(1-\eta)\ket2\!\bra2
    +3\eta(1-\eta)^2\ket1\!\bra1+(1-\eta)^3\ket0\!\bra0$ for $\bm a = (a_1,a_2,a_3)$ with respect to the loss parameter $\eta$. Precomputed bounds on threshold values for witnesses of the form $\hat \Omega_{\bm a}= a_1 \ket1\!\bra1 + a_2 \ket2\!\bra2 + a_3 \ket3\!\bra3 $ can be found in \cite{codes}. We use these values to find the witness $\hat\Omega_{\bm a,\alpha}$ giving the maximum lower bound $\delta_{\bm a,\alpha}^{\le} =\Tr(\hat\Omega_{\bm a,\alpha}\,\rho_{3,\eta})-\omega_{\bm a}^{\ge}$ on the violation $\delta_{\bm a,\alpha}$, for different values of the loss parameter $\eta$. These optimised lower bounds are represented in red. In blue is the maximal violation that can be detected using the witnesses $\hat \Omega_{(1,0,1),\alpha}=\hat D(\alpha)(\ket1\!\bra1+\ket3\!\bra3)\hat D^\dag(\alpha)$~\cite{chabaud2020certification}. In violet is the maximal violation that can be detected using the more naive witness $\hat \Omega_{(0,0,1),\alpha}=D(\alpha)\ket3\!\bra3D^\dag(\alpha)$. Note that $\rho_{3,\eta}$ has a nonnegative Wigner function for $\eta \geq 0.5$.
    }
		\label{fig:ch04_lossy}
	\end{center}
\end{figure}
Hereafter we discuss the heuristics for picking a good witness, with the theoretical example of the lossy $3$-photon Fock state:
\begin{align}
\begin{split}
\rho_{3,\eta} &:= (1-\eta)^3\ket3\!\bra3+3\eta(1-\eta)^2\ket2\!\bra2\\
&\;\;+3\eta^2(1-\eta)\ket1\!\bra1+\eta^3\ket0\!\bra0,
\end{split}
\end{align}
where $0\le\eta\le1$ is the loss parameter. Setting $\eta=0$ gives $\rho_{3,\eta}=\ket3\!\bra3$ while setting $\eta=1$ gives $\rho_{3,\eta}=\ket0\!\bra0$. This state has a nonnegative Wigner function for $\eta\geq\frac12$. 
The fidelities of $\rho_{3,\eta}$ with displaced Fock states $\hat D(\alpha)\ket l$ are given by:
\begin{equation}\label{eq:ch04_fiderho3etal}
    \begin{aligned}
        F(\rho_{3,\eta},\hat D(\alpha)&\ket l)\\
        &=(1-\eta)^3|\braket{3|\hat D(\alpha)|l}|^2\\
        &+3\eta(1-\eta)^2|\braket{2|\hat D(\alpha)|l}|^2\\
        &+3\eta^2(1-\eta)|\braket{1|\hat D(\alpha)|l}|^2\\
        &+\eta^3|\braket{0|\hat D(\alpha)|l}|^2.    
    \end{aligned}
\end{equation}
where the coefficients of the displacement operator in Fock basis are given in Eq.~\eqref{eq:ch01_coefD}.
In an experimental scenario, the state would be unknown and these fidelities should be estimated using samples from a homodyne or heterodyne detection of the state translated by $\alpha$, and estimating the fidelities with Fock states~\cite{chabaud2020efficient,chabaud2020certification}. 

Following the heuristic detailed in the previous section, we have determined suitable Wigner negativity witnesses for $50$ values of the loss parameter $\eta$ between $0$ and $0.5$ as follows: for each value $\eta$, we have computed numerically the values of the fidelities in Eq.~\eqref{eq:ch04_fiderho3etal} for $l=1,2,3$, and for displacement parameters $\alpha=q/10+ip/10$ for all $q,p\in\llbracket0,10\rrbracket$. Using these values, we have computed the expectation value of the witnesses $\hat\Omega_{\bm a,\alpha}$ for multiple choices of $\bm a=(a_1,a_2,a_3)$ with $\max_ia_i=1$. We have used the corresponding precomputed bounds $\omega_{\bm a}^{\ge}$ on the threshold values of $\hat\Omega_{\bm a,\alpha}$ in \cite{codes} to determine the witness leading to the maximal lower bound $\delta_{\bm a,\alpha}^{\le} :=\Tr(\hat\Omega_{\bm a,\alpha}\,\rho_{3,\eta})-\omega_{\bm a}^{\ge}$ on the violation $\delta_{\bm a,\alpha}=\Tr(\hat\Omega_{\bm a,\alpha}\,\rho_{3,\eta})-\omega_{\bm a}$ over the choice of $(\bm a,\alpha)$.

We have represented these violations for each value of the loss parameter $\eta$ in Fig.~\ref{fig:ch04_lossy}. For all values of $\eta$, we find that the optimal displacement parameter is $\alpha=0$. On the other hand, we find different optimal choices of $\bm a$ for different values of $\eta$. To illustrate the usefulness of the optimisation over the choice of witnesses parametrised by $(\bm a,\alpha)$, we have also represented the violations obtained when using the witnesses $\hat\Omega_{(1,0,1),\alpha}=\hat D(\alpha)(\ket1\!\bra1+\ket3\!\bra3)\hat D^\dag(\alpha)$\index{Displacement operator!CV} from~\cite{chabaud2020certification} for all values of $\eta$. In that setting, the violations obtained quantify how hard it is to detect the Wigner negativity of the state: a larger violation implies that a less precise estimate of the witness expectation value is needed to witness Wigner negativity.
In particular, we obtain that our optimised witnesses always provide a greater violation to detect negativity than the previous witnesses which will result in an easier experimental detection. 
We also represented the violation obtained when using the more naive witnesses $\hat\Omega_{(0,0,1),\alpha}=\hat D(\alpha)\ket3\!\bra3\hat D^\dag(\alpha)$ and we see that it is only useful when the loss parameter is smaller than $0.25$, while the optimised witnesses may detect negativity of the state $\rho_{3,\eta}$ up to $\eta=0.5$---when the Wigner functions becomes nonnegative---provided the estimates of the fidelities are precise enough.\index{Wigner negativity}

Overall, this procedure only amounts to a simple classical post-processing of samples from homodyne or heterodyne detection and yields a good witness for detecting Wigner negativity.

\subsection{Numerical implementation}
\label{subsec04:numerical}

Here we discuss numerical implementations of the semidefinite programs \refprog{lowerSDP} and \refprog{upperSDP}. All codes are available \href{https://archive.softwareheritage.org/swh:1:dir:d98f70e386783ef69bf8c2ecafdb7b328b19b7ec}{here} \cite{codes}.\index{Semidefinite program}

We implemented the semidefinite programs with Python through the interface provided by PICOS~\cite{sagnol2012picos}. We first used the solver Mosek \cite{mosek} to solve these problems but, while the size of the semidefinite programs remains relatively low for small values of $n$ and $m$, binomial terms grow rapidly and numerical precision issues arise quickly (usually around $m=12$, $n \le m$ on a laptop).  
The linear constraints involving $Q_{ij}$ in the semidefinite programs come from a polynomial equality (see Lemma~\ref{lemma:ch04_pospolyR}). While polynomial equalities are usually written in the canonical basis, a first trick is to express them in a different basis---for instance the basis $(1,\frac X{1!},\frac {X^2}{2!}, \dots)$---to counterbalance the binomial terms. 

However, this may not be sufficient to probe larger values of $m$. Instead, we used the solver SDPA-GMP~\cite{nakata2010numerical,fujisawa2002sdpa} which allows arbitrary precision arithmetic. While much slower, this solver is dedicated to solve problems requiring a lot of precision. Because our problems remain rather small, time efficiency is not an issue and this solver is particularly well-suited. All problems were initially solved on a regular laptop as warning flags on optimality were raised before the problems were too large. A high-performance computer\footnote{DELL PowerEdge R440, 384 Gb RAM, Intel Xeon Silver 4216 processor, 64 threads from LIP6. } handling floating point arithmetic more accurately was later used to compute further ranks in the hierarchy. 

Using the semidefinite programs \refprog{upperSDP} and \refprog{lowerSDP} for values of $m$ up to around $30$ and $\bm a=(0,0,\dots,0,1)$ (where the size $n$ of the vector $\bm a$ is ranging from $1$ to $10$), we have obtained upper and lower bounds for the threshold values of Wigner negativity witnesses corresponding to fidelities with Fock states from $1$ to $10$, reported in Table~\ref{tab:ch04_Fockbounds}.

We also computed upper and lower bounds on the threshold values of witnesses of the form:
\begin{equation}
    \hat \Omega_{(a_1,a_2,a_3)} = \sum_{k=1}^3a_k\ket k\!\bra k,
\end{equation}
where $\forall i \in \{1,2,3\},\, 0 \leq a_i \leq 1$ and $\max_i a_i =1$. We focused on these particular witnesses for experimental considerations as it is challenging to obtain fidelities with higher Fock states. We fix one coefficient equal to $1$ and vary each other $a_i$ from $0$ to $1$ with a step of $0.1$. The resulting bounds on the threshold values can be found in \cite{codes} and in \cite{Chabaud2021witnessingwigner}. 

We now turn to the mathematical proofs of our results, i.e., that the threshold values in \refeq{ch04_threshold} can be upper bounded and lower bounded by the optimal values of the converging hierarchies of semidefinite programs \refprog{upperSDP}$_{m\ge n}$ and \refprog{lowerSDP}$_{m\ge n}$, respectively.

The following section is rather technical as we dive into infinite-dimensional optimisation techniques to derive the hierarchies of semidefinite programs and prove their convergence. Some readers may want to skip directly to Section~\ref{sec04:multi}.

\section{Infinite-dimensional optimisation}
\label{sec04:opti}

\begin{figure}[t]
	\begin{center}
		\begin{tikzpicture}[scale=0.56]

\node[inner sep=0pt] (A) at (-0.8,-1) {};
\node[inner sep=0pt] (B) at (-0.8,8) {};
\node[inner sep=0pt] (C1) at (-0.8,0) {};
\node[inner sep=0pt] (C2') at (-0.8,2.75) {};
\node[inner sep=0pt] (C2) at (-0.8,4.25) {};
\node[inner sep=0pt] (C3) at (-0.8,7) {};

\draw[|-|] (-0.8,-1) -- (-0.8,-1);
\draw[-|] (-0.8,-1) -- (-0.8,0);
\draw[-|] (-0.8,0) -- (-0.8,2.75);
\draw[dashed] (-0.8,2.85) -- (-0.8,4.2);
\draw[|-|] (-0.8,4.25) -- (-0.8,7);
\draw[-|] (-0.8,7) -- (-0.8,8);

\node[left] (zero) at (A) {0};
\node[left] (one) at (B) {1};
\node[left] (omegainf) at (C1) {$\omega_n^{m,\le}$};
\node[left] (omega) at (C2) {$\omega_n^{L^2}$};
\node[left] (omega) at (C2') {$\omega_n^\mathcal{S}$};
\node[left] (omegasup) at (C3) {$\omega_n^{m,\ge}$};

\hypersetup{linkcolor=black}
\node[inner sep=0pt] (SDPsup) at (2,7) {\Large \ref{prog:upperSDPn}};
\node[inner sep=0pt] (D-SDPsup) at (10,7) {\Large \ref{prog:upperDSDPn}};

\node[inner sep=0pt] (LP) at (2,4.25) {\Large \ref{prog:LP}};
\node[inner sep=0pt] (D-LP) at (10,4.25) {\Large \ref{prog:DLP}};

\node[inner sep=0pt] (LPS) at (2,2.75) {\Large \ref{prog:LPS}};
\node[inner sep=0pt] (D-LPS) at (10,2.75) {\Large \ref{prog:DLPS}};

\node[inner sep=0pt] (SDPinf) at (2,0) {\Large \ref{prog:lowerSDPn}};
\node[inner sep=0pt] (D-SDPinf) at (10,0) {\Large \ref{prog:lowerDSDPn}};
\hypersetup{linkcolor=\lkcolor}

\node[inner sep=0pt] (eqsup) at ($.54*(SDPsup)+.46*(D-SDPsup)$) 
{$\substack{\text{Corollary}~\ref{th:ch04_sdupper} \\ =\joinrel=}$};

\node[inner sep=0pt] (eq) at ($.54*(LP)+.46*(D-LP)$) 
{$\substack{\text{Corollary}~\ref{th:ch04_sdLP} \\ =\joinrel=}$};

\node[inner sep=0pt] (eq) at ($.54*(LPS)+.46*(D-LPS)$) 
{$\substack{\text{Corollary}~\ref{th:ch04_sdLP} \\ =\joinrel=}$};

\node[inner sep=0pt] (eqinf) at ($.54*(SDPinf)+.46*(D-SDPinf)$) 
{$\substack{\text{Theorem}~\ref{th:ch04_sdlower} \\ =\joinrel=}$};

\node[inner sep=0pt] (CVlu) at ($.5*(SDPsup)+.5*(LP)$)
{$\substack{\text{Theorem~\ref{th:ch04_upperCV}} \\\hspace{20pt} \big\downarrow m\rightarrow \infty}$};

\node[inner sep=0pt] (CVld) at ($.5*(SDPinf)+.5*(LPS)$) 
{$\substack{\hspace{20pt} \big\uparrow m\rightarrow \infty \\ \text{Theorem~\ref{th:ch04_lowerCV}}}$};

\node[inner sep=0pt] (CVru) at ($.5*(D-SDPsup)+.5*(D-LP)$) 
{$\substack{\text{Theorem~\ref{th:ch04_upperCV}} \\\hspace{20pt} \big\downarrow m\rightarrow \infty}$};

\node[inner sep=0pt] (CVrd) at ($.5*(D-SDPinf)+.5*(D-LPS)$) 
{$\substack{\hspace{20pt} \big\uparrow m\rightarrow \infty \\ \text{Theorem~\ref{th:ch04_lowerCV}}}$};

\end{tikzpicture}
		\hypersetup{linkcolor=black}
		\caption{Hierarchy of semidefinite relaxations converging to the linear program \ref{prog:LP} and hierarchy of semidefinite restrictions converging to the linear program \ref{prog:LPS}, together with their dual programs. The upper index $m$ denotes the level of the relaxation or restriction. On the left are the associated optimal values. The equal sign denotes strong duality, i.e., equality of optimal values, and the arrows denote convergence of the corresponding sequences of optimal values. Note that the question of the closing the gap between the values of \ref{prog:LP} and \ref{prog:LPS} is left open.
		}
		\hypersetup{linkcolor=\lkcolor}
		\label{fig:ch04_structure}
	\end{center}
\end{figure}
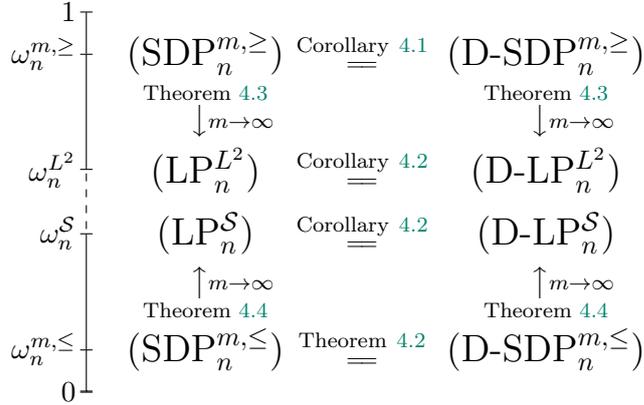

In this section we use infinite-dimensional optimisation techniques:\index{Linear program}
\begin{enumerate*}[label=(\roman*)]
    \item to phrase the computation of the witness threshold value introduced in \refeq{ch04_threshold} as an infinite-dimensional linear program in Section~\ref{subsec04:LP},
    \item to derive two hierarchies of finite-dimensional semidefinite programs that upper bound and lower bound the threshold value in Section~\ref{subsec04:SDP}, and
    \item to show, in Section~\ref{subsec04:CVproof}, that the sequence of upper bounds converges to the threshold value computed over $L^2(\R_ +)$ functions and the sequence of lower bounds converges to the threshold value computed over Schwartz functions (see~\refig{ch04_structure}). Given the technicalities of the proofs above, we sketch them in Section~\ref{subsec04:sketch} before detailing them in the following sections.
\end{enumerate*}

As a convention, except if specifically mentioned, we will use the terminology `relaxation' and `restriction' from the point of view of the primal program. We will refer to the hierarchy of semidefinite programs providing the upper bounds as a \textit{hierarchy of relaxations} because the obtained SDP programs are indeed relaxations of the primal program (while they are restrictions of the dual program). Likewise, we will refer to the hierarchy providing the lower bounds as a \textit{hierarchy of restrictions}.\index{Semidefinite program}

For clarity, we treat the case where the witnesses are given by the fidelity\index{Fidelity} with a single Fock state, corresponding to the case where one entry of the vector $\bm a$ is equal to $1$ and all the other entries are $0$. The generalisation to linear combinations of fidelities with Fock states is straightforward by linearity.

\subsection{Sketch of the proofs}
\label{subsec04:sketch}
This section aims at giving an overview of the rather technical proofs that follow. 
Our reasoning can be split into the following steps:

\subsubsection*{Obtaining an infinite-dimensional linear program and its dual}

\begin{itemize}
    \item We exploit the phase-space rotational invariance of Fock states in Lemma~\ref{lemma:ch04_random_deph} to express the computation of threshold values \refeq{ch04_threshold} on states that are diagonal in the Fock basis.
    \item This provides a maximisation problem which can be rephrased as an infinite-dimensional linear program \refprog{LP} (resp. \refprog{LPS}) on the space of square integrable functions $L^2(\R_+)$ (resp. Schwartz functions $\mathcal S(\R_+)$). We denote $\omega_n^{L^2}$ its optimal value (resp. $\omega_n^\mathcal S$).
    \item By duality, we recast this as an optimisation problem on finite-signed measures which is given by the dual infinite-dimensional linear program \refprog{DLP} (resp. \refprog{DLPS}). 
\end{itemize}

\subsubsection*{Hierarchy of relaxations}
See upper part of figure~\ref{fig:ch04_structure}.
\begin{itemize}
    \item We relax the program \refprog{LP}: instead of optimising over positive functions, we optimise over functions that have a positive inner product with positive polynomials of fixed degree $m$. The intuition for this procedure is given by Theorem~\ref{th:ch04_RHLaguerre}. This degree fixes a level within a hierarchy of relaxations. These constraints can be cast as a positive semidefinite constraint with Lemma~\ref{lemma:ch04_momentmatrix} and we thus obtain a hierarchy of semidefinite programs (see \refprog{upperSDPn}).
    \item For each level $m$, we derive the dual program \refprog{upperDSDPn} and show that strong duality holds in Theorem \ref{th:ch04_sdupper} by finding a strictly feasible solution of \refprog{upperSDPn}.
    \item We prove the convergence of the hierarchy of semidefinite relaxations towards \refprog{DLP}\footnote{By `convergence of the hierarchy of semidefinite relaxations towards \refprog{DLP}' we mean that the sequence obtained by taking the values of the programs in the hierarchy, and indexed by this hierarchy, converges to the value of \refprog{DLP}} in Theorem~\ref{th:ch04_upperCV} by first showing that the feasible set of \refprog{upperSDPn} is compact. Then, we perform a diagonal extraction on a sequence of optimal solutions of \refprog{upperSDPn} and we finally show that this provides a feasible solution of \refprog{LP} proving that $\underset{m\rightarrow+\infty}{\lim}\omega^{m,\ge}_n=\omega_n^{L_2}$. 
\end{itemize}

\subsubsection*{Convergence of hierarchy of restrictions}\index{Linear program}\index{Semidefinite program}
See bottom part of figure~\ref{fig:ch04_structure}.
\begin{itemize}
    \item We restrict the program \refprog{LP} (or equivalently the program \refprog{LP}): instead of optimising over positive functions, we optimise over positive polynomials of fixed degree $m$. Again, this degree fixes a level within a hierarchy of restrictions. Using the fact that univariate positive polynomials are sum-of-squares which can be written as a semidefiniteness constraint, we obtain a hierarchy of semidefinite programs (see \refprog{lowerSDPn}).
    \item For each level $m$, we derive the dual program \refprog{lowerDSDPn} and show that strong duality holds in Theorem \ref{th:ch04_sdlower} by finding a strictly feasible solution of \refprog{lowerSDPn}.
    \item We prove the convergence of the hierarchy towards \refprog{DLPS} in Theorem~\ref{th:ch04_lowerCV} by first showing that the feasible set of \refprog{lowerDSDPn} is compact. This is highly nontrivial since it requires exhibiting an analytical feasible solution of \refprog{lowerSDPn} which is difficult in general and in the present case. Then, we perform a diagonal extraction on a sequence of optimal solutions of \refprog{lowerDSDPn}. A technicality arises as it does not necessarily provide a feasible solution of \refprog{DLP} which is why we introduce the linear program expressed over Schwartz functions.
    We then show that diagonal extraction indeed provides a feasible solution of \refprog{DLPS} 
    proving that $\underset{m\rightarrow+\infty}{\lim}\omega^{m,\le}_n=\omega_n^{S}$. 
\end{itemize}

\subsection{Function spaces}
\label{subsec04:functionspaces}

We now review some function spaces which appear in the following sections, together with a few notations.

The half line of nonnegative real numbers is denoted $\R_+$. The space of real square-integrable functions over $\R_+$ is denoted $L^2(\R_+)$ and is equipped with the usual inner product:
\begin{equation}\label{eq:ch04_braket}
\braket{f,g}=\int_{\R_+}{f(x)g(x)dx},
\end{equation}
for $f,g\in L^2(\R_+)$.
This space is isomorphic to the space of square-summable real sequences indexed by $\N$ denoted $l^2(\N)$, by considering the expansion in a countable basis. Such a basis is given, e.g., by the Laguerre functions~\cite{szego1959orthogonal} (see \refeq{ch01_LaguerreFunction}),
modified here by a $(-1)^k$ prefactor to correspond to Fock state Wigner functions. We recall here its expression
\begin{equation}\label{eq:ch04_Lagf}
    \mathcal L_k(x):=(-1)^kL_k(x)e^{-\frac x2},
\end{equation}
for all $k\in\mathbb N$ and all $x\in\mathbb R_+$, where $L_k(x)=\sum_{l=0}^k\frac{(-1)^l}{l!}\binom klx^l$ is the $k^{\text{th}}$ Laguerre polynomial. \index{Laguerre!function} \index{Laguerre!polynomial}
These functions form an orthonormal basis: for all $p,q\in\mathbb N$, $\braket{\mathcal L_p,\mathcal L_q}=\delta_{pq}$.

The space $L^2(\R_+)$ is also isomorphic to its dual space ${L^2}'(\R_+)$: via
the
Radon--Nikodym theorem~\cite{Nikodym1930}\index{Radon-Nikodym theorem}
elements of ${L^2}'(\R_+)$ can be identified by the Lebesgue measure on $\R_+$ times the corresponding function in $L^2(\R_+)$.

We write $\mathcal S(\R_+)$ the space of Schwartz functions over $\R_+$, i.e., the space of $C^{\infty}$ functions that go to $0$ at infinity faster than any inverse polynomial, as do their derivatives. $\mathcal S'(\R_+)$ is its dual space, the space of tempered distributions over $\R_+$. 
$\mathcal S(\R_+) \subset {L^2}(\R_+)$ is dense in ${L^2}(\R_+)$.
We denote the space of rapidly decreasing real sequences by $\mathcal S(\mathbb N)$ (sequences that go to $0$ at infinity faster than any inverse polynomial), together with its dual space of slowly increasing real sequences $\mathcal S'(\mathbb N)$ (sequences that are upper bounded by a polynomial). The spaces $\mathcal S(\R_+)$ and $\mathcal S(\mathbb N)$ are isomorphic: any Schwartz function over $\R_+$ can be expanded uniquely in the basis of Laguerre functions with a rapidly decreasing sequence of coefficients. Similarly, the spaces $\mathcal S'(\R_+)$ and $\mathcal S'(\mathbb N)$ are also isomorphic: any tempered distribution over $\R_+$ can be written uniquely as a formal series of Laguerre functions with a slowly increasing sequence of coefficients~\cite{guillemot1971developpements}.
We extend the definition of the duality $\braket{\dummy,\dummy}$ in~\refeq{ch04_braket} to these spaces.

In order to denote nonnegative elements of these spaces, we will use the notations $L^2_+(\R_+)$, ${L_+^2}'(\R_+)$, $\mathcal S_+(\R_+)$ and $\mathcal S_+'(\R_+)$. A distribution $\mu$ in ${L_+^2}'(\R_+)$ (resp.\ in $\mathcal S_+'(\R_+)$) satisfies: $\forall f\in L^2_+(\R_+)$ (resp.\ $\forall f\in \mathcal S_+(\R_+)$), $\braket{\mu,f}\ge0$.

For all $m\in\mathbb N$, we define the following space of truncated series of Laguerre functions over $\R_+$:
\begin{equation}
     \mathcal R_m(\R_+):=\text{span}_{\R}\{\mathcal L_k\}_{0\le k\le m},
\end{equation}
which is equal to the set of real polynomials over $\R_+$ of degree less or equal to $m$ multiplied by the function $x\mapsto e^{-\frac x2}$. We denote by $\mathcal R_{m,+}(\R_+)$ its subset of nonnegative elements.

For all $\bm s = (s_k)_k \in\R^{\N}$, we define the associated formal series of Laguerre functions:
\begin{equation}\label{eq:ch04_isomorphism}
    f_{\bm s}:=\sum_{k\ge0}s_k\mathcal L_k,
\end{equation}
with the (formal) relation:
\begin{equation}\label{eq:ch04_coefLag}
    s_k=\braket{f_{\bm s},\mathcal L_k},
\end{equation}
for all $k\in\mathbb N$. We refer to $\bm s$ as the sequence of Laguerre moments of $f_{\bm s}$.\index{Moment sequence} We extend this definition to finite sequences by completing these sequences with zeros.
For $m\in\mathbb N$, we also define the matrix $A_{\bm s}$ (thus omitting the dependence in $m$ as it is always clear from the context) by
\begin{equation}\label{eq:ch04_momentmatrix}
    (A_{\bm s})_{0\le i,j\le m} := \begin{cases} 
      \sum\limits_{k=0}^l s_k \binom lkl! &\text{if } i+j=2l, \\
      0 &\text{otherwise.}
   \end{cases}
\end{equation}
$A_{\bm s}$ can be seen as the Laguerre moment matrix\index{Moment matrix} of the measure $f_{\bm s}$.
In what follows, we use standard techniques relating to the Stieltjes moment problem~\cite{reed1975ii}, which seeks conditions for a real sequence $\bm\nu=(\nu_k)_{k\in\N}\in\R^{\N}$ to be the sequence of moments $\int_{\R_+}x^kd\nu(x)$ of a nonnegative distribution $\nu$ over $\R_+$. We adapt these techniques to the basis of Laguerre functions\index{Laguerre!function}, rather than the canonical basis. We start by proving a change of basis result:

\begin{lemma}\label{lemma:ch04_changeofbasis}
Let $\bm\mu,\bm\nu\in\R^{\N}$. For all $m\in\N$, the following conditions are equivalent:
\begin{enumerate}[label=(\roman*)]
\item $\displaystyle \forall k\in\llbracket0,m\rrbracket,\quad\mu_k=\sum_{l=0}^k\nu_l\frac{(-1)^{k+l}}{l!}\binom kl$ \label{it:ch04_lemmaCoB1},
\item $\displaystyle \forall l\in\llbracket0,m\rrbracket,\quad\nu_l=\sum_{k=0}^l\mu_k\binom lkl!$.\label{it:ch04_lemmaCoB2}
\end{enumerate}
\end{lemma}

\noindent As a direct consequence, we retrieve the formula:
\begin{equation}\label{eq:ch04_Lagtox}
    x^l=\sum_{k=0}^l(-1)^k\binom lkl!L_k(x),
\end{equation}
for all $l\in\N$ and all $x\in\R_+$.

\begin{proof}
\ref{it:ch04_lemmaCoB1}$\Rightarrow$\ref{it:ch04_lemmaCoB2}: suppose that
\begin{equation}\label{eq:ch04_mutonuchange}
    \forall k\in\llbracket0,m\rrbracket,\quad\mu_k=\sum_{p=0}^k\nu_p\frac{(-1)^{k+p}}{p!}\binom kp.
\end{equation}
\begin{equation}\label{eq:ch04_derivationmutonu}
    \begin{aligned}
        \forall l\in\llbracket0,m\rrbracket \sum_{k=0}^l\mu_k\binom lkl!&=\sum_{k=0}^l\sum_{p=0}^k\nu_p\frac{(-1)^{k+p}}{p!}\binom kp\binom lkl!\\
        &=\sum_{p=0}^l\nu_p\frac{l!}{p!}\binom lp\sum_{k=p}^l(-1)^{k-p}\binom{l-p}{k-p}\\
        &=\sum_{p=0}^l\nu_p\frac{l!}{p!}\binom lp\sum_{q=0}^{l-p}(-1)^q\binom{l-p}q\\
        &=\nu_l,
    \end{aligned}
\end{equation}
where we used~\refeq{ch04_mutonuchange} in the first line and the binomial theorem in the last line which imposes $l=p$.

\ref{it:ch04_lemmaCoB2}$\Rightarrow$\ref{it:ch04_lemmaCoB1}: suppose that
\begin{equation}\label{eq:ch04_nutomuchange}
    \forall l\in\llbracket0,m\rrbracket,\quad\nu_l=\sum_{p=0}^l\mu_p\binom lpl!.
\end{equation}
\begin{equation}
    \begin{aligned}
        \forall k\in\llbracket0,m\rrbracket, \quad \sum_{l=0}^k\nu_l\frac{(-1)^{k+l}}{l!}\binom kl&=\sum_{l=0}^k\sum_{p=0}^l\mu_p(-1)^{k+l}\binom lp\binom kl\\
        &=\sum_{p=0}^k\mu_p(-1)^{k+p}\binom kp\sum_{l=p}^k(-1)^{l-p}\binom{k-p}{l-p}\\
        &=\sum_{p=0}^k\mu_p(-1)^{k+p}\binom kp\sum_{q=0}^{k-p}(-1)^q\binom{k-p}q\\
        &=\mu_k,
    \end{aligned}
\end{equation}
where we used~\refeq{ch04_nutomuchange} in the first line and the binomial theorem in the last line which imposes $k=p$.

\end{proof}
We prove a similar result that we will use later for proving that a pair of semidefinite programs are indeed dual programs.
\begin{lemma}\label{lemma:ch04_revert} Let $\bm u,\bm v\in\R^{m+1}$. The following propositions are equivalent:
\begin{enumerate}[label=(\roman*)]
\item $\displaystyle \forall k\in\llbracket0,m\rrbracket,\quad u_k=\sum_{l=0}^mv_l\binom lkl!$, \label{it:ch04_lemmaCoB21}
\item $\displaystyle \forall l\in\llbracket0,m\rrbracket,\quad v_l=\sum_{k=l}^m \frac{(-1)^{l+k}}{l!} \binom kl u_k$. \label{it:ch04_lemmaCoB22}
\end{enumerate}
\end{lemma}

\begin{proof} The proof is similar to that of Lemma~\ref{lemma:ch04_changeofbasis}. Note that we could start the first sum at $l=k$ since $\binom lk = 0 $ for $l < k$ but for convenience we start it at $l=0$.

\ref{it:ch04_lemmaCoB21}$\Rightarrow$\ref{it:ch04_lemmaCoB22}: suppose that
\begin{equation}\label{eq:mutoQ}
    \forall k\in\llbracket0,m\rrbracket,\quad u_k=\sum_{p=0}^mv_l\binom pkp!.
\end{equation}

\begin{equation}
    \begin{aligned}
        \forall l\in\llbracket0,m\rrbracket \frac{(-1)^l}{l!}\sum_{k=l}^m(-1)^k\binom kl u_k&= \frac{(-1)^l}{l!}\sum_{k=l}^m(-1)^k\binom kl\sum_{p=0}^mv_p\binom pkp!\\
        &=\sum_{p=0}^mv_p\frac{p!}{l!}\sum_{k=l}^m(-1)^{k+l}\binom pk\binom kl\\
        &=\sum_{p=l}^mv_p\frac{p!}{l!}\sum_{k=l}^p(-1)^{k+l}\binom pk\binom kl\\
        &=\sum_{p=l}^mv_p\frac{p!}{l!}\sum_{k=l}^p(-1)^{k+l}\frac{p!k!}{k!(p-k)!l!(k-l)!}\\
        &=\sum_{p=l}^mv_p\frac{p!}{l!}\binom pl\sum_{q=0}^{p-l}(-1)^q\binom{p-l}q\\
        &=v_l,
    \end{aligned}
\end{equation}
where we used~\refeq{mutoQ} in the first line, the fact that $\binom pk=0$ if $k>p$ in the third line, $q:=k-l$ in the fifth line, and the binomial theorem in the last line which imposes $p=l$.

\ref{it:ch04_lemmaCoB22}$\Rightarrow$\ref{it:ch04_lemmaCoB21}: suppose that
\begin{equation}\label{eq:Qtomu}
    \forall l\in\llbracket0,m\rrbracket,\quad v_l=\frac{(-1)^l}{l!}\sum_{p=l}^m(-1)^p\binom pl u_p.
\end{equation}
\begin{equation}
    \begin{aligned}
        \forall k\in\llbracket0,m\rrbracket \sum_{l=0}^mv_l\binom lkl!&=\sum_{l=0}^m\frac{(-1)^l}{l!}\sum_{p=l}^m(-1)^p\binom pl u_p\binom lkl!\\
        &=\sum_{p=0}^mu_p(-1)^p\sum_{l=0}^p(-1)^l\binom pl\binom lk\\
        &=\sum_{p=k}^mu_p(-1)^p\sum_{l=k}^p(-1)^l\binom pl\binom lk\\
        &=\sum_{p=k}^mu_p(-1)^{p-k}\binom pk\sum_{q=0}^{p-k}(-1)^q\binom{p-k}q\\
        &=u_k,
    \end{aligned}
\end{equation}
where we used~\refeq{Qtomu} in the first line, the fact that $\binom lk=0$ if $k>l$ in the third line, $q:=l-k$ in the fourth line, and the binomial theorem in the last line which imposes $p=k$.

\end{proof}

\noindent Now we turn our attention to a modified Stieltjes moment condition. 
\begin{theorem}\label{th:ch04_RHLaguerre}
Let $\bm\mu=(\mu_k)_{k\in\N}\in\R^{\N}$. The sequence $\bm\mu$\index{Moment sequence} is the sequence of Laguerre moments $\int_{\R_+}\mathcal L_k(x)d\mu(x)$\index{Laguerre!function} of a nonnegative distribution $\mu$ supported on $\R_+$ if and only if
\begin{equation}
    \forall m\in\N,\forall g\in\mathcal R_{m,+}(\R_+),\;\braket{f_{\bm\mu},g}\ge0.
\end{equation}
\end{theorem}

\noindent This result is based on the well-known Riesz--Haviland theorem~\cite{riesz1923probleme,haviland1936momentum}\index{Riesz!Haviland theorem} (see Theorem~\ref{th:ch01_RieszHaviland}). The Riesz functional\index{Riesz!functional} $L_{\bm \nu}$ for a sequence $\bm\nu=(\nu_l)_{l\in\N}\in\R^{\N}$ is defined in Definition~\ref{def:ch01_RieszFunctional}.

\begin{proof}
Let $\bm\mu=(\mu_k)_{k\in\N}\in\R^{\N}$, and suppose that the sequence $\bm\mu$ is the sequence of Laguerre moments $\int_{\R_+}\mathcal L_k(x)d\mu(x)$ of a nonnegative distribution $\mu$ supported on $\R_+$.\index{Laguerre!function}
Let $m\ge0$ and let $g=\sum_{k=0}^mg_k\mathcal L_k\in\mathcal R_{m,+}(\R_+)$. The distribution $\mu$ is nonnegative, so by definition $\braket{\mu,g}\ge0$. Moreover,
\begin{equation}
    \begin{aligned}
        \braket{f_{\bm\mu},g}&=\sum_{k=0}^m{\mu_kg_k}\\
        &=\int_{\R_+}\sum_{k=0}^m{g_k\mathcal L_kd\mu}\\
        &=\braket{\mu,g}.
    \end{aligned}
\end{equation}
Hence, for all $m\in\N$ and all $g\in\mathcal R_{m,+}(\R_+)$, $\braket{f_{\bm\mu},g}\ge0$. 

Conversely, let $\bm\mu=(\mu_k)_{k\in\N}\in\R^{\N}$,\index{Moment sequence} and suppose that for all $m\in\N$ and all $g\in\mathcal R_{m,+}(\R_+)$, $\braket{f_{\bm\mu},g}\ge0$.
We define the sequence $\bm\nu=(\nu_l)_{l\in\N}\in\R^{\N}$ by
\begin{equation}\label{eq:ch04_nu_ldef}
    \nu_l:=\sum_{k=0}^l\mu_k\binom lkl!,
\end{equation}
for all $l\in\N$. 

Let $m\in\N$ and let $P(x)=\sum_{l=0}^mp_lx^l$ be a nonnegative polynomial over $\R_+$. By~\refeq{ch04_Lagtox}, for all $x\in\R_+$,
\begin{equation}\label{eq:ch04_PinLag}
    \begin{aligned}
        P(x)&=\sum_{l=0}^mp_l\sum_{k=0}^l(-1)^k\binom lkl!L_k(x)\\
        &=\sum_{k=0}^m(-1)^kL_k(x)\left(\sum_{l=k}^mp_l\binom lkl!\right).
    \end{aligned}
\end{equation}
Let $g_P(x):=P(x)e^{-\frac x2}$, for $x\in\R_+$. We have $g_P\in\mathcal R_{m,+}(\R_+)$, so $\braket{f_{\bm\mu},g_P}\ge0$. Moreover, with~\refeq{ch04_PinLag}
\begin{equation}
    \begin{aligned}
        \braket{f_{\bm\mu},g_P}&=\sum_{k=0}^m\mu_k\left(\sum_{l=k}^mp_l\binom lkl!\right)\\
        &=\sum_{l=0}^m\left(\sum_{k=0}^l\mu_k\binom lkl!\right)p_l\\
        &=L_{\bm\nu}(P)
    \end{aligned}
\end{equation}
where we used~\refeq{ch04_nu_ldef} and the definition of the Riesz functional\index{Riesz!functional} from Definition~\ref{def:ch01_RieszFunctional} in the last line. 
In particular, $L_{\bm\nu}(P)\ge0$, and this holds for all nonnegative polynomials $P$ over $\R_+$. By the Riesz--Haviland theorem (Theorem~\ref{th:ch01_RieszHaviland})\index{Riesz!Haviland theorem}, this implies that $\bm\nu$ is the sequence of moments\index{Moment sequence} of a nonnegative distribution $\nu$ supported on $\R_+$.

Furthermore, we have that for all $k\in\N$:
\begin{equation}
    \begin{aligned}
        \mu_k&=\sum_{l=0}^k\nu_l\frac{(-1)^{k+l}}{l!}\binom kl\\
        &=\sum_{l=0}^k\frac{(-1)^{k+l}}{l!}\binom kl\int_{\R_+}x^ld\nu(x)\\
        &=\int_{\R_+}(-1)^k\sum_{l=0}^k\frac{(-1)^l}{l!}\binom klx^ld\nu(x)\\
        &=\int_{\R_+}(-1)^kL_k(x)d\nu(x)\\
        &=\int_{\R_+}\mathcal L_k(x)e^{\frac x2}d\nu(x)\\
    \end{aligned}
\end{equation}
where we used Lemma~\ref{lemma:ch04_changeofbasis} in the first line. Hence, $\bm\mu$ is the sequence of Laguerre moments of the distribution $\mu(x):=e^{\frac x2}\nu(x)$ supported on $\R_+$, which is nonnegative since $\nu$ is nonnegative.

\end{proof}

\subsection{Computing the threshold value as a linear program}
\label{subsec04:LP}

In this section, we phrase the computation of the witness threshold value introduced in \refeq{ch04_threshold} as an infinite-dimensional linear program, in the case where one entry of the vector $\bm a$ is equal to $1$ and all the other entries are $0$, the generalisation being straightforward by linearity.

Formally, we fix hereafter $n\in\mathbb N^*$ and we look for the witnesses threshold value $\omega_n$\footnote{Here we write generically the threshold value as $\omega_n$ while we use the more precise notation $\omega_n^{L^2}$ (resp. $\omega_n^{\mathcal S}$) to refer to its computation in the space of square integrable functions (resp. Schwartz functions).}
defined as
\begin{equation}\label{eq:ch04_formal_problem}
  \omega_n:=\sup_{\substack{\rho\in\mathcal D(\mathscr H)\\W_\rho\ge0}}\braket{n|\rho|n}.
\end{equation}
This is the maximal values such that for all states $\rho\in\mathcal D(\mathscr H)$:
\begin{equation}
  \braket{n|\rho|n}>\omega_n\quad\Rightarrow\quad\exists\alpha\in\mathbb C,\;W_\rho(\alpha)<0.
\end{equation}
Let $\mathcal C(\mathscr H)$ be the set of states that are invariant under phase-space rotations:
\begin{equation}
    \mathcal C(\mathscr H) := \enset{ \sigma\in\mathcal D(\mathscr H) :
    \forall\varphi\in[0,2\pi], \, \eu^{\im\varphi \hat{n}}\sigma\eu^{-\im\varphi \hat{n}}=\sigma },
\end{equation}
where $\hat n=\hat a^\dag\hat a$ is the number operator.
The witnesses corresponding to the fidelity\index{Fidelity} with a single Fock state (and linear combination of Fock states) feature a rotational symmetry in phase space, which we exploit in the following lemma. 

\begin{lemma}\label{lemma:ch04_random_deph}
The threshold value in~\refeq{ch04_formal_problem} can be expressed as
\begin{equation} \label{eq:ch04_WitnessPhaseInv}
  \omega_n=\sup_{\substack{\sigma\in\mathcal C(\mathscr H)\\W_\sigma\ge0}}\braket{n|\sigma|n}.
\end{equation}
\end{lemma}

\begin{proof}
Let $\rho\in\mathcal D(\mathscr H)$. We start by applying a random dephasing to the state $\rho$:
\begin{equation}
    \sigma = \int_0^{2\pi}\frac{\dd\varphi}{2\pi}\eu^{\im\varphi \hat{n}}\rho\eu^{-\im\varphi \hat{n}}\in \mathcal C(\mathscr H).
\end{equation}
The random dephasing does not change the fidelity\index{Fidelity} with any Fock state because of the rotational symmetry in phase space of the latter (see \refeq{ch01_WignerFockstates}), that is\index{Wigner negativity}
\begin{equation}
    \forall n \in \N, \, \braket{n|\sigma|n} = \braket{n|\rho|n}.
\end{equation} 
Moreover, it can only decrease the maximum negativity of the Wigner function. Indeed for all $\alpha\in\mathbb C$,
\begin{equation}\label{eq:ch04_inequality_dephasing}
    \begin{aligned}
        W_{\sigma}(\alpha)
        &=\int_0^{2\pi} \frac{\dd\varphi}{2\pi}W_{e^{\im \varphi \hat{n}} \rho  e^{- \im \varphi \hat{n}}}(\alpha) \\
        &=\int_0^{2\pi} \frac{\dd\varphi}{2\pi}W_{\rho}(\alpha \eu^{\im\varphi}) \\
        &\ge \min_{\varphi\in[0,2\pi]}W_{\rho}(\alpha \eu^{\im\varphi}) \\
        &\ge \min_{\beta\in\mathbb C}W_{\rho}(\beta), 
    \end{aligned} 
\end{equation}
and taking the minimum over all $\alpha\in\mathbb C$ then gives
\begin{equation}
    \min_{\alpha\in\mathbb C}W_{\sigma}(\alpha)\ge\min_{\beta\in\mathbb C}W_{\rho}(\beta).
\end{equation}
In particular, applying a random dephasing to a Wigner positive state yields a Wigner positive mixtures of Fock states, which is invariant under phase-space rotations. Hence, we can restrict without loss of generality to states that are invariant under phase-space rotations when looking for the maximum fidelity\index{Fidelity} of Wigner positive states with a given Fock state $\ket n$.
\end{proof}

\noindent Lemma~\ref{lemma:ch04_random_deph} ensures that the supremum in \refeq{ch04_formal_problem} can be computed over states that have a rotational symmetry in phase space. Such states $\sigma$ can be expanded diagonally in the Fock basis:
\begin{equation}
    \sigma = \sum_{k=0}^{\infty} F_k \ket{k}\!\bra{k},
\end{equation}
with the normalisation condition $\sum_kF_k=1$ and $0 \leq F_k \leq 1$ for all $k\in\N$ assuring that $\sigma$ is a positive semidefinite operator.
By linearity of the Wigner function: 
\begin{equation}
    \forall \alpha \in \C,\; W_{\sigma}(\alpha) = \sum_k F_k W_k(\alpha),
\end{equation}
where $W_k$ is the Wigner function of the $k^{th}$ Fock state~\cite{kenfack2004negativity}: 
\begin{equation}
    \forall \alpha \in \C, \, W_k(\alpha) = \frac{2}{\pi} \mathcal L_k(4|\alpha|^2),
\end{equation}
with $\mathcal L_k$ the $k^{\text{th}}$ Laguerre function, defined in~\refeq{ch04_Lagf}.\index{Laguerre!function} As noted before, Fock states are invariant under phase-space rotations: their Wigner function only depends on the amplitude of the phase-space point considered. We fix $x= 4 \vert \alpha \vert^2 \in \R^+$ hereafter.

\subsubsection*{Computation of the threshold value over square-integrable functions.}
We first consider the computation of the threshold value in \refeq{ch04_WitnessPhaseInv} over $L^2(\R_+)$ functions and we will denote the corresponding threshold value by $\omega_n^{L^2}$. With Lemma \ref{lemma:ch04_random_deph}, the computation of $\omega_n^{L^2}$ can thus be expressed as the following infinite-dimensional linear program:
\leqnomode
\begin{flalign*}\index{Linear program}
    \label{prog:LP}
    \tag*{(LP$_n^{L^2}$)}
    \hspace{3cm} \left\{
        \begin{aligned}
            & \quad \text{Find }  (F_k)_{k \in \N} \in \ell^2(\N) \\
            & \quad \text{maximising } F_n \\
            & \quad \text{subject to} \\
            & \hspace{1cm} \begin{aligned}
            & \sum_k F_k = 1  \\
            & \forall k \in \N, \;F_k \geq 0 \\
            & \forall x \in \R_+, \;\sum_k F_k\mathcal L_k(x) \geq 0.
            \end{aligned}
        \end{aligned}
    \right. &&
\end{flalign*}
\reqnomode
The first constraint ensures unit trace of the corresponding state $\sigma$, the second one ensures that its fidelity\index{Fidelity} with each Fock state is nonnegative and thus that the state is a proper positive semidefinite operator, and the last one ensures that its Wigner function $W_{\sigma}$ is nonnegative. Note that $\omega_n>0$ for all $n\in\N^*$, by considering a mixture of $\ket0$ and $\ket n$ with the vacuum component close enough to $1$. For convenience when considering the dual formulation, we use the superscript $L$ instead of $\ell$ in the label of the program. Even though it is formulated as an optimisation over sequences from $\ell^2$, we can equivalently see it as an optimisation over $L^2(\R_+)$ functions as $L^2(\R_+)$ and $\ell^2$ are isomorphic.

The dual linear program reads:
\leqnomode
\begin{flalign*}\index{Linear program}
    \label{prog:DLP}
    \tag*{(D-LP$_n^{L^2}$)}
    \hspace{3cm} \left\{
        \begin{aligned}
            & \quad \text{Find } y \in \R \text{ and } \mu \in {L^2}'(\R_+)\\
            & \quad \text{minimising } y\\
            & \quad \text{subject to}  \\
            & \hspace{1cm} \begin{aligned}
            & \forall k \neq n \in \N,\; y \geq \int_{\R_+}{\mathcal L_k}{d\mu}  \\
            & y \geq 1 + \int_{\R_+}{\mathcal L_n}{d\mu} \\
            & \forall f \in L^2_+(\R_+), \; \langle \mu,f \rangle \geq  0.
            \end{aligned}
        \end{aligned}
        \right. &&
\end{flalign*}
\reqnomode

\subsubsection*{Retrieving the canonical form of infinite-dimensional LP} 
Here we follow Subsection~\ref{subsec01:LP} to retrieve the standard form of infinite-dimensional linear program as presented in \cite[IV--(6.1)]{barvinok02} and see why programs \refprog{LP} and \refprog{DLP} are indeed dual programs. 

Recall that via expansion on a basis of $L^2(\R_ +)$, $\ell^2(\N)$ and $L^2(\R_ +)$ are isomomorphic and that the spaces $L^2(\R_ +)$ and ${L^2}'(\R_ +)$ are isomomorphic by the Radon--Nikodym theorem\index{Radon-Nikodym theorem}. Let us introduce the spaces:
\begin{itemize}
    \item $E_1 = \ell^2(\N) \times L^2(\R_ +)$. 
    \item $F_1 = \ell^2(\N) \times {L^2}'(\R_ +)$ the dual space of $E_1$.
    \item $E_2 = \R \times L^2(\R_ +)$.
    \item $F_2 = \R \times {L^2}'(\R_ +)$ the dual space of $E_2$.
\end{itemize}
We also define the dualities $\langle \dummy , \dummy \rangle_1 : E_1 \times F_1 \longrightarrow \R$ and $\langle \dummy , \dummy \rangle_2 : E_2 \times F_2 \longrightarrow \R$ as follows:
\begin{align}
    & \forall e_1 = ((u_k),f) \in E_1,\, \forall f_1 = ((v_k),\mu) \in F_1, & & \langle e_1,f_1 \rangle_1 \defeq \sum_k u_k v_k + \int_{\R_ +}{f}{d\mu},  \\ 
    & \forall e_2 = (x,f) \in E_2,\, \forall f_2 = (y,\mu) \in F_2, & & \langle e_2,f_2 \rangle_2 \defeq  xy + \int_{\R_ +}{f}{d\mu}.
\end{align}
Let $\fdec{A}{E_1}{E_2}$ be the following linear transformation:
\begin{equation}
    \forall e_1=((u_k),f) \in E_1, \quad A(e_1) \defeq \left(\sum_k u_k,\; x \in \R_ + \mapsto f(x) - \sum_k u_k \mathcal L_k(x)\right),
\end{equation}
and $\fdec{A^*}{F_2}{F_1}$ be defined as:
\begin{equation}
    \forall f_2=(y,\mu) \in F_2, \quad A^*(f_2) \defeq \left( (y-\int_{\R_ +}{\mathcal L_k}{d\mu})_{k \in \N},\; \mu \right).
\end{equation}
We can easily verify that $A^*$ is the dual transformation of $A$, i.e., $\forall e_1 \in E_1,\, \forall f_2 \in F_2$ we have $\langle A(e_1),f_2 \rangle_2 = \langle e_1, A^*(f_2) \rangle_1 $.

Recall that $L^2_+(\R_+)$ is the cone of nonnegative functions in $L^2(\R_+)$ and $\ell^2_+$ the cone of sequences in $\ell^2$ with nonnegative coefficients.
We will optimise in the convex cones $K_1 = \ell^2_+ \times L^2_+(\R_+) \subset E_1$ and $K_2 = \{ 0 \}$. The dual cones are then respectively: $K_1^* = \{f_1 \in F_1: \forall e_1 \in K_1, \; \langle e_1,f_1 \rangle \geq 0 \}$ and $K_2^* = F_2$. \index{Cone}

We can now rewrite the problem \refprog{LP} as a standard linear program in convex cones. We choose the vector function in the objective to be $c_n = ( (\delta_{kn})_k, \mathbf{0}) \in F_1$ and we also set $b = (1,\mathbf{0}) \in E_2$ for the constraints. The standard form of~\refprog{LP} in the sense of \cite{barvinok02} can be written as follows:
\leqnomode
\begin{flalign*} \index{Linear program}
    \tag*{(LP$_n^{L^2}$)}
    \hspace{3cm} \left\{
    \begin{aligned}
            & \quad \text{Find } e_1 \in E_1 \\
            & \quad \text{maximising } \langle e_1,c_n \rangle_1 \\
            & \quad \text{subject to:}  \\
            & \hspace{1cm} \begin{aligned}
            & A(e_1) = b  \\
            & e_1 \geq_{K_1} 0 \Mdot
            \end{aligned} 
    \end{aligned}
    \right. &&
\end{flalign*}
This is expressed with an equality constraint with a slack variable rather than an inequality constraint (see the canonical program in \ref{subsec01:LP}). Both formulation are equivalent. 
The standard form of the dual can be expressed as follows:
\begin{flalign*}
    \tag*{(D-LP$_n^{L^2}$)}
    \hspace{3cm}\left\{
    \begin{aligned}
        & \quad \text{Find } f_2 \in F_2 \\
        & \quad \text{minimising } \langle b,f_2 \rangle_2 \\
        & \quad \text{subject to:}  \\
        & \hspace{1cm} \begin{aligned}
        & A^*(f_2) \geq_{K_1^*} c_n \Mdot
        \end{aligned}
    \end{aligned}
    \right. &&
\end{flalign*}
\reqnomode
which can be expanded as program \refprog{DLP}. Note the same derivation hold if we optimise over Schwartz functions by considering $E_1 = \mathcal S(\N) \times \mathcal S(\R_+)$, $E_2 = \mathcal S'(\N) \times \mathcal S'(\R_+)$, $F_1 = \R \times \mathcal S(\R_+)$ and $F_2 =  \R \times \mathcal S'(\R_+)$.

\subsubsection*{Strong duality} \index{Strong duality!of linear programs}
This will follow immediately from the proof of convergence we provide later. As a sanity check, we use standard tools for proving strong duality as presented in~\ref{subsec01:LP} with Theorem~\ref{th:ch01_strongduality}. 

There is no duality gap between programs \refprog{LP} and \refprog{DLP} if there is a primal feasible plan and if the convex cone 
\begin{equation}
    \begin{aligned}
        \mathcal{K} &= \Big\{ \Big( A(e_1), \langle e_1,c \rangle_1 \Big) : e_1 \in K_1   \Big\} \\
        &= \Big\{ \Big( \sum_k u_k, x \in \R_ + \mapsto f(x) - \sum_k u_k \mathcal L_k(x), F_n \Big) : ((u_k),f) \in K_1 \Big\}
    \end{aligned}
\end{equation}
is closed in $E_2 \oplus \R$ (for the weak topology).

\begin{proof}
The null sequence and the null function provides a feasible plan for the primal problem.

Next, we consider a sequence $(e_{1j})_j = (((u_k^j)_k)_j\footnote{Because we are dealing with a sequence of sequences, we use the upper index to refer to the embracing sequence.},(f_j)_j) \in K_1^{\N} = \ell^2(\N)\times L^2_+(\R_+)^{\N} $ and we want to show that the accumulation point $(b,g,a) = \lim_{j \rightarrow \infty} (A(e_{1j}), \langle e_{1j},c \rangle_1) $ belongs to $\mathcal{K}$ where $a,b \in \R$ and $g \in L^2(\R_+)$. 

For all $j \in \N$, $(u_k^j)_k \in \ell^2$ and for all $k \in \N$, $u_k^j$ is bounded. 
Thus, for all $k \in \N$, the sequence $(u_k^j)_j$ is bounded and via diagonal extraction there exists $\phi: \N \rightarrow \N$ strictly increasing such that $(u_k^j)_{\phi(j)}$ converges. 
We denote $\tilde{u}_k$ its limit. Since $\ell^2(\N)$ is closed, the sequence $(\tilde{u}_k)_k$ belongs to $\ell^2(\N)$ and we have $b = \sum_k \tilde{u}_k$ and $a=\tilde{u}_n$. 

Now $f_j - \sum_k u_k^j \mathcal L_k \longrightarrow g$ so that $f_j \longrightarrow g + \sum_k \tilde{u}_k \mathcal L_k \in L^2_+(\R_+)$ since $L^2_+(\R_+)$ is closed.
Thus, for $\tilde{e}_1 = ((\tilde{u}_k)_k, g + \sum_k \tilde{u}_k \mathcal L_k) \in K_1$, $(b,g,a) = (A(\tilde{e}_1), \langle \tilde{e}_1,c \rangle_1)$ and $(b,g,a) \in \mathcal{K}$.
\end{proof}

Note that the same proof does not hold for the problems expressed over Schwartz functions as $\mathcal S(\N)$ and $\mathcal S_+(\R_+)$ are not closed. We will rely on the proof of convergence of the lower bounding hierarchy of semidefinite programs to prove strong duality in this case. 

\subsubsection*{Computation of the threshold over Schwartz functions} 
We also consider a restriction or \refprog{LP} by optimising over Schwartz functions rather than square integrable functions as $\mathcal S(\R_+) \subset L^2(\R_+)$. The primal program becomes: 
\leqnomode
\begin{flalign*}\index{Linear program}
    \label{prog:LPS}
        \tag*{(LP$_n^{\mathcal S}$)}
        \hspace{3cm} \left\{
        \begin{aligned}
            & \quad \text{Find } (F_k)_{k \in \N} \in \mathcal S(\N) \\
            & \quad \text{maximising } F_n \\
            & \quad \text{subject to} \\
            & \hspace{1cm} \begin{aligned}
            & \sum_k F_k = 1  \\
            & \forall k \in \N, \;F_k \geq 0\\
            & \forall x \in \R_+, \;\sum_k F_k\mathcal L_k(x) \geq 0,
            \end{aligned}
        \end{aligned}
        \right. &&
\end{flalign*}
We denote its value by $\omega_n^\mathcal S$.
Because it is a restriction of \refprog{LP}, we have that $\omega_n^{L^2} \geq \omega_n^\mathcal S$.
Its dual linear program can be expressed as:
\begin{flalign*}
    \label{prog:DLPS}
    \tag*{(D-LP$_n^{\mathcal S}$)} \index{Linear program}
    \hspace{3cm} \left\{
        \begin{aligned}
            & \quad \text{Find } y \in \R \text{ and } \mu \in\mathcal S'(\R_+) \\
            & \quad \text{minimising } y \\
            & \quad \text{subject to}  \\
            & \hspace{1cm} \begin{aligned}
            & \forall k \neq n \in \N,\; y \geq \int_{\R_+}{\mathcal L_k}{d\mu}  \\
            & y \geq 1 + \int_{\R_+}{\mathcal L_n}{d\mu} \\
            & \forall f \in \mathcal S_+(\R_+), \; \langle \mu,f \rangle \geq  0.
            \end{aligned}
        \end{aligned}
        \right. &&
    \end{flalign*}
\reqnomode
Note there is a rather strong structure with Wigner functions. In \cite{hernandezrapidly2021} it is shown that the Wigner function only need to decay towards infinity faster than any polynomial to be a Schwartz function (\ie it also implies that all its derivatives decay towards infinity faster than any polynomial).

\subsubsection*{Analytical solutions}
Even without strong duality (it is not proven in the case of Schwartz functions), weak duality of linear programming already ensures that the optimal value $\omega_n^{L^2}$ of \refprog{LP} (resp. $\omega_n^{\mathcal S}$ of \refprog{LPS}) is upper bounded by the optimal value of \refprog{DLP} (resp. \refprog{DLPS}). Hence, a possible way of solving the optimisation \refprog{LP} is to exhibit a feasible solution for \refprog{LP} and a feasible solution for \refprog{DLP} that have the same value. 

For $n=1$, choosing $(F_k)_{k\in\mathbb N}=(\frac12,\frac12,0,0,\dots)$ gives a feasible solution for~(LP$_1^{L^2}$)
(resp. (LP$_1^{\mathcal S}$))
with the value $\frac12$, while choosing $(y,\mu)=(\frac12,\frac12\delta(x))$, where $\delta$ is the Dirac delta function\footnote{Technically $\delta\notin{L^2}'(\R_+)$, but the result holds by considering a sequence of functions converging to a Dirac delta.} 
over $\R_+$, 
gives a feasible solution for~(D-LP$_1^{L^2}$) 
(resp. (D-LP$_1^{\mathcal S}$))
with the value $\frac12$. 
This shows that $\omega_1^{L^S} = \omega_1^{\mathcal S} =\frac12$.

Similarly, for $n=2$, choosing $(F_k)_{k\in\mathbb N}=(\frac12,0,\frac12,0,0,\dots)$ gives a feasible solution for~(LP$_2^{L^2}$) 
(resp. LP$_2^{\mathcal S}$)
with the value $\frac12$, while choosing $(y,\mu)=(\frac12,\frac e2\delta(x-2))$ gives a feasible solution for~(D-LP$_2$)
(resp. (D-LP$_2^{\mathcal S}$)), up to a conjecture\footnote{We checked numerically the corresponding constraints $|L_k(2)|\le1$ for $k$ up to $10^3$ and, considering asymptotic behaviors, we conjecture that these hold for all $k\ge0$.}, with the value $\frac12$. This shows that $\omega_2^{L^2} = \omega_2^{S} = \frac12$.

While this approach is sensible for small values of $n$, finding optimal analytical solutions for higher values of $n$ seems highly nontrivial. Moreover, the infinite number of variables prevents us from performing the optimisation \refprog{LP} numerically. A natural workaround is to find finite-dimensional relaxations or restrictions of the original problem---thus providing upper and lower bounds for the optimal value $\omega_n^{L^2}$, respectively. This is the approach we follow in the next section.

\subsection{Hierarchies of semidefinite programs}
\label{subsec04:SDP}

Semidefinite programming is a convex optimisation technique in the cone of positive semidefinite matrices. For more details, see a brief review in Subsection~\ref{subsec01:SDP}.\index{Cone}

\subsubsection{Preliminaries}

In this section, we introduce preliminary technical lemmas.

We recall the following standard result, which comes from the fact that any univariate polynomial nonnegative over $\R$ can be written as a sum-of-squares:

\begin{lemma}[\cite{hilbert1888darstellung}]\label{lemma:ch04_pospolyR}
Let $p\in\mathbb N$ and let $P$ be a univariate polynomial of degree $2p$. Let $\bm \vrm(x)=(1,x,\dots,x^p)$ be the vector of monomials. Then, $P$ is nonnegative over $\R$ if and only if there exists a real $(p+1)\times(p+1)$ positive semidefinite matrix $Q$ such that for all $x\in\mathbb R$,
\begin{equation}
    P(x)=\bm \vrm(x)^TQ\bm \vrm(x).
\end{equation}
\end{lemma}
\noindent This is the univariate case of Proposition~\ref{prop:ch01_sosdecomposition}. From this lemma we deduce the following characterisation of nonnegative polynomials over $\R_+$:

\begin{lemma}\label{lemma:ch04_pospolyR+}
nonnegative polynomials on $\R_+$ can be written as sums of polynomials of the form $\sum_{l=0}^px^l\sum_{i+j=2l}y_iy_j$, where $p\in\mathbb N$ and $y_i\in\mathbb R$, for all $0\le i\le p$.
\end{lemma}
\begin{proof}
Let $P$ be a univariate polynomial of degree $p$ which is nonnegative on $\R_+$.
Writing $\bm \vrm(x)=(1,x,\dots,x^p)$, the polynomial $x\mapsto P(x^2)$ of degree $2p$ is nonnegative on $\R$, so by Lemma~\ref{lemma:ch04_pospolyR} there exists a real positive semidefinite matrix $Q=(Q_{ij})_{0\le i,j\le p}$ such that for all $x\in\mathbb R$:
\begin{equation}
    \begin{aligned}
        P(x^2)&=\bm \vrm(x)^TQ\bm \vrm(x)\\
              &=\sum_{k=0}^{2p}x^k\sum_{i+j=k}Q_{ij}\\
              &=\sum_{l=0}^px^{2l}\sum_{i+j=2l}Q_{ij},
    \end{aligned}
\end{equation}
where the last line comes from the fact that $x\mapsto P(x^2)$ has no monomial of odd degree. Hence, for all $x\in\mathbb R_+$,
\begin{equation}
    P(x)=\sum_{l=0}^px^l\sum_{i+j=2l}Q_{ij}.
\end{equation}
$Q$ is a real $(p+1)\times(p+1)$ positive semidefinite matrix, so via Cholesky decomposition
\begin{equation}
    Q=\sum_{k=0}^p\bm y^{(k)}\bm y^{(k)T},
\end{equation}
where $\bm y^{(k)} \in\R^{p+1}$ for all $k\in\llbracket 0,p \rrbracket$. We finally obtain, for all $x\in\mathbb R_+$,
\begin{equation}
    \begin{aligned}
        P(x)&=\sum_{l=0}^px^l\sum_{i+j=2l}\sum_{k=0}^p\left(\bm y^{(k)}\bm y^{(k)T}\right)_{ij}\\
            &=\sum_{k=0}^p\left(\sum_{l=0}^px^l\sum_{i+j=2l} \bm y_i^{(k)} \bm y_j^{(k)}\right).
    \end{aligned}
\end{equation}
\end{proof}

\noindent Note that the characterisation in Lemma~\ref{lemma:ch04_pospolyR+} differs from that of Stieltjes~\cite{reed1975ii}, which expresses nonnegative polynomials over $\R_+$ as $x\mapsto A_1(x)+xA_2(x)$, where $A_1$ and $A_2$ are sums of squares. 
This slightly slows down the numerical resolution, which does not matter given the size of the programs considered. At the same time, this allows us to obtain more compact expressions for the semidefinite programs. 

We use the characterisation of Lemma~\ref{lemma:ch04_pospolyR+} to obtain the following crucial result: for $\bm s\in\R^{\N}$, the fact that the series $f_{\bm s}$ defined in~\refeq{ch04_isomorphism} has nonnegative scalar product with nonnegative truncated Laguerre series up to degree $m$ can be expressed as a positive semidefinite constraint involving the matrix $A_{\bm s}$ defined in~\refeq{ch04_momentmatrix}. Formally:

\begin{lemma}\label{lemma:ch04_momentmatrix}
Let $m\ge n$ and let $\bm s\in\R^{\N}$. The following propositions are equivalent:
\begin{enumerate}[label=(\roman*)]
\item $\forall g\in\mathcal R_{m,+}(\R_+),\;\braket{f_{\bm s},g}\ge0$,
\item $A_{\bm s}\succeq0$.
\end{enumerate}
\end{lemma}

\begin{proof}
By Lemma~\ref{lemma:ch04_pospolyR+}, any nonnegative polynomial over $\R_+$ of degree less or equal to $m$ can be expressed as a sum of polynomials of the form $\sum_{l=0}^mx^l\sum_{i+j=2l}y_iy_j$. We set the row vector $Y=(y_0,\dots,y_m)\in\mathbb R^{m+1}$. 
Hence, any nonnegative truncated Laguerre series (the elements of $\mathcal R_{m,+}(\R_+)$) can be expressed as as sum of terms of the form $e^{-\frac x2}\sum_{l=0}^mx^l\sum_{i+j=2l}y_iy_j$. 
By linearity, it is sufficient to check that the scalar products with one of these terms are nonnegative.

Recall that for all $k\in\mathbb N$ and $\bm s = (s_k)_k \in \R^\N$ (see \refeq{ch04_coefLag}) we have
\begin{equation}\label{eq:ch04_sk}
    s_k=\int_{\R_+}\mathcal L_k(x)f_{\bm s}(x)dx.
\end{equation}
Thus,
\begin{equation}
    \begin{aligned}
          A_{\bm s}\succeq0&\Leftrightarrow\forall Y\in\R^{m+1},\;Y^TA_{\bm s}Y\ge0\\
                           &\Leftrightarrow\forall Y\in\R^{m+1},\;\sum_{i,j=0}^my_iy_j(A_{\bm s})_{ij}\ge0\\
                           &\Leftrightarrow\forall Y\in\R^{m+1},\;\sum_{l=0}^m\sum_{i+j=2l}^my_iy_j\sum_{k=0}^ls_k\binom lkl!\ge0\\
                           &\Leftrightarrow\forall Y\in\R^{m+1},\;\int_{\R_+}\sum_{l=0}^m\sum_{i+j=2l}^my_iy_j\sum_{k=0}^l\binom lkl!\mathcal L_k(x)f_{\bm s}(x)dx\ge0\\
                           &\Leftrightarrow\forall Y\in\R^{m+1},\;\int_{\R_+}\sum_{l=0}^m\sum_{i+j=2l}^my_iy_j\sum_{k=0}^l(-1)^k\binom lkl!L_k(x)e^{-\frac x2}f_{\bm s}(x)dx\ge0\\
                           &\Leftrightarrow\forall Y\in\R^{m+1},\;\int_{\R_+}\left(e^{-\frac x2}\sum_{l=0}^mx^l\sum_{i+j=2l}^my_iy_j\right)f_{\bm s}(x)dx\ge0\\
                           &\Leftrightarrow\forall Y\in\R^{m+1},\;\left\langle f_{\bm s},x\mapsto e^{-\frac x2}\sum_{l=0}^mx^l\sum_{i+j=2l}^my_iy_j\right\rangle\ge0\\
                           &\Leftrightarrow\forall g\in\mathcal R_{m,+}(\R_+),\;\braket{f_{\bm s},g}\ge0,
    \end{aligned}
\end{equation}
where we used~\refeq{ch04_momentmatrix} in the third line, \refeq{ch04_sk} in the fourth line and~\refeq{ch04_Lagtox} in the sixth line.
\end{proof}

Using these results, we derive hierarchies of semidefinite relaxations and restrictions for the infinite-dimensional linear program \refprog{LP} in the following sections.

\subsubsection{Semidefinite restrictions for computing the threshold value}
\label{subsec04:SDPlower}

A trivial way to obtain a restriction of \refprog{LP} (or equivalently of \refprog{LPS}) is to impose $F_l=0$ for $l>m$, for some $m\ge n$. What is less trivial is that this yields a finite-dimensional semidefinite program.\index{Linear program} \index{Semidefinite program}
Indeed, the constraint~\eqref{eq:ch04_conditionpos} becomes
\begin{equation}
    \forall x\in\mathbb R_+,\;\sum_{k=0}^mF_k\mathcal L_k(x)\ge0,
\end{equation}
or equivalently:
\begin{equation} \label{eq:ch04_PosConstRestr}
    \forall x\in\mathbb R,\;\sum_{k=0}^m(-1)^kF_kL_k(x^2)\ge0,
\end{equation}
where we used~\refeq{ch04_Lagf}. Then, by Theorem~\ref{th:ch04_RHLaguerre}, instead of imposing Eq.~\eqref{eq:ch04_PosConstRestr}, one may equivalently require that:
\begin{equation}\label{eq:ch04_conditionposscalmod}
    \forall m\in \N,\, \forall g\in\mathcal R_{m,+}(\R_+),\;\braket{f_{\bm F},g}\ge0,
\end{equation}
The restriction at rank $m\ge n$ can be written as:
\leqnomode
\begin{flalign*}\index{Semidefinite program}
    \label{prog:LPrestr}
    \tag*{(LP$_n^{\text{restr},m})$}
    \hspace{3cm} \left\{
        \begin{aligned}
            & \quad \text{Find }  \bm F = (F_k)_{0 \leq k \leq m} \in \R^{m+1} \\
            & \quad \text{maximising } F_n \\
            & \quad \text{subject to} \\
            & \hspace{1cm} \begin{aligned}
            & \sum_{k=0}^m F_k = 1  \\
            & \forall k \le m, \;F_k \geq 0 \\
            & \forall p \in \N, \forall g\in\mathcal R_{p,+}(\R_+),\;\braket{f_{\bm F},g}\ge0.
            \end{aligned}
        \end{aligned}
    \right. &&
\end{flalign*}
\reqnomode

We now express this program more straightforwardly as a semidefinite program. 
By Lemma~\ref{lemma:ch04_pospolyR}, writing $\bm \vrm(x)=(1,x,\dots,x^m)$, Eq.~\eqref{eq:ch04_PosConstRestr} is equivalent to the existence of a positive semidefinite matrix $Q=(Q_{ij})_{0\le i,j\le m}$ such that for all $x\in\mathbb R$,
\begin{equation}\label{eq:ch04_FtoQ}
        \sum_{k=0}^m(-1)^kF_kL_k(x^2) = \bm \vrm(x)^TQ\bm \vrm(x) =\sum_{l=0}^m \left( \sum_{i+j=l}Q_{ij} \right) x^l.
\end{equation}
This is in turn equivalent to the linear constraints:
\begin{fleqn}
\begin{equation}
    \hspace{3cm}
    \begin{dcases}
        &\forall l \in \llbracket 1,m \rrbracket, \;\sum\limits_{i+j=2l-1}Q_{ij} = 0,\\
        &\forall l\le m,\sum\limits_{i+j=2l}Q_{ij} = \frac{(-1)^l}{l!}\sum\limits_{k=l}^{m} (-1)^k \binom kl F_k,\hspace{-1cm}
    \end{dcases}
\end{equation}
\end{fleqn}
by identifying the coefficients in front of each monomial in~\refeq{ch04_FtoQ}. Hence, the restriction of \refprog{LP} obtained by imposing $F_l=0$ for $l>m$, for a fixed $m\ge n$, is a semidefinite program given by:
\leqnomode
\begin{flalign*}
    \label{prog:lowerSDPn}
    \tag*{$(\text{SDP}^{m,\leq}_n)$}
    \hspace{3cm} \left\{
        \begin{aligned}
            & \quad \text{Find } Q\in\SymMatrices{m+1} \text{ and } \bm{F}\in\R^{m+1}\\
            & \quad \text{maximising } F_n \\
            & \quad \text{subject to} \\
            & \hspace{1cm} \begin{aligned}
                & \textstyle \sum_{k=0}^{m} F_k = 1  \\
                & \textstyle \forall k\le m, \; F_k \geq 0 \\                     
                & \textstyle \forall l \in \llbracket 1,m \rrbracket, \;\sum\limits_{i+j=2l-1}Q_{ij} = 0\\
                & \textstyle \forall l\le m,\sum\limits_{i+j=2l}Q_{ij} = \sum\limits_{k=l}^{m} \frac{(-1)^{k+l}}{l!}  \binom kl F_k \\
                & \textstyle Q \succeq 0.
             \end{aligned}
        \end{aligned}
        \right. &&
\end{flalign*}
\reqnomode
Let us denote its optimal value by $\omega^{m,\leq}_n$.
Each choice of $m$ leads to a different semidefinite restriction of \refprog{LP} and \refprog{LPS}, whose optimal value gets closer to $\omega_n^\mathcal S$ as $m$ increases (since a feasible solution at rank $m$ necessary provides a feasible solution for the program at rank $m+1$). The sequence $(\omega^{m,\leq}_n)_{m\ge n}$ is thus an increasing sequence and for all $m\ge n$, we have $\omega^{m,\leq}_n\le\omega_n^\mathcal S\le\omega_n^{L^2}$.

For each $m\ge n$, the program \refprog{lowerSDPn} has a dual semidefinite program which is given by:
\leqnomode
\begin{flalign*}\index{Semidefinite program}
   \label{prog:lowerDSDPn}
      \tag*{$(\text{D-SDP}^{m,\leq}_n)$}
      \hspace{3cm} \left\{
      \begin{aligned}
            & \quad \text{Find } A\in\SymMatrices{m+1}, \bm{\mu} \in \R^{m+1} \text{ and } y \in \R \\
            & \quad \text{minimising } y\\
            & \quad \text{subject to}\\
            & \hspace{1cm} \begin{aligned}
                & \textstyle y \geq 1 + \mu_n \\
                & \textstyle \forall k \le m,\; y \geq \mu_k \\
                & \textstyle \forall l\le m,\forall i+j=2l,\;A_{ij}=\sum\limits_{k=0}^l \mu_k \binom lkl!\\
                & \textstyle A \succeq 0.
            \end{aligned}
        \end{aligned}
        \right. &&
\end{flalign*}
\reqnomode

\subsubsection*{Retrieving the canonical form of SDP}

Here we write those SDP in the standard forms (as detailed in Subsection~\ref{subsec01:SDP}) to see why they are indeed dual programs. 
In the programs \refprog{SDPstdform} and \refprog{DSDPstdform} we can exchange \textit{minimisation} and \textit{maximisation} if we also change the sign of the semidefinite constraint for the dual program. We recall them below with this slight modification. For $M,N\in\mathbb N$, $\bm b=(b_1,\dots,b_M)\in\R^M$, $C\in\SymMatrices{N}$, and $B^{(i)}\in\SymMatrices{N}$ for all $i \in \llbracket 1,M \rrbracket$:
\leqnomode
\begin{flalign*}\index{Semidefinite program}
   \label{prog:SDPstdform_recall}
    \tag*{(SDP)}
    \hspace{3cm}\left\{
        \begin{aligned}
            & \quad \text{Find } X\in\SymMatrices{N} \\
            & \quad \text{maximising } \Tr(C^TX) \\
            & \quad \text{subject to:}  \\
            & \hspace{1cm} \begin{aligned}
                &  \forall i \in \llbracket 1,M \rrbracket,\quad\Tr(B^{(i)}X)=b_i \\
                &  X \succeq 0 \Mdot
            \end{aligned}
        \end{aligned}
    \right. &&
\end{flalign*}
\reqnomode
\leqnomode
\begin{flalign*}
   \label{prog:DSDPstdform_recall}
    \tag*{(D-SDP)}
    \hspace{3cm}\left\{
        \begin{aligned}
            & \quad \text{Find } \bm y\in\R^M \\
            & \quad \text{minimising } \bm b^T\bm y \\
            & \quad \text{subject to:}  \\
            & \hspace{1cm} \begin{aligned}
                & \sum_{i=1}^My_iB^{(i)} \succeq C \Mdot
            \end{aligned}
        \end{aligned}
    \right. &&
\end{flalign*}
\reqnomode

To put \refprog{lowerSDPn} in the standard form \refprog{SDPstdform_recall} we set $N=2\times(m+1)$ and $M=1+m+(m+1)$. For all $r\in\N^*$ and all $i,j\in\llbracket 1,r \rrbracket$, let $E^{(i,j)}_r$ be the $r\times r$ matrix whose $(i,j)$ entry is $1$ and all other entries are $0$. We set
\begin{equation}
    \begin{aligned}
    &X=\text{Diag}_{k=0,\dots,m}(F_k)\oplus Q\in\SymMatrices{N},\\
    &C=E^{(n,n)}_N=E^{(n,n)}_{m+1}\oplus\mymathbb 0_{m+1}\in\SymMatrices{N},\\
    &\bm b=(1,0,0,\dots,0)\in\R^M,\\
    &{B'}^{(0)}=\mymathbb 1_{m+1}\oplus\mymathbb 0_{m+1}\in\SymMatrices{N},\\
    &\forall l\in\llbracket 1,m \rrbracket,\quad {B'}^{(l)}=\mymathbb 0_{m+1}\oplus\left(\sum_{i+j=2l-1}E^{(i,j)}_{m+1}\right),\\
    &\forall l\in\llbracket 0,m \rrbracket,\quad B^{(l)}=\text{Diag}_{k=0,\dots,m}\left(-\frac{(-1)^{k+l}}{l!}\binom kl\right)\oplus\left(\sum_{i+j=2l}E^{(i,j)}_{m+1}\right),
    \end{aligned}
\end{equation}
with the convention $\binom kl=0$ when $l>k$. The matrix ${B'}^{(0)}$ corresponds to the constraint $\sum_{k=0}^mF_k=1$ and we denote the corresponding dual variable $y\in\R$. Similarly,  the matrices ${B'}^{(l)}$ correspond to the $m$ constraints $\sum_{i+j=2l-1}Q_{ij} = 0$, and we denote the corresponding dual variables $\nu_l'\in\R$. Finally, the matrices $B^{(l)}$ correspond to the $m+1$ constraints $\sum_{i+j=2l}Q_{ij} = \frac{(-1)^l}{l!}\sum_{k=l}^{m} (-1)^k  \binom kl F_k$, and we denote the corresponding dual variables $\nu_l\in\R$.

The standard form of the dual program \refprog{lowerDSDPn} thus reads:
\leqnomode
\begin{flalign*}\index{Semidefinite program}
   \label{prog:lowerDSDPn_std}
    \tag*{(D-SDP$_n^{m,\le}$)}
    \hspace{3cm}\left\{
        \begin{aligned}
            & \quad \text{Find } y\in\R \text{ and } \bm\nu,\bm\nu'\in\R^{m+1}\times\R^m \\
            & \quad \text{minimising } y \\
            & \quad \text{subject to:}  \\
            & \hspace{1cm} \begin{aligned}
            & \text{Diag}_{k=0,\dots,m}\left[y-\sum_{l=0}^k\nu_l\frac{(-1)^{k+l}}{l!}\binom kl\right]\\
            &\quad\oplus\left(\sum_{l=0}^m\sum_{i+j=2l}\nu_lE^{(i,j)}_{m+1}+\sum_{l=1}^m\sum_{i+j=2l-1}\nu_l'E^{(i,j)}_{m+1}\right) \succeq E^{(n,n)}_{m+1}\oplus\mymathbb 0_{m+1}.\hspace{-4cm}
            \end{aligned}       
        \end{aligned}
    \right. &&
\end{flalign*}
\reqnomode
Due to the block-diagonal structure of the matrices involved, the positive semidefinite constraint above is equivalent to the following constraints:
\begin{fleqn}
\begin{equation}
    \hspace{3cm}
    \left\{ \begin{aligned}
    &y\ge1+\sum_{l=0}^n\nu_l\frac{(-1)^{n+l}}{l!}\binom nl,\\
    &\forall k\in\llbracket0,m\rrbracket\setminus\{n\},\quad y\ge\sum_{l=0}^k\nu_l\frac{(-1)^{k+l}}{l!}\binom kl,\\
    &\left(\sum_{l=0}^m\sum_{i+j=2l}\nu_lE^{(i,j)}_{m+1}+\sum_{l=1}^m\sum_{i+j=2l-1}\nu_l'E^{(i,j)}_{m+1}\right)\succeq0.
    \end{aligned}\right.
\end{equation}
\end{fleqn}
Let us define $A=(A_{ij})_{0\le i,j\le m}$ by
\begin{equation}
    A:=\sum_{l=0}^m\sum_{i+j=2l}\nu_lE^{(i,j)}_{m+1}+\sum_{l=1}^m\sum_{i+j=2l-1}\nu_l'E^{(i,j)}_{m+1},
\end{equation}
or equivalently
\begin{equation}\label{eq:ch04_Atonu}
    A_{ij}=\begin{cases}\nu_l&\text{when }i+j=2l,\\\nu_l'&\text{when }i+j=2l-1,\end{cases}
\end{equation}
for all $i,j\in\llbracket0,m\rrbracket$.
For $k\in\llbracket0,m\rrbracket$, we also define
\begin{equation}
    \mu_k:=\sum_{l=0}^k\nu_l\frac{(-1)^{k+l}}{l!}\binom kl\in\R.
\end{equation}
By Lemma~\ref{lemma:ch04_changeofbasis}, the following conditions are equivalent:
\begin{enumerate}[label=(\roman*)]
\item $\displaystyle \forall k\in\llbracket0,m\rrbracket,\quad\mu_k=\sum_{l=0}^k\nu_l\frac{(-1)^{k+l}}{l!}\binom kl$,
\item $\displaystyle \forall l\in\llbracket0,m\rrbracket,\quad\nu_l=\sum_{k=0}^l\mu_k\binom lkl!$.
\end{enumerate}
With~\refeq{ch04_Atonu} we thus have
\begin{equation}
    A_{ij}=\sum_{k=0}^l\mu_k\binom lkl!\quad\text{when }i+j=2l,
\end{equation}
and we obtain the following expected expression for \refprog{lowerDSDPn}:
\leqnomode
\begin{flalign*}\index{Semidefinite program}
    \tag*{(D-SDP$_n^{m,\le}$)}\hspace{2.8cm}\left\{
        \begin{aligned}
            & \quad \text{Find } y,\bm\mu\in\R\times\R^{m+1} \text{ and } A\in\SymMatrices{m+1} \\
            & \quad \text{minimising } y \\
            & \quad \text{subject to:}  \\
            & \hspace{1cm} \begin{aligned}
            & y \geq 1 + \mu_n\\
            &\forall k \le m,\quad y \geq \mu_k\\
            & \forall l \le m,\forall i+j=2l,\quad A_{ij}=\sum_{k=0}^l\mu_k\binom lkl!\\
            & A \succeq 0.
            \end{aligned}
        \end{aligned}
    \right. &&
\end{flalign*}
\reqnomode
Note that the constraint $y\ge1+\mu_n$ implies the constraint $y\ge\mu_n$.

\subsubsection*{Strong duality} \index{Strong duality!of semidefinite programs}

We show that strong duality holds between the primal and the dual versions of this semidefinite program. In particular, numerical computations with either of these programs will yield the same value. 

\begin{theorem}\label{th:ch04_sdlower}\index{Strong duality!of semidefinite programs}
Strong duality holds between the programs \refprog{lowerSDPn} and \refprog{lowerDSDPn}.
\end{theorem}
\begin{proof}
We make use of Slater's condition for (finite-dimensional) semidefinite programs: strict feasibility of \refprog{lowerSDPn} implies strong duality between \refprog{lowerSDPn} and \refprog{lowerDSDPn}. \index{Strong duality!of semidefinite programs}

In order to obtain a strictly feasible solution, we define $Q \in \SymMatrices{m+1}$ and $\bm F=(F_0,\dots,F_m)\in\R^{m+1}$ by:
\begin{align}
    & Q \defeq \frac1{2^{m+1}-1}\text{Diag}_{k=0,\dots,m}(\frac1{k!}) \\
    & F_k \defeq \frac1{2^{m+1}-1} \binom{m+1}{k+1}
\end{align}
Then $Q \succ 0$ and $F_k>0$ for all $k \in \llbracket 0,m \rrbracket$. Moreover, we have
\begin{equation}
    \begin{aligned}
        \sum_{k=0}^m F_k &= \frac1{2^{m+1}-1}\sum_{k=0}^m \binom{m+1}{k+1} \\
        &= \frac1{2^{m+1}-1}\left(\sum_{k=0}^{m+1} \binom{m+1}k-1\right)\\
        &= 1.
    \end{aligned}
\end{equation}
We also have $\sum_{i+j=2l-1}Q_{ij} = 0$ for all $l\in\llbracket1,m\rrbracket$ since $Q$ is diagonal. Furthermore, for all $l\le m$,
\begin{equation}
    \sum_{i+j=2l}Q_{ij}=Q_{ll}=\frac1{2^{m+1}-1}\frac1{l!},
\end{equation}
and indeed 
\begin{equation}
    \begin{aligned}
        &\frac{(-1)^l}{l!}\sum\limits_{k=l}^{m} (-1)^k  \binom kl F_k\\
        &=\frac1{2^{m+1}-1}\frac1{l!}\sum_{k=l}^m(-1)^{k-l}\binom km\binom{m+1}{k+1}\\
        &=\frac1{2^{m+1}-1}\frac1{l!}\binom ml\sum_{q=0}^{m-l}(-1)^q\frac{m+1}{q+l+1}\binom{m-l}q\hspace{-1cm}\\
        &=\frac1{2^{m+1}-1}\frac1{l!},
    \end{aligned}
\end{equation}
where we used~\cite[(1.41)]{gould1972combinatorial} in the last line.
Therefore, $(Q,\bm F)$ is a strictly feasible solution of \refprog{lowerSDPn}, which implies strong duality.
\end{proof}

\subsubsection{Semidefinite relaxations for computing the threshold value}
\label{subsec04:SDPupper}

One way to obtain a relaxation of \refprog{LP} is to relax the constraint:\index{Linear program} \index{Semidefinite program}
\begin{equation}\label{eq:ch04_conditionpos}
    \forall x\in\mathbb R_+,\;f_{\bm F}(x)=\sum_kF_k\mathcal L_k(x)\ge0.    
\end{equation}
Instead, one may impose the weaker constraint:
\begin{equation}\label{eq:ch04_conditionposscal}
    \forall g\in\mathcal R_{m,+}(\R_+),\;\braket{f_{\bm F},g}\ge0,
\end{equation}
for some fixed $m\ge n$. This relaxation is motivated by Theorem~\ref{th:ch04_RHLaguerre} as it indicates that we will indeed have convergence. We will prove this formally later. Moreover the constraint in Eq.~\eqref{eq:ch04_conditionposscal} only concerns the coefficients $F_k$ for $0 \leq k \leq m$ and the other variables $F_l$ for $l > m$ are only constrained by $\sum_k F_k = 1$ and $F_l \ge 0$. Since $m \ge n$ and the objective is to maximise $F_n$ we can set all of the coefficients $F_l$ for $l>m$ to $0$ without loss of generality.
The relaxation at rank $m\ge n$ can thus be expressed as:
\leqnomode
\begin{flalign*}\index{Semidefinite program}
    \label{prog:LPrelax}
    \tag*{(LP$_n^{\text{relax},m})$}
    \hspace{3cm} \left\{
        \begin{aligned}
            & \quad \text{Find }  \bm F = (F_k)_{0 \leq k \leq m} \in \R^{m+1} \\
            & \quad \text{maximising } F_n \\
            & \quad \text{subject to} \\
            & \hspace{1cm} \begin{aligned}
            & \sum_{k=0}^m F_k = 1  \\
            & \forall k \le m, \;F_k \geq 0 \\
            &\forall g\in\mathcal R_{m,+}(\R_+),\;\braket{f_{\bm F},g}\ge0.
            \end{aligned}
        \end{aligned}
    \right. &&
\end{flalign*}
\reqnomode
It is interesting to relate this program to \refprog{LPrestr}. At first glance they seem identical. The only difference appears on the last constraint. In \refprog{LPrelax} we require that $f_{\bm F}$ has a nonnegative inner product with all nonnegative polynomials over $\R_+$ of degree less or equal than $m$ multiplied by the function $x \mapsto e^{- \frac x2}$. In \refprog{LPrestr} we impose the more restrictive constraint that $f_{\bm F}$ has a nonnegative inner product with all nonnegative square integrable functions $g$. Of course because $f_{\bm F}$ has an expansion on the Laguerre basis of fixed degree $m$, only the first $m+1$ coefficients of $g$ will contribute. This amounts to taking $g \in \mathcal R_{m}(\R_+)$ such that $g$ can be completed into a nonnegative $L^2(\R_+)$ function. 

By Lemma~\ref{lemma:ch04_momentmatrix}, the constraint in Eq.~\eqref{eq:ch04_conditionposscal} may in turn be expressed as a positive semidefinite constraint on the $(m+1)\times(m+1)$ matrix $A_{\bm F}$ defined in~\refeq{ch04_momentmatrix}. Each choice of $m$ thus leads to a different semidefinite program, whose optimal value gets closer to $\omega_n$ as $m$ increases (since the constraint~\eqref{eq:ch04_conditionposscal} gets stronger when $m$ increases).
This gives a hierarchy of finite-dimensional semidefinite relaxations for \refprog{LP}. The semidefinite relaxation of order $m$ is given by:
\leqnomode
\begin{flalign*}
    \label{prog:upperSDPn}
    \tag*{$(\text{SDP}^{m,\geq}_n)$}
    \hspace{3cm} \left\{
        \begin{aligned}
            & \quad \text{Find } A\in\SymMatrices{m+1} \text{ and } \bm F\in\R^{m+1} \\
            & \quad \text{maximising } F_n \\
            & \quad \text{subject to}\\
            & \hspace{1cm}\begin{aligned}
                        & \textstyle \sum_{k=0}^{m} F_k = 1  \\
                        & \textstyle \forall k\le m, \; F_k \geq 0 \\
                        & \textstyle \forall l\le m,\forall i+j=2l,\; A_{ij}=\sum\limits_{k=0}^l F_k \binom lkl!\hspace{-3cm}\\
                        & \textstyle \forall l\in\llbracket1,m\rrbracket,\forall i+j=2l-1,\;A_{ij}=0\hspace{-3cm}\\
                        & \textstyle A\succeq 0.
              \end{aligned}
        \end{aligned}
        \right. &&
\end{flalign*}
\reqnomode
Let us denote its optimal value by $\omega^{m,\geq}_n$. The sequence $(\omega^{m,\geq}_n)_{m\ge n}$ is a decreasing sequence and for all $m\ge n$, we have $\omega_n^{L^2}\le\omega^{m,\geq}_n$.

For each $m\ge n$, the program \refprog{upperSDPn} has a dual semidefinite program which is given by:
\leqnomode
\begin{flalign*}
    \label{prog:upperDSDPn}\index{Semidefinite program}
    \tag*{$(\text{D-SDP}^{m,\geq}_n)$}
    \hspace{3cm} \left\{
        \begin{aligned}
            & \quad \text{Find } Q \in \SymMatrices{m+1}, \bm{\mu} \in \R^{m+1} \text{ and } y \in \R \hspace{-3cm}\\
            & \quad \text{minimising } y \\
            & \quad \text{subject to} \\
            & \hspace{1cm} \begin{aligned} 
            & \textstyle y\ge1+\mu_n  \\
            & \textstyle \forall k\le m, \; y\ge \mu_k \\
            & \textstyle \forall l\le m, \sum\limits_{i+j=2l} Q_{ij} = \textstyle \sum\limits_{k=l}^{m} \frac{(-1)^{k+l}}{l!}\binom kl \mu_k\hspace{-2cm} \\
            & \textstyle Q \succeq 0.
            \end{aligned}
        \end{aligned}
        \right. &&
\end{flalign*}
\reqnomode

\subsubsection*{Retrieving the canonical form of SDP}

To put \refprog{upperSDPn} in the standard form \refprog{SDPstdform_recall} we set $N=2\times(m+1)$ and $M=1+(m+1)^2$. For all $r\in\N^*$ and all $i,j\in\llbracket 1,r \rrbracket$, recall that $E^{(i,j)}_r$ denote the $r\times r$ matrix whose $(i,j)$ entry is $1$ and all other entries are $0$. We set
\begin{equation}
    \begin{aligned}
    &X=\text{Diag}_{k=0,\dots,m}(F_k)\oplus A\in\SymMatrices{N},\\
    &C=E^{(n,n)}_N=E^{(n,n)}_{m+1}\oplus\mymathbb 0_{m+1}\in\SymMatrices{N},\\
    &\bm b=(1,0,0,\dots,0)\in\R^M,\\
    &B^{(0)}=\mymathbb 1_{m+1}\oplus\mymathbb 0_{m+1}\in\SymMatrices{N},\\
    &\forall i,j\in\llbracket0,m\rrbracket,\quad B^{(i,j)}=\begin{cases}\text{Diag}_{k=0,\dots,m}\left(-\binom lkl!\right)\oplus \left(\frac12E^{(i,j)}_{m+1}+\frac12E^{(j,i)}_{m+1}\right)&\text{when } i+j=2l,\\\mymathbb 0_{m+1}\oplus \left(\frac12E^{(i,j)}_{m+1}+\frac12E^{(j,i)}_{m+1}\right) &\text{otherwise,}\end{cases} \hspace{-2cm}
    \end{aligned}
\end{equation}
with the convention $\binom lk=0$ when $k>l$. The matrix $B^{(0)}$ corresponds to the constraint $\sum_{k=0}^mF_k=1$, and we denote the corresponding dual variable $y\in\R$. Similarly, the matrices $B^{(i,j)}$ correspond to the $(m+1)^2$ constraints defining the symmetric matrix $A$, and we denote the corresponding dual variables $Q_{ij}\in\R$, with $Q_{ij}=Q_{ji}$ for all $i,j\in\llbracket 0,m \rrbracket$. We write $Q=(Q_{ij})_{0\le i,j\le m}$.
The standard form of the dual program \refprog{upperDSDPn} thus reads:
\leqnomode
\begin{flalign*}
   \label{prog:upperDSDPn_std}
    \tag*{(D-SDP$_n^{m,\ge}$)}
    \hspace{3cm}\left\{
        \begin{aligned}
            & \quad \text{Find }  y\in\R \text{ and } Q\in\SymMatrices{m+1} \\
            & \quad \text{minimising } y \\
            & \quad \text{subject to:}  \\
            & \hspace{1cm} \begin{aligned}
            & \text{Diag}_{k=0,\dots,m}\left[y-\sum_{l=0}^m\sum_{i+j=2l}Q_{ij}\binom     lkl!\right]\oplus\frac12Q \succeq E^{(n,n)}_{m+1}\oplus\mymathbb 0_{m+1}.\hspace{-4cm}
            \end{aligned}
        \end{aligned}
    \right. &&
\end{flalign*}
\reqnomode 
Due to the block-diagonal structure of the matrices involved, the positive semidefinite constraint above is equivalent to the following constraints:
\begin{fleqn}
\begin{equation}
   \hspace{3cm} \left\{ \begin{aligned}
    &y\ge1+\sum_{l=0}^m\sum_{i+j=2l}Q_{ij}\binom lnl!,\\
    &\forall k\in\llbracket0,m\rrbracket\setminus\{n\},\quad y\ge\sum_{l=0}^m\sum_{i+j=2l}Q_{ij}\binom lkl!,\\
    &Q\succeq0.
    \end{aligned}\right.
\end{equation}
\end{fleqn}
For $k\in\llbracket0,m\rrbracket$, we define
\begin{equation}
    \mu_k:=\sum_{l=0}^m\sum_{i+j=2l}Q_{ij}\binom lkl!\in\R.
\end{equation}
We obtain the program:
\leqnomode
\begin{flalign*}
   \label{prog:upperDSDPn_std2}\index{Semidefinite program}
    \tag*{(D-SDP$_n^{m,\ge}$)}
    \hspace{3cm}\left\{
        \begin{aligned}
            & \quad \text{Find }  y\in\R,\, \bm \mu \in \R^{m+1} \text{ and } Q\in\SymMatrices{m+1} \\
            & \quad \text{minimising } y \\
            & \quad \text{subject to:}  \\
            & \hspace{1cm} \begin{aligned}
            & y\ge1+\mu_n\\
            & \forall k\in\llbracket0,m\rrbracket\setminus\{n\},\quad y\ge\mu_k\\
            & \forall k\in\llbracket0,m\rrbracket,\quad \mu_k=\sum_{l=0}^m\sum_{i+j=2l}Q_{ij}\binom lkl!\\
            & Q\succeq0.
            \end{aligned}
        \end{aligned}
    \right. &&
\end{flalign*}
\reqnomode

Combining Lemma~\ref{lemma:ch04_revert} for $u_k=\mu_k$ and $v_l=\sum_{i+j=2l}Q_{ij}$ for all $k,l\in\llbracket0,m\rrbracket$ with the previous expression of \refprog{upperDSDPn_std2} we finally obtain:
\leqnomode
\begin{flalign*}
    \tag*{(D-SDP$_n^{m,\ge}$)}\hspace{3cm}\left\{
        \begin{aligned}
            & \quad \text{Find }  y\in\R,\, \bm \mu \in \R^{m+1} \text{ and } Q\in\SymMatrices{m+1} \\
            & \quad \text{minimising } y \\
            & \quad \text{subject to:}  \\
            & \hspace{1cm} \begin{aligned}
            & y\ge1+\mu_n\\
            & \forall k\in\llbracket0,m\rrbracket\setminus\{n\},\quad y\ge\mu_k\\
            & \forall l\in\llbracket0,m\rrbracket,\quad\sum_{i+j=2l}Q_{ij}=\frac{(-1)^l}{l!}\sum_{k=l}^m(-1)^k\binom kl \mu_k\hspace{-2cm}\\
            & Q\succeq0,
            \end{aligned}
        \end{aligned}
    \right. &&
\end{flalign*}
\reqnomode
which is the required form. 

\subsubsection*{Strong duality}\index{Strong duality!of linear programs}

This proof is a direct consequence of the proof of Theorem~\ref{th:ch04_sdlower}.
\begin{corollary}\label{th:ch04_sdupper}
Strong duality holds between the programs \refprog{upperSDPn} and \refprog{upperDSDPn}.
\end{corollary}
\begin{proof}\index{Strong duality!of linear programs}
The program \refprog{upperSDPn} is a relaxation of \refprog{LP} and \refprog{lowerSDPn} is a restriction of \refprog{LP}, so \refprog{upperSDPn} is a relaxation of \refprog{lowerSDPn}. Hence, the strictly feasible solution of \refprog{lowerSDPn} derived in the proof of Theorem~\ref{th:ch04_sdlower} yields a strictly feasible solution $(A,\bm F)$ for \refprog{upperSDPn}: we set
$\bm F=(F_0,\dots,F_m)\in\R^{m+1}$, where for all $k \in \llbracket 0,m \rrbracket$, $F_k := \frac1{2^{m+1}-1} \binom{m+1}{k+1}$ and $A=A_{\bm F}\in\SymMatrices{m+1}$, where $A_{\bm F}$ is defined in~\refeq{ch04_momentmatrix}. 

With Slater's condition, this shows again that strong duality holds between the programs \refprog{upperSDPn} and \refprog{upperDSDPn}.
\end{proof}

\subsection{Convergence of the hierarchies of semidefinite programs}
\label{subsec04:CVproof}

From the previous sections, for $m\ge n$ the optimal values $\omega^{m,\geq}_n$ and $\omega^{m,\leq}_n$ of \refprog{upperSDPn} and \refprog{lowerSDPn} form decreasing and increasing sequences, respectively, which satisfy
\begin{equation}\label{eq:ch04_sandwich}
    0\le\omega^{m,\leq}_n\le\omega_n^{\mathcal S}\le\omega_n^{L^2}\le\omega^{m,\geq}_n\le1.
\end{equation}
Recall that $\omega_n^{L^2}$ is the optimal value of \refprog{LP} while $\omega_n^\mathcal S$ is the optimal value of \refprog{LPS}. These sequences thus both converge. The remaining question is whether $(\omega^{m,\geq}_n)_m$ converges to $\omega_n^{L^2}$ and $(\omega^{m,\leq}_n)_m$ converges to $\omega_n^{\mathcal S}$. 
In this section, we show that this is indeed the case.
The problem of proving that $\omega_n^{L^2} = \omega_n^{\mathcal S}$ is still open and we discuss this point later.

\subsubsection{Convergence of the sequence of upper bounds}

\begin{theorem}\label{th:ch04_upperCV}
The decreasing sequence of optimal values of \refprog{upperSDPn} converges to the optimal value of \refprog{LP} that is
\begin{equation}
    \lim_{m\rightarrow+\infty}\omega^{m,\ge}_n=\omega_n^{L^2}.
\end{equation}
\end{theorem}

\noindent In order to prove this theorem, we extract a limit from a sequence of optimal solutions of \refprog{upperSDPn}, for $m\ge n$, and we show using Theorem~\ref{th:ch04_RHLaguerre} that it provides a feasible solution of \refprog{LP}.

\begin{proof}
For all $m\ge n$, the feasible set of \refprog{upperSDPn} is non-empty (consider, e.g., $\bm F=(1,0,0,\dots,0)\in\R^{m+1}$). Moreover, due to the constraints $\sum_{k=0}^mF_k=1$ and $F_k\ge0$ for all $k\le m$, the feasible set of \refprog{upperSDPn} is compact. Hence, the program \refprog{upperSDPn} has feasible optimal solutions, for all $m\ge n$, by diagonal extraction.

The matrix $A$ in \refprog{upperSDPn} is entirely fixed by the choice of $\bm F$. Let $(\bm F^m)_{m\ge n}$ be a sequence of optimal solutions of \refprog{upperSDPn}, for $m\ge n$. 
For each $m\ge n$, we have by optimality that $F_n^m=\omega_n^{m,\ge}$, and the sequence $(F^m_n)_{m\ge n}$ converges. We complete each tuple $\bm F^m=(F_0^m,F_1^m,\dots,F_m^m)\in\R^{m+1}$ with zeros to obtain a sequence in $\R^{\N}$, which we still denote $\bm F^m=(F_0^m,F_1^m,\dots,F_m^m,0,0,\dots)\in\R^{\N}$ for simplicity.

Performing a diagonal extraction $\phi$ on the sequence of optimal solutions $(\bm F^m)_{m\ge n}$, we obtain a sequence of sequences $(\bm F^{\phi(m)})_{m\ge n}$ such that each sequence $(F^{\phi(m)}_k)_{m\ge n}$ converges when $m\rightarrow+\infty$, for all $k\in\N$. Let $F_k$ denote its limit, for each $k\in\N$. We write $\bm F=(F_k)_{k\in\N}\in\R^{\N}$ the sequence of limits. 

For all $m\ge n$, $F_k^{\phi(m)}\ge0$ for all $k\in\mathbb N$ and $\sum_kF_k^{\phi(m)}=1$, so taking $m\rightarrow+\infty$ we obtain $F_k\ge0$ for all $k\in\mathbb N$, and $\sum_kF_k\le1$. Moreover,
\begin{equation}\label{eq:ch04_Fnlim}
    F_n=\lim_{m\rightarrow+\infty}F_n^m=\lim_{m\rightarrow+\infty}\omega^{m,\ge}_n.
\end{equation}
For all $m\ge n$, we have $\omega^{m,\ge}_n\ge\omega_n^{L^2}$, so $F_n\ge\omega_n^{L^2}>0$. In particular, $\sum_kF_k>0$, so without loss of generality we may assume that $\sum_kF_k=1$ (otherwise we can always replace $F_k$ by $\frac{F_k}{\sum_lF_l} > F_k$).

Let $f_{\bm F}=\sum_kF_k\mathcal L_k\in L^2(\R_+)$. By construction we have (see program \refprog{LPrelax}):
\begin{equation}
    \forall m\ge n,\forall g\in\mathcal R_{m,+}(\R_+),\;\braket{f_{\bm F},g}\ge0.
\end{equation}
Hence, by Theorem~\ref{th:ch04_RHLaguerre}, $\bm F$ is the sequence of Laguerre moments of a nonnegative distribution over $\R_+$ (the Lebesgue measure times the function $f_{\bm F}$). \index{Moment sequence} \index{Measure} \index{Laguerre!function}
In particular,
\begin{equation}
    \forall x \in \R_+, \;f_{\bm F}(x)=\sum_k F_k\mathcal L_k(x) \geq 0.    
\end{equation}
With the constraints $F_k\ge0$ for all $k\in\mathbb N$, and $\sum_kF_k=1$, this implies that $\bm F$ is a feasible solution of \refprog{LP}, and in particular $F_n\le\omega_n^{L^2}$, since \refprog{LP} is a maximisation problem. Since we already had $F_n\ge\omega_n^{L^2}$ we obtain with~\refeq{ch04_Fnlim}:
\begin{equation}
    \lim_{m\rightarrow+\infty}\omega^{m,\ge}_n=\omega_n^{L^2},
\end{equation}
which concludes the proof.
\end{proof}
This immediately implies strong duality for programs \refprog{LP} and \refprog{DLP} because of the following remarks:
\begin{enumerate}[label=(\roman*)] 
\item we have weak duality between those programs so that the optimal value $\omega'^{L^2}_n$ of \refprog{DLP} upper bounds the optimal value of \refprog{LP} $\omega^{L^2}_n$ \ie $\omega^{L^2}_n \le \omega'^{L^2}_n$; \index{Strong duality!of semidefinite programs}
\item we have strong duality between \refprog{upperSDPn} and \refprog{upperDSDPn} by Corollary~\ref{th:ch04_sdupper};
\item \refprog{upperSDPn} is a relaxation of \refprog{LP} so that $\forall m,\; \omega^{m,\ge}_n \ge\omega_n^{L^2}$;
\item \refprog{upperDSDPn} is a restriction of \refprog{DLP} so that $\forall m,\; \omega^{m,\ge}_n \ge\omega'^{L^2}_n\ge\omega^{L^2}_n$;
\item we showed that the optimal value of the hierarchy \refprog{upperSDPn} converges to $\omega^{L^2}_n$ \ie $\lim_{m\rightarrow+\infty}\omega^{m,\ge}_n=\omega_n^{L^2}$.
\end{enumerate}
Thus necessarily $\omega'^{L^2}_n = \omega^{L^2}_n$.

\subsubsection{Convergence of the sequence of lower bounds}
\label{subsubsec04:CVowerbound} 
 
\begin{theorem}\label{th:ch04_lowerCV}
The increasing sequence of optimal values of \refprog{lowerSDPn} converges to the optimal value of \refprog{LP}:
\begin{equation}
    \lim_{m\rightarrow+\infty}\omega^{m,\le}_n=\omega_n^\mathcal S.
\end{equation}
\end{theorem}
 
\noindent As a sanity check, we will try to prove the harder result that it converges in order to $\omega_n^{L^2}$ to see where might be the gap. The proof is similar to that of Theorem~\ref{th:ch04_upperCV} using the dual programs: we attempt to construct a feasible optimal solution of \refprog{DLP} by extracting a limit from a sequence of optimal solutions of \refprog{lowerDSDPn}, for $m\ge n$ and then conclude using the strong duality between \refprog{lowerSDPn} and \refprog{lowerDSDPn}, which we proved in Theorem~\ref{th:ch04_sdlower}. However, it turns out that \refprog{DLP} may not have feasible optimal solutions in ${L^2}'(\R_+)$ (as anticipated with the analytical optimal solutions for $n=1$ and $n=2$ from Section~\ref{subsec04:LP}). \index{Linear program} \index{Semidefinite program}

To deal with this issue, we have extended the formulation of \refprog{DLP} to a larger space where it has feasible optimal solutions, namely the space of tempered distributions $\mathcal S'(\R_+)$ (see \refprog{DLPS}).

Note that the semidefinite restrictions \refprog{lowerSDPn} of \refprog{LP} are also restrictions of \refprog{LPS} for all $m\ge n$, while \refprog{LPS} is itself a restriction of \refprog{LP}, and \refprog{DLPS} is a relaxation of \refprog{DLP}. We denote by ${\omega_n'^{\mathcal S}}$ the optimal value of \refprog{DLPS}. Recall that the optimal value of \refprog{LPS} is denoted $\omega_n^{\mathcal S}$ and that of \refprog{LP} and \refprog{DLP} are denoted $\omega_n^{L^2}$ (by strong duality). By weak duality of linear programming we thus have
\begin{equation}\label{eq:ch04_chainopti}
    \omega_n^{m,\le}\le\omega_n^{\mathcal S}\le{\omega_n'^{\mathcal S}}\le\omega_n^{L^2},
\end{equation}
for all $m\ge n$. 

First we use a reformulation of \refprog{lowerDSDPn} over Schwartz space.
For all $m\ge n$:
\leqnomode
\begin{flalign*}
    \label{prog:newlowerDSDPn}
    \tag*{(D-SDP$_n^{m,\le}$)}
    \hspace{3cm} \left\{
        \begin{aligned}
            & \quad \text{\upshape Find } y \in \R \; \text{\upshape and } \bm\mu \in\mathcal S'(\N) \\
            & \quad \text{\upshape minimising } y \\
            & \quad \text{\upshape subject to}  \\
            & \hspace{1cm} \begin{aligned}
            & \forall k \neq n \in \N,\; y \ge\mu_k  \\
            & y \geq 1 + \mu_n \\
            & \forall g \in \mathcal R_{m,+}(\R_+), \; \langle f_{\bm\mu},g \rangle \geq  0\Mdot
            \end{aligned}
        \end{aligned}
        \right. &&
\end{flalign*}
\reqnomode
This amounts to develop $\mu \in \mathcal S'(\R_+)$ in \refprog{DLPS} in the Laguerre basis with a sequence $\bm \mu \in \mathcal S'(\N)$ and then relax the positivity constraint as before. Note that we only consider a sequence $\bm \mu \in \R^{m+1}$ in \refprog{lowerDSDPn} and that by completing it with 0 we obtain automatically a sequence from $\mathcal S'(\N)$.

Before proving Theorem~\ref{th:ch04_lowerCV}, we provide a nontrivial analytical solution to the primal program \refprog{lowerSDPn}.
Let us define $\bm F^n=(F_k^n)_{k\in\N}\in\R^{\N}$ by \\
$\bullet$ if $n$ is even:
\begin{equation}\label{eq:ch04_Fkforneven}
        \mspace{36  mu} F_k^n:=\begin{cases}\frac1{2^n}\binom k{\frac k2}\binom{n-k}{\frac{n-k}2}&\text{when }k\le n, k\text{ even},\\0&\text{otherwise},\end{cases}
\end{equation}
$\bullet$ if $n$ is odd:
\begin{equation}\label{eq:ch04_Fkfornodd}
        F_k^n:=\begin{cases} \frac1{2^n}\frac{\binom n{\floor{\frac n2}}\binom{\floor{\frac n2}}{\floor{\frac k2}}^2 }{\binom nk},&\text{when }k\le n,\\0&\text{otherwise}. \end{cases}
\end{equation}
In both cases,
\begin{equation}
        F_n^n =\frac1{2^n}\binom n{\floor{\frac n2}},
\end{equation}
and we have that:
\begin{equation}\label{eq:ch04_boundFnn}
    F_n^n \ge\frac1{n+1}
\end{equation}
where we used $\binom nj\le\binom n{\floor{\frac n2}}$ for all $j\in\llbracket0,n\rrbracket$, summed over $j$ from 0 to $n$. 
We extrapolated these analytical expressions from numerical values: running \refprog{lowerSDPn} for several values of $n$ and $m$ allowed us to deduce these sequences (we acknowledge here the great help from \href{https://oeis.org}{oeis.org}).

The proof of feasibility consists in checking that the constraints of \refprog{lowerSDPn} are satisfied by $\bm F^n$. To do so, we make use of Zeilberger's algorithm~\cite{zeilberger1991method}, a powerful algorithm for proving binomial identities. The proof of optimality for $m=n$ is obtained by deriving an analytical feasible solution of (D-SDP$_n^{n,\le}$) with the same optimal value. Note that the optimality of this solution does not play a role in the proof of convergence. A word of warning as this proof is rather long. We start by showing two results, corresponding to $n$ even and $n$ odd, respectively, where we make use of Zeilberger's algorithm, a powerful algorithm for proving binomial identities \cite{zeilberger1991method}.  Given a holonomic function, this algorithm outputs a recurrence relation that it satisfies, thus reducing the proof of identity between binomial expressions to the verification that the initialisation is correct. A Mathematica notebook is available for the implementation of Zeilberger's algorithm  \cite{codes}.

\begin{lemma}
\label{lemma:ch04_QdecFnkeven}
For $n \in \N$ even:
\begin{equation}
    \sum_{k=0}^{n} (-1)^k F_k^n L_k(x) = \sum_{l=0}^n x^l \sum_{i+j=2n-2l} p_{n-i} p_{n-j},
    \label{eq:ch04_polyeqeven}
\end{equation}
where:
\begin{align}\label{eq:ch04_coefpnk}
    p_{n-k}:=&\begin{cases} \sqrt{\frac{1}{2^n n!} \binom{n}{\frac{n}{2}} },&\text{when }k=0,\\
    (-1)^{\frac k2} 2^{\frac k2} (\frac k2)!  \binom{\frac{n}{2}}{\frac k2}^2 p_n,&\text{when }0<k\le n,\;k \text{ even },\\
    0,&\text{otherwise}.
\end{cases}
\end{align}
\end{lemma}

\noindent The coefficients $p_{n-k}$ (and $q_{n-k}$ later on) were found by hand when looking for an analytical sum-of-squares decomposition.

\begin{proof}
To prove the polynomial equality~\eqref{eq:ch04_polyeqeven}, we start by equating the coefficients in $x^{l}$ for $l \in \llbracket 0,n \rrbracket$ which gives:
\begin{equation}    \label{eq:ch04_claim_SOSeven}
    \begin{aligned}
        \frac{(-1)^l}{l!}\sum_{k=l}^n (-1)^k  \binom{k}{l}F_k^n 
        &= \sum_{i+j=2n-2l} p_{n-i} p_{n-j}\\
        &= \sum_{i=0}^{2n-2l} p_{n-i} p_{n-(2n-2l-i)}\\
        &= \sum_{i=0}^{n-l} p_{n-2i} p_{n-(2n-2l-2i)},
    \end{aligned}
\end{equation}
where we used $p_{n-i}=0$ for $n-i<0$ in the second line. 
These are equalities between holonomic functions of parameters $n$ and $l$ that are trivial for $l>n$.

\begin{itemize}[leftmargin=*]
\item For $l\le n$ even, because $F^n_k=0$ for $k$ odd, \refeq{ch04_claim_SOSeven} becomes:
\begin{equation}
    \sum_{k=\frac l2}^{\frac n2} \frac1{l!} \binom{2k}{l} F_{2k}^n = \sum_{i=0}^{n-l} p_{n-2i} p_{n-(2n-2l-2i)},
\end{equation}
that is, taking into account the parity of $l=2s$ and $n=2t$,
\begin{equation}
    \sum_{k=s}^{t} \frac1{(2s)!} \binom{2k}{2s} F_{2k}^{2t} = \sum_{i=0}^{2t-2s} p_{2t-2i} p_{2t-(4t-4s-2i)}.
    \label{eq:Zeil_neven_leven_pre}
\end{equation}
Inserting the expressions from~\refeq{ch04_Fkforneven} and~\refeq{ch04_coefpnk}, we thus have to check the identity:
\begin{equation}\label{eq:Zeil_neven_leven}
    \sum_{k=s}^t\frac1{2^{2t}(2s)!}\binom{2k}{2s}\binom{2k}{k}\binom{2t-2k}{t-k}= \sum_{i=0}^{2t-2s}\frac{i!(2t-2s-i)!}{2^{2s}(2t)!}\binom{2t}t\binom ti^2\binom t{2t-2s-i}^2
\end{equation}
for all $t\in\N$ and all $s\le t$ (with the convention $\binom kj=0$ for $j>k$).
We ran Zeilberger's algorithm to show that the right-hand side and the left-hand side of~\refeq{Zeil_neven_leven_pre} both satisfy the following recurrence relation, for all $s,t\in\N$:
\begin{equation}
    \begin{aligned}
    & 2(t+1)^2 S(s,t)+(-2s^2-4t^2+4st+5s-11t-8) S(s,t+1) \\
    & +(s-t-2) (2s-2t-3) S(s,t+2) =0.
    \end{aligned}
\end{equation}
It remains to check that the initialisation is correct. For all $s\in\N$, this recurrence relation in $t$ is of order $2$. Since the identities in~\refeq{Zeil_neven_leven} are trivial when $l>n$, i.e., $s>t$, we thus only need to check~\refeq{Zeil_neven_leven} for $(s,t)=(0,0)$, $(s,t)=(0,1)$, and $(s,t)=(1,1)$, which is straightforward: we obtain the values $1$, $1$ and $\frac14$ respectively, for both sides of~\refeq{Zeil_neven_leven}.
\item For $l\le n$ odd, \refeq{ch04_claim_SOSeven} becomes:
\begin{equation}
    - \sum_{k=\frac{l+1}{2}}^{\frac n2} \frac1{l!} \binom{2k}{l} F_{2k}^n = \sum_{i=0}^{n-l} p_{n-2i} p_{n-(2n-2l-2i)},
\end{equation}
that is, taking into account the parity of $l=2s+1$ and $n=2t$:
\begin{equation}
    - \sum_{k=s+1}^{t} \frac{1}{(2s+1)!} \binom{2k}{2s+1} F_{2k}^{2t} = \sum_{i=0}^{2t-2s-1} p_{2t-2i} p_{2t-(4t-4s-2i-2)}.
    \label{eq:Zeil_neven_lodd_pre}
\end{equation}
Inserting the expressions from~\refeq{ch04_Fkforneven} and~\refeq{ch04_coefpnk}, we thus have to check the identity:
\begin{equation}\label{eq:Zeil_neven_lodd}
    \begin{aligned}
        & \sum_{k=s+1}^{t} \frac1{2^{2t}(2s+1)!}\binom{2k}{2s+1}\binom{2k}k\binom{2t-2k}{t-k} \\
        = & \sum_{i=0}^{2t-2s-1}\frac{i!(2t-2s-i-1)!}{2^{2s+1}(2t)!}\binom{2t}t\binom ti^2\binom t{2t-2s-i-1}^2
    \end{aligned}
\end{equation}
for all $t\in\N$ and all $s\le t$ (with the convention $\binom kj=0$ for $j>k$).
Likewise, we ran Zeilberger's algorithm to show that the right-hand side and the left-hand side of~\refeq{Zeil_neven_lodd_pre} both satisfy the following recurrence relation, for all $s,t\in\N$:
\begin{equation}
\begin{aligned}
    & 2(t+1)^2 S(s,t) + (-2s^2-4t^2+4st+3s-9t-6) S(s,t+1) \\
    & +(s-t-1)(2s-2t-3) S(s,t+2) =0.
\end{aligned}
\end{equation}
It remains to check that the initialisation is correct. For all $s\in\N$, this recurrence relation in $t$ is of order $2$. Since the identities in~\refeq{Zeil_neven_lodd} are trivial when $l>n$, i.e., $s\ge t$, we thus only need to check~\refeq{Zeil_neven_lodd} for $(s,t)=(0,1)$, which is straightforward: we obtain the value $1$ for both sides of~\refeq{Zeil_neven_lodd}.
\end{itemize}

\end{proof}

Now we turn to the case where $n \in \N$ is odd. 
\begin{lemma}
\label{lemma:ch04_QdecFnkodd}
For $n \in \N$ odd:
\begin{equation}
    \sum_{k=0}^{n} (-1)^k F_k^n L_k(x) = \sum_{l=0}^n x^l \sum_{i+j=2l} q_i q_j
    \label{eq:ch04_polyeqodd}
\end{equation}
\end{lemma}
where:
\begin{align}
    q_{n-k}:=&\begin{cases} \sqrt{\frac{1}{2^n n!} \binom{n}{\floor{\frac{n}{2}}} },&\text{when }k=0,\\
    (-1)^{\frac k2}2^{\frac k2}(\frac k2)! \frac{n+1}{n-k+1}\binom{\floor{\frac{n}{2}}}{\frac k2}^2q_n,&\text{when }0<k< n,\;k \text{ even},\\
    0.&\text{otherwise}.
\label{eq:ch04_coefqnk}
\end{cases}
\end{align}

\begin{proof}

Unlike the case where $n$ is even, $\bm F^n$ is non-zero for $k\le n $ odd, and the expression of $F^n_k$ depends on the parity of $k$ (which is not linear in $k$). Thus we cannot use directly~\refeq{ch04_Fkfornodd} in Zeilberger's algorithm as we did in the previous lemma, hence the development below in order to obtain expressions that the algorithm can take as inputs. We fix $n$ odd and $l\le n$.

We start by equating coefficents in $x^l$ in~\refeq{ch04_polyeqodd}:
\begin{equation}
    \begin{aligned}
        \frac{(-1)^l}{l!}\sum_{k=l}^n (-1)^k  \binom{k}{l}F_k^n &= \sum_{i+j=2n-2l} q_{n-i} q_{n-j} \\
        &= \sum_{i=0}^{n-l} q_{n-2i} q_{n-(2n-2l-2i)}.
    \end{aligned}
\label{eq:ch04_claim_SOSodd}
\end{equation}
\begin{itemize}[leftmargin=*]
    \item For $l$ even, writing $l=2s$ and $n=2t+1$, the left-hand side of~\refeq{ch04_claim_SOSodd} becomes:
    \begin{equation}
        \begin{aligned}
            \frac{1}{(2s)!}\sum_{k=2s}^{2t+1} & (-1)^k  \binom{k}{2s}F_k^{2t+1} \\ &= \frac{1}{(2s)!}\sum_{k=0}^{2t+1-2s} (-1)^k  \binom{k+2s}{2s}F_{k+2s}^{2t+1}\\
            &= \frac{1}{(2s)!}\sum_{k=0}^{t-s} \left( \binom{2k+2s}{2s}F_{2k+2s}^{2t+1} - \binom{2k+2s+1}{2s}F_{2k+2s+1}^{2t+1} \right)\\
            &= \frac{\binom{2t+1}{t}}{2^{2t+1}(2s)!} \sum_{k=0}^{t-s} \left(  \frac{\binom{2k+2s}{2s} \binom{t}{k+s}^2}{\binom{2t+1}{2k+2s}} - \frac{\binom{2k+2s+1}{2s} \binom{t}{k+s}^2}{\binom{2t+1}{2k+2s+1}} \right)\\
            &= q_{2t+1}^2 \frac{(2t+1)!}{(2s)!} \sum_{k=0}^{t-s}  \frac{\binom{t}{k+s}^2 \binom{2k+2s}{2s}}{\binom{2t+1}{2k+2s}} \left( 1 - \frac{(2k+2s+1)^2}{(2k+1)(2t-2s-2k+1)} \right).
        \end{aligned}
    \end{equation}
    With~\refeq{ch04_coefqnk} and~\refeq{ch04_claim_SOSodd}, we thus have to check the identity:
    \begin{equation}
        \begin{aligned}
          \frac{(2t+1)!}{(2s)!} &\sum_{k=0}^{t-s}  \frac{\binom{t}{k+s}^2 \binom{2k+2s}{2s}}{\binom{2t+1}{2k+2s}} \left( 1 - \frac{(2k+2s+1)^2}{(2k+1)(2t-2s-2k+1)} \right) \\ 
          =  &\sum_{i=0}^{2t+1-2s} \frac{q_{2t+1-2i} q_{2t+1-(4t+2-4s-2i)}}{q_{2t+1}^2} \\
          = &-\sum_{i=0}^{2t+1-2s}2^{2t-2s+1}\frac{(2t+2)^2 i! (2t+1-2s-i)!}{(2t-2i+2)(4s+2i-2t)}\binom ti^2\binom t{2t-2s-i+1}^2 \hspace{-2cm}
        \end{aligned}
        \label{eq:ch04_Zeil_nodd_leven}
    \end{equation}
    for all $t\in\N$ and all $s\le t$ (with the convention $\binom kj=0$ for $j>k$).
    Zeilberger's algorithm certifies that the right-hand side and the left-hand side of~\refeq{ch04_Zeil_nodd_leven} both satisfy for all $s,t\in\N$:
    \begin{equation}
        \begin{aligned}
          -32 &(t+2)^3 (t+1)^2 (t+3) S(s,t) \\ 
          &+ 4(t+3)(t+2)(2s^2+4t^2-4st-7s+15t+14) S(s,t+1) \\
          & + (-2s+2t+5)(s-t-2)S(s,t+2) =0.
        \end{aligned}
    \end{equation}
    It remains to check that the initialisation is correct. For all $s\in\N$, this recurrence relation in $t$ is of order $2$. Since the identities in~\refeq{ch04_Zeil_nodd_leven} are trivial when $l>n$, i.e., $s>t$, we thus only need to check~\refeq{ch04_Zeil_nodd_leven} for $(s,t)=(0,0)$, $(s,t)=(0,1)$, and $(s,t)=(1,1)$. We obtain the values $0$, $0$ and $-8$ respectively, for both sides of~\refeq{ch04_Zeil_nodd_leven}.
    \item For $l$ odd, writing $l=2s+1$ and $n=2t+1$, the left-hand side of~\refeq{ch04_claim_SOSodd} becomes:
    \begin{equation}
        \begin{aligned}
            \frac{-1}{(2s+1)!}&\sum_{k=2s+1}^{2t+1}  (-1)^k  \binom{k}{2s+1}F_k^{2t+1} \\ 
            &=  \frac{-1}{(2s+1)!} \sum_{k=0}^{2t-2s} (-1)^{k+2s+1} \binom{k+2s+1}{2s+1} F_{k+2s+1}^{2t+1} \\
            &= \frac 1{(2s+1)!} \sum_{k=0}^{t-s} \left( - \binom{2k+2s}{2s+1} F_{2k+2s}^{2t+1} + \binom{2k+2s+1}{2s+1} F_{2k+2s+1}^{2t+1} \right) \\
            &= q_{2t+1}^2 \frac{(2t+1)!}{(2s+1)!} \sum_{k=0}^{t-s} \left( - \frac{\binom{2k+2s}{2s+1} \binom{t}{k+s}^2 }{ \binom{2t+1}{2k+2s} } + \frac{\binom{2k+2s+1}{2s+1} \binom{t}{k+s}^2 }{ \binom{2t+1}{2k+2s+1} } \right) \\
            &= q_{2t+1}^2 \frac{(2t+1)!}{(2s+1)!} \sum_{k=0}^{t-s} \frac{\binom{2k+2s}{2s+1} \binom{t}{k+s}^2 }{ \binom{2t+1}{2k+2s} } \left(-1 + \frac{(2k+2s+1)^2}{2k(2t-2k-2s+1)} \right),
        \end{aligned}
    \end{equation}
    where we used that $\binom{2k+2s}{2s+1}=0$ for $k=0$ in the third line. Note that when factorising we introduced an indeterminate form in the last line that Zeilberger's algorithm can resolve. This is necessary since the algorithm cannot deal with differences of binomial terms.  
    With~\refeq{ch04_coefqnk} and~\refeq{ch04_claim_SOSodd}, we thus have to check the identity:
    \begin{equation}\label{eq:ch04_Zeil_nodd_lodd}
        \begin{aligned}
          \frac{(2t+1)!}{(2s+1)!} & \sum_{k=0}^{t-s} \frac{\binom{2k+2s}{2s+1} \binom{t}{k+s}^2 }{ \binom{2t+1}{2k+2s} } \left(-1 + \frac{(2k+2s+1)^2}{2k(2t-2k-2s+1)} \right) \\ 
          &= \sum_{i=0}^{2t-2s} \frac{q_{2t+1-2i} q_{2t+1-(4t-4s-2i)} }{q_{2t+1}^2} \\
          &=\sum_{i=0}^{2t-2s}2^{2t-2s}\frac{(2t+2)^2 i! (2t-2s-i)!}{(2t-2i+2)(4s+2i-2t+2)}\binom ti^2\binom t{2t-2s-i}^2,
        \end{aligned}
    \end{equation}
    for all $t\in\N$ and all $s\le t$ (with the convention $\binom kj=0$ for $j>k$).
    Zeilberger's algorithm certifies that both the right-hand side and the left-hand side of~\refeq{ch04_Zeil_nodd_lodd} satisfy for all $s \le t$:
    \begin{equation}
        \begin{aligned}
          -32 &(t+2)^3 (t+1)^2 (t+3) S(s,t) \\ 
          &+ 4(t+3)(t+2)(2s^2+4t^2-4st-5s+13t+11) S(s,t+1) \\
          & + (-2s+2t+3)(s-t-2)S(s,t+2) =0.
        \end{aligned}
    \end{equation}
    It remains to check that the initialisation is correct. For all $s\in\N$, this recurrence relation in $t$ is of order $2$. Since the identities in~\refeq{ch04_Zeil_nodd_lodd} are trivial when $l>n$, i.e., $s>t$, we thus only need to check~\refeq{ch04_Zeil_nodd_lodd} for $(s,t)=(0,0)$, $(s,t)=(0,1)$, and $(s,t)=(1,1)$. We obtain the values $1$, $16$ and $1$ respectively, for both sides of~\refeq{ch04_Zeil_nodd_lodd}.
\end{itemize}

\end{proof}

\noindent Having derived these identities, we prove the lemma of interest before proving the theorem on the convergence of the lower bounding hierarchy:
\begin{lemma}
\label{lemma:ch04_feasible}
For all $m\ge n$, $\bm F^n$ is a feasible solution of \refprog{lowerSDPn}. Moreover, it is optimal when $m=n$.
\end{lemma}
\begin{proof}
The proof has three parts. In the first we focus on $n$ even, and in the second on $n$ odd. 
In the last part, we exhibit a feasible solution of \refprog{lowerDSDPn} for $m=n$ with the same optimal value $F_n^n$. 

\paragraph{$\bullet$ $n$ even, $m \ge n$:} we check that $\bm F^n$ defined in~\refeq{ch04_Fkforneven} satisfies all the constraints of \refprog{lowerSDPn}.

\begin{itemize}
    \item For all $k \in \N$, $F_k^n \ge 0$.
    \item We have
    \begin{equation}\label{eq:sumFkneven}
        \begin{aligned}
            \sum_{k=0}^{\infty} F_k^n &=  \sum_{\substack{k=0 \\ \text{even}}}^{n} \frac1{2^{n}}\binom k{\frac k2}\binom{n-k}{\frac{n-k}2}\\
            & = \sum_{k=0}^{\frac n2} \frac1{2^{n}}\binom{2k}{k}\binom{n-2k}{\frac n2 -k} \\
            & = 1,
        \end{aligned}
    \end{equation}
    where the last equality follows from \cite[(3.90)]{gould1972combinatorial}.
    \item We have to show that $x\mapsto\sum_{k=0}^{n} (-1)^k F_k^n L_k(x^2)$ is a positive function on $\R$.  
    Due to Lemma \ref{lemma:ch04_pospolyR}, we aim to find a sum-of-squares decomposition for this polynomial. Guided by numerical results, we look for a polynomial $P(x) \defeq \sum_{i=0}^{n} p_i x^i$ such that:
    \begin{equation}
        \begin{aligned}
          \sum_{k=0}^{n} (-1)^k F_k^n L_k(x^2) & = P^2(x^2) \\
          & = \left( \sum_{i=0}^n p_i x^{2i} \right)^2 \\
          & = \sum_{l=0}^n \left( \sum_{\substack{i+j=l \\ 0 \leq i,j \leq n}} p_i p_j \right) x^{2l},
        \end{aligned}
    \end{equation}
    and the sought coefficients are given by Lemma~\ref{lemma:ch04_QdecFnkeven}, which concludes the first part of the proof.
\end{itemize}

\paragraph{$\bullet$ $n$ odd, $m \ge n$:} similarly, we check that $\bm F^n$ defined in~\refeq{ch04_Fkfornodd} satisfies all the constraints of \refprog{lowerSDPn}.

\begin{itemize}
    \item For all $k \in \N$, $F_k^n \ge 0$.
    \item Writing $n=2t+1$, from~\refeq{ch04_Fkfornodd} we have
    \begin{equation}
        \begin{split}
            & \forall s\le t, \; F^{2t+1}_{2s} = \frac1{2^{2t+2}}\left(1-\frac{s}{t+1}\right)\binom{2s}s\binom{2t+2-2s}{t+1-s}  \\
            & \forall s\in\llbracket1,t+1\rrbracket, \; F^{2t+1}_{2s-1} =   \frac1{2^{2t+2}}\frac{s}{t+1}\binom{2s}s\binom{2t+2-2s}{t+1-s}.
        \end{split}
    \end{equation}
    In particular, for all $s\le t$ we have
    \begin{equation}
       F^{2t+1}_{2s-1}+F^{2t+1}_{2s}=F^{2t+2}_{2s},
    \end{equation}
    where $F^{2t+2}_{2s}$ is defined in \refeq{ch04_Fkforneven}. Since $F^{2t+2}_{2s+1}=0$ for all $s\le t$, and $F_k^n=0$ for all $k>n$, we have:
    \begin{equation}
        \begin{aligned}
            \sum_{k=0}^{\infty} F_k^n&=\sum_{k=0}^{\infty} F_k^{n+1} \\
            & = 1,
        \end{aligned}
    \end{equation}
    where we used~\refeq{sumFkneven}.
    \item We have to show that $x\mapsto\sum_{k=0}^{n} (-1)^k F_k^n L_k(x^2)$ is a positive function on $\R$.  
    Due to Lemma \ref{lemma:ch04_pospolyR}, we aim to find a sum-of-squares decomposition for this polynomial. Guided by numerical results, we look for a polynomial $Q(x) \defeq \sum_{i=0}^{n} q_i x^i$ such that:
\begin{equation}
    \begin{aligned}
        \sum_{k=0}^{n} (-1)^k F_k^n L_k(x^2) & = Q^2(x^2) \\
            & = \left( \sum_{i=0}^n q_i x^{2i} \right)^2 \\
            & = \sum_{l=0}^n \left( \sum_{\substack{i+j=l \\ 0 \leq i,j \leq n}} q_i q_j \right) x^{2l},
    \end{aligned}
\end{equation}
and the sought coefficients are given by Lemma~\ref{lemma:ch04_QdecFnkodd}, which concludes the second part of the proof.
\end{itemize}

\noindent We thus obtained a feasible solution of \refprog{LP} for all $n\in\N^*$, which is feasible for \refprog{lowerSDPn} for all $m\ge n$. 

\paragraph{$\bullet$ Optimality for $m=n$:} we now find an analytical solution of the dual (D-SDP$_n^{n,\le}$) with the same optimal value as the primal (SDP$_n^{n,\le}$), by finding a Cholesky decomposition for the matrix appearing in the dual program \refprog{lowerDSDPn} for $m=n$. The coefficients of the Cholesky decomposition are given by the triangular matrix $L$ with coefficients:
\begin{equation}
    \begin{split}
        \forall j \leq i, \quad & l_{2i,2j} = 2^i i! \binom ij \\ 
        \forall j \leq i, \quad & l_{2i+1,2j+1} = 2^{i+1/2}\frac{(i+1)!}{\sqrt{j+1}} \binom ij \\
        & l_{nn} = 0\\
        & l_{ij}=0\quad\text{otherwise}.
    \end{split}
\end{equation}
Once again, these analytical expressions were extrapolated from numerical values.
Then, $A = L L^T$ is a positive semidefinite matrix given by:
\begin{equation}
    A_{ij} = \sum_{k=0}^{\min(i,j)} l_{ik} l_{jk}.
\end{equation}
Now $l_{ij}$ is non-zero only when $i$ and $j$ have the same parity, so for all $k \in \llbracket 0,\min(i,j) \rrbracket$, $i$ and $j$ must have the same
parity than $k$ for $l_{ik} l_{jk}$ to be non-zero.

\begin{itemize}
    \item Suppose $i = 2i'$, $j = 2j'$, $i' \leq j'$ and $(i',j') \neq (n,n)$. Furthermore let $l=\frac{i+j}2$.
    \begin{equation}
      \begin{aligned}
        A_{ij} &= \sum_{\substack{k=0 \\ k \text{ even}}}^{i} l_{2i', k} l_{2j', k'}\\
        &= \sum_{k=0}^{i'} 2^{i'} i'! \binom{i'}k 2^{j'} j'! \binom{j'}k \\
        &= 2^l i'! (l-i')! \sum_{k=0}^{i'} \binom{i'}k\binom{l-i'}k \\ 
        &= 2^l l!.
        \end{aligned}
    \end{equation}
    \item Suppose $i = 2i'+1$, $j = 2j'+1$, $i' \leq j'$ and $(i',j') \neq (n,n)$. Furthermore let $l=\frac{i+j}2$.
    \begin{equation}
      \begin{aligned}
        A_{ij} &= \sum_{\substack{k=0 \\ k \text{ odd}}}^{i} l_{2i'+1 ,k} l_{2j'+1 ,k}\\
        &= \sum_{k=0}^{i'} 2^{i'+1/2} \frac{(i'+1)!}{\sqrt{k+1}} \binom{i'}k 2^{j'+1/2} \frac{(j'+1)!}{\sqrt{k+1}} \binom{j'}k \\
        &= 2^l i'! (l-i')! \sum_{k=0}^{i'} \binom{i'+1}{k+1}\binom{l-i'-1}k \\
        &= 2^l l!.      
    \end{aligned}
    \end{equation}
    \item Suppose $n = 2t$:
    \begin{equation}
      \begin{aligned}
        A_{nn} &= \sum_{k=0}^{t} l_{2t,2k}^2 \\
        &= 2^{2t} (t!)^2 \sum_{k=0}^{t-1} \binom tk^2\\
        &= 2^n (t!)^2 \left( \binom{2t}t - 1  \right) \\
        &= 2^n n! \left( 1- \binom n{\floor*{\frac n2}}^{-1}  \right).
    \end{aligned}
    \end{equation}
    \item Suppose $n = 2t+1$:
    \begin{equation}
      \begin{aligned}
        A_{nn} &= \sum_{\substack{k=0 \\ k \text{ odd}}}^{t} l_{2t+1,k}^2 \\
        &= 2^{2t+1} (t+1)!^2 \sum_{k=0}^{t-1} \frac{1}{k+1}\binom tk^2\\
        &= 2^n t! (t+1)! \sum_{k=0}^{t-1} \binom{t+1}{k+1}\binom tk \\
        &= 2^n t! (t+1)! \left( \binom{2t+1}t - 1  \right) \\
        &= 2^n n! \left( 1- \binom n{\floor*{\frac n2}}^{-1}  \right).
    \end{aligned}
    \end{equation}
\end{itemize}

\noindent In both cases, $A$ is indeed constructed as:
\begin{equation}
    (A_{\bm{\mu}})_{i,j} = \begin{cases} 
      \sum_{k=0}^l \mu_k \binom lk l! & \text{when } i+j=2l, \\
      0 & \text{otherwise},
\end{cases}
\end{equation}
for $\bm{\mu} = (F_n^n, F_n^n, \dots, F_n^n, 1 - F_n^n)$ with $F_n^n = \frac {1}{2^n} \binom n{\floor*{\frac n2}}$, and this provides a feasible solution of \refprog{lowerDSDPn} for $m=n$, with value $F_n^n$.
This shows the optimality of $\bm F^n$ for (SDP$_n^{n,\le}$) (and the fact that strong duality holds between the programs (SDP$_n^{n,\le}$) and (D-SDP$_n^{n,\le}$), which we already knew from Theorem~\ref{th:ch04_sdlower}).

\end{proof}

\noindent For \refprog{lowerDSDPn}, we see numerically that the optimal solution is the same then for (D-SDP$_n^{n,\le}$), for a few values of $m$ greater than $n$. However, this is no longer the case for higher values, for example when $n=3$ and $m=10$.

We also obtain the following analytical lower bound for the optimal value of \refprog{LP}:
\begin{equation}
    \omega_n^{L^2}\ge \omega_n^{n,\le}=\frac1{2^n}\binom n{\floor{\frac n2}}\underset{n\rightarrow+\infty}\sim\sqrt{\frac2{\pi n}},
\end{equation}
which is superseded by numerical bounds when $n\ge3$ (see Table~\ref{tab:ch04_Fockbounds}).

We now combine the reformulation of program \refprog{newlowerDSDPn} and Lemma~\ref{lemma:ch04_feasible} to prove Theorem~\ref{th:ch04_lowerCV}.

\begin{proof}
The feasible set of \refprog{newlowerDSDPn} is non-empty, by considering the null sequence, which achieves value $1$. Without loss of generality, we add the constraint $y\le1$ in \refprog{DLPS} and \refprog{newlowerDSDPn}.

Let $m\ge n$ and let $(y,\bm\mu)\in\R\times S'(\N)$ be a feasible solution of \refprog{newlowerDSDPn}. 
The constraint $\braket{f_{\bm\mu},x\mapsto e^{-\frac x2}}\ge0$ implies $\mu_0\ge0$ and thus $y\ge0$. Without loss of generality, we may set $\mu_k=0$ for $k>m$, since these coefficients are only constrained by $\mu_k\le y\le1$. We also have $\mu_k\le1$ for all $k\in\N$.

By Lemma~\ref{lemma:ch04_feasible}, $\bm F^l$ is a feasible solution of (SDP$_l^{m,\le}$) for all $l\le m$, so in particular $f_{\bm F^l}=\sum_{k=0}^lF_k^l\mathcal L_k\in\mathcal R_{m,+}(\R_+)$. Hence, $\bm\mu$ must satisfy the constraint $\braket{f_{\bm\mu},f_{\bm F^l}}\ge0$, which gives
\begin{equation}
    \sum_{k=0}^l\mu_kF_k^l\ge0,
\end{equation}
for all $l\le m$.
Thus we have, for all $l\in\llbracket1,m\rrbracket$,
\begin{equation}\label{eq:ch04_mucompact}
    \begin{aligned}
        \mu_l&\ge-\frac1{F_l^l}\sum_{k=0}^{l-1}\mu_kF_k^l\\
        &\ge-\frac1{F_l^l}\sum_{k=0}^{l-1}F_k^l\\
        &=1-\frac1{F_l^l}\\
        &\ge-l,
    \end{aligned}
\end{equation}
where we used $F_l^l>0$ in the first line, $\mu_k\le1$ and $F_k^l\ge0$ in the second line, $\sum_{k=0}^lF_k^l=1$ in the third line and~\refeq{ch04_boundFnn} in the last line. With $\mu_k\le1$, we obtain $|\mu_k|\le k$ for $k\in\N^*$, and thus $|\mu_k|\le k+1$\footnote{Critically this bounds does not depend on the level $m$ of the hierarchy. } for all $k\in\mathbb N$. Hence, the feasible set of \refprog{newlowerDSDPn} is compact and the program \refprog{newlowerDSDPn} has feasible optimal solutions for all $m\ge n$, by diagonal extraction.

Let $(y^m,\bm\mu^m)_{m\ge n}$ be a sequence of optimal solutions of \refprog{newlowerDSDPn}, for $m\ge n$. By Theorem~\ref{th:ch04_sdlower}, we have strong duality between the programs \refprog{lowerSDPn} and \refprog{newlowerDSDPn}, so the optimal value of \refprog{newlowerDSDPn} is given by $\omega_n^{m,\le}$, for all $m\ge n$. By optimality $y^m=\omega_n^{m,\le}$, for all $m\ge n$, and the sequence $(y^m)_{m\ge n}$ converges. 

Performing a diagonal extraction $\phi$ on the sequence $(\bm\mu^m)_{m\ge n}$, we obtain a sequence of sequences $(\bm\mu^{\phi(m)})_{m\ge n}$ such that each sequence $(\mu^{\phi(m)}_k)_{m\ge n}$ converges when $m\rightarrow+\infty$, for all $k\in\N$. Let $\mu_k$ denote its limit, for each $k\in\N$. We write $\bm \mu=(\mu_k)_{k\in\N}\in\R^{\N}$ the sequence of limits. We also write 
\begin{equation}\label{eq:ch04_ylim}
    y:=\lim_{m\rightarrow+\infty}y^m=\lim_{m\rightarrow+\infty}\omega_n^{m,\le}.
\end{equation}
For all $m\ge n$, we have $\omega_n^{\phi(m),\le}\ge\mu_k^{\phi(m)}$ for all $k\in\mathbb N$ and $\omega_n^{\phi(m),\le}\ge1+\mu_n^{\phi(m)}$, so taking $m\rightarrow+\infty$ we obtain $y\ge\mu_k$ for all $k\in\mathbb N$ and $y\ge1+\mu_n$. By~\refeq{ch04_chainopti}, we have $\omega^{m,\le}_n\le\omega_n^{\mathcal S}$ for all $m\ge n$, so $y\le\omega_n^{\mathcal S}$.

\index{Moment sequence}
Moreover, $|\mu_k^{\phi(m)}|\le k+1$ for all $k\in\mathbb N$, so taking $m\rightarrow+\infty$ we obtain $|\mu_k|\le k+1$ for all $k\in\mathbb N$, which implies that $\bm\mu\in\mathcal S'(\N)$~\cite{guillemot1971developpements}. Let $f_{\bm\mu}=\sum_k\mu_k\mathcal L_k\in \mathcal S'(\R_+)$. We have 
\begin{equation}
    \mu_k=\braket{f_{\bm\mu},\mathcal L_k}.
\end{equation}
By construction we also have:
\begin{equation}
    \forall m\ge n,\forall g\in\mathcal R_{m,+}(\R_+),\;\braket{f_{\bm \mu},g}\ge0.
\end{equation}
By Theorem~\ref{th:ch04_RHLaguerre}, this implies that the distribution $\mu:=f_{\bm\mu}(x)\in\mathcal S'(\R_+)$ is nonnegative, i.e.,
\begin{equation}
    \forall f \in \mathcal S_+(\R_+), \;\braket{f_{\bm \mu},f}\ge0.    
\end{equation}
With the constraints $y\ge\mu_k$ for all $k\in\mathbb N$ and $y\ge1+\mu_n$, we have that $(y,\mu)$ is a feasible solution of \refprog{DLPS}, and in particular $y\ge{\omega_n'^{\mathcal S}}$, since \refprog{DLPS} is a minimisation problem. Since $y\le\omega_n^{\mathcal S}$ we obtain with~\refeq{ch04_chainopti} and~\refeq{ch04_ylim}:
\begin{equation}
    \lim_{m\rightarrow+\infty}\omega^{m,\le}_n=\omega_n^{\mathcal S}={\omega_n'^{\mathcal S}}.
\end{equation}
\end{proof}
 
\noindent As a direct corollary of the proofs of Theorem~\ref{th:ch04_upperCV} and Theorem~\ref{th:ch04_lowerCV} (in the same spirit than the remark after the proof of Theorem~\ref{th:ch04_upperCV}), we obtain the following strong duality result:\index{Strong duality!of linear programs}
 
\begin{corollary}\label{th:ch04_sdLP}
Strong duality holds between the programs \refprog{LPS} and \refprog{DLPS} and between the programs \refprog{LP} and \refprog{DLP}.
\end{corollary}

We thus have shown the convergence of the semidefinite hierarchies of upper and lower bounds: $(\omega^{m,\ge}_n)_{m\ge n}$ towards the optimal value of \refprog{LP} and
$(\omega^{m,\le}_n)_{m\ge n}$ towards the opimal value of \refprog{LPS}.
By linearity, these results generalise straightforwardly to the case of witnesses corresponding to linear combinations of fidelities with displaced Fock states:
\begin{align}
    & \lim_{m\rightarrow+\infty}\!\omega^{m,\le}_{\bm a}=\omega_{\bm a}^\mathcal S, \\
    & \lim_{m\rightarrow+\infty}\!\omega^{m,\ge}_{\bm a}=\omega_{\bm a}^{L^2},
\end{align}
for all $n\in\N^*$ and all $\bm a\in[0,1]^n$ where, as an example, $\omega_{\bm a}^{L^2}$ is the value of the following program: 
\leqnomode
\begin{flalign*}\index{Linear program}
    \label{prog:LP_a}
        \tag*{($\text{LP}_{\bm a}^{L^2}$)}
        \hspace{3cm} \left\{
        \begin{aligned}
            & \quad \text{Find }  (F_k)_{k \in \N} \in \ell^2(\N) \\
            & \quad \text{maximising } \sum_{k=1}^n a_k F_k \\
            & \quad \text{subject to} \\
            & \hspace{1cm} \begin{aligned}
            & \sum_k F_k = 1 \\
            & \forall k \in \N, \quad F_k \geq 0 \\
            & \forall x \in \R_+, \quad \sum_k F_k \mathcal L(x) \geq 0,
            \end{aligned}
        \end{aligned}
    \right. &&
\end{flalign*}
and its dual program reads:
\begin{flalign*}\index{Linear program}
    \label{prog:DLP_a}
    \tag*{($\text{D-LP}_{\bm a}^{L^2}$)}
    \hspace{3cm} \left\{
        \begin{aligned}
            & \quad \text{Find }  y \in \R \text{ and } \mu \in {L^2}'(\R_+) \\
            & \quad \text{minimising } y \\
            & \quad \text{subject to} \\
            & \hspace{1cm} \begin{aligned}
            & \forall k \le n \in \N, \quad y \geq a_k + \int_{\R_+}{\mathcal L_k}{d\mu} \\
            & \forall k > n \in \N, \quad y \geq \int_{\R_+}{\mathcal L_k}{d\mu} \\
            & \forall f \in L^2_+(\R_+), \quad\langle \mu,f \rangle=\int_{\R_+}fd\mu \geq  0.
            \end{aligned}
        \end{aligned}
    \right. &&
\end{flalign*}
\reqnomode
The relaxation is given by \refprog{upperSDP} and the restriction is given by \refprog{lowerSDP}.

\subsubsection{Discussion on the possible gap between $\omega_n^{L^2}$ and $\omega_n^{\mathcal S}$}
In practice, we have two hierarchies providing numerical lower bounds and upper bounds on the threshold value. A natural question that now arises is the following: is there a gap between the optimal values of \refprog{LP} and \refprog{LPS}? This is left as an open question but we provide several leads. 

One potential direction would be to prove there is convergence at a finite rank in the hierarchy (see \cite{nie2013certifying}). This would automatically implies the equivalence of the formulation in the space of square integrable functions and in the space of Schwartz functions.

Another direction would be to prove that the upper hierarchy actually converges towards the optimal value or \refprog{LPS} rather than \refprog{LP}. However the limit $(F_k)_k$ of optimal feasible solutions for \refprog{upperSDPn} is only constrained by $\sum_k F_k =1$ which gives that $(F_k)_k$ belongs to $\ell^2$ but not to $\mathcal S(\N)$ in general. 

Another lead would be to prove the other direction \ie that the upper lower hierarchy converges towards the optimal value of \refprog{LP} rather than \refprog{LPS}. But as we saw in the proof above, the analytical feasible solution only ensures that for all $k\in \N$, $\vert \mu_k \vert \le k+1$ and so that $(\mu)_k \in \mathcal S '(\N)$. To prove a better bound on $\mu_k$ so that $(\mu)_k \in \ell^2(\N)$, we would need to exhibit another analytical solution which seems very difficult.

Finally, we also considered proving that any feasible solution of \refprog{LP} (resp. \refprog{DLPS}) can be approximated by a feasible solution of \refprog{LPS} (resp. \refprog{DLP}). However the constraints present in the programs prevent us from using standard density results.

\section{Witnessing multimode Wigner negativity}
\label{sec04:multi}

In this section we discuss the generalisation of our Wigner negativity witnesses to the more challenging multimode setting.
Hereafter, $M$ denotes the number of modes. 

\subsection{Multimode notations}

We use the multi-index notations presented in Eq.~\ref{eq:ch00_multiindex}. A multivariate polynomial of degree $p\in\mathbb N$ may then be written in the compact form $P(\bm x)=\sum_{|\bm l|\le p}p_{\bm l}\bm x^{\bm l}$, where the sum is over all the tuples $\bm l\in\N^M$ such that $|\bm l|\le m$, also known as the weak compositions of the integer $m$. There are $s(m)$ such tuples. In what follows, we will also consider sums over all the tuples $\bm l\in\N^M$ such that $\bm l\le\bm k$, for $\bm k=(k_1,\dots,k_M)\in\N^M$. There are $\pi_{\bm k}$ such tuples.
In particular, for all $\bm x\in\R_+^M$ and all $\bm k\in\N^M$,
\begin{equation}
    L_ {\bm k}(\bm x)=\sum_{\bm l\le\bm k}\frac{(-1)^{|\bm l|}}{\bm l!}\binom{\bm k}{\bm l}\bm x^{\bm l}\quad\text{and}\quad \bm x^{\bm k}=\sum_{\bm l\le\bm k}(-1)^{|\bm l|}\binom{\bm k}{\bm l}\bm k!L_{\bm l}(\bm x).
\end{equation}
Note that there are several ways of considering relaxations or restrictions of fixed degree for the multimode case. The `triangle' approach is to fix a degree $m \in \N$ and consider monomials of degrees $\bm l \in \N^M$ such that $\vert \bm l \vert \le m$. Another way---the `rectangle' approach--- is to fix a degree $\bm m \in \N^M$ and consider monomials of degrees $\bm l \in \N^M$ such that $\bm l \le \bm m$. 
The former is usually used in literature (see~\cite{lasserre10}) while we will need the latter for proving the convergence of the lower bounding hierarchy. These two procedures are interleaved and the convergence of one will provide the convergence of the other. 

We now extend a few definitions from the single-mode case.
For $\bm s=(s_{\bm k})_{\bm k\in\N^M}\in\R^{\N^M}$, we define the associated formal series of multivariate Laguerre functions:\index{Laguerre!function}
\begin{equation}
    f_{\bm s}:=\sum_{\bm k}s_{\bm k}\mathcal L_{\bm k},
\end{equation}
where $\bm s$ is the so-called sequence of Laguerre moments of $f_{\bm s}$. 
For $m\in\mathbb N$, we define the associated $s(m)\times s(m)$ moment matrix $A_{\bm s}$ by
\begin{equation}
    (A_{\bm s})_{\bm i,\bm j} = \begin{cases} 
      \sum_{\bm k\le\bm l}s_{\bm k}\binom{\bm l}{\bm k}\bm l! &\text{when }\bm i+\bm j=2\bm l, \\
      0 &\text{otherwise,}
   \end{cases}
\end{equation}
where $\bm i,\bm j\in\N^M$ with $|\bm i|\le m$ and $|\bm j|\le m$.

The multimode Laguerre functions $(\mathcal L_{\bm k})_{\bm k\in\N^M}$ form an orthonormal basis of the space $L^2(\R_+^M)$ of real square-integrable functions over $\R_+^M$ equipped with the usual scalar product:
\begin{equation}\label{eq:ch04_braketmulti}
\braket{f,g}=\int_{\R_+^M}{f(\bm x)g(\bm x)d\bm x},
\end{equation}
for $f,g\in L^2(\R_+^M)$. We denote by $L^2_+(\R_+^M)$ its subset of nonnegative elements. 
The space $L^2(\R_+^M)$ is isomorphic to its dual space ${L^2}'(\R_+^M)$: elements of ${L^2}'(\R_+^M)$ can be identified by the Lebesgue measure on $\R_+^M$ times the corresponding function in $L^2(\R_+^M)$. The space $L^2(\R_+^M)$ is also isomorphic to $\ell^2(\N^M)$ via expansion on the Laguerre basis. 

Moreover, the elements of the space $\mathcal S(\R_+^M)$ of Schwartz functions over $\R_+^M$, i.e., the space of $C^{\infty}$ functions that go to $0$ at infinity faster than any inverse polynomial, as do their derivatives, can be written as series of Laguerre functions with a sequence indexed by $\N^M$ of rapidly decreasing coefficients $\mathcal S(\N^M)$ (which go to $0$ at infinity faster than any inverse $M$-variate polynomial). Its dual space $\mathcal S'(\R_+^M)$ of tempered distributions over $\R_+^M$ is characterised as the space of formal series of Laguerre functions with a slowly increasing sequence of coefficients $\mathcal S'(\N^M)$ (sequences that are upper bounded by an $M$-variate polynomial)~\cite{guillemot1971developpements}.
We also extend the definition of the duality $\braket{\dummy,\dummy}$ in~\refeq{ch04_braketmulti} to these spaces.

For all $m\in\mathbb N$, the set of series of Laguerre functions over $\R_+^M$ truncated at $m$ is denoted $\mathcal R_m(\R_+^M)$. This is the set of $M$-variate polynomials $P(\bm x)=\sum_{|\bm k|\le m}p_{\bm k}\bm x^{\bm k}$ of degree at most $m$ multiplied by the function $\bm x\mapsto e^{-\frac{\bm x}2}$. 
Let $\mathcal R_{m,+}(\R_+^M)$ denotes its subset of nonnegative elements where the polynomial $P$ is such that $\bm x\mapsto P(\bm x^2)$ has a sum-of-squares decomposition. 

Similarly, for $\bm m\in\N^M$, the set of truncated series of Laguerre functions over $\R_+^M$ with monomials smaller than $\bm m$ is denoted $\mathcal R_{\bm m}(\R_+^M)$ (with a bold $m$ subscript). This is the set of $M$-variate polynomials $P(\bm x)=\sum_{\bm k\le\bm m}p_{\bm k}\bm x^{\bm k}$, multiplied by the function $\bm x\mapsto e^{-\frac{\bm x}2}$. Let $\mathcal R_{\bm m,+}(\R_+^M)$ denotes its subset of nonnegative elements where the polynomial $P$ is such that $\bm x\mapsto P(\bm x^2)$ has a sum-of-squares decomposition. 

\subsection{Multimode Wigner negativity witnesses}

The single-mode Wigner negativity witnesses defined in Eq.~\eqref{eq:ch04_witnessOmega} are naturally generalised to\index{Displacement operator!CV}
\begin{equation}\label{eq:ch04_witnessOmegamulti}
    \hat\Omega_{\bm a,\bm\alpha}:=\smashoperator{\sum_{\bm{1}\le\bm k\le\bm n}}
      a_{\bm k}\hat D(\bm\alpha)\ket{\bm k}\!\bra{\bm k}\hat D^\dag(\bm\alpha),
\end{equation}
for $\bm n=(n_1,\dots,n_M)\in\mathbb N^M\setminus\{\bm0\}$, $\bm a=(a_{\bm k})_{\bm{1}\le\bm k\le\bm n}\in[0,1]^{n_1\cdots n_M}$, 
with $\max_{\bm k}a_{\bm k}=1$, and $\bm\alpha\in\mathbb C^M$. Similar to the single-mode case, these POVM elements\index{POVM} are weighted sums of multimode displaced Fock states projectors, and their expectation value for a quantum state $\bm\rho\in\mathcal D(\mathscr H^{\otimes M})$ is given by
\begin{equation}
    \Tr(\hat\Omega_{\bm a,\bm\alpha}\bm\rho)
    =\smashoperator{\sum_{\bm{1}\le\bm k\le\bm n}}
       a_{\bm k}F\left(\hat D^\dag(\bm\alpha)\bm\rho\hat D(\bm\alpha),\ket{\bm k}\right)
\end{equation}
where $F$ is the fidelity.\index{Fidelity} However, unlike in the single-mode case however, estimating this quantity with homodyne or heterodyne detection by direct fidelity estimation is no longer efficient when the number of modes becomes large. Instead, one may use \textit{robust lower bounds on the multimode fidelity} from~\cite{chabaud2020efficient} which can be obtained efficiently with homodyne or heterodyne detection. A lower bound on the estimated experimental multimode fidelity will allow to detect Wigner negativity if it is larger than an upper bound on the threshold value associated to a given witness.\index{Fidelity}

These lower bounds are obtained as follows: given a target multimode Fock state $\ket{\bm n}=\ket{n_1}\otimes\cdots\otimes\ket{n_M}$ and multiple copies of an $M$-mode experimental state $\bm \rho$, measure all single-mode subsystems of $\bm\rho$ and perform fidelity estimation with each corresponding single-mode target Fock state. That is, the samples obtained from the detection of the $i^{th}$ mode of $\bm\rho$ are used for single-mode fidelity estimation with the Fock state $\ket{n_i}$. Let $F_1,\dots,F_M$ be the single-mode fidelity estimates obtained and let $\tilde F(\bm\rho,\ket{\bm n}):=1-\sum_{i=1}^M(1-F_i)$. Then~\cite{chabaud2020efficient},
\begin{equation}\label{eq:ch04_robustfide}
    \hspace{-0.15cm}1\!-\!M(1\!-\!F(\bm\rho,\ket{\bm n}))\le\tilde F(\bm\rho,\ket{\bm n})\le F(\bm\rho,\ket{\bm n}).
\end{equation}
In particular, $\tilde F$ provides a good estimate of the multimode fidelity $F$ whenever $F$ is not too small. 
The same procedure is followed in the case of target \textit{displaced} Fock states, with classical translations of the samples in order to account for the displacement parameters.

To each witness $\hat\Omega_{\bm a,\bm\alpha}$ is associated its threshold value:
\begin{equation}\label{eq:ch04_threshold_multi}
    \omega_{\bm a}\footnote{Again we write the threshold value as $\omega_{\bm a}$ generically and we will use the more precise notation $\omega_{\bm a}^{L^2}$ (resp. $\omega_{\bm a}^{\mathcal S}$) to refer to the optimisation over square integrable functions (resp. Schwartz functions) on $\R^M$.}
    :=\sup_{\substack{\bm\rho\in \mathcal D(\mathscr H^{\otimes M})\\W_{\bm\rho} \geq 0}}\Tr\!\left(\hat\Omega_{\bm a,\bm\alpha}\,\bm\rho\right).
\end{equation}
With Eq.~\eqref{eq:ch04_robustfide}, if the value of the bound $\tilde F$ obtained experimentally is greater than $\omega_{\bm a}$ (or an upper bound on it) then the multimode state $\bm\rho$ has a negative Wigner function. 

With the same arguments as in the single-mode case, the multimode Wigner negativity\index{Wigner negativity} witnesses in Eq.~\eqref{eq:ch04_witnessOmegamulti} form a complete family and preserve the interpretation from Lemma~\ref{lemma:ch04_operational}: the violation of the threshold value provides a lower bound on the distance to the set of multimode states with nonnegative Wigner function. However, the limited robustness of the bound $\tilde F$ may affect the performance of the witnesses in practical scenarios, in particular for witnesses that are sums of different projectors. Still, we show in Section~\ref{subsec04:example_multi} the applicability of the method with a genuinely multimode example.

We first generalise the single-mode semidefinite programming approach for approximating the threshold values to the multimode case.

\subsection{Infinite-dimensional linear programs}

By linearity, we restrict our analysis to the case of Wigner negativity witnesses that are projectors onto a single multimode Fock state $\ket{\bm n}$, for $\bm n\in\mathbb N^M\setminus\{\bm0\}$. We thus consider the computation of
\begin{equation}
    \omega_{\bm n}=\sup_{\substack{\bm\rho\in\mathcal D(\mathscr H^{\otimes M})\\W_{\bm\rho}\ge0}}\braket{\bm n|\bm\rho|\bm n}.
\end{equation}
A similar reasoning to the single-mode case shows that the computation of the corresponding threshold value in Eq.~\eqref{eq:ch04_threshold_multi} may be rephrased as the following infinite-dimensional linear program:
\leqnomode
\begin{flalign*}\index{Linear program}
    \label{prog:LP_multi}
        \tag*{(LP$_{\bm n}^{L^2}$)}
        \hspace{3cm} \left\{
        \begin{aligned}
            & \quad \text{Find }  (F_{\bm k})_{\bm k \in \N^M} \in \ell^2(\N^M) \\
            & \quad \text{maximising } F_{\bm n} \\
            & \quad \text{subject to} \\
            & \hspace{1cm} \begin{aligned}
            & \sum_{\bm k} F_{\bm k} =  1  \\
            & \forall \bm k \in \N^M, \;F_{\bm k} \geq  0  \\
            & \forall \bm x \in \R_+^M, \; \sum_{\bm k} F_{\bm k}\mathcal L_{\bm k}(\bm x) \geq  0,
            \end{aligned}
        \end{aligned}
        \right. &&
\end{flalign*}
where the optimisation is over square-summable real sequences indexed by elements of $\mathbb N^M$---or equivalently square integrable functions over $\R_+^M$. Its dual linear program reads
\begin{flalign*}
    \label{prog:DLP_multi}
    \tag*{(D-LP$_{\bm n}^{L^2}$)}
    \hspace{3cm} \left\{
        \begin{aligned}
            & \quad \text{Find } y \in \R \text{ and } \mu \in {L^2}'(\R_+^M)\\
            & \quad \text{minimising } y\\
            & \quad \text{subject to}  \\
            & \hspace{1cm} \begin{aligned}
            & \forall \bm k \neq \bm n \in \N^M,\;  y \geq \int_{\R_+^M}{\mathcal L_{\bm k}}{d\mu}  \\
            & y \geq  1 + \int_{\R_+^M}{\mathcal L_{\bm n}}{d\mu} \\
            & \forall f \in L^2_+(\R_+^M), \; \langle \mu,f \rangle \geq  0.
            \end{aligned}
        \end{aligned}
        \right. &&
\end{flalign*}
\reqnomode
We denote their optimal value $\omega_{\bm n}^{L^2}$---as in the single mode case, we have strong duality which can be proven using the same proof with standard infinite-dimensional optimisation techniques as in Subsection~\ref{subsec04:LP}. This will also be obtained from the convergence of the upper bounding hierarchy of semidefinite programs. 
We also introduce the programs over Schwartz functions:
\leqnomode
\begin{flalign*}
    \label{prog:LPS_multi}
        \tag*{(LP$_{\bm n}^{\mathcal S}$)}
        \hspace{3cm} \left\{
        \begin{aligned}
            & \quad \text{Find }  (F_{\bm k})_{\bm k \in \N^M} \in \mathcal S(\N^M) \\
            & \quad \text{maximising } F_{\bm n} \\
            & \quad \text{subject to} \\
            & \hspace{1cm} \begin{aligned}
            & \sum_{\bm k} F_{\bm k} =  1  \\
            & \forall \bm k \in \N^M, \;F_{\bm k} \geq  0  \\
            & \forall \bm x \in \R_+^M, \; \sum_{\bm k} F_{\bm k}\mathcal L_{\bm k}(\bm x) \geq  0,
            \end{aligned}
        \end{aligned}
        \right. &&
\end{flalign*}
and
\begin{flalign*}\index{Linear program}
    \label{prog:DLPS_multi}
    \tag*{(D-LP$_{\bm n}^{\mathcal S}$)}
    \hspace{3cm} \left\{
        \begin{aligned}
            & \quad \text{Find } y \in \R \text{ and } \mu \in {\mathcal S}'(\R_+^M)\\
            & \quad \text{minimising } y\\
            & \quad \text{subject to}  \\
            & \hspace{1cm} \begin{aligned}
            & \forall \bm k \neq \bm n \in \N^M,\;  y \geq \int_{\R_+^M}{\mathcal L_{\bm k}}{d\mu}  \\
            & y \geq  1 + \int_{\R_+^M}{\mathcal L_{\bm n}}{d\mu} \\
            & \forall f \in L^2_+(\R_+^M), \; \langle \mu,f \rangle \geq  0.
            \end{aligned}
        \end{aligned}
        \right. &&
\end{flalign*}
\reqnomode
We denote their optimal value $\omega_{\bm n}^{S}$.

We will then derive two hierarchies of semidefinite programs, lower bounding and upper bounding the threshold value. While our proof of convergence of the single-mode hierarchy of upper bounds towards $\omega_{\bm n}^{L^2}$ transfers easily to the multimode setting, the proof of convergence of the hierarchy of lower bounds towards $\omega_{\bm n}^\mathcal S$ requires the analytical expression of feasible solutions for each level of the hierarchy. We show how to construct such solutions in the multimode case using products of single-mode feasible solutions---this requires introducing an equivalent hierarchy of restrictions, where constraints are expressed on polynomials of $M$ variables with the degree in each individual variable being less or equal to $m$, rather than on polynomials of degree $m$ (that is, constraints of the form $\bm k \leq m \bm 1$ where $m \bm 1=(m,\dots,m)\in\N^M$ rather than $\vert \bm k \vert \leq m$). Along the way, we also prove strong duality of the programs involved. This is the subject of the next subsections. 

\subsubsection{Approximating the threshold value}

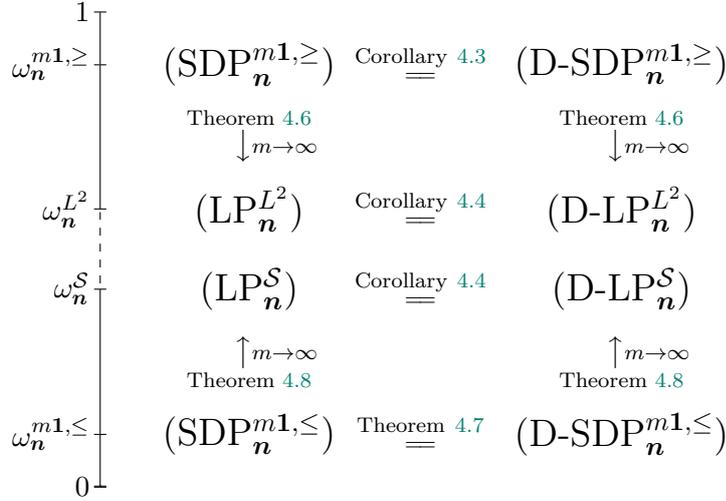
\begin{figure}[ht]
	\begin{center}
		\begin{tikzpicture}[scale=0.70]

\node[inner sep=0pt] (A) at (-0.8,-1) {};
\node[inner sep=0pt] (B) at (-0.8,8) {};
\node[inner sep=0pt] (C1) at (-0.8,0) {};
\node[inner sep=0pt] (C2') at (-0.8,2.75) {};
\node[inner sep=0pt] (C2) at (-0.8,4.25) {};
\node[inner sep=0pt] (C3) at (-0.8,7) {};

\draw[|-|] (-0.8,-1) -- (-0.8,-1);
\draw[-|] (-0.8,-1) -- (-0.8,0);
\draw[-|] (-0.8,0) -- (-0.8,2.75);
\draw[dashed] (-0.8,2.85) -- (-0.8,4.2);
\draw[|-|] (-0.8,4.25) -- (-0.8,7);
\draw[-|] (-0.8,7) -- (-0.8,8);

\node[left] (zero) at (A) {0};
\node[left] (one) at (B) {1};
\node[left] (omegainf) at (C1) {$\omega_{\bm n}^{m\bm1,\le}$};
\node[left] (omega) at (C2) {$\omega_{\bm n}^{L^2}$};
\node[left] (omega) at (C2') {$\omega_{\bm n}^{\mathcal S}$};
\node[left] (omegasup) at (C3) {$\omega_{\bm n}^{m\bm1,\ge}$};

\hypersetup{linkcolor=black}
\node[inner sep=0pt] (SDPsup) at (2,7) {\Large \ref{prog:upperSDPnalt1_multi}};
\node[inner sep=0pt] (D-SDPsup) at (9,7) {\Large \ref{prog:upperDSDPnalt1_multi}};

\node[inner sep=0pt] (LP) at (2,4.25) {\Large \ref{prog:LP_multi}};
\node[inner sep=0pt] (D-LP) at (9,4.25) {\Large \ref{prog:DLP_multi}};

\node[inner sep=0pt] (LPS) at (2,2.75) {\Large \ref{prog:LPS_multi}};
\node[inner sep=0pt] (D-LPS) at (9,2.75) {\Large \ref{prog:DLPS_multi}};

\node[inner sep=0pt] (SDPinf) at (2,0) {\Large \ref{prog:lowerSDPnalt1_multi}};
\node[inner sep=0pt] (D-SDPinf) at (9,0) {\Large \ref{prog:lowerDSDPnalt1_multi}};
\hypersetup{linkcolor=\lkcolor}

\node[inner sep=0pt] (eqsup) at ($.54*(SDPsup)+.46*(D-SDPsup)$) {$\substack{\text{Corollary}~\ref{th:ch04_sduppermulti} \\ =\joinrel=}$};

\node[inner sep=0pt] (eq) at ($.54*(LP)+.46*(D-LP)$) {$\substack{\text{Corollary}~\ref{th:ch04_sdLPmulti} \\ =\joinrel=}$};

\node[inner sep=0pt] (eq) at ($.54*(LPS)+.46*(D-LPS)$) {$\substack{\text{Corollary}~\ref{th:ch04_sdLPmulti} \\ =\joinrel=}$};

\node[inner sep=0pt] (eqinf) at ($.54*(SDPinf)+.46*(D-SDPinf)$) {$\substack{\text{Theorem}~\ref{th:ch04_sdlowermulti} \\ =\joinrel=}$};

\node[inner sep=0pt] (CVlu) at ($.5*(SDPsup)+.5*(LP)$) {$\substack{\text{Theorem~\ref{th:ch04_upperCVmulti}} \\\hspace{20pt} \big\downarrow m\rightarrow \infty}$};

\node[inner sep=0pt] (CVld) at ($.5*(SDPinf)+.5*(LPS)$) {$\substack{\hspace{20pt} \big\uparrow m\rightarrow \infty \\ \text{Theorem~\ref{th:ch04_lowerCVmulti}}}$};

\node[inner sep=0pt] (CVru) at ($.5*(D-SDPsup)+.5*(D-LP)$) {$\substack{\text{Theorem~\ref{th:ch04_upperCVmulti}} \\\hspace{20pt} \big\downarrow m\rightarrow \infty}$};

\node[inner sep=0pt] (CVrd) at ($.5*(D-SDPinf)+.5*(D-LPS)$) {$\substack{\hspace{20pt} \big\uparrow m\rightarrow \infty \\ \text{Theorem~\ref{th:ch04_lowerCVmulti}}}$};

\end{tikzpicture}
		\hypersetup{linkcolor=black}
		\caption{Multimode hierarchies of semidefinite relaxations and restrictions converging to the linear program \ref{prog:LP_multi}, together with their dual programs. The upper index $m$ denotes the level of the relaxation or restriction. On the left are the associated optimal values. The equal sign denotes strong duality, i.e., equality of optimal values, and the arrows denote convergence of the corresponding sequences of optimal values. The hierarchies \ref{prog:upperSDPn_multi} and \ref{prog:lowerSDPn_multi} in the main text are different 
		from the ones appearing in the figure, but equivalent by Lemma~\ref{lemma:ch04_equiv}. The question of whether $\omega_{\bm n}^{L^2} = \omega_{\bm n}^{\mathcal S}$ is left open.}
		\label{fig:structuremulti}
	\end{center}
	\hypersetup{linkcolor=\lkcolor}
\end{figure}

Hereafter, we state without proofs the technical results used to derive the semidefinite programs and their dual programs. They are straightforward generalisation from their single-mode version by using multi-index notations and will be given without proofs.

As mentioned there are two natural ways to obtain relaxations and restrictions by replacing constraints on nonnegative functions by constraints on nonnegative polynomials: either by considering polynomials $P(\bm x)=\sum_{|\bm k|\le m}p_{\bm k}\bm x^{\bm k}$ of degree at most $m$ for $m\in\N$ (`triangle' procedure), or by considering polynomials $P(\bm x)=\sum_{\bm k\le \bm m}p_{\bm k}\bm x^{\bm k}$ with monomials smaller than $\bm m$ for $m\in\N^M$ (`rectangle' procedure). Note that when $M=1$ these two are equal. 

Crucially, all the results below containing conditions of the form $|\bm k|\le m$ for $m\in\N$ are also valid when replaced by conditions of the form $\bm k\le\bm m$ for $\bm m=(m_1,\dots,m_M)\in\N^M$, with the same proofs, by replacing $s(m)$ by $\pi_{\bm m}=\prod_{i=1}^M(m_i+1)$.

\begin{lemma}[Equivalent of Lemma~\ref{lemma:ch04_pospolyR}]\label{lemma:ch04_pospolyRmulti}\index{Sum-of-squares polynomial}
Let $p\in\mathbb N$ and let $P$ be a multivariate polynomial of degree $2p$. Let $\bm \vrm(\bm x)=(\bm x^{\bm k})_{|\bm k|\le p}$ be the vector of monomials. Then, $P$ has a sum-of-squares decomposition if and only if there exists a real $s(p)\times s(p)$ positive semidefinite matrix $Q$ such that for all $x\in\R^M$,
\begin{equation}
    P(\bm x)=\bm \vrm(\bm x)^TQ\bm \vrm(\bm x).
\end{equation}
\end{lemma}

\begin{lemma}[Generalisation of Lemma~\ref{lemma:ch04_pospolyR+}]
Let $P$ be a nonnegative polynomial over $\R_+^M$ such that $\bm x\mapsto P(\bm x^2)$ has a sum-of-squares decomposition.\index{Sum-of-squares polynomial}
Then, $P$ can be written as a sum of polynomials of the form $\sum_{|\bm l|\le p}\bm x^{\bm l}\sum_{\bm i+\bm j=2\bm l}y_{\bm i}y_{\bm j}$ for $p\in\N$ and $y_{\bm i}\in\mathbb R$ for all $\bm i\in\mathbb N^M$ such that $|\bm i|\le p$.
\end{lemma}

\begin{lemma}[Generalisation of Lemma~\ref{lemma:ch04_momentmatrix}]\label{lemma:ch04_momentmatrixmulti}
Let $m\in\N$ and let $\bm s=(s_{\bm k})_{\bm k\in\N^M}\in\R^{\N^M}$. The following propositions are equivalent:
\begin{enumerate}[label=(\roman*)]
\item $\forall g\in\mathcal R_{m,+}(\R_+^M),\;\braket{f_{\bm s},g}\ge0$,
\item $A_{\bm s}\succeq0$.
\end{enumerate}
\end{lemma}

In the single-mode case, we obtained hierarchies of SDP relaxations and restrictions for~\refprog{LP} by replacing constraints involving nonnegative functions by constraints involving nonnegative polynomials $P$ of fixed degree. 
We then exploited the existence of a sum-of-squares decomposition for nonnegative monovariate polynomials.
In the multimode setting, the polynomials involved are multivariate, so that the set of nonnegative polynomials over $\R$ of a given degree may be strictly larger than the set of sum-of-square polynomials~\cite{hilbert1888darstellung}. Instead, we replace directly constraints involving nonnegative functions over $\R_+$ by constraints involving nonnegative polynomials $P$ of fixed degree such that $\bm x\mapsto P(\bm x^2)$\footnote{We write $\bm x^2$ in short for $\bm x^{2 \bm 1} = (x_1^2,\dots,x_M^2)$.} has a sum-of-squares decomposition, implying that the multimode semidefinite relaxations and restrictions are possibly looser than their single-mode counterparts. 

Moreover, the dimension of the semidefinite programs increases exponentially with the level of the hierarchy $m$, as the number of $M$-variate monomials of degree less or equal to $m$ is given by $s(m)$. This implies that the semidefinite programs remain tractable only for a constant number of levels.

In spite of these observations, and following similar steps to the single-mode case (see Subsection~\ref{subsec04:SDP}), we use Lemma~\ref{lemma:ch04_momentmatrixmulti} to obtain the following semidefinite \textit{relaxations} providing upperbounds on the threshold value:
\leqnomode
\begin{flalign*}\index{Semidefinite program}
    \label{prog:upperSDPn_multi}
    \tag*{$(\text{SDP}^{m,\geq}_{\bm n})$}
    \hspace{3cm} \left\{
        \begin{aligned}
            & \quad \text{Find } A=(A_{\bm i\bm j})_{|\bm i|,|\bm j|\le m}\in\SymMatrices{s(m)} \text{ and } \bm F=(F_{\bm k})_{|\bm k|\le m}\in\R^{{s(m)}} \hspace{-3cm}\\
            & \quad \text{maximising } F_{\bm n} \\
            & \quad \text{subject to}\\
            & \hspace{1cm}\begin{aligned}
                        & \textstyle \sum_{|\bm k|\le m} F_{\bm k} =  1  \\
                        & \textstyle \forall |\bm k|\le m, \; F_{\bm k} \geq  0 \\
                        & \textstyle \forall|\bm l|\le m,\forall\bm i\!+\!\bm j\!=\!2\bm l,\; A_{\bm i\bm j}=\sum_{\bm k\le\bm l}F_{\bm k}\binom{\bm l}{\bm k}\bm l!\\
                        & \textstyle \forall|\bm r|\le2m,\bm r\!\neq\!2\bm l,\forall|\bm l|\le m,\forall\bm i\!+\!\bm j\!=\!\bm r,\;A_{\bm i\bm j}=0\\
                        & \textstyle A\succeq 0,
              \end{aligned}
        \end{aligned}
        \right. &&
\end{flalign*}
for all $m\ge|\bm n|$. We denote its optimal value by $\omega_{\bm n}^{m,\ge}$.
The corresponding dual programs are given by:
\leqnomode
\begin{flalign*}
    \label{prog:upperDSDPn_multi}
    \tag*{(D-SDP$_{\bm n}^{m,\ge}$)}
    \hspace{3cm}\left\{
        \begin{aligned}
            & \quad \text{Find } Q\in\SymMatrices{s(m)},\; y \in \R \text{ and } \bm\mu\in\R^{s(m)} \\
            & \quad \text{minimising } y \\
            & \quad \text{subject to}\\
            & \hspace{1cm} \begin{aligned}
            & y\ge1+\mu_{\bm n}\\
            & \forall |\bm k|\le m,\bm k\neq\bm n,\quad y\ge\mu_{\bm k}\\
            & \forall|\bm l|\le m,\quad\sum_{\bm i+\bm j=2\bm l}Q_{\bm i\bm j}=\sum_{\bm k\ge\bm l}\frac{(-1)^{|\bm k|+|\bm l|}}{\bm l!}\binom{\bm k}{\bm l}\mu_{\bm k}\\
            & Q\succeq0.
            \end{aligned}
        \end{aligned}
    \right. &&
\end{flalign*}
\reqnomode

Similarly, using Lemma~\ref{lemma:ch04_pospolyRmulti}, the semidefinite restrictions providing lower bounds for the threshold value are given by
\leqnomode
\begin{flalign*}\index{Semidefinite program}
   \label{prog:lowerSDPn_multi}
    \tag*{$(\text{SDP}^{m,\leq}_{\bm n})$}
    \hspace{3cm} \left\{
        \begin{aligned}
            & \quad \text{Find } Q\in \SymMatrices{s(m)} \text{ and } \bm F\in\R^{{s(m)}} \\
            & \quad \text{maximising } F_{\bm n} \\
            & \quad \text{subject to }\\
            & \hspace{1cm} \begin{aligned}
                        & \textstyle \sum_{|\bm k|\le m} F_{\bm k} =  1  \\
                        & \textstyle \forall |\bm k|\le m, \; F_{\bm k} \geq  0 \\
                        & \textstyle \forall |\bm l|\le m,\sum_{\bm i+\bm j=2\bm l}Q_{\bm i\bm j}=\sum_{\bm k\ge\bm l} \frac{(-1)^{|\bm k|+|\bm l|}}{\bm l!}\binom{\bm k}{\bm l}F_{\bm k}\\
                        & \textstyle \forall |\bm r|\le 2m,\bm r\neq2\bm l,\forall |\bm l|\le m,\sum_{\bm i+\bm j=\bm r}Q_{\bm i\bm j}=0\\
                        & \textstyle Q\succeq 0,
              \end{aligned}
        \end{aligned}
        \right. &&
\end{flalign*}
\reqnomode
for all $m\ge|\bm n|$. We denote its optimal value by $\omega_{\bm n}^{m,\le}$.
The corresponding dual programs are given by:
\leqnomode
\begin{flalign*}
   \label{prog:lowerDSDPn_multi}
    \tag*{$(\text{D-SDP}^{m,\leq}_{\bm n})$}
    \hspace{3cm}\left\{
        \begin{aligned}
            & \quad \text{Find } A\in\SymMatrices{s(m)},\; y \in \R \text{ and } \bm\mu\in\R^{s(m)}  \\
            & \quad \text{minimising } y \\
            & \quad \text{subject to }\\
            & \hspace{1cm} \begin{aligned}
            & y\ge1+\mu_{\bm n}\\
            & \forall |\bm k|\le m,\bm k\neq\bm n,\quad y\ge\mu_{\bm k}\\
            & \forall|\bm l|\le m,\forall\bm i+\bm j=2\bm l,\quad A_{\bm i\bm j}=\sum\limits_{\bm k\le\bm l}\mu_{\bm k}\binom{\bm l}{\bm k}\bm l!\\
            & A \succeq 0,
            \end{aligned}
        \end{aligned}
    \right. &&
\end{flalign*}
\reqnomode
for all $m\ge|\bm n|$. Like in the single-mode case, note that without loss of generality the condition $y\le1$, and thus $\mu_{\bm k}\le1$ for all $\bm k$, can be added to the optimisation, since setting $A=0$, $y=1$ and $\bm\mu=0$ gives a feasible solution with objective value 1.

These are the relaxations and restrictions of~\refprog{LP_multi} obtained by considering polynomials of degree less or equal to $m$, where the optimisation is over matrices and vectors indexed by elements of $\mathbb N^m$ with sum of coefficients lower that $m$. 
Alternatively, we may also consider the relaxations and restrictions obtained by considering polynomials with monomials smaller than $\bm m\in\N^M$, where the optimisation is over matrices and vectors indexed by elements of $\mathbb N^m$ lower or equal to $\bm m$. Recalling the notation $\pi_{\bm m}=\prod_{i=1}^M(m_i+1)$, the corresponding semidefinite \textit{relaxations} are given by
\leqnomode
\begin{flalign*}\index{Semidefinite program}
   \label{prog:upperSDPnalt_multi}
    \tag*{$(\text{SDP}^{\bm m,\geq}_{\bm n})$}
    \hspace{3cm}\left\{
        \begin{aligned}
            & \quad \text{Find } A\in\SymMatrices{\pi_{\bm m}} \text{ and } \bm F\in\R^{\pi_{\bm m}} \\
            & \quad \text{maximising } F_{\bm n} \\
            & \quad \text{subject to }\\
            & \hspace{1cm} \begin{aligned}
            & \sum_{\bm k\le\bm m} F_{\bm k} =  1  \\
            & \forall\bm k\le\bm m, \quad F_{\bm k} \geq  0 \\
            & \forall\bm l\le\bm m,\forall\bm i+\bm j=2\bm l,\quad A_{\bm i\bm j}=\sum\limits_{\bm k\le\bm l}F_{\bm k}\binom{\bm l}{\bm k}\bm l!\\
            & \forall\bm r\le 2\bm m,\bm r\!\neq\!2\bm l,\forall\bm l\le \bm m,\forall\bm i+\bm j=\bm r,\;A_{\bm i\bm j}=0 \\
            & A \succeq 0.
            \end{aligned}
        \end{aligned}
    \right. &&
\end{flalign*}
\reqnomode
for all $\bm m\ge\bm n$. We denote its optimal value by $\omega_{\bm n}^{\bm m,\ge}$.
The corresponding dual programs are given by:
\leqnomode
\begin{flalign*}
    \label{prog:upperDSDPnalt_multi}
    \tag*{(D-SDP$_{\bm n}^{\bm m,\ge}$)}\hspace{3cm}\left\{
        \begin{aligned}
            & \quad \text{Find } Q\in\SymMatrices{\pi_{\bm m}},\; y \in \R \text{ and } \bm\mu\in\R^{\pi_{\bm m}}  \\
            & \quad \text{minimising } y \\
            & \quad \text{subject to }\\
            & \hspace{1cm} \begin{aligned}
            & y\ge1+\mu_{\bm n}\\
            & \forall\bm k\le\bm m,\bm k\neq\bm n,\quad y\ge\mu_{\bm k}\\
            & \forall\bm l\le\bm m,\quad\sum_{\bm i+\bm j=2\bm l}Q_{\bm i\bm j}=\sum_{\bm k\ge\bm l}\frac{(-1)^{|\bm k|+|\bm l|}}{\bm l!}\binom{\bm k}{\bm l}\mu_{\bm k}\hspace{-2cm}\\
            & Q\succeq0,
            \end{aligned}
        \end{aligned}
    \right. &&
\end{flalign*}
\reqnomode
for all $\bm m\ge\bm n$.
Similarly, the semidefinite \textit{restrictions} are given by:
\leqnomode
\begin{flalign*}
    \label{prog:lowerSDPnalt_multi}
    \tag*{(SDP$_{\bm n}^{\bm m,\le}$)}\hspace{3cm}\left\{
        \begin{aligned}
            & \quad \text{Find } Q\in\SymMatrices{\pi_{\bm m}} \text{ and } \bm F\in\R^{\pi_{\bm m}}  \\
            & \quad \text{maximising } F_{\bm n} \\
            & \quad \text{subject to }\\
            & \hspace{1cm} \begin{aligned}
            & \sum_{\bm k\le\bm m} F_{\bm k} =  1\\
            & \forall\bm k\le\bm m, \quad F_{\bm k} \geq  0\\
            & \forall\bm l\le\bm m,\quad\sum_{\bm i+\bm j=2\bm l}Q_{\bm i\bm j}=\sum_{\bm k\ge\bm l} \frac{(-1)^{|\bm k|+|\bm l|}}{\bm l!}\binom{\bm k}{\bm l}F_{\bm k}\\
            & \forall \bm r\le 2\bm m,\bm r\neq2\bm l,\forall\bm l\le\bm m,\quad\sum_{\bm i+\bm j=\bm r}Q_{\bm i\bm j}=0\\
            & Q\succeq0,
            \end{aligned}
        \end{aligned}
    \right. &&
\end{flalign*}
\reqnomode
for all $\bm m\ge\bm n$. We denote its optimal value by $\omega_{\bm n}^{\bm m,\le}$.
The corresponding dual programs are given by:
\leqnomode
\begin{flalign*}\index{Semidefinite program}
   \label{prog:lowerDSDPnalt_multi}
    \tag*{$(\text{D-SDP}^{\bm m,\leq}_{\bm n})$}
    \hspace{3cm}\left\{
        \begin{aligned}
            & \quad \text{Find } A\in\SymMatrices{\pi_{\bm m}},\; y \in \R \text{ and } \bm\mu\in\R^{\pi_{\bm m}}  \\
            & \quad \text{minimising } y \\
            & \quad \text{subject to }\\
            & \hspace{1cm} \begin{aligned}
            & y\ge1+\mu_{\bm n}\\
            & \forall\bm k\le\bm m,\bm k\neq\bm n,\quad y\ge\mu_{\bm k}\\
            & \forall\bm l\le\bm m,\forall\bm i+\bm j=2\bm l,\quad A_{\bm i\bm j}=\sum\limits_{\bm k\le\bm l}\mu_{\bm k}\binom{\bm l}{\bm k}\bm l!\\
            & A \succeq 0,
            \end{aligned}
        \end{aligned}
    \right. &&
\end{flalign*}
\reqnomode
for all $\bm m\ge\bm n$. 

The programs~\refprog{upperSDPn_multi} and~\refprog{lowerSDPn_multi} respectively provide hierarchies of relaxations and restrictions of~\refprog{LP_multi}, since the set of $M$-variate polynomials of degree $m$ is included in the set of $M$-variate polynomials of degree $m+1$.
On the other hand, there is no natural ordering in $\mathbb N^M$ of the relaxations~\refprog{upperSDPnalt_multi} or the restrictions~\refprog{lowerSDPnalt_multi} (consider for instance $\bm m=(2,1)$ and $\bm m'=(1,2)$). In order to obtain proper hierarchies of semidefinite programs, we thus consider the subset of these programs where the tuple $\bm m$ is of the form $m\bm1=(m,\dots,m)\in\N^M$, for $m\in\N$. We have $\pi_{m\bm1}=(m+1)^M$, and the \textit{relaxations} are then given by
\leqnomode
\begin{flalign*}
   \label{prog:upperSDPnalt1_multi}\index{Semidefinite program}
    \tag*{$(\text{SDP}^{m\bm1,\geq}_{\bm n})$}
    \hspace{3cm}\left\{
        \begin{aligned}
            & \quad \text{Find } A\in\SymMatrices{(m+1)^M} \text{ and } \bm F\in\R^{(m+1)^M} \\
            & \quad \text{maximising } F_{\bm n} \\
            & \quad \text{subject to }\\
            & \hspace{1cm} \begin{aligned}
            & \sum_{\bm k\le m\bm1} F_{\bm k} =  1  \\
            & \forall\bm k\le m\bm1, \quad F_{\bm k} \geq  0 \\
            & \forall\bm l\le m\bm1,\forall\bm i+\bm j=2\bm l,\quad A_{\bm i\bm j}=\sum\limits_{\bm k\le\bm l}F_{\bm k}\binom{\bm l}{\bm k}\bm l!\\
            & \forall\bm r\le2m\bm 1,\bm r\!\neq\!2\bm l,\forall\bm l\le m\bm1,\forall\bm i+\bm j=\bm r,\;A_{\bm i\bm j}=0\\
            & A \succeq 0.
            \end{aligned}
        \end{aligned}
    \right. &&
\end{flalign*}
\reqnomode
for $m\ge\max_in_i$. We denote its optimal value by $\omega_{\bm n}^{m\bm1,\ge}$.
The corresponding dual programs are given by:
\leqnomode
\begin{flalign*}
    \label{prog:upperDSDPnalt1_multi}
    \tag*{(D-SDP$_{\bm n}^{m\bm1,\ge}$)}
    \hspace{3cm}\left\{
        \begin{aligned}
            & \quad \text{Find } Q\in\SymMatrices{(m+1)^M},\; y \in \R \text{ and } \bm\mu\in\R^{(m+1)^M}  \\
            & \quad \text{minimising } y \\
            & \quad \text{subject to }\\
            & \hspace{1cm} \begin{aligned}
            & y\ge1+\mu_{\bm n}\\
            & \forall\bm k\le m\bm1,\bm k\neq\bm n,\quad y\ge\mu_{\bm k}\\
            & \forall\bm l\le m\bm1,\quad\sum_{\bm i+\bm j=2\bm l}Q_{\bm i\bm j}=\sum_{\bm k\ge\bm l}\frac{(-1)^{|\bm k|+|\bm l|}}{\bm l!}\binom{\bm k}{\bm l}\mu_{\bm k}\\
            & Q\succeq0,
            \end{aligned}
        \end{aligned}
    \right. &&
\end{flalign*}
\reqnomode
for $m\ge\max_in_i$.
Similarly, the \textit{restrictions} are given by:
\leqnomode
\begin{flalign*}\index{Semidefinite program}
    \label{prog:lowerSDPnalt1_multi}
    \tag*{(SDP$_{\bm n}^{m\bm1,\le}$)}
    \hspace{3cm}\left\{
        \begin{aligned}
            & \quad \text{Find } Q\in\SymMatrices{(m+1)^M} \text{ and } \bm F\in\R^{(m+1)^M}  \\
            & \quad \text{maximising } F_{\bm n} \\
            & \quad \text{subject to }\\
            & \hspace{1cm} \begin{aligned}
            & \sum_{\bm k\le m\bm1} F_{\bm k} =  1\\
            & \forall\bm k\le m\bm1, \quad F_{\bm k} \geq  0\\
            & \forall\bm l\le m\bm1,\quad\sum_{\bm i+\bm j=2\bm l}Q_{\bm i\bm j}=\sum_{\bm k\ge\bm l} \frac{(-1)^{|\bm k|+|\bm l|}}{\bm l!}\binom{\bm k}{\bm l}F_{\bm k}\\
            & \forall \bm r\le2m\bm1,\bm r\neq2\bm l,\forall\bm l\le m\bm1,\quad\sum_{\bm i+\bm j=\bm r}Q_{\bm i\bm j}=0\\\
            & Q\succeq0,
            \end{aligned}
        \end{aligned}
    \right. &&
\end{flalign*}
\reqnomode
for $m\ge\max_in_i$. We denote its optimal value by $\omega_{\bm n}^{m\bm1,\le}$.
The corresponding dual programs are given by:
\leqnomode
\begin{flalign*}
   \label{prog:lowerDSDPnalt1_multi}
    \tag*{$(\text{D-SDP}^{m\bm1,\leq}_{\bm n})$}
    \hspace{3cm}\left\{
        \begin{aligned}
            & \quad \text{Find } A\in\SymMatrices{(m+1)^M},\; y \in \R \text{ and } \bm\mu\in\R^{(m+1)^M}  \\
            & \quad \text{minimising } y \\
            & \quad \text{subject to }\\
            & \hspace{1cm} \begin{aligned}
            & y\ge1+\mu_{\bm n}\\
            & \forall\bm k\le m\bm1,\bm k\neq\bm n,\quad y\ge\mu_{\bm k}\\
            & \forall\bm l\le m\bm1,\forall\bm i+\bm j=2\bm l,\quad A_{\bm i\bm j}=\sum\limits_{\bm k\le\bm l}\mu_{\bm k}\binom{\bm l}{\bm k}\bm l!\\
            & A \succeq 0,
            \end{aligned}
        \end{aligned}
    \right. &&
\end{flalign*}
\reqnomode
for $m\ge\max_in_i$. 

The programs~\refprog{upperSDPnalt1_multi} and~\refprog{lowerSDPnalt1_multi}  are respectively relaxations and restrictions of~\refprog{LP_multi} obtained by considering polynomials of individual degree in each variable less or equal to $m$. These programs respectively provide hierarchies of relaxations and restrictions of~\refprog{LP_multi}, since the set of $M$-variate polynomials with monomials lower than $m\bm1$ is included in the set of $M$-variate polynomials with monomials lower than $(m+1)\bm1$. 

Note that these hierarchies of programs obtained by setting $\bm m$ of the form $m\bm1$ capture the behaviour of all bounds that can be obtained from the more general family of programs indexed by $\bm m=(m_1,\dots,m_M)$, since $\bm m\le(\max_im_i)\bm1$, i.e., the bound obtained by considering the program indexed by $(\max_im_i)\bm1$ supersedes the bound obtained by considering the program indexed by $\bm m$. Formally, for all $\bm m=(m_1,\dots,m_M)\in\N^M$,
\begin{equation}
    \omega_{\bm n}^{\bm m,\ge}\ge\omega_{\bm n}^{(\max_im_i)\bm1,\ge}\quad\text{and}\quad\omega_{\bm n}^{\bm m,\le}\le\omega_{\bm n}^{(\max_im_i)\bm1,\le}.
\end{equation}
Finally, we show that both ways of defining the hierarchies---\ie with $m\in \N$ corresponding to programs \refprog{lowerSDPn_multi} and \refprog{upperSDPn_multi} and with $m \bm1 \in \N^M$ corresponding to programs \refprog{lowerSDPnalt1_multi} and \refprog{upperSDPnalt1_multi}---are equivalent:

\begin{lemma}\label{lemma:ch04_equiv}
For all $m\in\N$,
\begin{equation}\label{eq:ch04_interupper}
    \omega_{\bm n}^{m,\ge}\ge\omega_{\bm n}^{m\bm1,\ge}\ge\omega_{\bm n}^{Mm,\ge},
\end{equation}
and
\begin{equation}\label{eq:ch04_interlower}
    \omega_{\bm n}^{m,\le}\le\omega_{\bm n}^{m\bm1,\le}\le\omega_{\bm n}^{Mm,\le}.
\end{equation}
\end{lemma}

\begin{proof} For all $m\in\N$ we have (for $M>1$):
\begin{equation}
    \enset{\bm k \in \N^M :\, |\bm k|\le m}
    \; \subset \;
    \enset{\bm k \in \N^M :\, \bm k\le m\bm1}
    \;\subset\;
    \enset{\bm k \in \N^M :\, |\bm k|\le Mm}.
\end{equation}
Geometrically
We thus obtain the corresponding inclusions between sets of $M$-variate polynomials:
\begin{enumerate*}[label=(\roman*)] \item\label{item:ch04_inclusion1}$M$-variate polynomials of degree less or equal to $m$ have all their monomials lower than $m\bm1$, and \item\label{item:ch04_inclusion2} all $M$-variate polynomials with monomials lower than $m\bm1$ have degree less or equal to $Mm$. \end{enumerate*}
Hence,
\begin{equation}\label{eq:ch04_inclusion}
    \mathcal R_{m,+}(\R_+^M)\overset{\text{\ref{item:ch04_inclusion1}}}{\subset}\mathcal R_{m\bm1,+}(\R_+^M)\overset{\text{\ref{item:ch04_inclusion2}}}{\subset}\mathcal R_{Mm,+}(\R_+^M).
\end{equation}
As a consequence, \refprog{upperSDPn_multi} is a relaxation of~\refprog{upperSDPnalt1_multi} which is itself a relaxation of $(\text{SDP}_{\bm n}^{Mm,\ge})$, and \refprog{lowerSDPn_multi} is a restriction of~\refprog{lowerSDPnalt1_multi} which is itself a restriction of $(\text{SDP}_{\bm n}^{Mm,\le})$.
\end{proof}

\noindent This result implies that the two versions of the hierarchies of relaxations are interleaved (Eq.~\eqref{eq:ch04_interupper}), and that the two versions of the hierarchies of restrictions are also interleaved (Eq.~\eqref{eq:ch04_interlower}). As such, for any bound obtained with one version of the hierarchy at some fixed level, a better bound can be obtained with the other version at some other level. While this means that the hierarchies are equivalent, note that in practice it may be simpler to solve numerically the version where the parameter space is smaller.

\subsection{Convergence of the multimode hierarchies}
\label{subsec:ch04_app_multiCV}

For $\bm n=(n_1,\dots,n_M)\in\N^M$, the sequences $(\omega_{\bm n}^{m,\ge})_{m\ge|\bm n|}$ and $(\omega_{\bm n}^{m\bm1,\ge})_{m\ge\max_in_i}$ (resp.\ the sequences $(\omega_{\bm n}^{m,\le})_{m\ge|\bm n|}$ and $(\omega_{\bm n}^{m\bm1,\le})_{m\ge\max_in_i}$) are decreasing (resp.\ increasing) sequences, lower bounded (resp.\ upper bounded) by $\omega_{\bm n}^{L^2}$ (resp. $\omega_{\bm n}^{\mathcal S})$). Hence, these sequences are converging. 

We show in what follows that $(\omega_{\bm n}^{m\bm1,\ge})_{m\ge\max_in_i}$ (resp.\ $(\omega_{\bm n}^{m\bm1,\le})_{m\ge\max_in_i}$) converges to $\omega_{\bm n}^{L^2}$ (resp. $\omega_{\bm n}^{\mathcal S}$). By Lemma~\ref{lemma:ch04_equiv}, this implies that the sequence $(\omega_{\bm n}^{Mm,\ge})_{m\ge\max_in_i}$ (resp.\ $(\omega_{\bm n}^{Mm,\le})_{m\ge\max_in_i}$) also converges to $\omega_{\bm n}$. Since this is a subsequence of the converging sequence $(\omega_{\bm n}^{m,\ge})_{m\ge|\bm n|}$ (resp.\ $(\omega_{\bm n}^{m,\le})_{m\ge|\bm n|}$), it implies that the sequence $(\omega_{\bm n}^{m,\ge})_{m\ge|\bm n|}$ (resp.\ $(\omega_{\bm n}^{m,\le})_{m\ge|\bm n|}$) also converges to $\omega_{\bm n}^{L^2}$ (resp. $\omega_{\bm n}^{\mathcal S}$).

With similar proofs to the single-mode case using multi-index notations, we obtain the following result:

\begin{theorem}[Generalisation of Theorem~\ref{th:ch04_RHLaguerre}]
Let $\bm\mu=(\mu_{\bm k})_{\bm k\in\N^M}\in\R^{\N^M}$. Then, $\bm\mu$ is the sequence of Laguerre moments $\int_{\R_+^M}\mathcal L_{\bm k}(\bm x)d\mu(\bm x)$ of a nonnegative distribution $\mu$ supported on $\R_+^M$ if and only if $\forall m\in\N,\forall g\in\mathcal R_{m,+}(\R_+^M),\;\braket{f_{\bm\mu},g}\ge0$.
\end{theorem}

\noindent The proof of this theorem is identical to the univariate case, with the use of Riesz--Haviland theorem over $\R_+^M$~\cite{haviland1936momentum} rather than $\R_+$. \index{Riesz!Haviland theorem}

With Eq.~\eqref{eq:ch04_inclusion}, the proof of convergence of the multimode hierarchy of upper bounds is then obtained directly from its single-mode counterpart using multi-index notations:

\begin{theorem}[Generalisation of Theorem~\ref{th:ch04_upperCV}]\label{th:ch04_upperCVmulti}
The decreasing sequence of optimal values $\omega^{m\bm1,\ge}_{\bm n}$ of~\refprog{upperSDPnalt1_multi} converges to the optimal value $\omega_{\bm n}^{L^2}$ of \refprog{LP_multi}:
\begin{equation}
    \lim_{m\rightarrow+\infty}\omega^{m\bm1,\ge}_{\bm n}=\omega_{\bm n}^{L^2}.
\end{equation}
\end{theorem}

\noindent With Lemma~\ref{lemma:ch04_equiv}, we also obtain
\begin{equation}
    \lim_{m\rightarrow+\infty}\omega^{m,\ge}_{\bm n}=\omega_{\bm n}^{L^2}.
\end{equation}

On the other hand, the proof of convergence of the single-mode hierarchy of lower bounds crucially exploits analytical feasible solutions of the programs~\refprog{lowerSDPn} in order to obtain two results:

\begin{itemize}
    \item Strong duality between programs~\refprog{lowerSDPn} and~\refprog{lowerDSDPn} (Theorem~\ref{th:ch04_sdlower}).
    \item The fact that the feasible set of~\refprog{lowerDSDPn} is compact with coefficients bounded independently of $m$ (Eq.~\eqref{eq:ch04_mucompact}) which exploits an analytical feasible solution of the primal program. 
\end{itemize}

\noindent In what follows, we generalise these two results to the multimode setting by obtaining multimode analytical feasible solutions from products of single-mode ones. 

\begin{lemma}\label{lemma:ch04_productsol}
For all $m,n\in\mathbb N$ with $m\ge n$, suppose that $Q(m,n)\in\SymMatrices{m+1}$ and $\bm F(m,n) = (F_i(m,n))_{0 \le i\le m} \in\R^{m+1}$ are feasible solutions of~\refprog{lowerSDPn}. Let $\bm m=(m_1,\dots,m_M)\in\N^M$ and $\bm n=(n_1,\dots,n_M)\in\N^M$ with $\bm m\ge\bm n$. Define $Q:=Q(m_1,n_1)\otimes\dots\otimes Q(m_M,n_M)\in\SymMatrices{\pi_{\bm m}}$ and $\bm F=(F_{\bm k})_{\bm k\le\bm m}\in\R^{\pi_{\bm m}}$, where for all $\bm k=(k_1,\dots,k_M)\le\bm m$, $F_{\bm k}:=\prod_{i=1}^MF_{k_i}(m_i,n_i)$. 

Then, $(Q,\bm F)$ is a feasible solution of \refprog{lowerSDPnalt_multi}. Moreover, if $(Q(m_i,n_i),\bm F(m_i,n_i))$ is strictly feasible for all $i=1,\dots,M$ then $(Q,\bm F)$ is a strictly feasible solution of \refprog{lowerSDPnalt_multi}.
\end{lemma}

\begin{proof}
With the notations of the Lemma, we show the feasibility of $(Q,\bm F)$ (resp.\ strict feasibility).
We immediately have $Q\succeq0$, $F_{\bm k}\ge0$ (resp.\ $Q\succ0$, $F_{\bm k}>0$) for all $\bm k\le\bm m$, and $Q_{\bm i\bm j}=\prod_{p=1}^MQ_{i_pj_p}(m_p,n_p)$ for all $\bm i=(i_1,\dots,i_M)\le\bm m$ and $\bm j=(j_1,\dots,j_M)\le\bm m$. Hence, for all $\bm r=(r_1,\dots,r_M)\le2\bm m$,
\begin{equation}
    \begin{aligned}
        \sum_{\bm i+\bm j=\bm r}Q_{\bm i\bm j}&=\sum_{i_1+j_1=r_1,\dots,i_M+j_M=r_M}\prod_{p=1}^MQ_{i_pj_p}(m_p,n_p)\\
        &=\prod_{p=1}^M\sum_{i_p+j_p=r_p}Q_{i_pj_p}(m_p,n_p).
    \end{aligned}
\end{equation}
In particular, if $\bm r\neq2\bm l$ for all $\bm l\le\bm m$, then at least one coefficient $r_p$ is odd, and the corresponding sum gives $0$ since $(Q(m_p,n_p),\bm F(m_p,n_p))$ is feasible for $(\text{SDP}_{n_p}^{m_p,\le})$. In that case, $\sum_{\bm i+\bm j=\bm r}Q_{\bm i\bm j}=0$. Otherwise, for all $\bm l=(l_1,\dots,l_M)\le\bm m$,
\begin{equation}
    \begin{aligned}
        \sum_{\bm i+\bm j=2\bm l}Q_{\bm i\bm j}&=\prod_{p=1}^M\sum_{i_p+j_p=2l_p}Q_{i_pj_p}(m_p,n_p)\\
        &=\prod_{p=1}^M\sum_{k_p\ge l_p} \frac{(-1)^{k_p+l_p}}{l_p!}\binom{k_p}{l_p}F_{k_p}(m_p,n_p)\\
        &=\sum_{l_1\le k_1\le m_1,\dots,l_M\le k_M\le m_M}\prod_{p=1}^M \frac{(-1)^{k_p+l_p}}{l_p!}\binom{k_p}{l_p}F_{k_p}(m_p,n_p)\\
        &=\sum_{\bm k\ge\bm l} \frac{(-1)^{|\bm k|+|\bm l|}}{\bm l!}\binom{\bm k}{\bm l}F_{\bm k},
    \end{aligned}
\end{equation}
where we used the feasibility of $(Q(m_p,n_p),\bm F(m_p,n_p))$ in the second line. Finally, 
\begin{equation}
    \begin{aligned}
        \sum_{\bm k\le\bm m}F_{\bm k}&=\sum_{k_1\le m_1,\dots,k_M\le m_M}\prod_{i=1}^MF_{k_i}(m_i,n_i)\\
        &=\prod_{i=1}^M\sum_{k_i=0}^{m_i}F_{k_i}(m_i,n_i)\\
        &=1,
    \end{aligned}
\end{equation}
since $\sum_{k=0}^mF_{k}(m,n)=1$ for all $m,n\in\N$ with $m\ge n$. This shows that $(Q,\bm F)$ is a feasible solution of~\refprog{lowerSDPnalt_multi} (resp.\ strictly feasible).
\end{proof}

\noindent A direct consequence of this construction is the following result:

\begin{theorem}[Generalisation of Theorem~\ref{th:ch04_sdlower}]\label{th:ch04_sdlowermulti}
Strong duality holds between the programs \refprog{lowerSDPnalt_multi} and \refprog{lowerDSDPnalt_multi}.
\end{theorem}

\begin{proof}
The proof of Theorem~\ref{th:ch04_sdlower} gives a strictly feasible solution $(Q(m,n),\bm F(m,n))$ of \refprog{lowerSDPn} for all $m\ge n$. By Lemma~\ref{lemma:ch04_productsol}, the program \refprog{lowerSDPnalt_multi} thus has a strictly feasible solution, for all $\bm m\ge\bm n$. By Slater's condition, this implies that strong duality holds between the programs \refprog{lowerSDPnalt_multi} and \refprog{lowerDSDPnalt_multi}.
\end{proof}

\noindent In particular, strong duality holds between the programs \refprog{lowerSDPnalt1_multi} and \refprog{lowerDSDPnalt1_multi}. Note that the multimode generalisation of Corollary~\ref{th:ch04_sdupper} is a direct consequence of Theorem~\ref{th:ch04_sdlowermulti}:

\begin{corollary}[Generalisation of Corollary~\ref{th:ch04_sdupper}]\label{th:ch04_sduppermulti}
Strong duality holds between the programs \refprog{upperSDPnalt_multi} and \refprog{upperDSDPnalt_multi}.
\end{corollary} \index{Strong duality!of semidefinite programs}

\begin{proof}
the strictly feasible solution of \refprog{lowerSDPnalt_multi} derived in the proof of Theorem~\ref{th:ch04_sdlowermulti} yields a strictly feasible solution for \refprog{upperSDPnalt_multi}. 
With Slater's condition, this shows again that strong duality holds between the programs \refprog{upperSDPnalt_multi} and \refprog{upperDSDPnalt_multi}.
\end{proof}

\noindent In particular, strong duality holds between the programs \refprog{upperSDPnalt1_multi} and \refprog{upperDSDPnalt1_multi}.

We recall the following definition from Subsection~\ref{subsec04:CVproof}: for all $n\in\mathbb N$, $\bm F^n=(F_k^n)_{k\in\N}\in\R^{\N}$ where\\
$\bullet$ if $n$ is even:
\begin{equation}
        \mspace{36mu} F_k^n:=\begin{cases}\frac1{2^n}\binom k{\frac k2}\binom{n-k}{\frac{n-k}2}&\text{when }k\le n, k\text{ even},\\0&\text{otherwise},\end{cases}
\end{equation}
$\bullet$ if $n$ is odd:
\begin{equation}
        F_k^n:=\begin{cases} \frac1{2^n}\frac{\binom n{\floor{\frac n2}}\binom{\floor{\frac n2}}{\floor{\frac k2}}^2 }{\binom nk},&\text{when }k\le n,\\0&\text{otherwise}. \end{cases}
\end{equation}
Let us define, for all $\bm n=(n_1,\dots,n_M)\in\mathbb N^m$, $\bm F^{\bm n}=(F_{\bm k}^{\bm n})_{k\in\N^M}\in\R^{\N^M}$ where
\begin{equation}\label{eq:ch04_defFbm}
    \mspace{-35mu} F_{\bm k}^{\bm n}:=\begin{cases}
      \prod_{i=1}^MF_{k_i}^{n_i} &\text{when }\bm k\le\bm n, \\
      0 &\text{otherwise.}
      \end{cases}
\end{equation}
By~\eqref{eq:ch04_boundFnn}, for all $n\in\mathbb N$, $F_n^n\ge\frac1{n+1}$, so for all $\bm n=(n_1,\dots,n_M)\in\mathbb N^M$,
\begin{equation}\label{eq:ch04_boundFnnmulti}
    F_{\bm n}^{\bm n}\ge\frac1{\pi_{\bm n}}.
\end{equation}
Like in the single-mode case, the program~\refprog{lowerSDPnalt_multi} is equivalent to
\leqnomode
\begin{flalign*}\index{Semidefinite program}
    \label{prog:lowerSDPnalt_multiapp2}
    \tag*{(SDP$_{\bm n}^{\bm m,\le}$)}\hspace{3cm}\left\{
        \begin{aligned}
            & \quad \text{Find } \bm F\in\R^{\pi_{\bm m}}  \\
            & \quad \text{maximising } F_{\bm n} \\
            & \quad \text{subject to }\\
            & \hspace{1cm} \begin{aligned}
            & \sum_{\bm k\le\bm m} F_{\bm k} =  1\\
            & \forall\bm k\le\bm m, \, F_{\bm k} \geq  0\\
            & f_{\bm F} \in\mathcal R_{\bm m,+}(\R_+^M),
            \end{aligned}
        \end{aligned}
    \right. &&
\end{flalign*}\index{Laguerre!function}
\reqnomode
with $f_{\bm F} = \sum_{\bm k\le\bm m}F_{\bm k}\mathcal L_{\bm k}$ since $\bm F \in \R^{\pi_{\bm m}}$. Its dual program is given by
\leqnomode
\begin{flalign*}
   \label{prog:lowerDSDPnalt_multiapp2}
    \tag*{$(\text{D-SDP}^{\bm m,\leq}_{\bm n})$}
    \hspace{3cm}\left\{
        \begin{aligned}
            & \quad \text{Find } y,\bm\mu\in\R\times\R^{\pi_{\bm m}}  \\
            & \quad \text{minimising } y \\
            & \quad \text{subject to }\\
            & \hspace{1cm} \begin{aligned}
            & y\ge1+\mu_{\bm n}\\
            & \forall\bm k\le\bm m,\bm k\neq\bm n,\, y\ge\mu_{\bm k}\\
            & \forall g\in\mathcal R_{\bm m,+}(\R_+^M),\, \left\langle f_{\bm \mu},g\right\rangle\ge0,
            \end{aligned}
        \end{aligned}
    \right. &&
\end{flalign*}
\reqnomode
for all $\bm m\ge\bm n$, where $f_{\bm \mu} = \sum_{\bm k\le\bm m}\mu_{\bm k}\mathcal L_{\bm k}$. Moreover, adding the condition $\mu_{\bm k}\le1$ for all $\bm k\le\bm m$ does not change the optimal value of the program. We enforce this condition in what follows.
With Lemma~\ref{lemma:ch04_feasible} and Lemma~\ref{lemma:ch04_productsol}, we thus obtain the following result:

\begin{lemma}[Generalisation of Lemma~\ref{lemma:ch04_feasible}]\label{lemma:ch04_feasiblemulti}
For all $\bm m\ge\bm n$, $\bm F^{\bm m}$ (defined in Eq.~\eqref{eq:ch04_defFbm}) is a feasible solution of~\refprog{lowerSDPnalt_multiapp2}.
\end{lemma}

\noindent In particular, for all $\bm m\in\N^M$, $\sum_{\bm k\le\bm m}F_{\bm k}^{\bm m}\mathcal L_{\bm k}\in\mathcal R_{\bm m,+}(\R_+^M)$. For $\bm m,\bm n\in\N^M$ with $\bm m\ge\bm n$, let $\bm\mu\in\R^{\pi_{\bm m}}$ be a feasible solution of~\refprog{lowerDSDPnalt_multiapp2}. Then, for all $\bm l\le\bm m$
\begin{equation}
    \left\langle\sum_{\bm k\le\bm m}\mu_{\bm k}\mathcal L_{\bm k},\sum_{\bm k\le\bm l}F_{\bm k}^{\bm l}\mathcal L_{\bm k}\right\rangle\ge0,
\end{equation}
so that
\begin{equation}
    \sum_{\bm k\le\bm l}\mu_{\bm k}F_{\bm k}^{\bm l}\ge0.
\end{equation}
Hence, for all $\bm l\ge\bm m$,
\begin{equation} \label{eq:ch04_boundmumulti}
    \begin{aligned}
        \mu_{\bm l}&\ge-\frac1{F_{\bm l}^{\bm l}}\sum_{\substack{\bm k\le\bm l\\\bm k\neq\bm l}}\mu_{\bm k}F_{\bm k}^{\bm l}\\
        &\ge-\frac1{F_{\bm l}^{\bm l}}\sum_{\substack{\bm k\le\bm l\\\bm k\neq\bm l}}F_{\bm k}^{\bm l}\\
        &=1-\frac1{F_{\bm l}^{\bm l}}\\
        &\ge1-\pi_{\bm l},
    \end{aligned}
\end{equation}
where we used $F_{\bm l}^{\bm l}>0$ in the first line, $\mu_{\bm k}\le1$ and $F_{\bm k}^{\bm l}\ge0$ in the second line, $\sum_{\bm k\le\bm l}F_{\bm k}^{\bm l}=1$ in the third line, and Eq.~\eqref{eq:ch04_boundFnnmulti} in the last line. Eq.~\eqref{eq:ch04_boundmumulti} is the multimode generalisation of Eq.~\eqref{eq:ch04_mucompact}. It is essential to prove the convergence of the lower bounding hierarchy as \begin{enumerate*}[label=(\roman*)] \item it proves that the feasible set of \refprog{lowerSDPnalt1_multi} is compact and \item it shows that the sequence of optimal solutions obtained by diagonal extraction belongs to $\mathcal S(\N^M)$.\end{enumerate*}

With these additional results, the proof of convergence of the multimode hierarchy of lower bounds \refprog{lowerSDPnalt1_multi} is then obtained directly from its single-mode counterpart using multi-index notations:

\begin{theorem}[Generalisation of Theorem~\ref{th:ch04_lowerCV}]\label{th:ch04_lowerCVmulti}
The increasing sequence of optimal values $\omega^{m\bm1,\le}_{\bm n}$ of \refprog{lowerSDPnalt1_multi} converges to the optimal value $\omega_{\bm n}^\mathcal S$ of \refprog{LPS}:
\begin{equation}
    \lim_{m\rightarrow+\infty}\omega^{m\bm1,\le}_{\bm n}=\omega_{\bm n}^\mathcal S.
\end{equation}
\end{theorem}

\noindent With Lemma~\ref{lemma:ch04_equiv}, we also obtain
\begin{equation}
    \lim_{m\rightarrow+\infty}\omega^{m,\le}_{\bm n}=\omega_{\bm n}^\mathcal S.
\end{equation}
Like in the single-mode case, Theorem~\ref{th:ch04_upperCVmulti} and Theorem~\ref{th:ch04_lowerCVmulti} imply strong duality between the linear programs:

\begin{corollary}[Generalisation of Corollary~\ref{th:ch04_sdLP}]\label{th:ch04_sdLPmulti}
Strong duality holds between the programs \refprog{LP_multi} and \refprog{DLP_multi} and between programs \refprog{LPS_multi} and \refprog{DLPS_multi}.
\end{corollary}\index{Strong duality!of linear programs}

\subsection{Multimode example}
\label{subsec04:example_multi}

To illustrate the usefulness of our Wigner negativity witnesses in the multimode setting, we consider a lossy Fock state over two-modes:
\begin{equation}
    \begin{aligned}
        \bm\rho_{1,1,\eta}:=&(1-\eta)^2\ket1\!\bra1\otimes\ket1\!\bra1
        +\eta(1-\eta)\ket1\!\bra1\otimes\ket0\!\bra0\\
        &\;\;+\eta(1-\eta)\ket0\!\bra0\otimes\ket1\!\bra1
        +\eta^2\ket0\!\bra0\otimes\ket0\!\bra0,
    \end{aligned}
\end{equation}
with loss parameter $0\le\eta\le1$. Setting $\eta=0$ gives $\bm\rho_{1,1,\eta}=\ket1\!\bra1\otimes\ket1\!\bra1$ while setting $\eta=1$ gives $\bm\rho_{1,1,\eta}=\ket0\!\bra0\otimes\ket0\!\bra0$. This state has a nonnegative Wigner function for $\eta\geq\frac12$.

We also consider the multimode Wigner negativity witness $\ket1\!\bra1\otimes\ket1\!\bra1$, which is a projector onto the Fock state $\ket1\otimes\ket1$. Solving numerically the corresponding hierarchy~\refprog{lowerSDPn_multi} up to $m=3$, we obtain the lower bound $0.266$ and solving the hierarchy~\refprog{upperSDPn_multi} up to $m=10$, we obtain the upper bound $0.320$. \index{Semidefinite program}

A direct consequence of the numerical lower bound is that tensor product states are not the closest among Wigner positive states to tensor product states with a negative Wigner function. Indeed, the maximum achievable fidelity\index{Fidelity} with the state $\ket1\otimes\ket1$ using Wigner positive tensor product states is equal to the square of the maximum achievable fidelity\index{Fidelity} with the state $\ket1$ using single-mode Wigner positive states, that is $0.5^2=0.25<0.266$.

\begin{figure}[t]
	\begin{center}
		\includegraphics[width=.7\columnwidth]{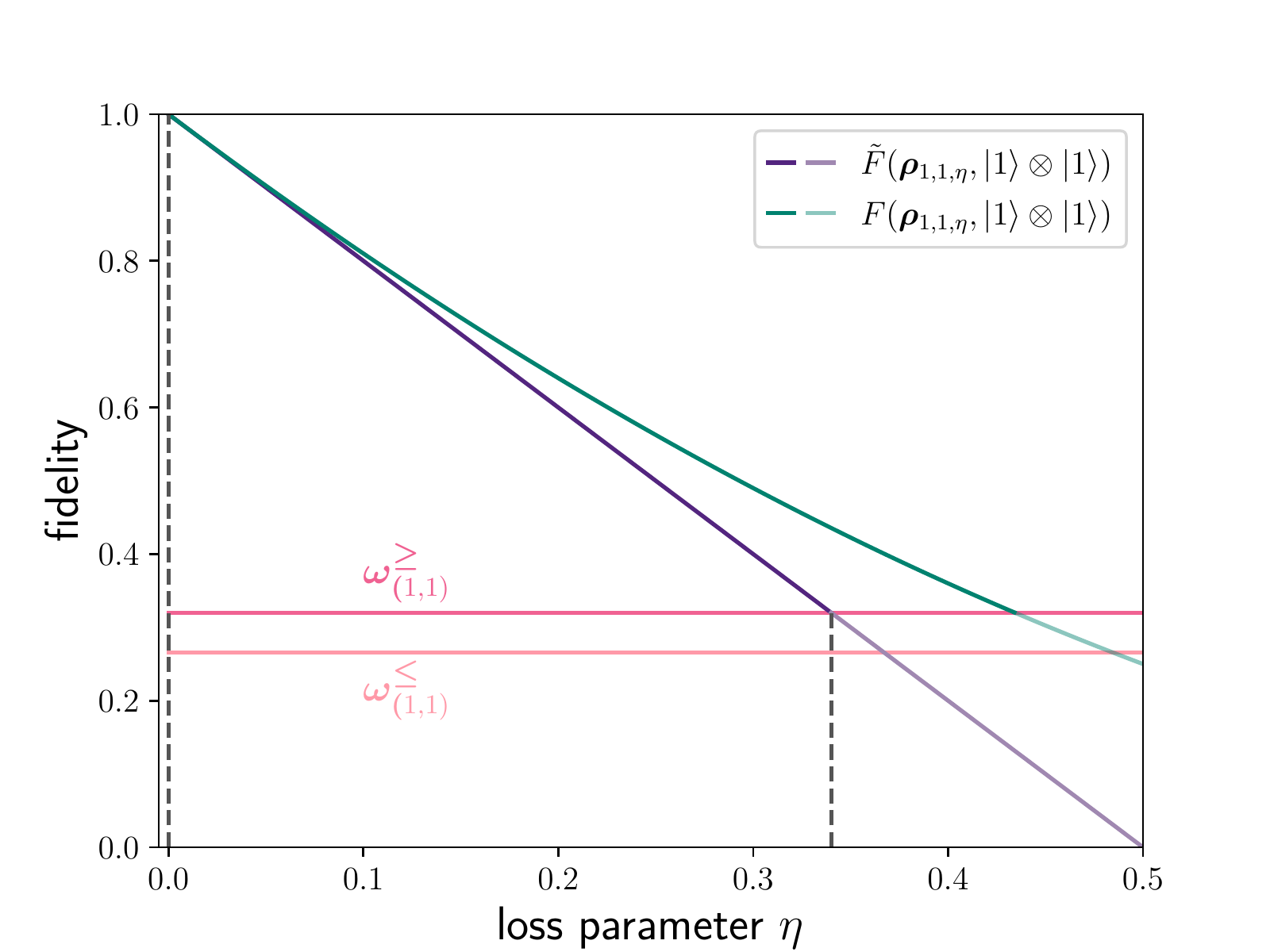}
		\caption{Witnessing Wigner negativity of the lossy Fock state $\bm\rho_{1,1,\eta}$ over two modes using the witness $\ket1\!\bra1\otimes\ket1\!\bra1$. The threshold value for that witness is upper bounded by $0.320$ and lower bounded by $0.266$.  The dashed grey line delimits the interval of loss parameter values where the witness can be used to detect Wigner negativity of $\bm\rho_{1,1,\eta}$ efficiently, i.e., when the robust bound $\tilde F(\bm\rho_{1,1,\eta},\ket1\otimes\ket1)$ (violet curve) on the fidelity\index{Fidelity} from Eq.~\eqref{eq:ch04_tildeF11} is above the witness upper bound (pink line). When it is below the witness lower bound (light pink line), we are guaranteed that the witness cannot be used to detect Wigner negativity of the state. The fidelity\index{Fidelity} $F(\bm\rho_{1,1,\eta},\ket1\otimes\ket1)$ is also depicted above (green curve). Note that $\bm\rho_{1,1,\eta}$ has a nonnegative Wigner function for $\eta\ge0.5$.}\index{Fidelity}
		\label{fig:multi}
	\end{center}
\end{figure}

We now use the upper bound to witness the Wigner negativity of the state $\bm\rho_{1,1\eta}$ (see Fig.~\ref{fig:multi}).
The fidelity\index{Fidelity} between $\bm\rho_{1,1,\eta}$ and $\ket1\otimes\ket1$ is given by $F(\bm\rho_{1,1,\eta},\ket1\otimes\ket1)=(1-\eta)^2$, for all $0\le\eta\le1$. This fidelity\index{Fidelity} is above the upper bound $0.320$ on the threshold value of the witness $\ket1\!\bra1\otimes\ket1\!\bra1$ when $\eta\le0.434$. 

However, in practice one would not obtain a precise estimate of the fidelity\index{Fidelity} efficiently, but rather a robust lower bound on the fidelity\index{Fidelity} computed from single-mode fidelities, which satisfies Eq.~\eqref{eq:ch04_robustfide}. In the worst case, the estimate obtained is closer to $1-2(1- F(\bm\rho_{1,1,\eta},\ket1\otimes\ket1))$ than to $F(\bm\rho_{1,1,\eta},\ket1\otimes\ket1)$. When the value of this robust lower bound is greater than the threshold value of the witness, this implies that the state has a negative Wigner function. 

In the present case, the two single-mode reduced states of $\bm\rho_{1,1,\eta}$ are the same, given by
\begin{equation}
    \Tr_2(\bm\rho_{1,1,\eta})=(1-\eta)\ket1\!\bra1+\eta\ket0\!\bra0,
\end{equation}
so the single-mode fidelities with $\ket1$ are equal for each mode and given by $1-\eta$. Hence, the robust lower bound on $F(\bm\rho_{1,1,\eta},\ket1\otimes\ket1)$ is given by
\begin{equation}\label{eq:ch04_tildeF11}
    \tilde F(\bm\rho_{1,1,\eta},\ket1\otimes\ket1)=1-2\eta.
\end{equation}
It is above the upper bound $0.320$ on the threshold value of the witness $\ket1\!\bra1\otimes\ket1\!\bra1$ when $\eta\le0.340$.\index{Fidelity}

This example highlights the use of efficient and robust lower bounds on multimode fidelities rather than fidelity\index{Fidelity} estimates~\cite{chabaud2020efficient}, in conjunction with our family of multimode witnesses to detect Wigner negativity of realistic experimental states.

\section{Discussion and open problems}
\label{sec04:conclusion}

Characterising quantum properties of physical systems is an important step in the development of quantum technologies, and negativity of the Wigner function, a necessary resource for any quantum computational speedup, is no exception.
In this chapter, we have derived a complete family of Wigner negativity witnesses which provide an operational quantification of Wigner negativity, both in the single-mode and multimode settings. 
In the context of quantum optical information processing, the main application of our method is in experimental scenarios, where it leads to robust and efficient certification of negativity of the Wigner function. \index{Wigner function} \index{Wigner negativity}

What is more, our witnesses also delineate the set of quantum states with positive Wigner function, and it would be interesting to understand whether additional insights on this set can be obtained using these witnesses. This is of particular importance for information theoretic notions like the recently defined quantum Wigner entropy \cite{vanherstraetenquantum2021} where it is argued that this is the natural measure in order to characterise quantum uncertainty in phase-space even though it is limited to Wigner positive states. 

Hierarchies of semidefinite programs (in particular with non-commutative variables~\cite{navascues2008convergent,klep2020optimization,klep2021sparse})\index{Semidefinite program} have found many recent applications in quantum information theory.
From an infinite-dimensional linear program, we were able to use numerically both a hierarchy of upper bounds and a hierarchy of lower bounds---thus obtaining a certificate for the optimality of these bounds by looking at their difference---whereas this only works in specific cases for the Lasserre hierarchy of upper bounds~\cite{lasserre2011new}.\index{Lasserre hierarchy} \index{Linear program} Can we find other interesting cases where we can exploit both hierarchies? Moreover, we obtained an analytical sequence of lower bounds for the threshold value of the program \refprog{LP}. Can we also get an analytical sequence of upper bounds? In particular, we anticipate that Fock states $\ket n$ get further away from the set of states having a positive Wigner function as $n$ increases and that $\omega_n = \mathcal{O}(\frac 1{\sqrt{n}})$ as $n\rightarrow+\infty$.
The question of whether the gap between the optimisation over square integrable functions and Schwartz functions is also left open.

\dobib

\clearemptydoublepage


\chapter{Quantum advantage in Information Retrieval}
\label{chap:informationretrieval}

\begin{figure}[htbp]
        \centering
        \includegraphics[scale=1.8]{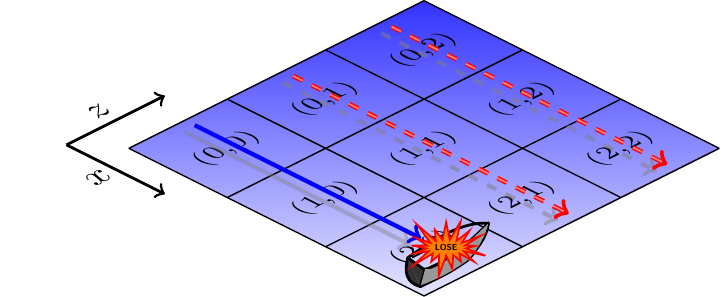}
        \caption{The Torpedo Game is a pacificist alternative to \textit{Battleship} where the aim is to avoid sinking Alice's ship, depicted here in dimension $3$.}
        \label{fig:TorpedoCartoon}
\end{figure}

\lettrine{W}{hile} the previous chapters focused on continuous-variable systems, we investigate here information retrieval tasks in discrete-variable systems that require Wigner negativity and contextuality to outperform the corresponding classical task.
A well-known information retrieval task is random access coding which involves the encoding of a random input string into a
shorter message string. \index{Contextuality!DV} \index{Wigner negativity}
The encoding should be such that any element of the
input string can be retrieved with high probability from the message string.
Such tasks have long been studied as examples in which the communication of
quantum information can provide advantage, \ie enhanced performance,
over classical information, e.g.\
\cite{ambainis1999dense,pawlowski2009information,spekkens2009preparation,grudka2014popescu,tavakoli2015quantum,chailloux2016optimal,aguilar2018connections,farkas2019self,Galvao2001CommunicationComplexity}.
However, random access coding concerns only one kind of information retrieval. 
In this chapter we define a general notion of information retrieval tasks and introduce another example called the Torpedo Game.
It is similar to random access coding, but with additional requirements involving the retrieval of relative information about elements of the input string.
Taking a geometric perspective it may also be viewed as a pacifist
version of the popular strategy game \textit{Battleship} (see \refig{TorpedoCartoon}).
Quantum strategies can be implemented in prepare-and-measure scenarios, and
outperform classical strategies for the Torpedo Games with bit and trit inputs.
In particular, quantum perfect strategies exist in the trit case and provide a
greater quantum advantage than for the comparable random access coding task
\cite{tavakoli2015quantum}.

Critically, optimal quantum strategies emerge from an analysis in terms of the discrete
Wigner function and exploits maximum Wigner negativity which has been shown to be necessary for computational speedup \cite{galvao2005discrete,Galvao2006ClassicalityDiscreteWigner,Mari2012}.
Yet while negativity is necessary for advantage in the Torpedo game, it does not seem to be sufficient.
To more precisely pinpoint the source of quantum advantage we must
look further.
One natural candidate would be preparation contextuality \cite{Spekkens2005}, another signature of non-classicality that has been linked to quantum random access codes in numerous studies \cite{spekkens2009preparation,chailloux2016optimal,ambainis2019parity}.
It has been shown to be necessary for advantage in a restricted class of random access codes subject to an obliviousness constraint \cite{hameedi2017communication,saha2019state}. \index{Contextuality!sequential}
Here however, we focus on a different characteristic called sequential contextuality \cite{mansfield2018quantum}.
It indicates the absence of a hidden-variable model\index{Hidden-variable model!DV} respecting the sequential structure of a given protocol.
It has already been used to explain the promotion of linear computation with access to a qubit quantum resource to the complexity class $P$ in the setting of $l2$-MBQC \cite{mansfield2018quantum,AndersBrowne2009,Galvao2017ContextualCorr}.
Subject to an assumption of bounded-memory,
we find that this characteristic is necessary and sufficient for quantum advantage, not just in random access coding but in any information retrieval task.
Moreover, we show that it quantifies the degree of advantage that can be achieved.

As recalled in Section~\ref{sec01:sheaf}, contextuality can exhibit itself at the level of probability distributions (e.g.\ quantum violations of the CHSH inequality \cite{CHSH1969}) but also at the level of the supports of these distributions in some cases. \index{CHSH model}
In other words contextuality can be inferred by a series of logical deductions about which events are possible or not, e.g.\ Hardy's paradox \cite{Hardy1992,Hardy1993}.
This situation has come to be known as logical contextuality \cite{abramsky2011sheaf}.
In the most extreme cases, known as strong contextuality \cite{abramsky2011sheaf}, every possible event triggers such a paradox \cite{mansfield2017consequences}, e.g.\ Popescu-Rohrlich box violations\index{PR box} of the CHSH inequality \cite{popescu1994quantum}.
For qutrits the quantum perfect strategies we introduce for the Torpedo game display analogous contextuality, and hence paradoxes, of this strongest form in a prepare-and-measure scenario.

Section~\ref{sec05:irtasks} gives an overview of information retrieval tasks including random access coding and the Torpedo Game. Section~\ref{sec05:DWF} provides background on discrete Wigner functions.\index{Wigner function} Section~\ref{sec05:optimalstrat} deals with optimal classical and quantum strategies for the Torpedo Game. Finally, Section~\ref{sec05:contextuality} establishes the relationship between sequential contextuality and quantum advantage in bounded-memory information retrieval tasks. This chapter is based on \cite{emeriau2020quantum}.

\section{Information retrieval tasks}
\label{sec05:irtasks}

\subsection{Random access codes}
\label{subsec05:RAC}

For $n,m \in \N^*$ with $m<n$, an $(n,m)_2$ {random access code (RAC)}---sometimes denoted $n \rightarrow m$---is a communication task in which one aims to encode information about a random $n$-bit input string into an $m$-bit message, in such a way that any one of the input bits may be retrieved from the message with high probability.
A $(n,m)_2$ {quantum random access code (QRAC)} instead encodes the input into an $m$-qubit (quantum) message state.

Such tasks may be considered as two-party cooperative games in which the first party, Alice, receives a random input string from a referee.
Alice encodes information about this in a message that is communicated to the second party, Bob.
The referee then asks Bob to retrieve the value of the bit at a randomly chosen position in the input string.
We will assume that the referee's choices are made uniformly at random.

For instance, for the $(2,1)_2$ RAC game \cite{ambainis1999dense} an optimal classical strategy is for Alice to directly communicate one of the input bits to Bob.
If asked for this bit, Bob can always return the correct answer, otherwise Bob guesses and will provide the correct answer with probability $\frac{1}{2}$.
Thus the game has a classical value of $$\theta^C_{2\rightarrow 1} = \frac{1}{2} \left( 1 + \frac{1}{2} \right) = \frac{3}{4}\, .$$

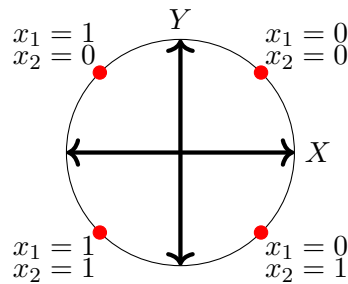
\begin{figure}[ht!]
    \centering
    \begin{tikzpicture}[scale=.5]

\draw (0,0) circle (3cm);
\draw [<->,ultra thick] (-3,0) -- (3,0);
\draw [<->,ultra thick] (0,-3) -- (0,3);

\node [right] (X) at (3,0) {$X$};
\node [above] (Y) at (0,3) {$Y$};

\node (i00) at (2.121,2.121) {};
\node (texte000) at ($(i00)+(1.2,1)$) {$x_1=0$};
\node (texte001) at ($(i00)+(1.2,.4)$) {$x_2=0$};
\draw [fill=red,red] (i00) circle (5pt);

\node (i01) at (2.121,-2.121) {};
\node (texte010) at ($(i01)+(1.2,-.4)$) {$x_1=0$};
\node (texte011) at ($(i01)+(1.2,-1)$) {$x_2=1$};
\draw [fill=red,red] (i01) circle (5pt);

\node (i10) at (-2.121,2.121) {};
\node (texte100) at ($(i10)+(-1.2,1)$) {$x_1=1$};
\node (texte101) at ($(i10)+(-1.2,.4)$) {$x_2=0$};
\draw [fill=red,red] (i10) circle (5pt);

\node (i11) at (-2.121,-2.121) {};
\node (texte110) at ($(i11)+(-1.2,-.4)$) {$x_1=1$};
\node (texte111) at ($(i11)+(-1.2,-1)$) {$x_2=1$};
\draw [fill=red,red] (i11) circle (5pt);

\end{tikzpicture}
    \caption{The four red dots correspond to the four states $\ket{\psi_{x_1,x_2}}$ defined in Eq.~\eqref{eq:ch05_QRAC21_optimalstate} depicted as points on the equator of the Bloch sphere.}
    \label{fig:ch05_QRAC21}
\end{figure}

Quantum strategies can outperform this classical bound.
An optimal quantum strategy is for Alice to communicate the qubit state
\begin{equation}
    \ket{\psi_{x_1,x_2}} = \frac{1}{\sqrt{2}} \left( \ket{0} +  \frac{1}{\sqrt{2}} \left(  (-1)^{x_1} + (-1)^{x_2} i \right) \ket{1} \right)
    \label{eq:ch05_QRAC21_optimalstate}
\end{equation}
where $(x,z)$ is the input bit-string they have received.
Bob measures in the $X$ basis when asked for $x_1$ and in the $Y$ basis when asked for $x_2$ (see \refig{ch05_QRAC21}).
If they obtain the $+1$ eigenvalue they return the value $1$ and if they obtain the $-1$ eigenvalue they return $0$.
This yields a quantum value for the game of $$\theta^Q_{2\rightarrow 1} = \cos^2\left(\frac{\pi}{8}\right) \approx 0.85 \, .$$

\subsection{General Information Retrieval Tasks}

One may also consider more general communication scenarios.
In an $(n,m)_d$ \textit{communication scenario} the input is a random string of $n$ dits and the message is a string of $m$ (qu)dits, for $d \geq 2$.
(Q)RAC tasks have previously been considered in such scenarios, e.g.\ in \cite{tavakoli2015quantum,casaccino2008extrema}.

However, we also wish to accommodate for a much wider range of retrieval tasks regarding the input $d$its.
An \textit{information retrieval task} in an $(n,m)_d$ communication scenario is specified by a tuple $\langle Q, \{w_q\}_{q \in Q} \rangle$ where
\begin{itemize}
\item $Q$ is a finite set of \textit{questions};
\item The $w_q : \mathbb{Z}_d^n \rightarrow \mathbb{Z}_d$ are \textit{winning relations} which pick out the good answers to question $q$ given an input string in $\mathbb{Z}_d^n$. Note that there may be more than one good answer, or none. It is assumed that inputs and outputs are endowed with the structure of the commutative ring $\mathbb{Z}_d$.
\end{itemize}

Standard $(n,m)_d$ (Q)RACs are recovered when the questions ask precisely for the respective input dits.
In that case the winning relations $w_{i}$ reduces simply to projectors onto the respective dits of the input string.
However, other interesting tasks arise when the questions also concern relative information about the input string, in the form of parities or linear combinations modulo $d$ of the input dits.
A similar generalisation for $d=2$, using functions rather than relations, has been independently proposed in \cite{doriguello2020quantum}. Below we introduce an example called the Torpedo Game which is distinct from random access coding. We show that Wigner negativity is necessary to outperform the best classical strategy. We later link this advantage to sequential contextuality \cite{mansfield2018quantum}.

\subsection{The Torpedo Game}
\label{subsec05:torpedo_game}

\noindent
Of particular interest here is an information retrieval task for $(2,1)_d$ communication scenarios where we fix the dimension to be $d=2$ or $d=3$.
We take the game perspective and refer to the task as the dimension $d$ {Torpedo Game} (see \refig{TorpedoCartoon} in dimension 3).
Let $x$ and $z$ be the two input bits or trits.
There are $3$ questions in $Q = \{ \infty, 0, 1\}$ dimension $2$ and $4$ questions $Q = \{ \infty, 0, 1,2\}$ in dimension $3$. The labelling comes from a geometric interpretation to be elaborated upon shortly.
Winning relations for the Torpedo Game are given by
\begin{align}
\begin{split}
w_\infty (x,z) &= \{ a \in \mathbb{Z}_d \mid a \neq x \} \\
w_0 (x,z) &= \{ a \in \mathbb{Z}_d \mid a \neq -z \} \\
w_1 (x,z) &= \{ a \in \mathbb{Z}_d \mid a \neq x-z \} \\
w_{2} (x,z) &= \{ a \in \mathbb{Z}_d \mid a \neq 2x-z \} \, .
\end{split}
\label{eq:ch05_winning_conditions}
\end{align}
All arithmetic is modulo $d=2$ or 3 depending on the considered dimension. The last winning relation only features in the dimension 3 Torpedo game.

For $d=2$, the Torpedo Game is equivalent to a $(2,1)_2$ (Q)RAC, but with an additional question.
Bob may be asked to retrieve either one of the individual input dits, or to retrieve relative information about them in the form of their parity $x \oplus z$. Note that in dimension 2, the winning relations are actually functions (there is only one good answer per question).

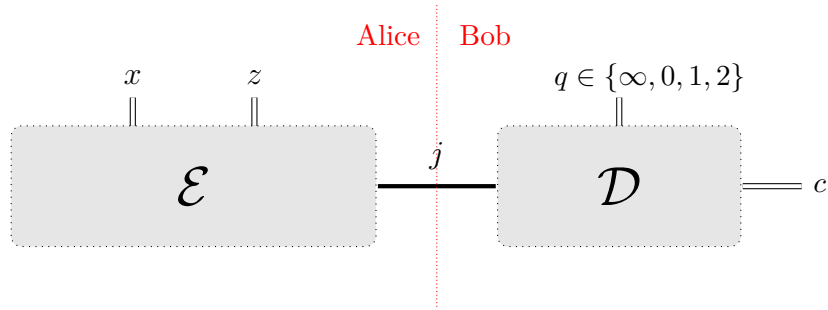
\begin{figure}[htbp]
    \centering
    \scalebox{\myscale}{\def\tkzscl{0.8}

\begin{tikzpicture}[scale=\tkzscl]

\node (up) at (0,1) {};
\node (down) at (0,-1) {};
\node (left) at (-1,0) {};
\node (right) at (1,0) {};

\node (a1) at (0,0) {};
\node (a2) at (6,2) {};
\node [inner sep=.0pt] (a3) at (6,1) {};

\node (b1) at (8,0) {};
\node (b2) at (12,2) {};
\node [inner sep=.0pt] (b3) at (8,1) {};

\draw[rounded corners,dotted,fill=gray!20] (a1) rectangle (a2);
\node (am) at (2,1) {};
\node (E) at ($(am)+(right)$) {\huge $\mathcal{E}$};
\node [inner sep=0 pt] (d1) at ($(am)+1.5*(up)$) {};
\node [inner sep=0 pt] (d2) at ($(am)+1.5*(up)+2*(right)$) {};
\draw [double distance=1.5pt] (d1) to ($(d1)+.5*(down)$);
\draw [double distance=1.5pt] (d2) to ($(d2)+.5*(down)$);
\node (d1texte) at ($(d1)+.3*(up)$) {\large $x$};
\node (d2texte) at ($(d2)+.3*(up)$) {\large $z$};

\draw[rounded corners,dotted,fill=gray!20,dotted] (b1) rectangle (b2);
\node (bm) at (10,1) {\huge $\Dc$};
\node [inner sep=0 pt] (j) at ($(bm)+1.5*(up)$) {};
\draw [double distance=1.5pt] (j) to ($(j)+.5*(down)$);
\node (jtexte) at ($(j)+.3*(up)+.5*(right)$) {$q \in \{\infty,0,1,2 \}$};
\node [inner sep=0 pt] (o) at ($(bm)+2*(right)$) {};
\draw [double distance=1.5pt] (o) to ($(o)+(right)$);
\node (otexte) at ($(o)+1.3*(right)$) {$c$};

\draw[ultra thick] (a3) to (b3); 
\draw[densely dotted, red] ($.5*(a2)+.5*(b1)+3*(up)$) to ($.5*(a2)+.5*(b1)+2*(down)$);
\node (Alice) at ($.5*(a3)+.5*(b3)+2.5*(up)+.8*(left)$) {\textcolor{red}{Alice}};
\node (Bob) at ($.5*(a3)+.5*(b3)+2.5*(up)+.8*(right)$) {\textcolor{red}{Bob}};
\node (psi) at ($.5*(a2)+.5*(b1)+.5*(up)$) {\large $j$};

\end{tikzpicture}}
    \caption{Prepare-and-measure protocol for the dimension 3 Torpedo Game: Alice receives trits $x$ and $z$ and sends a single message (qu)trit $j$ via the encoding $\Ec$. Bob is asked a question ${q \in \{ \infty,0,1,2 \}}$, performs decoding $\Dc$, and outputs $c$ which should satisfy the winning conditions given by $w_q (x,z)$ with high probability.} 
    \label{fig:ch05_QRAC_pres}
\end{figure}

The Torpedo Game may be framed as a cooperative, pacifist alternative to the popular game \textit{Battleship},\index{Phase-space!DV}
in which Alice and Bob, finding themselves on opposing sides in a context of naval warfare, wish to subvert the conflict and cooperate to avoid casualities while not directly disobeying orders.
The input dits received by Alice designates the coordinates in which Alice is ordered by their commander to position their one-cell ship on the affine plane of order $d$ for $d=2$ or $d=3$.
We may think of the affine plane as a toric $d \times d$ grid, with $x$ designating the row and $z$ the column.
For instance in \refig{ch05_d3_torpedo_firing} we identify the top edge with the bottom edge and the left edge with the right edge.
Bob is a naval officer on the opposing side who is ordered by their commander to shoot a torpedo along a line of the grid with slope specified by $q \in Q$.
The $\infty$ question requires Bob to shoot along some row, and the $0$ question requires Bob to shoot along some column, etc.
However, Bob retains the freedom to choose which row, or column, or diagonal of given slope, as the case may be.
In other terms, upon receiving $q$ Bob must shoot along a lines $q x - z = c$ (if $q \neq \infty$) or $x = c$ (if $q = \infty$) but is free to choose the constant $c$.

Alice and Bob wish to coordinate a strategy for avoiding casualities, while still obeying their explicit orders.
Alice may communicate a single (qu)dit to Bob---greater communication may risk revealing their position should it be intercepted---and based on this Bob must choose their $c$ in such a way that they avoid Alice's ship.

\section{The discrete Wigner function and information retrieval}
\label{sec05:DWF}
  
\begin{figure}[htbp]
    \centering
    \scalebox{\myscale}{\begin{tikzpicture}[scale=.19]

\draw [thick, fill=gray!20,step=3] (0,0) grid (9,9) rectangle (0,0);
\draw [->,ultra thick, red] (1.5,7.5) -- (8,7.5);

\draw [fill=gray!20] (3,-4) rectangle (6,-1);
\draw [fill=blue!70] (3,-3) rectangle (6,-2);
\draw [thick, fill=gray!20] (3,-4) grid (6,-1);

\draw [fill=gray!20] (-1,-4) rectangle (2,-1);
\draw [fill=blue!70] (-1,-4) rectangle (2,-3);
\draw [thick, fill=gray!20] (-1,-4) grid (2,-1);

\draw [fill=gray!20] (7,-4) rectangle (10,-1);
\draw [fill=blue!70] (7,-2) rectangle (10,-1);
\draw [thick, fill=gray!20] (7,-4) grid (10,-1);

\draw [thick, fill=gray!20,step=3,xshift=1cm] (12,0) grid (21,9) rectangle (12,0);
\draw [ultra thick, red, <-] (14.5,1.5) to (14.5,8);

\draw [fill=gray!20] (12,-4) rectangle (15,-1);
\draw [fill=blue!70] (12,-4) rectangle (13,-1);
\draw [thick, fill=gray!20] (12,-4) grid (15,-1);

\draw [fill=gray!20] (16,-4) rectangle (19,-1);
\draw [fill=blue!70] (17,-4) rectangle (18,-1);
\draw [thick, fill=gray!20] (16,-4) grid (19,-1);

\draw [fill=gray!20] (20,-4) rectangle (23,-1);
\draw [fill=blue!70] (22,-4) rectangle (23,-1);
\draw [thick, fill=gray!20] (20,-4) grid (23,-1);

\draw [thick, fill=gray!20,step=3] (27,0) grid (36,9) rectangle (27,0);
\draw [ultra thick, red, <-] (34.5,1.5) to (28.5,7.5);

\draw [fill=gray!20] (26,-4) rectangle (29,-1);
\draw [fill=blue!70] (26,-4) rectangle (27,-3);
\draw [fill=blue!70] (27,-2) rectangle (28,-1);
\draw [fill=blue!70] (28,-3) rectangle (29,-2);
\draw [thick, fill=gray!20] (26,-4) grid (29,-1);

\draw [fill=gray!20] (30,-4) rectangle (33,-1);
\draw [fill=blue!70] (30,-2) rectangle (31,-1);
\draw [fill=blue!70] (31,-3) rectangle (32,-2);
\draw [fill=blue!70] (32,-4) rectangle (33,-3);
\draw [thick, fill=gray!20] (30,-4) grid (33,-1);

\draw [fill=gray!20] (34,-4) rectangle (37,-1);
\draw [fill=blue!70] (34,-3) rectangle (35,-2);
\draw [fill=blue!70] (35,-4) rectangle (36,-3);
\draw [fill=blue!70] (36,-2) rectangle (37,-1);
\draw [thick, fill=gray!20] (34,-4) grid (37,-1);

\draw [thick, fill=gray!20,step=3,xshift=2cm] (39,0) grid (48,9) rectangle (39,0);
\draw [ultra thick, red, <-] (42.5,1.5) to (49,7.5);

\draw [fill=gray!20] (40,-4) rectangle (43,-1);
\draw [fill=blue!70] (40,-4) rectangle (41,-3);
\draw [fill=blue!70] (41,-3) rectangle (42,-2);
\draw [fill=blue!70] (42,-2) rectangle (43,-1);
\draw [thick, fill=gray!20] (40,-4) grid (43,-1);

\draw [fill=gray!20] (44,-4) rectangle (47,-1);
\draw [fill=blue!70] (44,-3) rectangle (45,-2);
\draw [fill=blue!70] (45,-2) rectangle (46,-1);
\draw [fill=blue!70] (46,-4) rectangle (47,-3);
\draw [thick, fill=gray!20] (44,-4) grid (47,-1);

\draw [fill=gray!20] (48,-4) rectangle (51,-1);
\draw [fill=blue!70] (48,-2) rectangle (49,-1);
\draw [fill=blue!70] (49,-4) rectangle (50,-3);
\draw [fill=blue!70] (50,-3) rectangle (51,-2);
\draw [thick, fill=gray!20] (48,-4) grid (51,-1);

\end{tikzpicture}}
    \caption{The red arrows depict the directions or slopes ($\infty, 0, 1, 2$, respectively) along which Bob may be asked to shoot in the $d=3$ Torpedo Game. For each direction, Bob has three possibilities, depicted by the blue lines. In the affine plane of order 3, each of these groups of three blue cells forms a line.} \index{Phase-space!DV}
    \label{fig:ch05_d3_torpedo_firing}
\end{figure}
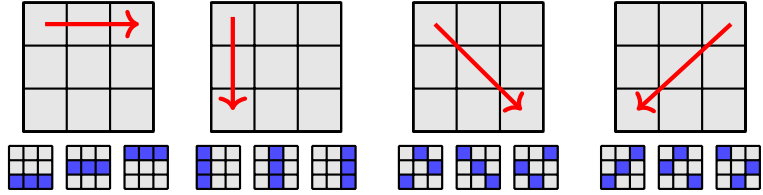

Below we work in dimension 3 because the construction we present below only provide a deterministic strategy for power-of-prime dimension Torpedo games. We will see that there exists a quantum strategy inspired by this construction that outperforms the best classical strategy for the dimension 2 Torpedo game though it is not deterministic. Wootters's geometric construction \cite{gibbons2004discrete,wootters1987wigner} of discrete Wigner functions (DWF) based on finite fields is useful for visualising our Torpedo Game as exemplified in Figure~\ref{fig:ch05_d3_torpedo_firing}, where each distinct orthonormal basis corresponds to a set of $3$ parallel (non-intersecting) lines. \index{Wigner function}

We recall from Subsection~\ref{subsec01:phasespace_qudits} (see Eq.~\eqref{eq:ch01_parityDV}) the expression of the phase point operator at the origin of phase space $\hat \Pi = \hat A_{0,0}$ (also called the parity operator\index{Parity operator}):\index{Parity operator!DV}
\begin{align*}
\hat A_{0,0} \defeq \sum_{k\in\mathbb{Z}_3} \ketbra{-k}{k},
\end{align*}
The other phase-point operators are found by conjugation with displacement operators (see Eq.~\eqref{eq:ch01_displacementDV}). The phase-point operator at point $(x,z) \in \Z_3^2$ reads:\index{Displacement operator!DV}
\begin{align}
\hat A_{x,z} \defeq \hat D_{x,z}\hat A_{0,0}\hat D_{x,z}^\dag\,. \label{eq:ch05_PPOdefn}
\end{align}
We recall the definition of the Wigner function (Eq.~\eqref{eq:ch01_Wignerdef}). An example of representation of the DWF in dimension 3 is given in Figure~\ref{fig:ch05_WigFigCartoon}. For a density matrix $\rho$ and for a phase-space point $(x,z) \in \Z_3^2$, it is given by:
\begin{equation}
    W_{\rho}(x,z) = \frac 1d \Tr(\hat A_{x,z} \rho) \Mdot
    \label{eq:ch05_Wignerdef_recall}
\end{equation}

Crucially phase operators may also be constructed from projectors onto mutually unbiased bases (MUBs) \cite{wootters1987wigner}. For $x,z \in \Z_3$:
\begin{equation}
    \hat A_{x,z} = \Pi_\infty^x + \Pi_0^{-z} + \Pi_1^{x-z} + \Pi_{2}^{2x-z} - \Id
    \label{eq:ch05_phasepoint_projMUBs}
\end{equation}
where $\Pi_q^{i}$ is the projector corresponding to dit value $i$ in Bob's $q^\text{th}$ measurement setting for $q \in \{\infty,0,1,2\}$. \index{PVM}

The eigenvectors of phase point operators are the objects of interest. The maximizing eigenvectors of the phase point operators in Eq.~\eqref{eq:ch05_PPOdefn} (and additional ones from different choices of DWF) were used in Casaccino \textit{et al.} \cite{casaccino2008extrema} as the encoded messages of a $(d+1,1)_d$ QRAC (for any odd power-of-prime dimension $d$). This is natural given the use of MUBs in constructing DWFs, and prominence of MUBs in the QRAC literature \cite{aguilar2018connections}. 
If Alice receives input $\pmb{k}=(k_1,k_2,\ldots,k_{d+1})\in\mathbb{Z}_d^{d+1}$ that they encodes in  $\rho_{\pmb{k}}$ and transmits to Bob, then the average probability of success for the Casaccino \textit{et al.} QRAC is\index{Wigner negativity} 
\begin{align}
		\frac{1}{(d+1)d^{d+1}}\sum_{\pmb{k}\in\mathbb{Z}_d^{d+1}}\Tr\left[\rho_{\pmb{k}}(\Pi_1^{k_1}+\ldots+ \Pi_{d+1}^{k_{d+1}})\right] \Mdot \label{eq:ch05_GalvaoQRAC}
\end{align}
where $\Pi_q^{i}$ is the projector corresponding to dit value $i$ in Bob's $q$-th measurement setting.
Since phase point operators can also be constructed using sums of projectors onto MUBs (see Eq.~\eqref{eq:ch05_phasepoint_projMUBs}),
the use of a maximizing eigenvector of a phase point operator for $\rho_{\pmb{k}}$ is natural to maximize Eq.~\eqref{eq:ch05_GalvaoQRAC}.

Here we instead make use of the \textit{minimizing} eigenvectors of phase point operators. The rationale for this is two-fold (i) these eigenvectors display remarkable geometric properties with respect to the measurements in (their constituent) mutually unbiased bases, and (ii) negativity is the hallmark of non-classicality which has already been identified with contextuality (with the previously mentioned caveat that an additional ``spectator'' subsystem was required). These will be seen to lead to a perfect quantum strategy for the Torpedo Game, in which the goal is to avoid certain answers.

Any state in the $-1$ eigenspace of phase point operators defined in Eq.~\eqref{eq:ch05_PPOdefn} has an outcome that is forbidden \cite{van2011noise,bengtsson2012kochen} 
in each of a complete set of MUBs. For example, let 
\begin{equation}
    \ket{\psi_{0,0}}=\frac{\ket{1}-\ket{2}}{\sqrt{2}}
\end{equation} satisfying 
\begin{equation}
    \hat A_{0,0}\ket{\psi_{0,0}}=-\ket{\psi_{0,0}}.
\end{equation}
This state, which is indeed an eigenvector of $A_{0,0}$ with eigenvalue -1, obeys 
\begin{equation}
    \Tr(\Pi_q^0 \ketbra{\psi_{0,0}}{\psi_{0,0}})=0
\end{equation}
where $\Pi^{0}_q$ is the projector on the $0^\text{th}$ eigenvector in the $q^\text{th}$ basis for $q \in \{\infty,0,1,2\}$. More specifically, $\Pi^{0}_q$ is the projector corresponding to the $\omega^{0}=+1$ eigenvector of the $q^\text{th}$ displacement operator from $\left\{\hat D_{0,1},\hat D_{1,0},\hat D_{1,1},\hat D_{1,2}\right\}$.\index{Displacement operator!DV}
These displacement operators have eigenvectors leading to mutually unbiased measurement bases $q\in\{\infty,0,1,2\}$ respectively. For $x,z \in \Z_3$, the related states $\ket{\psi_{x,z}}=\hat D_{x,z}\ket{\psi_{0,0}}$, which are eigenstates $\hat A_{x,z}\ket{\psi_{x,z}}=-\ket{\psi_{x,z}}$,
obey
\begin{equation}
	\Tr\left[(\Pi_{\infty}^{x}+\Pi_0^{-z}+\Pi_1^{x-z}+\Pi_{2}^{2x-z})\ketbra{\psi_{x,z}}{\psi_{x,z}}\right]=0\,. \label{eq:ch05_KeyFact}
\end{equation}
Eq.~\eqref{eq:ch05_KeyFact} implies that the probability of the relevant outcome (outcome $x$ in the first basis, $-z$ in the second basis, etc.) in each of the MUBs is zero (see the winning relations of the Torpedo game in Eq.~\ref{eq:ch05_winning_conditions}). The general expression in Eq.~\eqref{eq:ch05_KeyFact} for any odd power-of-prime $d$ is proven in \cite{howard2015classical,appleby2008spectra}.

\section{Optimal strategies for the Torpedo Game}
\label{sec05:optimalstrat}

Here we gather the optimal classical, quantum and (in one case) post-quantum strategies for the Torpedo Game. The dimension 3 Torpedo game is notable for the existence of a perfect quantum strategy due to the fact that there exist $d+1=4$ MUBs for this dimensions (it is true in general for odd power-of-prime dimensions though there also exists a perfect classical strategy for these dimensions). The dimension 2 Torpedo Game also has a quantum advantage though the quantum strategy is not perfect. The classical optimum is also established rigorously for dimension 2 and 3. We obtain a quantum advantage in the two cases. 

\subsection{Optimal quantum and post-quantum strategies}
\label{subsec05:qstrat}

\begin{figure}[ht!]
\centering
\includegraphics[width=0.5\textwidth]{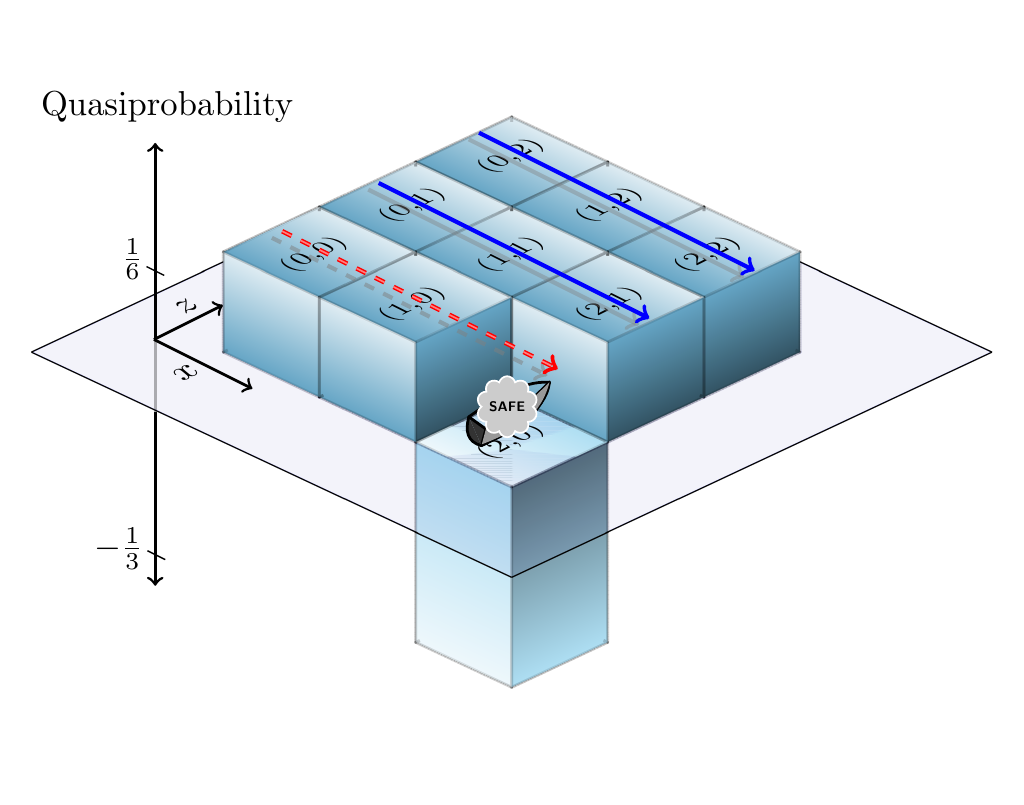}
        \caption{The perfect quantum strategy can be understood by plotting the discrete Wigner function (see Eq.~\eqref{eq:ch01_Wignerdef}) of the message state sent by Alice. In the above qutrit case, Alice is in coordinate $(x,z)=(2,0)$, so Alice sends Bob the state $\ket{\psi_{2,0}}$ whose Wigner function is $-1/3$ at coordinate $(2,0)$ and $1/6$ otherwise. Bob measures this state along any of the four allowed directions, wherein the probability of each outcome is given by sum of quasiprobabilities along the corresponding line. Hence, the only outcomes with non-zero probability of occurring correspond to lines not passing through $(2,0)$. Whichever outcome Bob sees, they may fire their torpedo along the corresponding line, safe in the knowledge that it will not intersect Alice's ship. In this figure the solid blue lines correspond to the possible outcomes for the $q=0$ direction, but the same argument holds for all the other directions.
        }\index{Wigner negativity}
        \label{fig:ch05_WigFigCartoon}
\end{figure}

\paragraph{Quantum perfect strategy for the dimension 3 Torpedo Game.} 
From Eq.~\eqref{eq:ch05_KeyFact} it follows that there is a perfect quantum strategy for the dimension 3 Torpedo game:
\begin{enumerate}
    \item Upon receiving trits $x$ and $z$ Alice sends the following state to Bob:
        \begin{equation}
            \ket{\psi_{x,z}} = \hat D_{x,z} \ket{\psi_{0,0}} = \hat D_{x,z} \left( \frac {\ket{1} - \ket{2}}{\sqrt{2}}\right). 
            \label{eq:ch05_PerfectState}
        \end{equation}\index{Displacement operator!DV}
    \item Bob receives $\ket{\psi_{x,z}}$ and is asked a question $q \in \{\infty,0, 1,2 \}$.
    They measure the state in the MUB corresponding to $q$ and outputs the dit corresponding to the measurement outcome.
\end{enumerate}
This quantum strategy wins the Torpedo Game deterministically, \ie with probability $1$. In Figure \ref{fig:ch05_WigFigCartoon} we also provide geometric intuition for why this strategy is perfect in dimension 3.

\paragraph{Optimal quantum strategy for the dimension 2 Torpedo Game.}
An analogous strategy to the qutrit case can be employed for the qubit Torpedo Game, using message states $\ket{\psi_{x,z}} = \Xp^x\Zp^z \ket{\psi_{0,0}}$ where
$\ketbra{\psi_{0,0}}{\psi_{0,0}}=\frac{1}{2}\left(\mathbb{I}-(\Xp+\Yp+\Zp)/\sqrt{3}\right)$. For $d=2$, while this does not constitute a perfect strategy it still achieves an advantage over classical strategies.
In fact, it turns out to be an optimal strategy: this strategy achieves a winning probability of approximately 0.79 and we show that this is optimal. As explained in Subsection~\ref{subsec01:quadratic}, this problem is very hard in general (NP-hard) since it is a nonconvex bilinear quadratic problem.
To show it achieves 0.79, first we can leverage the fact that the $(3,1)_2$ (Q)RAC attributed to Isaac Chuang is at least as hard to win as the Torpedo Game. This $(3,1)_2$ (Q)RAC has 3 independent inputs while the dimension 2 Torpedo game can be viewed as a modified $(3,1)_2$ (Q)RAC where the last input is the parity of the two first input bits. 
The $(3,1)_2$ (Q)RAC is then a harder task and it has been shown to have an optimal quantum value of $\frac{1}{2}\left(1+\frac{1}{\sqrt{3}}\right)$. Thus we get a lower bound of $\frac{1}{2}\left(1+\frac{1}{\sqrt{3}}\right) \approx 0.79$ on the optimal quantum value. To obtain a matching upper bound, we implemented numerically the NPA hierarchy \cite{navascues2008convergent} which is a hierarchy of semidefinite programs converging from the exterior to the correlations arising from quantum systems. 
Because the message sent from Alice to Bob is of finite dimension we relied mostly on \cite{navascues2015characterizing} which permits a characterisation of correlations arising from finite-dimensional quantum systems. 
We found a matching upper bound proving that indeed $\theta_{d=2}^Q \approx 0.79$.

\paragraph{Perfect Post-quantum Strategy for for the dimension 2 Torpedo Game.}

\begin{figure}[ht!]
        \centering
        \includegraphics[scale=0.2]{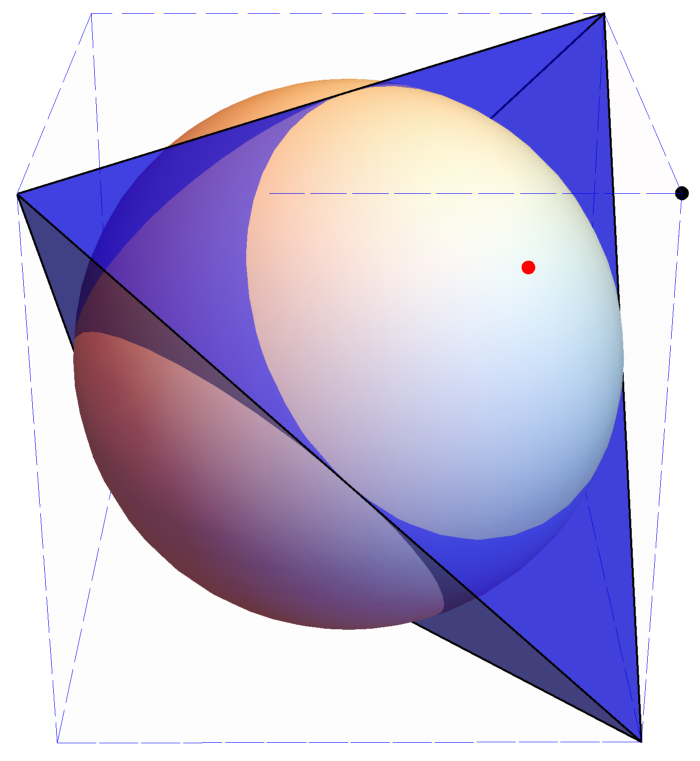}
        \caption{The qubit version of the Torpedo Game has a perfect strategy when allowed access to post-quantum ``states''. The red point on the surface of the Bloch sphere represents the optimal message state $\ket{\psi_{0,0}}$, achieving of a value of $0.79$ for the Torpedo game. The black point representing $\frac{1}{2}\left(\mathbb{I}-(X+Y+Z)\right)$ is not a valid density matrix, but achieves a value of 1 in the Torpedo Game. }
        \label{fig:ch05_Postquantum}
\end{figure}

\sloppy The average probability of success for the Casaccino et al.~QRAC, see Eq.~\eqref{eq:ch05_GalvaoQRAC}, can be maximized by using a post-quantum ``state'' of the form $\Pi_1^{k_1}+\Pi_2^{k_2}+\ldots+ \Pi_{d+1}^{k_{d+1}}-\mathbb{I}$ for any odd power-of-prime dimension $d$, where scare quotes reflect the fact that, although it is Hermitian and has unit trace, its spectrum is not necessarily nonnegative. In fact the ``state'' above is a phase point operator, $\hat A_{\pmb{k}}$, for one of Wootters' discrete Wigner functions. Since phase point operators obey $\Tr(\hat A\Id)=1$ and $\Tr(\hat A\hat A)=d$ then \index{PVM} $\frac{1}{(d+1)d^{d+1}}\sum_{\pmb{k}\in\mathbb{Z}_d^{d+1}}\Tr\left[\hat A_{\pmb{k}}(\hat A_{\pmb{k}}+\Id)\right] =1$. In other words, there is a perfect strategy by using post-quantum states. Seen in this way, phase point operators in a $(d+1,1)_d$ QRAC scenario are similar to Popescu-Rohrlich\index{PR box} \cite{popescu1994quantum} boxes in the CHSH scenario. As such, our Torpedo Game has a perfect strategy within quantum mechanics for all odd power-of-prime dimensions, by construction. In contrast, we saw that the qubit Torpedo Game only has quantum value of
roughly $0.79$.
To reach a perfect strategy, we may once again use a phase point operator as the non-physical ``state'' that Alice sends to Bob, see Figure~\ref{fig:ch05_Postquantum}.

\subsection{Optimal classical strategies}
\label{subsec05:classical_strat}

In what follows $d \in \N^*$ is arbitrary and we derive the classical value of a generic information retrieval task in a communication scenario $(2,1)_d$. We describe an encoding map $\Ec = \{p_{\mathcal{E}}(\cdot|x,z)\}_{x,z}$ as specifying a probability distribution over messages $j \in \mathbb{Z}_d$  for each combination of inputs $x,z \in \mathbb{Z}_d$.
Similarly a decoding map $\Dc = \{ p_{\mathcal{D}}(\cdot|j,q) \}_{j,q}$ specifies a probability distribution over outputs $c \in \mathbb{Z}_d$ for each combination of a message and question, $j \in \mathbb{Z}_d$ and $q \in Q$ respectively.

Combining an encoding
$\Ec$
and a decoding
$\Dc$
results in an empirical behaviour that we can write $e = \{ p_{e}(\cdot|x,z,q) \}_{x,z,q}$.
This is a set of probability distributions over outputs $c \in \mathbb{Z}_d$, one for each combination of the referee variables $x,z \in \mathbb{Z}_d$, $q \in Q$
such that
\begin{equation}
p_{e}(c |x,z,q) = \sum_{j \in \Z_d} \, p_{\mathcal{D}}( c |j,q) \, p_{\mathcal{E}}(j|x,z) \, .
    \label{eq:ch05_cl_proba_emp}
\end{equation}
By comparison, quantum mechanical empirical behaviours arise via the Born rule that is $p_e(c|x,z,q) = \Tr(\rho_{x,z}\Pi_q^c)$.\index{Born rule}

Assuming the referee variables to be uniformly distributed, a strategy has a winning probability given in terms of its empirical probabilities as
\[
\frac{1}{d^2(d+1)} \sum_{x,z,q} p_e (w_q(x,z) \mid x,z,q) \, .
\]
The classical value of the Torpedo Game can thus be expressed as
\begin{equation}
    \theta^C_{d} = \Max{\Ec,\Dc}
    \bigg[ \frac{1}{d^2(d+1)} \sum_{x,z,q}
    \, p_e ( w_q(x,z) \mid x,z,q) \bigg] \, .
    \label{eq:ch05_cval}
\end{equation}

To evaluate this expression note that it suffices to consider deterministic encodings and decodings.
In the presence of shared randomness,
nondeterministic strategies can always be obtained as convex combinations of deterministic ones and the expression is convex linear \cite{gallego2010device}.
Furthermore, for each encoding there exists a decoding that is optimal with respect to it.
This fact was also observed for one-way communication tasks with messages of bounded dimension in \cite{saha2019state}.
Thus it is possible to evaluate the classical value by enumerating over deterministic encodings only.

\begin{proposition}
The classical value of an information retrieval task in a $(2,1)_d$ communication scenario can be expressed as a maximum over encodings as
\begin{equation}
	\theta^C=
	\max_{\Ec}
	\bigg[
	\frac{1}{d^2(d+1)}\sum_{j,q}
	\max_c \sum_{\substack{ (x,z) \text{ s.t.} \\ c \in w_q(x,z)}} p_{\mathcal{E}}(j|x,z)
	 \bigg] .
\label{eq:ch05_classical_formula}
\end{equation}
\end{proposition}

\begin{proof}
Starting from Eq.~\eqref{eq:ch05_cval},
\begin{align*}
    \theta^C &= \Max{\Ec,\Dc}
    \bigg[ \frac{1}{d^2(d+1)} \, \sum_{x,z,q}
    \, p_e ( w_q(x,z) \mid x,z,q) \bigg] \\
    &= \Max{\Ec,\Dc}
    \bigg[ \frac{1}{d^2(d+1)} \, \sum_{x,z,q} \, \sum_{c \in w_q(x,z)}
    \, p_e ( c \mid x,z,q) \bigg] \\
    &= \Max{\Ec,\Dc}
    \bigg[ \frac{1}{d^2(d+1)} \, \sum_{q,c} \, \sum_{\substack{(x,z) \text{s.t. } \\ c \in w_q(x,z)}}
    \, p_e ( c \mid x,z,q) \bigg] \\
    &= \Max{\Ec,\Dc}
    \bigg[ \frac{1}{d^2(d+1)} \, \sum_{j,q,c} \, \sum_{\substack{(x,z) \text{s.t. } \\ c \in w_q(x,z)}}
    \, p_\Dc (c | j,q) \, p_\Ec ( j \mid x,z) \bigg] \\
    &= \Max{\Ec} \bigg[ \frac{1}{d^2(d+1)} \, \sum_{j,q} \, \max_c \sum_{\substack{(x,z) \text{s.t. } \\ c \in w_q(x,z)}}
    \, p_\Ec ( j \mid x,z) \bigg] \, ,
\end{align*}
where the last line follows by using a deterministic decoding that is optimal with respect to the encoding.
\end{proof}

A useful way of representing any deterministic encoding is as a colouring of the $d \times d$ affine plane using no more than $d$ colours.
Observe that a deterministic encoding can alternatively be expressed as a function $f_\Ec : \mathbb{Z}_d \times \mathbb{Z}_d \rightarrow \mathbb{Z}_d$, where $f_\Ec(x,z)$ is the message dit to be sent (with probability $1$) given inputs $x,z$.
Thinking of the inputs as coordinates in the $d \times d$ affine plane
a deterministic encoding is equivalent to a partition of the plane into no more than $d$ equivalence classes,
or a colouring using no more than $d$ colours.

\paragraph{Optimal classical strategies for $d=2$ and $d=3$ Torpedo Game.}
In general there are $d^{d^2}$ partitions of a $d \times d$ grid.
For low dimensions the expression in Eq.~\eqref{eq:ch05_classical_formula} can be evaluated by exhaustive search over partitions.
For dimension $2$ and $3$ we find
\begin{equation}
\label{eq:ch05_bounds}
    \theta^{C}_{d=2} = \frac{3}{4} \quad \text{ and } \quad \theta^{C}_{d=3} = \frac{11}{12} \, .
\end{equation}
Example of strategies that attain these values are depicted below in Figure \ref{fig:ch05_cl_d2} and in Figure \ref{fig:ch05_cl_d3}.

    \begin{figure}[htbp]
        \centering
        \scalebox{\myscale}{\begin{tikzpicture}[scale=1]

        \draw [thick, fill=blue!40] (0,0) grid (2,2) rectangle (0,0);
        \draw [thick, fill=purple!70] (0,0) rectangle (1,1);

\end{tikzpicture}}
        \caption{
        An optimal classical strategy for the $d=2$ Torpedo Game. Alice uses their bit of communication to indicate in which class of the partition that they find themself. Classes are represented here by colours.
        }
        \label{fig:ch05_cl_d2}
    \end{figure}
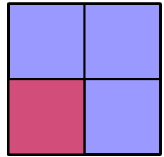

    \begin{figure}[htpb]
        \centering
        \begin{tikzpicture}[scale=.6]
	
	\draw [thick, fill=blue!40] (0,0) grid (3,3) rectangle (0,0);
	
	\draw [thick, fill=purple!70] (0,0) rectangle (1,1);
	\draw [thick, fill=purple!70] (1,0) rectangle (2,1);
	\draw [thick, fill=purple!70] (2,1) rectangle (3,2);
	\draw [thick, fill=purple!70] (2,1) rectangle (3,2);

	\draw [thick, fill=green!30] (0,2) rectangle (1,3);
	\draw [thick, fill=green!30] (1,2) rectangle (2,3);
	\draw [thick, fill=green!30] (1,1) rectangle (2,2);
	
	\end{tikzpicture}
	
\bigskip
\begin{tikzpicture}[scale=.15]


\draw [fill=gray!20] (-1,-4) rectangle (2,-1);
\draw [fill=purple!70] (-1,-2) rectangle (2,-1);
\draw [thick, fill=gray!20] (-1,-4) grid (2,-1);

\draw [fill=gray!20] (7,-4) rectangle (10,-1);
\draw [fill=green!30] (7,-4) rectangle (10,-3);
\draw [thick, fill=gray!20] (7,-4) grid (10,-1);

\draw [fill=gray!20] (3,-4) rectangle (6,-1);
\draw [fill=black!70] (3,-3) rectangle (6,-2);
\draw [thick, fill=gray!20] (3,-4) grid (6,-1);
\node (T1) at (4.5,-6) {$\neg x$};


\draw [fill=gray!20] (16,-4) rectangle (19,-1);
\draw [fill=blue!40] (17,-4) rectangle (18,-1);
\draw [thick, fill=gray!20] (16,-4) grid (19,-1);

\draw [fill=gray!20] (12,-4) rectangle (15,-1);
\draw [fill=black!70] (12,-4) rectangle (13,-1);
\draw [thick, fill=gray!20] (12,-4) grid (15,-1);

\draw [fill=gray!20] (20,-4) rectangle (23,-1);
\draw [fill=green!30] (22,-4) rectangle (23,-1);
\draw [thick, fill=gray!20] (20,-4) grid (23,-1);
\node (T1) at (17.5,-6) {$\neg (-z)$};


\draw [fill=gray!20] (29,-4) rectangle (32,-1);
\draw [fill=purple!70] (29,-2) rectangle (30,-1);
\draw [fill=purple!70] (30,-3) rectangle (31,-2);
\draw [fill=purple!70] (31,-3) rectangle (32,-4);
\draw [thick, fill=gray!20] (29,-4) grid (32,-1);
\node (T1) at (30.5,-6) {$\neg (x-z)$};

\draw [fill=gray!20] (25,-4) rectangle (28,-1);
\draw [fill=blue!40] (25,-4) rectangle (26,-3);
\draw [fill=blue!40] (26,-2) rectangle (27,-1);
\draw [fill=blue!40] (27,-3) rectangle (28,-2);
\draw [thick, fill=gray!20] (25,-4) grid (28,-1);

\draw [fill=gray!20] (33,-4) rectangle (36,-1);
\draw [fill=green!30] (33,-3) rectangle (34,-2);
\draw [fill=green!30] (34,-4) rectangle (35,-3);
\draw [fill=green!30] (35,-2) rectangle (36,-1);
\draw [thick, fill=gray!20] (33,-4) grid (36,-1);


\draw [fill=gray!20] (46,-4) rectangle (49,-1);
\draw [fill=black!70] (46,-4) rectangle (47,-3);
\draw [fill=black!70] (47,-3) rectangle (48,-2);
\draw [fill=black!70] (48,-2) rectangle (49,-1);
\draw [thick, fill=gray!20] (46,-4) grid (49,-1);

\draw [fill=gray!20] (38,-4) rectangle (41,-1);
\draw [fill=purple!70] (38,-3) rectangle (39,-2);
\draw [fill=purple!70] (39,-2) rectangle (40,-1);
\draw [fill=purple!70] (40,-4) rectangle (41,-3);
\draw [thick, fill=gray!20] (38,-4) grid (41,-1);

\draw [fill=gray!20] (42,-4) rectangle (45,-1);
\draw [fill=blue!40] (42,-2) rectangle (43,-1);
\draw [fill=blue!40] (43,-4) rectangle (44,-3);
\draw [fill=blue!40] (44,-3) rectangle (45,-2);
\draw [thick, fill=gray!20] (42,-4) grid (45,-1);
\node (T1) at (43.5,-6) {$\neg (2x-z)$};

\end{tikzpicture}
        \caption{
        An optimal classical strategy for the $d=3$ Torpedo Game.
        Alice uses their dit of communication to indicate in which equivalence class (represented by same coloured cells) of the large grid partition they find themselves.
        The smaller grids (cf.\ Fig.~\ref{fig:ch05_d3_torpedo_firing}) show where Bob chooses to shoot, given a direction and a colour.
        For the first direction, when asked to shoot horizontally in the grid, notice that Bob may avoid Alice with certainty if Alice is in either of the red or green partitions.
        Lines that avoid Alice with certainty are depicted in the corresponding colour,
        whereas black lines intersect with Alice's position with probability $\frac{1}{3}$.
        Overall, this strategy wins the Torpedo Game with probability
$\frac{1}{4}(\frac{8}{9}+\frac{8}{9}+1+\frac{8}{9}) = \frac{11}{12}$.
        }
        \label{fig:ch05_cl_d3}
    \end{figure}

\subsection{Comparison of quantum and classical game values}
Recall the optimal quantum values established in Subsection~\ref{subsec05:qstrat},
\begin{equation}
    \theta^{Q}_ {d = 2} \simeq 0.79 \quad \text{ and } \quad \theta^{Q}_{d \geq 3}= 1 \, . \label{eq:ch05_QuantumValues}
\end{equation}
Comparing these with the classical bounds from Subsection~\ref{subsec05:classical_strat} we obtain the ratios
\begin{equation}
    \frac{\theta^{Q}_ {d = 2}}{\theta^{C}_ {d = 2}} \simeq 1.053 \quad \text{ and } \quad \frac{\theta^{Q}_ {d = 3}}{\theta^{C}_ {d = 3}} \simeq 1.091 \, .
\end{equation}
By comparison, it was shown in \cite{tavakoli2015quantum} that the classical and quantum values of the $(4,1)_3$ (Q)RAC are $\frac{16}{27}$ and $0.637$, respectively, giving a ratio of $\theta^{Q}_ {d = 3} / \theta^{C}_ {d = 3} \simeq 1.075$.
Accordingly, we note that the $d=3$ Torpedo Game admits a greater quantum-over-classical advantage than the standard random access coding task whose optimal QRAC also exploits the $4$ mutually unbiased bases available in dimension $3$.


\subsection{Dimensional witness}
The Torpedo Game can be used as a dimensional witness \cite{Brunner2008TestingDimension} for qubits and qutrits. In the following, we modify slightly the setting of the game to allow the message to be of arbitrary dimension. In particular, we no longer require that the message between Alice and Bob is of the same dimension as the inputs. For instance, we will allow Alice to send a (qu)trit while they receive two input bits. We will thus specify the \textit{dimension of the inputs} as well as the \textit{dimension of the message}. Questions are defined as before and are fixed by the dimension of the inputs: $d+1$ questions for inputs of dimension $d$.

\begin{table}[ht!]
    \centering
        \begin{tabular}{cc|cc}
            \toprule \toprule
            \begin{tabular}{c}
                 Input  \\
                 dimension
            \end{tabular} & 
            \begin{tabular}{c}
                 Message  \\
                 dimension
            \end{tabular}& $\theta^C$ & $\theta^Q$ \\ \midrule
            2 & 2 & 0.75 & 0.789 \\
            2 & 3 & 0.833 & 0.875 \\[10pt]
            3 & 2 & 0.833 & $> 0.867$ \\
            3 & 3 & 0.917 & 1 \\
            \bottomrule \bottomrule
        \end{tabular}
    \caption{Classical and quantum optimal values of the Torpedo Game when we allow the message dimension to differ from that of the inputs. Classical values were computed by exhaustive search. Quantum values were obtained by a combination of a seesaw algorithm and implementation of the NV hierarchy \cite{navascues2015characterizing} through the interface QDimSum \cite{tavakoli2019enabling}. Note that only the seesaw algorithm succeeded for the trit-input, qubit-message Torpedo Game as the NV hierarchy does not perform well with POVMs.}\index{NV hierarchy}
    \label{tab:ch05_dimension_witness}
\end{table}

Following Table \ref{tab:ch05_dimension_witness}, we can use the Torpedo Game to discriminate between qubits and qutrits. Moreover these witnesses can distinguish between classical and quantum systems of the same dimension.

\section{Sequential contextuality}
\label{sec05:contextuality}\index{Contextuality!sequential}
        
When quantum advantage is observed in a bounded-memory information retrieval task like a (Q)RAC task or the Torpedo Game, it highlights a difference between the information carrying capacities of qudits compared to dits for a fixed dimension.
It can be remarked that such a difference is a consequence of the different geometries of the respective state spaces.
In this section, however, we seek a sharper, quantified analysis of the source of the advantage in terms of contextuality.

We will be using the notion of sequential contextuality that was introduced in \cite{mansfield2018quantum}
to extend structural treatments of Bell-Kochen-Specker contextuality \cite{abramsky2011sheaf} to sequential operational scenarios.\index{Contextuality!DV}
As such, it is a behavioural characteristic that can arise in experiments involving sequences of operations.
While \cite{mansfield2018quantum} was concerned specifically with sequences of transformations,
here we take a broader view that also includes the operations of preparation and measurement.
In the special case of prepare-and-measure scenarios, sequential noncontextuality
recovers a natural notion of classicality in terms of realisability by hidden-variables.
For instance in the bounded-memory regime we are interested in, sequential contextuality
also matches the characteristic introduced by \.Zukowski in \cite{zukowski2014temporal}.

We note that sequential contextuality\index{Contextuality!sequential} is distinct from the notion of contextuality due to Spekkens \cite{Spekkens2005},
as discussed in \cite{mansfield2018quantum}.
It is also distinct from the analyses of \cite{kirchmair2009state,amselem2009state,lapkiewicz2011experimental,guhne2010compatibility},
which sought to close potential loopholes created by sequentiality of measurements in experimental tests of the more traditional Bell--Kochen--Specker form of contextuality.
To determine its precise relationship with the latter analyses is however an interesting open question.

The study of contextuality arose in quantum foundations,
where a major theme is the attempt to understand empirical behaviours that may appear non-intuitive from a classical perspective, e.g.\ the EPR paradox \cite{einsteincan1935}.\index{EPR argument}
The typical approach is to look for a description of physical systems at a deeper level than the quantum one at which more classically intuitive properties may be restored.
Such a description is usually formalised as a hidden-variable model\index{Hidden-variable model!DV} for the behaviour (sometimes also referred to as an ontological model \cite{Spekkens2005}).
The great significance of the celebrated no-go theorems of quantum foundations,
like Bell's Theorem \cite{belleinstein1964} and the Bell--Kochen--Specker Theorem \cite{bell1966,kochen1975problem},
was to prove that certain `non-classical' features of the empirical behaviours of quantum systems are necessarily inherited by any underlying model under minimal assumptions.

Non-classical features of quantum systems like contextuality are also increasingly investigated for their practical utility.
For instance, contextuality of the Bell--Kochen--Specker kind was shown to be a prerequisite for quantum speed-up \cite{howard2014contextuality} and to quantify quantum-over-classical advantage in a variety of information processing tasks \cite{abramsky2017contextual}.

Bell--Kochen--Specker contextuality essentially concerns the statistics that arise under varied measurements on a physical system.
In contrast, our notion of contextuality concerns the statistics that arise from sequences of operations -- preparations, transformations and measurements -- all of which can vary.
As a behavioural feature, sequential contextuality signifies the absence of any hidden-variable model that would preserve a compositional description of operations performed in sequence.
In other words, sequential \emph{non}contextuality requires a hidden-variable model in which each operation has an independent, modular description (as a transformation on hidden variables) such that to describe a sequence of operations one simply composes their hidden-variable descriptions.
We provide a more rigorous mathematical description in the following subsections.\index{Contextuality!sequential}

Rather than focusing on characteristics that must be inherited by all hidden-variable models,\index{Hidden-variable model!DV}
as is common in foundational works,
we also take a practical perspective and shift focus to characteristics that must be inherited by bounded-memory models --
a constraint that matches the information retrieval problem at hand.
In this respect, the significance of sequential contextuality in what follows can also be viewed through a practical rather than a foundational lens, as a characteristic that quantifies quantum advantage.

\subsection{Empirical behaviours and hidden-variable Models}
\label{subsec05:emp_behaviours}

Recall from Section \ref{subsec05:classical_strat} that any strategy for a general information retrieval task gives rise to an empirical behaviour $e = \{ p_e (\cdot | i,q) \}_{i,q}$ (see Eq.~\eqref{eq:ch05_cl_proba_emp}).\index{Empirical model!DV}
In other words, for each combination of input string $i \in \mathbb{Z}_d^n$ and question $q \in Q$
there is a resulting probability distribution over outputs.
This is true regardless of whether the strategy is classical, quantum, post-quantum, or other (like generalised probabilistic theories \cite{BarrettGPT}).
The combination of input string and question fully specifies the precise operations that are performed in the sequence.
This is what we refer to as the context,
just as a context in a (Bell--Kochen--Specker) measurement scenario specifies a set compatible measurements to be performed jointly.
We have also chosen a formal description of empirical behaviour that echoes the formalism of empirical models for measurement scenarios in \cite{abramsky2011sheaf} (see Subsection~\ref{sec01:sheaf}).

Given an empirical behaviour, one can ask whether it can be simulated by a bounded-memory hidden-variable model.
In particular we will be interested in models that respect the sequential structure of the strategy,
bearing in mind that the inputs and questions specify a sequence of operations: either preparations, transformations, or measurements.
Our main focus is on prepare-and-measure scenarios, in which sequences arise from a combination of a preparation and a measurement.
To match the constraints of information retrieval tasks, memory is bounded by the dimension of the message string. This is further motivated by the Holevo bound \cite{holevo1973bounds}, according to which one can faithfully retrieve no more than $n$ dits of classical information from $n$ qudits.

In a bounded-memory model, the hidden variable is restricted to take values in $\mathbb{Z}_d$ with $d$ fixed by the communication scenario.
A state preparation $P$ is modelled by a probability distribution $p_\pi(\cdot \mid P)$ over the hidden-variable space $\mathbb{Z}_d$.
Similarly, a {measurement} $M$ is modelled by a family of probability distributions $\{ p_\mu(\cdot \mid \lambda, M) \}_\lambda$ over the outcome space, which, to match the communication scenario, is also $\mathbb{Z}_d$.
A hidden-variable model---e.g.\ for a prepare-and-measure sequence of operations $M_q \circ P_{x,z}$ with inputs $x,z \in \Z_d$ and a question $q\in Q$, as in \refig{ch05_QRAC_pres}---simulates an empirical behaviour as
\begin{equation}
p_e(\cdot \mid x,z,q ) = \\ \sum_{\lambda} \, p_\mu(\cdot \mid \lambda, M_q) \cdot p_\pi(\lambda \mid P_{x,z}) \, .
    \label{eq:ch05_pamprealise}
\end{equation}
Each operation is thus modelled in sequence as an operation on the hidden-variable space.\index{Empirical model!DV} \index{Hidden-variable model!DV}

With Eq.~\eqref{eq:ch05_pamprealise} in mind, the bounded-memory classical strategies of Section~\ref{sec05:optimalstrat} can also be interpreted as bounded-memory hidden-variable models themselves.
To see this, note that in Eq.~\eqref{eq:ch05_cl_proba_emp} encoding corresponds to hidden-variable preparation and decoding to hidden-variable measurement.

The above description makes contact with the strategies and empirical behaviours of Section~\ref{sec05:optimalstrat} and it will be convenient for the remainder of this section to use a simplified notation that meets the one from \cite{mansfield2018quantum}.
Let us work in the real vector space $\Z_d$ with basis given by the hidden-variable states.
For all preparations $P_{x,z}$, the probability distribution $p_\pi(\cdot \mid P_{x,z})$ will be more concisely denoted as a probability vector $\pmb{\lambda}_{x,z}$ in $\Z_d$. Deterministic encoding will result in vectors $\pmb{\lambda}_{x,z}$ indicating in which partition $x,z$ belong to.
For measurements, we express $\{ p_\mu(\cdot \mid \lambda, M) \}_\lambda$ more concisely as, for each question, a choice of a $d\times d$ left stochastic matrix\footnote{A left-stochastic matrix is a square matrix with each column summing to 1. It is furthermore binary if each column consists of a single 1 and 0 elsewhere.}
$T_q$ acting on $\bm \lambda_{x,z}$.
If we also denote by $e_{x,z,q} := p_e(\cdot \mid x,z,q)$ the empirical probability vector over outcomes $\mathbb{Z}_d$, then Eq.~\eqref{eq:ch05_pamprealise} can be rewritten for each outcome $c \in \Z_d$ in simplified notation as the dot product
\begin{equation}
    e_{x,z,q}(c) = T_q \bm{\lambda}_{x,z} \cdot \bm \delta_c \, ,
    \label{eq:ch05_pamrealise}
\end{equation}
where $\bm \delta_c$ is the $d$-dimensional vector filled with zeros except for the $c^\text{th}$ coordinate which equals 1.

As mentioned in Subsection~\ref{subsec05:classical_strat}, it is sufficient to consider deterministic encodings and decodings to achieve optimal strategies. As in \cite{saha2019state}, if the message is of dimension $d$, if there are $\vert Q\vert$ possible questions, and if the outcome is a dit, then there are $d^{\vert Q \vert d}$ different deterministic decoding strategies. This number can be calculated by enumerating the possible choices of $ T_q $ for each question. There are $d^d$ possible binary left stochastic matrices hence $d^{d \vert Q \vert}$ deterministic decoding strategies. 

As our focus is on prepare-and-measure scenarios, we have not discussed hidden-variable modelling
of transformations. This can be found in \cite{mansfield2018quantum}.
For our main example of an information retrieval task we have also focused on the prepare-and-measure version of the Torpedo Game. Note, however, that it can be equivalently expressed in a sequential scenario with fixed preparation and measurement. 

\subsection{Sequential contextuality in information retrieval tasks}
\index{Contextuality!sequential}

Here we introduce a different notion of contextuality than the one presented in Section~\ref{sec01:sheaf}. It concerns the sequence of operations that might be applied to a system.
An empirical behaviour is sequential noncontextual if it admits a hidden-variable model that:
(i) preserves a modular sequential description of operations, and
(ii) the hidden-variable representation of operations is context-independent. We refer readers to \cite{mansfield2018quantum}
for a detailed presentation of this concept. 

These assumptions have been implicitly built into the above definition of hidden-variable models displayed in Eq.~\eqref{eq:ch05_pamrealise}.
For (i), note that each operation has an individual description at the hidden-variable level.\index{Contextuality!sequential} \index{Hidden-variable model!DV}
For example, to obtain predictions for a prepare-and-measure experiment we compose the individual hidden-variable descriptions of the preparation and of the measurement, as in Eq.~\eqref{eq:ch05_pamrealise}.
And for (ii), note for example that regardless of which context the preparation $P_{x,z}$ appears in it should be modelled by the same vector $\bm{\lambda}_{x,z}$.
One could relax these assumptions, in which case it would become trivial to find a hidden-variable model for any behaviour,
but it would also entail giving up the intuitive sense of what the model means.

If an empirical behaviour does not admit a sequential noncontextual hidden-variable model it is said to be sequential contextual.
In this chapter we will only be considering sequential contextuality with respect to bounded-memory models, though the definition may be applied more generally.

A useful intuition for sequential contextuality is that, within the memory constraints, for any faithful model of the behaviour, the whole (the description of the context) is more than the composition of its parts (the descriptions of the individual operations).
A contextual model would always involve additional memory and communication to track the context, which would be outside of the constraints of the task --
involving, e.g., a contextuality demon analogous to Maxwell's demon in thermodynamics.
Indeed it was shown in \cite{henaut2018tsirelson} that a related characteristic incurs a simulation cost as measured by Landauer erasure.

\subsection{Quantified Contextual Advantage in Information Retrieval Tasks}

The following proposition can be understood as a no-go theorem stating that winning the Torpedo Game deterministically for $d=2$ and $d=3$ is incompatible with the assumptions of sequential noncontexutality and bounded memory. If such a performance is observed then one is forced to abandon at least one of the assumptions, and we note that the Holevo bound gives an argument that perhaps the noncontextuality assumption is the weaker of these.

\begin{proposition}
\label{pro:ssc}
For $d=2$ and $d=3$, strong sequential contextuality with respect to bounded memory is necessary and sufficient to win the Torpedo Game deterministically.
\end{proposition}

\begin{proof}
Suppose a bounded-memory hidden-variable model realises an empirical model that wins the Torpedo Game deterministically.
Input-question combinations $(x,z,q)$ label the contexts.
Recall that the winning relation is $\omega_q(x,z)$,
and the winning condition for the Torpedo Game is
\begin{align}
p_e(c \not\in {\omega}_q(x,z) \mid x,z,q ) = 0. \label{eq:pam_sequential_context}
\end{align}

Using notations introduced in Subsection~\ref{subsec05:emp_behaviours}, the hidden-variable model must specify probability vectors $\left\{\bm{\lambda}_{x,z}\right\}_{x,z \in \Z_d}$ and left stochastic matrices $\left\{ T_q \right\}_{q \in Q}$ such that
\begin{equation}
    T_q \bm{\lambda}_{x,z} \cdot \bm \delta_c= 0\, ,
    \label{eq:ch05_linear_eq_prop}
\end{equation}
for all $x,z \in \Z_d, \forall q \in Q$ and $c \not\in \omega_q(x,z)$.

As mentioned in Subsection~\ref{subsec05:classical_strat}, it suffices to consider deterministic strategies. Eq.~\eqref{eq:ch05_linear_eq_prop} reduces to a set of binary linear equations (36 equations for $d=3$) that any sequentially noncontextual realisation must jointly satisfy. 

This cannot be possible since it would provide a perfect classical strategy for the $d=2$ and $d=3$ Torpedo Games, violating the optimal bounds Eq.~\eqref{eq:ch05_bounds} that were obtained by exhaustive search.
On the other hand, it is always possible to obtain a contextual realisation, by taking context-wise solutions to Eq.~\eqref{eq:ch05_linear_eq_prop}: e.g.\ where the choice of $\pmb{\lambda}_{x,z}$ is not only a function of $x$ and $z$, but also of $q$.\index{Hidden-variable model!DV}

It can further be observed that if any fraction of an empirical model $e$ can be described noncontextually, i.e.\ $\NCF(e) = p >0$, then with an average probability at least $p$ the empirical model $e$ fails in the Torpedo Game. \index{Contextuality!sequential}
Therefore, to win the Torpedo Game deterministically requires strong contextuality.

\end{proof}

An explicit noncontextual memory-bounded hidden-variable model that fails to fully realise the empirical predictions but that satisfies the maximum of 33 out of 36 constraints---thus reaching the maximum probability $\frac{11}{12}$ for the Torpedo Game with input trits---from Eq.~\eqref{eq:ch05_linear_eq_prop} for $d=3$ is the following.
The state vectors are
\begin{align*}
& \bm{\lambda}_{0,0}=\bm{\lambda}_{0,1}=\bm{\lambda}_{1,1}=\bm \delta_0 \, , \\
& \bm{\lambda}_{1,0}=\bm{\lambda}_{0,2}=\bm{\lambda}_{2,2}=\bm \delta_1 \, , \\
& \bm{\lambda}_{2,0}=\bm{\lambda}_{2,1}=\bm{\lambda}_{1,2}=\bm \delta_2 \, .
\end{align*}
The measurement left-stochastic matrices are
\begin{align*}
T_\infty = \begin{pmatrix}
0&0&1\\
0&1&0\\
1&0&0
\end{pmatrix}\, , \; 
T_0 = \begin{pmatrix}
0&0&1\\
0&1&0\\
1&0&0
\end{pmatrix}, \; 
T_1 = \begin{pmatrix}
0&1&0\\
0&0&1\\
1&0&0
\end{pmatrix}, \; 
T_2 = \begin{pmatrix}
0&0&1\\
0&1&0\\
1&0&0
\end{pmatrix} \Mdot
\end{align*}
This corresponds to the strategy depicted in \refig{ch05_cl_d3} with the green section corresponding to sending the message $\bm \delta_0$, the blue one to $\bm \delta_1$ and the red one to $\bm \delta_2$.

We also obtain the following more general result, of which Proposition \ref{pro:ssc} is a special case. This is an extension of \cite[Theorem 3]{abramsky2017contextual} to the case of information retrieval tasks with sequential contextuality rather than measurement contextuality.
\begin{theorem}
\label{thm:sc}
Given any information retrieval task and strategy with empirical behaviour $e$,
\begin{equation*}
    \varepsilon \geq \textsf{\upshape NCF}(e) \, \nu \,
\end{equation*}
where $\varepsilon$ is the probability of failure, averaged over inputs and questions, $\NCF(e)$ is the bounded-memory noncontextual fraction of $e$ with memory size $d$ fixed by the scenario, and $\nu := 1- \theta^C$ measures the hardness of the task, $\theta^C$ being the classical value.
\end{theorem} \index{Empirical model!DV}

\begin{proof}
The empirical behaviour can be decomposed as:
\begin{equation*}\index{Contextual fraction!DV}
    e = \NCF(e) e^{\text{NC}} + \CF(e) e'
\end{equation*}
where $e'$ is necessarily strongly contextual. From this convex decomposition, we obtain that the probability of success using the empirical model $e$ reads:
\begin{equation*}
    p_{S,e} = \NCF(e) p_{S,e^{\text{NC}}} + \CF(e) p_{S,e'}
\end{equation*}
where $p_{S,e^{\text{NC}}}$ and $p_{S,e'}$ are the average probabilities associated with empirical models $e^{\text{NC}}$ and $e'$ respectively. At best, $e'$ wins with probability $1$ and thus:
\begin{align*}
    p_{S,e} & \leq \NCF(e) p_{S,e^{\text{NC}}} + \CF(e) \\
    \varepsilon & \geq \NCF(e) \varepsilon_{e^{\text{NC}}} 
\end{align*}
and $\varepsilon_{e^{\text{NC}}} = 1 - p_{S,e^{\text{NC}}}$ is the average probability of failure associated to $e^{\text{NC}}$.
Since $e^{\text{NC}}$ is noncontextual, we know that the minimum probability of failure is $\nu = 1 - \theta^C$, where $\theta^C$ is the classical value of the game.
Then $\varepsilon_{e^{\text{NC}}} \ge \nu$, from which we obtain the desired inequality:
\begin{equation*}
    \varepsilon \geq \NCF(e) \nu
\end{equation*}

\end{proof}

This provides a quantifiable relationship between quantum advantage and sequential contextuality\index{Contextuality!sequential}.
Inequalities of this {form} are also known to arise for a variety of other informational tasks that admit quantum advantage,
with hardness measures and notions of non-classicality adapted to the particular task \cite{abramsky2017contextual,mansfield2018quantum,linde}.

\section{Discussion and open problems}
\label{sec05:conclusion}

We have formalised a class of information retrieval tasks in communication scenarios, of which the much-studied problem of (quantum) random access coding is a special case. We showed that quantum-over-classical advantage is explained by quantum (sequential) contextuality.
We have identified a distinct information retrieval task that we have presented as the Torpedo Game,
which admits a greater quantum-over-classical advantage than the comparable QRAC for qutrits by exploiting Wigner negativity. Remarkably, the qutrit torpedo strategy is maximally contextual, meaning that no fraction of it can be explained by an underlying noncontextual model.
By choosing measurements that are associated with a particular discrete Wigner function, Wigner negativity seems a necessary ingredient for a quantum strategy to perform better than a classical one as the optimal quantum strategy relies on it. 
However post-quantum strategies (e.g. using phase point operators) might be optimal while remaining Wigner positive. 
A more thorough investigation of the precise relationship between negativity and advantage is an interesting direction to pursue.

To obtain perfect quantum strategies for the Torpedo Game we have derived a prepare-and-measure scenario for which quantum mechanics exhibits logically paradoxical behaviour (with respect to noncontextual hidden variable assumptions).
More generally, we have identified this as a characteristic that quantifies quantum advantage for any bounded-memory information retrieval task.
\index{Contextuality!sequential}
\index{Wigner negativity}

In the specific case of random access coding,
some works have imposed obliviousness constraints as part of the task as opposed to bounded-memory.
These restrict what information the receiver can be allowed to infer about the input string.
Whereas preparation contextuality is known to be necessary and sufficient for quantum advantage in oblivious tasks \cite{hameedi2017communication,saha2019state},
we have shown sequential contextuality to be necessary and sufficient characteristic for bounded-memory tasks.

Importantly, a nonlocal version of the Torpedo Game was derived after our corresponding paper in \cite{TavakoliExp2021}, where it was also implemented experimentally. 
This example was employed to certify quantum correlations allowing to test for nonlocality, steering and quantum state tomography in a single experiment. Based on this nonlocality experiment, entanglement in prepare-and-measure scenarios was further studied in \cite{pauwels2021entanglement}.

\begin{figure}[ht!]
        \centering
        \begin{tikzpicture}[scale=.45]

\draw [thick, fill=blue!40] (0,0) grid (5,5) rectangle (0,0);

\draw [thick, fill=brown!60] (0,0) rectangle (1,1);
\draw [thick, fill=brown!60] (1,0) rectangle (2,1);
\draw [thick, fill=brown!60] (0,2) rectangle (1,3);
\draw [thick, fill=brown!60] (0,4) rectangle (1,5);
\draw [thick, fill=brown!60] (4,1) rectangle (5,2);

\draw [thick, fill=green!30] (0,1) rectangle (1,2);
\draw [thick, fill=green!30] (2,2) rectangle (3,3);
\draw [thick, fill=green!30] (2,3) rectangle (3,4);
\draw [thick, fill=green!30] (3,0) rectangle (4,1);
\draw [thick, fill=green!30] (3,1) rectangle (4,2);

\draw [thick, fill=cyan!70] (2,0) rectangle (3,1);
\draw [thick, fill=cyan!70] (1,2) rectangle (2,3);
\draw [thick, fill=cyan!70] (1,4) rectangle (2,5);
\draw [thick, fill=cyan!70] (3,4) rectangle (4,5);
\draw [thick, fill=cyan!70] (4,3) rectangle (5,4);

\draw [thick, fill=purple!70] (1,1) rectangle (2,2);
\draw [thick, fill=purple!70] (3,2) rectangle (4,3);
\draw [thick, fill=purple!70] (3,3) rectangle (4,4);
\draw [thick, fill=purple!70] (2,4) rectangle (3,5);
\draw [thick, fill=purple!70] (4,4) rectangle (5,5);

\end{tikzpicture}
\bigskip

\begin{tikzpicture}[scale=.18]

\draw[fill = green!30] (0,4) -- (0,5) -- (1,4);
\draw[fill = green!30] (1,4) -- (1,5) -- (2,4);
\draw[fill = green!30] (2,4) -- (2,5) -- (3,4);
\draw[fill = green!30] (3,4) -- (3,5) -- (4,4);
\draw[fill = green!30] (4,4) -- (4,5) -- (5,4);
\draw[fill = blue!40] (0,5) -- (1,5) -- (1,4);
\draw[fill = blue!40] (1,5) -- (2,5) -- (2,4);
\draw[fill = blue!40] (2,5) -- (3,5) -- (3,4);
\draw[fill = blue!40] (3,5) -- (4,5) -- (4,4);
\draw[fill = blue!40] (4,5) -- (5,5) -- (5,4);
\draw[fill= purple!70] (0,0) rectangle (5,1);
\draw[fill= cyan!70] (0,1) rectangle (5,2);
\draw[fill= brown!60] (0,3) rectangle (5,4);
\node[below] (notx) at (2.5,0) {\scriptsize{$\neg (x)$}};
\draw (0,0) grid (5,5);

\draw[fill = purple!70] (7,0) -- (8,0) -- (7,1);
\draw[fill = purple!70] (7,1) -- (8,1) -- (7,2);
\draw[fill = purple!70] (7,2) -- (8,2) -- (7,3);
\draw[fill = purple!70] (7,3) -- (8,3) -- (7,4);
\draw[fill = purple!70] (7,4) -- (8,4) -- (7,5);
\draw[fill = cyan!70] (7,1) -- (8,1) -- (8,0);
\draw[fill = cyan!70] (7,2) -- (8,2) -- (8,1);
\draw[fill = cyan!70] (7,3) -- (8,3) -- (8,2);
\draw[fill = cyan!70] (7,4) -- (8,4) -- (8,3);
\draw[fill = cyan!70] (7,5) -- (8,5) -- (8,4);
\draw[fill = green!30] (8,0) rectangle (9,5);
\draw[fill = blue!40] (10,0) rectangle (11,5);
\draw[fill = brown!60] (9,0) rectangle (10,5);
\draw (7,0) grid (12,5);
\node[below] (notz) at (9.5,0) {\scriptsize{$\neg (-z)$}};

\draw[fill = blue!40] (14,0) rectangle (15,1);
\draw[fill = blue!40] (15,4) rectangle (16,5);
\draw[fill = blue!40] (16,3) rectangle (17,4);
\draw[fill = blue!40] (17,2) rectangle (18,3);
\draw[fill = blue!40] (18,1) rectangle (19,2);
\draw[fill = cyan!70] (14,1) rectangle (15,2);
\draw[fill = cyan!70] (15,0) rectangle (16,1);
\draw[fill = cyan!70] (16,4) rectangle (17,5);
\draw[fill = cyan!70] (17,3) rectangle (18,4);
\draw[fill = cyan!70] (18,2) rectangle (19,3);
\draw[fill = green!30] (14,2) rectangle (15,3);
\draw[fill = green!30] (15,1) rectangle (16,2);
\draw[fill = green!30] (16,0) rectangle (17,1);
\draw[fill = green!30] (17,4) rectangle (18,5);
\draw[fill = green!30] (18,3) rectangle (19,4);
\draw[fill = brown!60] (14,3) rectangle (15,4);
\draw[fill = brown!60] (15,2) rectangle (16,3);
\draw[fill = brown!60] (16,1) rectangle (17,2);
\draw[fill = brown!60] (17,0) rectangle (18,1);
\draw[fill = brown!60] (18,4) rectangle (19,5);
\draw[fill = purple!70] (14,4) rectangle (15,5);
\draw[fill = purple!70] (15,3) rectangle (16,4);
\draw[fill = purple!70] (16,2) rectangle (17,3);
\draw[fill = purple!70] (17,1) rectangle (18,2);
\draw[fill = purple!70] (18,0) rectangle (19,1);
\draw (14,0) grid (19,5);
\node[below] (notxz) at (16.5,0) {\scriptsize{$\neg (x-z)$}};

\draw[fill = brown!60] (21,3) rectangle (22,4);
\draw[fill = brown!60] (22,1) rectangle (23,2);
\draw[fill = brown!60] (23,4) rectangle (24,5);
\draw[fill = brown!60] (24,2) rectangle (25,3);
\draw[fill = brown!60] (25,0) rectangle (26,1);
\draw[fill = green!30] (21,4) rectangle (22,5);
\draw[fill = green!30] (22,2) rectangle (23,3);
\draw[fill = green!30] (23,0) rectangle (24,1);
\draw[fill = green!30] (24,3) rectangle (25,4);
\draw[fill = green!30] (25,1) rectangle (26,2);
\draw[fill = purple!70] (21,0) rectangle (22,1);
\draw[fill = purple!70] (22,3) rectangle (23,4);
\draw[fill = purple!70] (23,1) rectangle (24,2);
\draw[fill = purple!70] (24,4) rectangle (25,5);
\draw[fill = purple!70] (25,2) rectangle (26,3);
\draw[fill = blue!40] (21,1) rectangle (22,2);
\draw[fill = blue!40] (22,4) rectangle (23,5);
\draw[fill = blue!40] (23,2) rectangle (24,3);
\draw[fill = blue!40] (24,0) rectangle (25,1);
\draw[fill = blue!40] (25,3) rectangle (26,4);
\draw[fill = cyan!70] (21,2) rectangle (22,3);
\draw[fill = cyan!70] (22,0) rectangle (23,1);
\draw[fill = cyan!70] (23,3) rectangle (24,4);
\draw[fill = cyan!70] (24,1) rectangle (25,2);
\draw[fill = cyan!70] (25,4) rectangle (26,5);
\draw (21,0) grid (26,5);
\node[below] (notxz2) at (23.5,0) {\scriptsize{$\neg (2x-z)$}};

\draw[fill = green!30] (28,2) rectangle (29,3);
\draw[fill = green!30] (29,4) rectangle (30,5);
\draw[fill = green!30] (30,1) rectangle (31,2);
\draw[fill = green!30] (31,3) rectangle (32,4);
\draw[fill = green!30] (32,0) rectangle (33,1);
\draw[fill = brown!60] (28,1) rectangle (29,2);
\draw[fill = brown!60] (29,3) rectangle (30,4);
\draw[fill = brown!60] (30,0) rectangle (31,1);
\draw[fill = brown!60] (31,2) rectangle (32,3);
\draw[fill = brown!60] (32,4) rectangle (33,5);
\draw[fill = blue!40] (28,0) rectangle (29,1);
\draw[fill = blue!40] (29,2) rectangle (30,3);
\draw[fill = blue!40] (30,4) rectangle (31,5);
\draw[fill = blue!40] (31,1) rectangle (32,2);
\draw[fill = blue!40] (32,3) rectangle (33,4);
\draw[fill = purple!70] (28,3) rectangle (29,4);
\draw[fill = purple!70] (29,0) rectangle (30,1);
\draw[fill = purple!70] (30,2) rectangle (31,3);
\draw[fill = purple!70] (31,4) rectangle (32,5);
\draw[fill = purple!70] (32,1) rectangle (33,2);
\draw[fill = cyan!70] (28,4) rectangle (29,5);
\draw[fill = cyan!70] (29,1) rectangle (30,2);
\draw[fill = cyan!70] (30,3) rectangle (31,4);
\draw[fill = cyan!70] (31,0) rectangle (32,1);
\draw[fill = cyan!70] (32,2) rectangle (33,3);
\draw (28,0) grid (33,5);
\node[below] (notxz3) at (30.5,0) {\scriptsize{$\neg (3x-z)$}};

\draw[fill = cyan!70] (35,2) rectangle (36,3);
\draw[fill = cyan!70] (36,3) rectangle (37,4);
\draw[fill = cyan!70] (37,4) rectangle (38,5);
\draw[fill = cyan!70] (38,0) rectangle (39,1);
\draw[fill = cyan!70] (39,1) rectangle (40,2);
\draw[fill = blue!40] (35,0) rectangle (36,1);
\draw[fill = blue!40] (36,1) rectangle (37,2);
\draw[fill = blue!40] (37,2) rectangle (38,3);
\draw[fill = blue!40] (38,3) rectangle (39,4);
\draw[fill = blue!40] (39,4) rectangle (40,5);
\draw[fill = purple!70] (35,1) rectangle (36,2);
\draw[fill = purple!70] (36,2) rectangle (37,3);
\draw[fill = purple!70] (37,3) rectangle (38,4);
\draw[fill = purple!70] (38,4) rectangle (39,5);
\draw[fill = purple!70] (39,0) rectangle (40,1);
\draw[fill = brown!60] (35,3) rectangle (36,4);
\draw[fill = brown!60] (36,4) rectangle (37,5);
\draw[fill = brown!60] (37,0) rectangle (38,1);
\draw[fill = brown!60] (38,1) rectangle (39,2);
\draw[fill = brown!60] (39,2) rectangle (40,3);
\draw[fill = green!30] (35,4) rectangle (36,5);
\draw[fill = green!30] (36,0) rectangle (37,1);
\draw[fill = green!30] (37,1) rectangle (38,2);
\draw[fill = green!30] (38,2) rectangle (39,3);
\draw[fill = green!30] (39,3) rectangle (40,4);
\draw (35,0) grid (40,5);
\node[below] (notxz4) at (37.5,0) {\scriptsize{$\neg (4x-z)$}};

\end{tikzpicture}
        \caption{
        A perfect classical strategy for the $d=5$ Torpedo Game. 
        As in Figure \ref{fig:ch05_cl_d3}, cells of the same color belong to the same partition.
        The lines that avoid Alice are depicted below for every question Bob can be asked.
        }
        \label{fig:ch05_cl_d5}
    \end{figure}
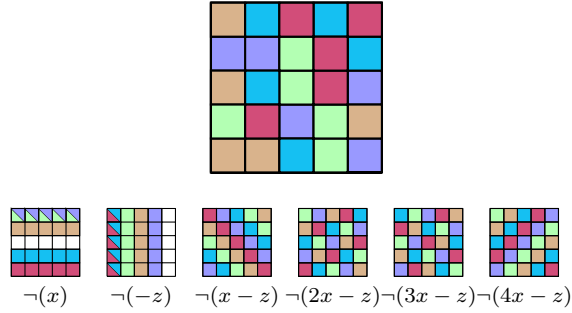

We only restricted the analysis to dimensions 2 and 3 because as the dimension increases, the Torpedo Game gets easier as the number of right answers to each questions grows linearly with the dimension (there are $d-1$ correct answers in dimension $d$). The classical optimum can only be established rigorously for small dimensions, owing to the proliferation of possible hidden variable assignments as the dimension increases. We have, however, found perfect classical strategies, i.e.\ strategies that win with probability $1$, for $d=4$ up to $d=23$ by randomly generating grids (see Figure \ref{fig:ch05_cl_d5} for an example in dimension 5).
We thus conjecture that the Torpedo Game has a perfect classical strategy for dimension 4 and above. 

However, we briefly comment on a possible generalisation to higher dimensions. 
In order to re-instate a quantum-over-classical advantage, as we had in dimensions two and three, we may modify the Torpedo game to make it harder to win classically. Note that the following modifications have no effect on the quantum values, which remain $\theta^Q_{d\geq 3}=1$.
As previously noted in \cite{gross2006hudson,van2011noise}, the eigenvectors of phase point operators Eq.~\eqref{eq:ch05_PPOdefn} are degenerate: a $+1$ eigenspace of dimension $\frac{d+1}{2}$ and a $-1$ eigenspace of dimension $\frac{d-1}{2}$.
Thus it is possible to enlarge Alice's input from $d^2$ to $\frac{d^2(d-1)}{2}$. Formally, let $0\leq \ell < \frac{d-1}{2}$, so Alice sends Bob $\ket{\psi_{x,z,\ell}}=X^xZ^z \left(\ket{\ell+1}+\ket{-(\ell+1)}\right){/\sqrt{2}}$, instead of just $\ket{\psi_{x,z,\ell=0}}:=\ket{\psi_{x,z}}$ as before. The modification changes a single relation from $w_\infty (x,z) =  \{ a \in \mathbb{Z}_d \mid a \neq x \} $ to
\begin{align*}
	w_\infty(x,z,\ell)	&=\{a\in\mathbb{Z}_d|a \in \{x+\ell+1,x-\ell-1\}\},
\end{align*}
whereas the remaining conditions persist i.e, $w_q(x,z,\ell)=w_q(x,z)$ in Eq.~\eqref{eq:ch05_winning_conditions} for other questions $q \in \{0,1,\ldots,d-1\}$. It seems reasonable that such a game, with more restrictive winning conditions, should be harder to win classically. Indeed, we were unable to find any perfect classical strategy by sampling, although we cannot rule out its existence since we were unable to exhaustively check all classical strategies. More generally, we have motivated how our perfect quantum strategies for this information retrieval task arise from a remarkable geometric feature of maximally negative states (c.f.~Eq.~\eqref{eq:ch05_KeyFact}), and we expect that this insight can be further mined for obtaining other examples of quantum advantage.

\dobib

\clearemptydoublepage


\chapter*{Conclusion and outlook}
\label{chap:ccl}
\addcontentsline{toc}{chapter}{Conclusion and outlook}
\chaptermark{Conclusion and outlook}

Since the seminal papers of Feynman \cite{Feynman1982} and Deutsch \cite{deutsch1985quantum}, quantum information scientists have asked some fundamental questions: are there computational problems that are provably hard for classical computers but which are efficiently solvable on a quantum computer? what really makes a quantum computer non-classical? what is the best route towards scalable fault-tolerant quantum computers? Unable to answer any of these, in this thesis we have considered how contextuality and Wigner negativity play a fundamental role as non-classical features.\footnotemark 
\footnotetext{These ironic few last sentences are a pastiche of the introduction in \cite{ross2021hodge}.}

In this dissertation, we have introduced the first robust framework for contextuality in continuous variables along the lines of the discrete-variable one introduced in~\cite{abramsky2011sheaf}. The Fine--Abramsky--Brandenburger (FAB) theorem extends to continuous-variables, even in the case where there are an uncountably infinite set of measurement labels. The contextual fraction~\cite{abramsky2017contextual} can be defined with infinite-dimensional linear programming and it can be approximated through Lasserre-type of semidefinite relaxations~\cite{lasserre10}. \index{Linear program} \index{Lasserre hierarchy} \index{Contextual fraction!CV}

Thanks to the FAB theorem, we have generalised the seminal result from~\cite{howard2014contextuality}---namely that Wigner negativity corresponds to measurement contextuality with respect to Pauli measurements---to the continuous-variable realm.\index{FAB theorem!CV} These first two contributions have closed a portion of the gap between discrete variables and continuous variables by showing how some important concepts and results extend between them. 

We have introduced a reliable method for witnessing Wigner negativity that can be efficiently implemented experimentally. Each witness is associated with a threshold value that can be computed using infinite-dimensional linear programming. Relaxing and restricting these programs yield two converging hierarchies of semidefinite programs that provide tighter and tighter upper and lower bounds on the threshold values for increasing ranks in the hierarchies.\index{Wigner negativity} \index{Contextuality!CV} Importantly, we have shown how well the hierarchies perform numerically on realistic examples.

We have then sought to extend the range of tasks for which non-classicality is known to be a useful resource, by tying it to and using it to design information retrieval tasks with a quantum-over-classical advantage.
We have introduced a game---the Torpedo Game---where a quantum advantage can be derived for systems of dimensions 2 and 3. Crucially, the optimal quantum strategies make use of quantum states with maximally negative discrete Wigner functions. To pinpoint the source of quantum advantage, we have shown that, subject to an assumption of bounded memory, sequential contextuality is necessary and sufficient for quantum advantage in any information retrieval task and in particular the Torpedo Game.\index{Contextuality!DV} \index{Wigner function} 
A nonlocal version of the Torpedo Game was later derived in \cite{TavakoliExp2021}, where it was experimentally tested to certify quantum correlations, allowing to test for nonlocality, steering and quantum state tomography in a single experiment. 

\bigskip

As contextuality may prove to be an important resource for quantum computation, the framework for continuous-variable contextuality may be very relevant with the promise of continuous-variable computation with for instance the generation of photonic cluster states \cite{Nielsen2004,Browne2005,Kok2007} along with Gottesman--Kitaev--Preskill type codes \cite{GKP2001}.

As shown in this dissertation, optimisation theory can lead to interesting results that help answering questions arising in quantum information. 
It has also been used to obtain information about the ground-state energy of a Hamiltonian \cite{brandao2013product} or to obtain bounds on the maximum quantum violations of Bell inequalities \cite{navascues2015characterizing}.\index{Bell inequality}
It is natural to wonder what other problems from quantum information might benefit from optimisation theory and in particular from nonlinear polynomial optimisation with noncommutative variables \cite{Navascues2007bounding,klep2020optimization}.

Perhaps a less obvious outlook is the converse question, that is to wonder whether optimisation theory may benefit from quantum information from a theoretical point of view\footnotemark.
\footnotetext{Of course having a quantum computer might help greatly for solving some optimisation problems but this is not what we are wondering here.}
In order to solve a Global Moment Problem (see \refprog{GMPstdform}) for \textit{sparse instances} with the Lasserre hierarchy, the original problem is split into several sub-problems \cite{weisser2018sparse,klep2021sparse}.
Solving a sparse GMP thus amounts to solving smaller instances of the problem \ie finding optimal measures on reduced space and then combining the solutions into a global solution. Of course this can only be implemented if we are certified that the solutions of the smaller problems can be combined---or glued---consistently. In words used throughout this dissertation, this requires that the optimal measures on sub-problems can be extended consistently to a global probability measure.
Because the smaller optimisation problems can pick out any solution of the corresponding feasible space, being able to glue them consistently requires that the splitting into sub-problems corresponds to a noncontextual scenario for otherwise a consistent global measure may not exist. 
Interestingly, the problem of gluing measures consistently was first introduced by Vorob'ev \cite{vorob1962consistent} and of course it bears a close relationship to noncontextuality \cite{Barbosa14}. 
It is also extensively studied for optimal transport \cite{villani2009optimal}.\index{Extendability}

It is expected that results linking contextuality and/or Wigner negativity to advantage will continue to emerge. 
We have provided some answers in this dissertation, but it will be interesting to see what other tools, results, and insights can be transferred from discrete variables to continuous variables and vice versa.

The equivalence between contextuality and Wigner negativity in continuous variables for at least two systems was shown using all possible Weyl measurements. Of course the set of all possible Weyl measurements is a continuum and one has to wonder if the equivalence still holds in a realistic experimental setting where only a finite number of measurements can be implemented. 
One problem that may arise is that the fact that the Wigner function is the unique quasiprobability distribution yielding the correct marginal distributions can only be established using all possible Weyl measurements \cite{bertrand1987tomographic,blass2020negative}. \index{Wigner negativity} \index{Contextuality!CV} \index{Pauli measurement}

From this equivalence, it is also very tempting to understand the precise relationship between measures of quantumness. In particular, is there a direct link between the negativity volume and the contextual fraction? 
\index{Contextual fraction!CV}
This dissertation may provide a jumping off point for understanding how to derive a general resource theory of non-classicality along the lines of the resource theory of contextuality presented in \cite{abramsky2019comonadic}, for which all of these measures would be non-classicality monotones.

Also we know that the violation of the threshold value associated to a Wigner negativity witness introduced in Chapter~\ref{chap:witnessing} provide a lower bound on the distance to the set of Wigner positive states. One may wonder if there is a link between this distance and the contextual fraction so that the aforementioned violation may provide a lower bound on the contextual fraction.

Different and more fine-grained notions of contextuality or non-classicality can be useful for example in obtaining an ever finer understanding of quantum advantages. Sequential contextuality is one example that we have shown to provide a good characterisation of advantage in information retrieval. We expect that this 
this notion and other finer or more generalised notions will continue to give rise to communication or computation tasks with a quantum-over-classical advantage.  

Finally, it is still an open question to understand precisely and quantifiably what is 
the non-classical phenomenon at the root of recent,and much mediatised, demonstrations of quantum computational advantage \cite{arute2019quantum,zhong2020quantum}.
This dissertation might provide partial answers to start answering this very challenging question. 

\dobib

\backmatter


\scriptsize{
\bibliographystyle{alphamod}
\bibliography{biblio}
}

\printindex

\end{document}